\documentclass[11pt,twoside,american]{report}
\usepackage[T1]{fontenc}
\usepackage[utf8x]{inputenc}
\usepackage{geometry}
\usepackage{pdfpages}

\usepackage{fancyhdr}
\usepackage{babel}
\usepackage{calc}
\usepackage{setspace}
\usepackage{amsmath}
\usepackage{mathrsfs}
\usepackage{amssymb}
\usepackage{graphicx}
\usepackage{physics}
\usepackage{color}
\usepackage{comment}
\usepackage{tikz}
\usepackage{standalone}
\usepackage[unicode=true,pdfusetitle,bookmarks=true,bookmarksnumbered=false,bookmarksopen=false,breaklinks=false,backref=false,colorlinks=false]{hyperref}

\usepackage{slashed}

\usepackage{nextpage}
\newcommand\myclearpage{\cleartooddpage[\thispagestyle{empty}]}

\usepackage[fit]{truncate}
\usepackage{xcolor}
\usepackage{cite}

\usepackage{doi}

\usepackage[comma]{natbib}
\setcitestyle{numbers,square}

\begin{document}
\pagenumbering{gobble}

\begin{center}
{\huge Universidade Federal do ABC\\
\vspace{1cm}
Pós-Graduação em Física}
\par\end{center}

\vspace{2cm}

\begin{center}
{\huge Iara Naomi Nobre Ota}
\par\end{center}
\vspace{2cm}

\begin{center}
{\huge{}Black hole spectroscopy: prospects for testing the nature of black holes with gravitational wave observations}{\huge\par}
\par\end{center}

\begin{center}
\vspace{3cm}
\par\end{center}

\begin{center}
\vspace{3cm}
Santo André - SP
\par\end{center}

\begin{center}
2022
\par\end{center}

\begin{center}
{\large{}\thispagestyle{empty}}{\large\par}
\par\end{center}

\begin{center}
\pagebreak{}
\par\end{center}

\pagenumbering{gobble}

\begin{center}
Iara Naomi Nobre Ota
\par\end{center}

\vspace{2cm}

\begin{center}
{\huge{}Espectroscopia de buracos negros: perspectivas para testar a natureza dos buracos negros com observações de ondas gravitacionais}{\huge\par}
\par\end{center}

\begin{center}
\vspace{2.5cm}
\par\end{center}

\begin{center}
\begin{minipage}[t]{0.6\columnwidth}%
\begin{center}
Tese apresentada ao Programa de Pós-Graduação em Física
da Universidade Federal do ABC (UFABC), como requisito parcial à obtenção
do título de Doutora em Física.
\par\end{center}%
\end{minipage}
\par\end{center}

\begin{center}
\vspace{1.5cm}
Orientadora: Profa. Dra. Cecilia Bertoni Martha Hadler Chirenti
\par\end{center}

\begin{center}
Co-orientador: Prof. Dr. Mauricio Richartz
\par\end{center}

\begin{center}
\vspace{2.5cm}
Santo André - SP
\par\end{center}

\begin{center}
2022
\par\end{center}

\begin{center}
{\large{}\thispagestyle{empty}}{\large\par}
\par\end{center}

\begin{center}
\pagebreak{}
\par\end{center}

\vfill{}

\begin{center}
\fbox{\begin{minipage}[t]{0.8\columnwidth}%
NOBRE OTA, Iara Naomi

\hspace{1cm}ESPECTROSCOPIA DE BURACOS NEGROS: perspectivas para testar a natureza dos buracos negros com observações de ondas gravitacionais / Iara Naomi Nobre Ota - Santo André,
Universidade Federal do ABC, 2022.

\vspace{0.5cm}

\hspace{1cm} 55 fls. \vspace{0.5cm}

\hspace{1cm}Orientadora: Cecilia Bertoni Martha Hadler Chirenti \\ \vspace{0.1cm}
\hspace{0.9cm}Co-orientador: Mauricio Richartz\vspace{0.5cm}

\hspace{1cm}Tese (doutorado) - Universidade
Federal do ABC, Programa de Pós-Graduação em Física, 2022\vspace{0.5cm}

\hspace{1cm}1. Buracos negros. 2. Ondas gravitacionais. 3. Teorema no-hair. 4. Modos quasi-normais.
I. NOBRE OTA, Iara Naomi. II. Programa de Pós-Graduação em Física,
2022. III. Espectroscopia de buracos negros: perspectivas para testar a natureza dos buracos negros com observações de ondas gravitacionais%
\end{minipage}}
\par\end{center}

\begin{center}
{\large{}\thispagestyle{empty}}{\large\par}
\par\end{center}

\begin{center}
\myclearpage
\par\end{center}


\begin{center}
\textbf{\large{}\rule[0.5ex]{1\columnwidth}{1pt}}{\large\par}
\par\end{center}

\begin{center}
\textbf{\Large{}Acknowledgements}{\large\par}
\par\end{center}

\vspace{0.5cm}

I am deeply grateful to my advisor, Professor Dr. Cecilia Chirenti, for all the valuable support and mentorship she gave me since my undergrad studies.
I would like thank my co-advisor Prof.
Dr. Mauricio Richartz for all the help and assistance.

A special thanks to all my friends Fabricio, Enesson, Helena and Nathalia, for valuable discussions and friendship.
In particular, I am indebted to Lucas, for all the extra help he gave me.

I thank Bruno, for all the love, care and friendship.

I thank my parents, Andr\'{e} and In\^{e}s, for all the encouragement and support, my siblings, Laura and N\'{a}dia and Iuri, for all the care.

This thesis was supported by grant 2018/21286-3 of the S\~{a}o Paulo Research Foundation (FAPESP) and by the Federal University of ABC.
This work was also supported in part by the Simons Foundation through the Simons Foundation Emmy Noether Fellows Program at Perimeter Institute.
I acknowledge travel funding support from Funda\c{c}\~{a}o Norte-rio-grandense de Pesquisa e Cultura (FUNPEC).
This study was financed in part by the Coordena\c{c}\~{a}o de Aperfei\c{c}oamento de Pessoal de N\'{\i}vel Superior - Brasil (CAPES) - Finance Code 001.
Resources supporting this work were partially provided by the NASA High-End Computing (HEC) Program through the NASA Center for Climate Simulation (NCCS) at Goddard Space Flight Center.

\begin{center}
{\large{}\thispagestyle{empty}}{\large\par}
\par\end{center}

\begin{center}
\myclearpage
\par\end{center}


\begin{center}
\textbf{\large{}\rule[0.5ex]{1\columnwidth}{1pt}}{\large\par}
\par\end{center}

\begin{center}
\textbf{\Large{}Resumo}{\large\par}
\par\end{center}

\vspace{0.5cm}

As ondas gravitacionais fornecem informações sobre a natureza do espaço-tempo e evidências da existência de buracos negros.
O buraco negro resultante de uma fusão de um binário de buracos negros emite ondas gravitacionais na forma de modos quasi-normais, cujo espectro, que depende apenas das propriedades do mesmo, é conhecido como as ``digitais'' do buraco negro.
Os modos quasi-normais podem ser usados para testar quão bem um buraco negro astrofísico pode ser descrito pela geometria de Kerr.
Cada modo é parametrizado por três índices: os números harmônicos $(\ell, m)$ e o índice $n$, o qual indica o modo fundamental ($n = 0$) e os modos superiores ($n = 1,2,3,\ldots$).
A espectroscopia de buracos negros propõe utilizar o espectro de modos quasi-normais para testar o teorema no-hair.
Neste trabalho, investigamos as perspectivas de se realizar espectroscopia de buracos negros.
Por meio da análise de simulações de relatividade numérica, investigamos a contribuição de modos subdominantes, além do modo dominante $(2,2,0)$.
Mostramos que o primeiro modo superior $(2,2,1)$ tem amplitude maior ou comparável com a amplitude dos outros modos harmônicos mais relevantes.
Para detectores atuais e futuros, obtivemos os horizontes de espectroscopia de buracos negros, que indicam a distância máxima em que um evento pode estar para que dois ou mais modos sejam detectados.
Para binários com razão entre as massas pequena, os modos $(2,2,1)$ e $(3,3,0)$ são os modos secundário e terciário e, para o caso de razão entre as massas grande, os modos $(3,3,0)$ e $(4,4,0)$ são os modos subdominantes mais relevantes para detecção.
Nosso trabalho indica que há boas perspectivas para a detecção de modos subdominantes com detectores futuros.
As taxas de eventos para o LIGO são muito menores, porém não são impeditivas.

\vspace{0.5cm}

\textbf{Palavras-chave: buracos negros, ondas gravitacionais, teorema no-hair, modos quasi-normais}

\begin{center}
{\large{}\thispagestyle{empty}}{\large\par}
\par\end{center}

\begin{center}
\myclearpage
\par\end{center}


\begin{center}
\textbf{\large{}\rule[0.5ex]{1\columnwidth}{1pt}}{\large\par}
\par\end{center}

\begin{center}
\textbf{\Large{}Abstract}{\Large\par}
\par\end{center}

\vspace{0.5cm}

Gravitational waves provide direct information about the nature of spacetime and the existence of black holes.
The remnant of a binary black hole merger emits gravitational waves in the form of quasinormal modes, whose spectrum is known as the ``fingerprints'' of a black hole, as it depends only on the properties of the remnant.
The quasinormal modes can be used to test how closely an astrophysical black hole matches the Kerr geometry.
Each mode is parameterized by three indices: the harmonic numbers $(\ell, m)$ and the overtone index $n$, that labels the fundamental mode ($n = 0$) and the overtones ($n = 1,2,3,\ldots$).
Black hole spectroscopy is the proposal to use the detection of multiple quasinormal modes to test the no-hair theorem.
In this work, we investigate the prospects for performing black hole spectroscopy.
The $(2,2,0)$ is the dominant mode, and we analyze the contribution of the most relevant subdominant modes in numerical relativity simulations.
We show that the overtone mode $(2,2,1)$ has an amplitude higher or comparable to the amplitude of the most relevant higher harmonic modes.
For current and future gravitational wave detectors, we compute the black hole spectroscopy horizon, which is the maximum distance of an event up to which two or more quasinormal modes can be detected.
For low mass ratio binaries,  the secondary and tertiary modes are the $(2,2,1)$ and $(3,3,0)$, respectively, and, for the large mass ratio case, the $(3,3,0)$ and $(4,4,0)$ are the most relevant subdominant modes for detection.
Our work indicates promising prospects for the detection of subdominant modes with future gravitational wave detectors.
The event rate for LIGO is much smaller, but not prohibitively so.

\vspace{0.5cm}

\textbf{Keywords: black holes, gravitational waves, no-hair theorem, quasinormal modes}

\begin{center}
{\large{}\thispagestyle{empty}}{\large\par}
\par\end{center}

\begin{center}
\myclearpage
\par\end{center}


\tableofcontents{}

\clearpage
\ifodd\value{page}\else
  \thispagestyle{empty}
\fi

%

\pagenumbering{arabic}
\chapter{Introduction}
\label{ch:introduction}

In 2015, the first \emph{direct} detection of gravitational waves was achieved by the Laser Interferometer Gravitational-Wave Observatory (LIGO)~\cite{LIGOScientific:2016aoc}.
The importance of this observation was recognized with the Physics Nobel Prize in 2017, awarded to Rainer Weiss, Kip Thorne and Barry Barish.
The existence of gravitational waves is a major consequence of the theory of General Relativity that contrasts with the Newtonian theory of gravity, but their direct observation is far more important than a proof of the success of Einstein's theory.
In fact, gravitational waves have already been measured indirectly before that: the Nobel Prize in Physics of 1993 was awarded to Russell Hulse and Joseph Taylor for the discovery of the binary pulsar PSR 1913+16, whose orbit decays in agreement with the loss of energy due to gravitational wave emission~\cite{Taylor:1982zz, Weisberg:2004hi}.

It is now common to hear that "gravitational waves opened a new window to the universe".
Similarly to analysing the Sun's electromagnetic radiation to determine its structure and composition, properties of a source are determined by the emitted gravitational waves.
For instance, the neutron star equation of state is still unknown, and new information extracted from the gravitational waveform will increase our understanding of the interior composition of such stars~\cite{LIGOScientific:2018cki}.

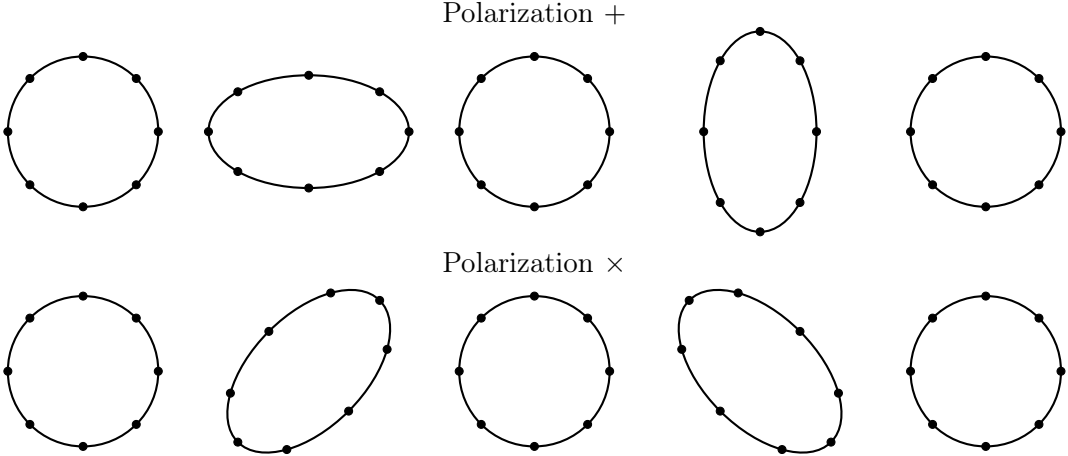
\begin{figure}[hbt!]
  \centering
  {Polarization $+$}
  {\begin{tikzpicture}
\draw[color = black, thick](0,0) circle (1);

	\draw[fill] (1,0) circle  [radius=1.5pt];
	\draw[fill] (0.70710678118, 0.70710678118) circle  [radius=1.5pt];
	\draw[fill] (0,1) circle  [radius=1.5pt];
	\draw[fill] (-0.70710678118, 0.70710678118) circle  [radius=1.5pt]; 
	\draw[fill] (-1,0) circle  [radius=1.5pt];
	\draw[fill] (-0.70710678118, -0.70710678118) circle  [radius=1.5pt]; 
	\draw[fill] (0,-1) circle  [radius=1.5pt];
	\draw[fill] (0.70710678118, -0.70710678118) circle  [radius=1.5pt]; 
\draw[color = black, thick] (3,0) ellipse (1.33333333333 and 0.75);
	\draw[fill] (3 + 1.33333333333,0) circle  [radius=1.5pt];
	\draw[fill] (3 + 1.33333333333*0.70710678118, 0.75*0.70710678118) circle  [radius=1.5pt];
	\draw[fill] (3, 0.75) circle  [radius=1.5pt];
	\draw[fill] (3 + 1.33333333333*-0.70710678118, 0.75*0.70710678118) circle  [radius=1.5pt]; 
	\draw[fill] (3 -1.33333333333,0) circle  [radius=1.5pt];
	\draw[fill] (3 + 1.33333333333*-0.70710678118, 0.75*-0.70710678118) circle  [radius=1.5pt]; 
	\draw[fill] (3,-0.75) circle  [radius=1.5pt];
	\draw[fill] (3 + 1.33333333333*0.70710678118, 0.75*-0.70710678118) circle  [radius=1.5pt]; 

\draw[color = black, thick](6,0) circle (1);

	\draw[fill] (7,0) circle  [radius=1.5pt];
	\draw[fill] (6 + 0.70710678118, 0.70710678118) circle  [radius=1.5pt];
	\draw[fill] (6,1) circle  [radius=1.5pt];
	\draw[fill] (6 -0.70710678118, 0.70710678118) circle  [radius=1.5pt]; 
	\draw[fill] (6 -1,0) circle  [radius=1.5pt];
	\draw[fill] (6 -0.70710678118, -0.70710678118) circle  [radius=1.5pt]; 
	\draw[fill] (6,-1) circle  [radius=1.5pt];
	\draw[fill] (6 + 0.70710678118, -0.70710678118) circle  [radius=1.5pt]; 	

\draw[color = black, thick] (9 ,0) ellipse (0.75 and 1.33333333333);
	\draw[fill] (9 + 0.75,0) circle  [radius=1.5pt];
	\draw[fill] (9 + 0.75*0.70710678118, 1.33333333333*0.70710678118) circle  [radius=1.5pt];
	\draw[fill] (9, 1.33333333333) circle  [radius=1.5pt];
	\draw[fill] (9 + 0.75*-0.70710678118, 1.33333333333*0.70710678118) circle  [radius=1.5pt]; 
	\draw[fill] (9 - 0.75,0) circle  [radius=1.5pt];
	\draw[fill] (9 + 0.75*-0.70710678118, 1.33333333333*-0.70710678118) circle  [radius=1.5pt]; 
	\draw[fill] (9,-1.33333333333) circle  [radius=1.5pt];
	\draw[fill] (9 + 0.75*0.70710678118, 1.33333333333*-0.70710678118) circle  [radius=1.5pt];

\draw[color = black, thick](12 ,0) circle (1);

	\draw[fill] (12 + 1,0) circle  [radius=1.5pt];
	\draw[fill] (12 + 0.707111178118, 0.707111178118) circle  [radius=1.5pt];
	\draw[fill] (12 ,1) circle  [radius=1.5pt];
	\draw[fill] (12 -0.707111178118, 0.707111178118) circle  [radius=1.5pt]; 
	\draw[fill] (12 -1,0) circle  [radius=1.5pt];
	\draw[fill] (12 -0.707111178118, -0.707111178118) circle  [radius=1.5pt]; 
	\draw[fill] (12 ,-1) circle  [radius=1.5pt];
	\draw[fill] (12 + 0.707111078118, -0.707101078118) circle  [radius=1.5pt]; 	

\end{tikzpicture}}
  {Polarization $\times$}
  {\begin{tikzpicture}
\draw[color = black, thick](0,0) circle (1);

	\draw[fill] (1,0) circle  [radius=1.5pt];
	\draw[fill] (0.70710678118, 0.70710678118) circle  [radius=1.5pt];
	\draw[fill] (0,1) circle  [radius=1.5pt];
	\draw[fill] (-0.70710678118, 0.70710678118) circle  [radius=1.5pt]; 
	\draw[fill] (-1,0) circle  [radius=1.5pt];
	\draw[fill] (-0.70710678118, -0.70710678118) circle  [radius=1.5pt]; 
	\draw[fill] (0,-1) circle  [radius=1.5pt];
	\draw[fill] (0.70710678118, -0.70710678118) circle  [radius=1.5pt]; 
\draw[color = black, thick, rotate around={45:(3,0)}] (3,0) ellipse (1.33333333333 and 0.75);
	\draw[fill] (3 + 1.33333333333*0.70710678118, 1.33333333333*0.70710678118) circle  [radius=1.5pt];
	\draw[fill] (3 + 1.33333333333*0.70710678118*0.70710678118 - 0.75*0.70710678118*0.70710678118, 1.33333333333*0.70710678118*0.70710678118 + 0.75*0.70710678118*0.70710678118) circle  [radius=1.5pt];
	\draw[fill] (3 - 0.75*0.70710678118, 0.75*0.70710678118) circle  [radius=1.5pt];
	\draw[fill] (3 - 1.33333333333*0.70710678118*0.70710678118 - 0.75*0.70710678118*0.70710678118, -1.33333333333*0.70710678118*0.70710678118 + 0.75*0.70710678118*0.70710678118) circle  [radius=1.5pt]; 
	\draw[fill] (3 - 1.33333333333*0.70710678118, -1.33333333333*0.70710678118) circle  [radius=1.5pt];
	\draw[fill] (3 - 1.33333333333*0.70710678118*0.70710678118 + 0.75*0.70710678118*0.70710678118, -1.33333333333*0.70710678118*0.70710678118 - 0.75*0.70710678118*0.70710678118) circle  [radius=1.5pt]; 
	\draw[fill] (3 + 0.75*0.70710678118,- 0.75*0.70710678118) circle  [radius=1.5pt];
	\draw[fill] (3 + 1.33333333333*0.70710678118*0.70710678118 + 0.75*0.70710678118*0.70710678118, 1.33333333333*0.70710678118*0.70710678118 - 0.75*0.70710678118*0.70710678118) circle  [radius=1.5pt]; 

\draw[color = black, thick](6,0) circle (1);

	\draw[fill] (7,0) circle  [radius=1.5pt];
	\draw[fill] (6 + 0.70710678118, 0.70710678118) circle  [radius=1.5pt];
	\draw[fill] (6,1) circle  [radius=1.5pt];
	\draw[fill] (6 -0.70710678118, 0.70710678118) circle  [radius=1.5pt]; 
	\draw[fill] (6 -1,0) circle  [radius=1.5pt];
	\draw[fill] (6 -0.70710678118, -0.70710678118) circle  [radius=1.5pt]; 
	\draw[fill] (6,-1) circle  [radius=1.5pt];
	\draw[fill] (6 + 0.70710678118, -0.70710678118) circle  [radius=1.5pt]; 	

\draw[color = black, thick, rotate around={45:(9,0)}] (9,0) ellipse (0.75 and 1.33333333333);
	\draw[fill] (9 - 1.33333333333*0.70710678118, 1.33333333333*0.70710678118) circle  [radius=1.5pt];
	\draw[fill] (9 - 1.33333333333*0.70710678118*0.70710678118 + 0.75*0.70710678118*0.70710678118, 1.33333333333*0.70710678118*0.70710678118 + 0.75*0.70710678118*0.70710678118) circle  [radius=1.5pt];
	\draw[fill] (9 + 0.75*0.70710678118, 0.75*0.70710678118) circle  [radius=1.5pt];
	\draw[fill] (9 + 1.33333333333*0.70710678118*0.70710678118 + 0.75*0.70710678118*0.70710678118, -1.33333333333*0.70710678118*0.70710678118 + 0.75*0.70710678118*0.70710678118) circle  [radius=1.5pt]; 
	\draw[fill] (9 + 1.33333333333*0.70710678118, -1.33333333333*0.70710678118) circle  [radius=1.5pt];
	\draw[fill] (9 + 1.33333333333*0.70710678118*0.70710678118 - 0.75*0.70710678118*0.70710678118, -1.33333333333*0.70710678118*0.70710678118 - 0.75*0.70710678118*0.70710678118) circle  [radius=1.5pt]; 
	\draw[fill] (9 - 0.75*0.70710678118,- 0.75*0.70710678118) circle  [radius=1.5pt];
	\draw[fill] (9 - 1.33333333333*0.70710678118*0.70710678118 - 0.75*0.70710678118*0.70710678118, 1.33333333333*0.70710678118*0.70710678118 - 0.75*0.70710678118*0.70710678118) circle  [radius=1.5pt];

\draw[color = black, thick](12 ,0) circle (1);

	\draw[fill] (12 + 1,0) circle  [radius=1.5pt];
	\draw[fill] (12 + 0.707111178118, 0.707111178118) circle  [radius=1.5pt];
	\draw[fill] (12 ,1) circle  [radius=1.5pt];
	\draw[fill] (12 -0.707111178118, 0.707111178118) circle  [radius=1.5pt]; 
	\draw[fill] (12 -1,0) circle  [radius=1.5pt];
	\draw[fill] (12 -0.707111178118, -0.707111178118) circle  [radius=1.5pt]; 
	\draw[fill] (12 ,-1) circle  [radius=1.5pt];
	\draw[fill] (12 + 0.707111078118, -0.707101078118) circle  [radius=1.5pt]; 	

\end{tikzpicture}}
  \caption{Distortions in a free particle ring caused by a gravitational wave that propagates orthogonally to the ring plane.}
  \label{fig:polarization}
\end{figure}

In contrast to the other types of radiation, the spacetime is not just a background for the gravitational radiation, as the waves are oscillations of the spacetime itself.
Gravitational waves have two polarizations, are transversal and only the plane orthogonal to the propagation direction will be deformed by the gravitational radiation~\cite{maggiore-vol1, poisson-gravity}.
The polarizations are called \emph{plus} ($+$) and \emph{cross} ($\times$).
These names are related to the deformation of free particle rings caused by a passing orthogonal gravitational wave, as shown in Figure~\ref{fig:polarization}.
The polarizations are related to each other by a 45-degree rotation.
Moreover, the stretch in one direction of the circle is compensated by a squeeze in the orthogonal direction, which keeps the area inside the circle unchanged~\cite{maggiore-vol1, poisson-gravity}.

The intensity of the deformation depends on the gravitational wave source and on its distance to the observer.
Similarly to electromagnetic radiation, non-inertial movements in the source produce radiation.
In general, the electromagnetic radiation is dominated by the radiation emitted by variations in the dipole moment, as there is no monopole radiation due to charge conservation.
In the gravitational case, there are more conservation laws.
Due to conservation of the total mass-energy of the system (monopole moment), the center of mass-energy and total angular momentum of the system (dipole moments) there are no monopole nor dipole radiation terms for a mass-energy distribution $\rho(\vec{x})$, where $\vec{x}$ is the position vector.
The next term in the multipolar expansion is the quadrupole $I_{ij} = \int \rho(\vec{x}) x_i x_j d^3 x$~\cite{maggiore-vol1, poisson-gravity}, and there is no conservation law associated with the quadrupole moment.
Therefore, in general, the gravitational radiation is dominated by radiation emitted by variations in the quadrupole moment.

The variation $\Delta L$ of the length of an object with proper length $L$ caused by a gravitational wave is proportional to the second time derivative of the quadrupole moment~\cite{maggiore-vol1, poisson-gravity} and inversely proportional to the source distance $r$.
Using dimensional analysis, one can conclude that
\begin{equation}
\frac{\Delta L}{L} \propto \frac{G}{c^4 r} \frac{d^2I_{ij}}{dt^2},
\label{eq:delta-L}
\end{equation}
where $G$ is the gravitational constant and $c$ is the speed of light.

As $G/c^4 \sim 10^{-45}$, the length variation caused by gravitational waves is terribly small and only extremely ``violent'' events are detectable.
For this reason, the sensitivity needed to directly detect a gravitational wave was only achieved in 2015.
Take the example of the first detection GW150914\footnote{``GW'' indicates a \textbf{G}ravitational \textbf{W}ave event and the numbers indicate the detection date 20\textbf{15}-\textbf{09}-\textbf{14}.
With the increased sensitivity of the detectors, there are some days when there are more than one triggered event. Thus, a new notation was introduced, adding  after the date an underscore followed by the UTC time of the event~\cite{LIGOScientific:2020ibl}.}~\cite{LIGOScientific:2016aoc}.
This event is compatible with a binary black hole merger, with initial black hole masses $m_1 \sim 36 M_\odot$ and $m_2 \sim 31 M_\odot$ and effective spin compatible with zero, resulting in a remnant black hole with final mass $M_f \sim 63 M_\odot$ and dimensionless spin $a_f \sim 0.69$.
The event happened at a luminosity distance of $D_L \sim 440$ Mpc, which is approximately 1.4 billion light years away from the Earth.
The fractional length variation of LIGO's arms, which have length $L = 4$ km, was $\Delta L/ L \sim 10^{-21}$, which is smaller than the ratio between the radius of a proton and the elevation of Mount Everest!

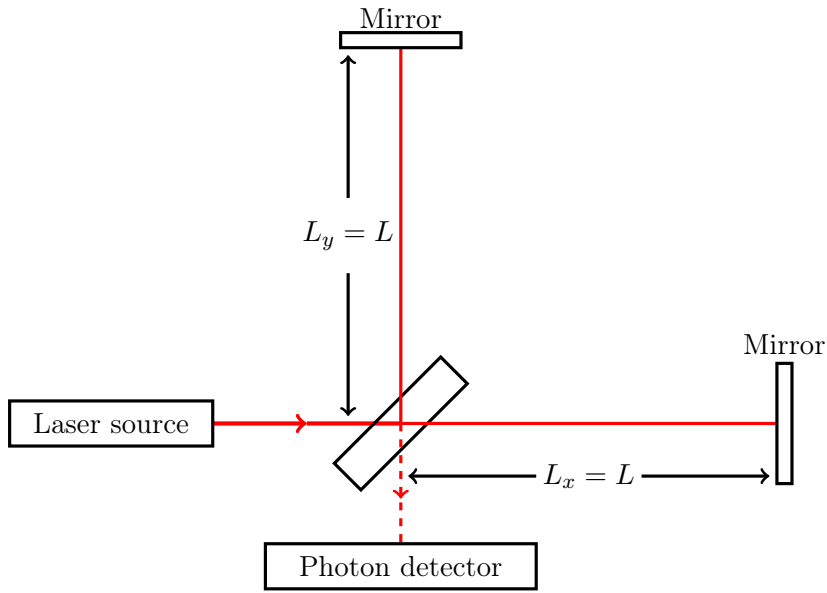
\begin{figure}[hbt!]
	\centering
	\begin{tikzpicture}

\draw[->,red, ultra thick]  (-2.5, 0) -- (-1.25,0);
\draw[red, ultra thick]  (-1.25, 0) -- (0,0);
\draw[very thick] (-5.2,-0.3) rectangle (-2.5,0.3) node[pos=.5] {Laser source };

\draw[rotate around = {45: (0,0)}, very thick](-1,-0.25) rectangle (1,0.25);

\draw[red, very thick] (0,0) -- (0,5); 
\draw[very thick] (-0.8,5) rectangle (0.8,5.2) node[pos=.5, above = 0.5] {Mirror};
\draw[<-, very thick] (-0.7, 0.1) -- (-0.7, 2);
\draw[->, very thick] (-0.7, 3) -- (-0.7, 4.9);
\draw[very thick] (-0.7, 2.5) node[] {$L_y = L$};

\draw[red, very thick] (0,0) -- (5,0); 
\draw[very thick] (5, -0.8) rectangle (5.2, 0.8) node[pos=.5, above = 22] {Mirror};
\draw[<-, very thick] (0.1, -0.7) -- (1.8,-0.7);
\draw[->, very thick] (3.2, -0.7) -- (4.9,-0.7);
\draw[very thick] (2.5, -0.7) node[] {$L_x = L$};

\draw[->, dashed, red, very thick]  (0, 0) -- (0,-1);
\draw[dashed,red, very thick] (0, -1.05) -- (0,-1.6);

\draw[very thick] (-1.8, -1.6) rectangle (1.8, - 2.2) node[pos=.5] {Photon detector};
\end{tikzpicture}
	\caption{Laser interferometer diagram.}
	\label{fig:interferometer}
\end{figure}

To date, laser interferometers are the only instruments sensitive enough to detect gravitational waves.
The gravitational waves are measured by the changes in the interference pattern of the laser.
As illustrated in Figure~\ref{fig:interferometer}, a laser beam is split into two orthogonal beams, which are reflected back by mirrors and recombined.
The mirrors are separated by the same distance $L$ and the laser beam is tuned to destructively interfere after it bounces back at the detector.
When a gravitational wave passes through the detector, the arms stretch and squeeze, resulting in a not completely destructive interference, which is related to the gravitational waveform.

We can see from equation~\eqref{eq:delta-L} that the variation $\Delta L$ is proportional to the length of the detector's arms $L$, that is, the sensitivity of the interferometer increases with its arm's length.
By comparing Figures~\ref{fig:polarization} and~\ref{fig:interferometer}, we can see that interferometers are more sensitive to specific propagation directions, and are even ``blind'' to certain directions and polarizations.
Take the example that Figures~\ref{fig:polarization} and~\ref{fig:interferometer} are in the same plane $x-y$, where the distortion in the arms will be fully due to the polarization $+$.
On the other hand, a wave propagating in the detector's plane in the direction that forms a 45-degree angle with the detector's arms will not cause any distortion in the arms' direction.
Moreover, considering again that the wave propagates orthogonal to the detector's plane, there is no way to distinguish between a wave entering and a wave leaving the plane.
These limitations are true even for a noiseless ideal detector.
For real detectors, there are many additional technical limitations, such as seismic noise and quantum effects of light~\cite{Abbott:2016xvh, aLIGO:2020wna}.
Regardless of the noise, a single ideal detector cannot obtain all the information about a gravitational source.
For example, unique determination of the direction to a source using time delays requires at least four detectors.

Currently, there are five laser interferometer detectors in operation~\cite{KAGRA:2013rdx, Shoemaker:2019bqt, LIGOScientific:2021djp}:
the two identical LIGO detectors, which have 4 km long arms and are located in Hanford and Livingston, in the USA\@.
The German detector GEO600 was built in Hannover, and it has 600 m long arms.
This detector is not sensitive enough to detect gravitational waves, however it is used to assist in technological improvements of the detectors network.
In Cascina, the Italian detector Virgo, which has 3 km long arms, is detecting gravitational waves since 2017.
The newest detector is the Japanese detector KAGRA, which also has 3 km arms, and was build in Hida, and started operating in 25 February 2020.
KAGRA is the first detector build underground, in the Kamioka Observatory inside the Mozumi Mine.
It is also the first detector to use cryogenic technology.
In the future, another LIGO detector is planned to be built in India.
We refer to the joint current detectors LIGO, Virgo, and KAGRA Collaboration as LVKC\@.

These detectors are only sensitive to a limited part of the frequency spectrum of gravitational radiation: they are able to detect waves in the frequency band of approximately 20 to 2000 Hz~\cite{Abbott:2016xvh,aLIGO:2020wna}.
The sources in this range are compact binary mergers of stellar mass black holes and/or neutron stars,
rotating neutron stars and supernovae.
To date, only gravitational waves emitted by compact binary systems have been detected.
In the first two observing runs,  O1  (12 September 2015 to 19 January 2016) and O2 (30 November 2016 to 25 August 2017), gravitational waves from ten binary black hole mergers and one binary neutron star merger were detected~\cite{LIGOScientific:2018mvr}.
With the improvement of the detectors, the third observing run, which was divided in two parts O3a (1 April 2019 to 1 October 2019) and O3b (1 November 2019 to 27 March 2020) observed almost eight times more gravitational wave events than the two previous observing runs, adding 79 events to the catalog,
consisting in 73 binary black hole mergers, one binary neutron star merger and five events with very asymmetric initial masses that are compatible with either binary black hole or neutron star-black hole mergers~\cite{LIGOScientific:2020ibl,LIGOScientific:2021usb, LIGOScientific:2021djp}.

There are plans to expand the frequency range and sensitivity of the current detectors and to build new instruments.
To decrease these limitations, there are plans to make improvements in the current LIGO instruments~\cite{LIGOScientific:T1700231} and build new gravitational wave detectors.
For greater sensitivity in the same frequency range of the current detectors, besides the improvements in the instruments, there are proposed projects for improved third generation (3G) ground-based detectors.
The Einstein Telescope (ET)~\cite{Punturo:2010zz} is the European proposed detector, which will be build underground, and will consist in a set of six interferometers that forms an equilateral triangular shape, each arm will be 10 km long.
As it is an underground project, the seismic noise should be suppressed and the detector will be able to detect frequencies as low as 1 Hz.
The Cosmic Explorer (CE)~\cite{LIGOScientific:2016wof}, in the USA, is planned to be built on the surface, but will have two arms with a ninety-degree opening angle.
There are two possible versions: one with 40 km arms (in length) and the other with 20 km arms.
The consortium is considering to build both versions in sites apart from each other (one in each site).
Because it is on the surface, CE will not be as sensitive as ET for low frequencies, but it is expected to be very sensitive for higher frequencies, which will advance the detection of less massive events, such as radiation from rotating neutron stars.
There is also an Australian proposal for a gravitational wave detector: Neutron Star Extreme Matter Observatory (NEMO).
NEMO will be optimized to study nuclear physics with merging neutron stars, that is, it will be most sensitive in the high frequency band.

The Laser Interferometer Space Antenna (LISA)~\cite{amaroseoane2017laser} is an accepted project for a space-based detector and is one of the main research joint missions between the European Space Agency and NASA, with a planned launch in mid-2030s.
It will consist in three satellites orbiting the Sun, which will maintain a near-equilateral triangular formation.
The satellites will be separated by millions of kilometers, and will consist of high precision interferometers.
The scale of the arms will allow the measurements of long period (low frequency) waves,
which are unachievable by ground-based detectors, due to seismic and gravity gradient noise.
The mili Hertz sensitivity band of LISA will detect the coalescence and merger of supermassive black hole binaries ($M > 10^{4} M_{\odot}$), among other sources.
Similar to LISA there is the Chinese project TianQin~\cite{TianQin:2015yph}, which is also planned to launch in the 2030s.

The laser interferometers are not the only type of gravitational wave detectors.
The first detectors built~\cite{Weber:1960zz} are called resonant mass antennae, and consist of a large, solid metal object isolated from outside vibrations.
A passing gravitational wave could excite the body's resonant frequency.
Joseph Weber built a cylindrical bar detector in the 1960s~\cite{Weber:1960zz} and his claimed detection~\cite{Weber:1969bz} could not be achieved by a similar detector~\cite{Tyson:1973ra} and other more advanced resonant mass detectors.
Several resonant mass antennas were build and Brazil has a pioneer project.
The Mario Schenberg Gravitational Wave Antenna is a 65~cm diameter spherical mass suspended in a cryogenic vacuum enclosure~\cite{Aguiar:2002eq, Aguiar:2005tp, Aguiar:2008zz, Oliveira:2016gds}.
Like other resonant mass detectors, Mario Schenberg was not sensitive enough to detect gravitational waves, however, the development of the detector continues.

\begin{figure}[htb!]
  \centering
  \includegraphics[width=0.85\linewidth]{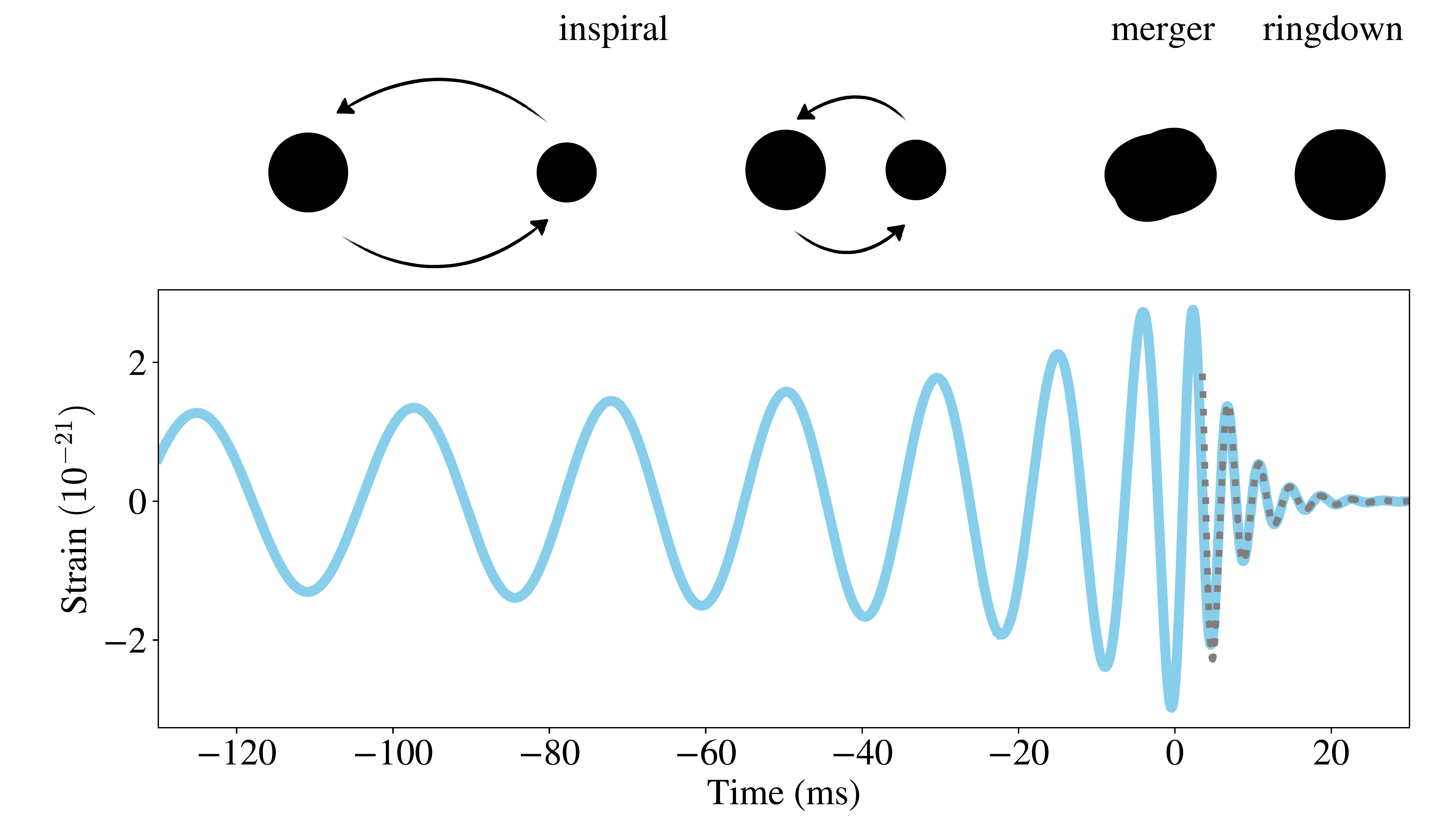}
  \caption{\emph{Top:} illustration of a binary black hole dynamical evolution and its three stages: inspiral, merger and ringdown.
  \emph{Bottom:} the blue curves represents the quadrupolar gravitational waveform emitted during the merger process of a nonspinning binary of a numerical simulation~\cite{Boyle:2019kee, SXS-catalog}.
  The fundamental quasinormal mode corresponding to the remnant black hole is plotted in dashed gray.}
  \label{fig:waveform_gw150914}
\end{figure}

The current detections show that binary black hole mergers are the main source of gravitational waves,
and black holes are expected to be the main source for LISA\@.
The gravitational waves emitted during the merger of a binary black hole system are divided in three stages, as illustrated in Figure~\ref{fig:waveform_gw150914}:
\begin{enumerate}
    \item \emph{inspiral}: the orbit of the binary black hole system is approximately Kleperian, and it shrinks due to the emission of gravitational waves;
    \item \emph{merger}: the black holes merge into a single black hole;
    \item \emph{ringdown}: the remnant black hole ``relaxes'' and the waveform amplitude decreases.
\end{enumerate}

The waveform of the inspiral can be analytically described by the post-Newtonian approximation~\cite{Blanchet:2013haa}, where the solutions of the Einstein field equations are expanded in factors of $1/c^n$.
During the merger stage, the weak field approximation is not valid and Numerical Relativity simulations are needed to precisely describe the waveform~\cite{Buonanno:2006ui}.
After the non-linear regime, the gravitational wave emitted by remnant black hole is well described by black hole perturbation theory and the excess of energy emitted can be described by the black hole \emph{quasinormal modes}~\cite{Buonanno:2006ui,Seidel_2004}.

Black hole perturbation theory was first studied by Regge and Wheeler~\cite{Regge:1957td}, when they analyzed the stability of the Schwarzschild metric under small perturbations.
The evolution of these perturbations was studied by Vishveshwara~\cite{Vishveshwara:1970zz}, who showed that, after some time interval, the gravitational waves are described by damped sinusoids, whose frequencies of oscillation and damping times depend only on the black hole parameters (the black hole mass, in the Schwarzschild case), and are independent of the initial perturbation.
Therefore, the quasinormal modes are considered the ``characteristic sounds'' of black holes.
In contrast to normal modes, these modes are not stationary, as the system is open, which is the reason why they are named ``quasinormal''~\cite{Press:1971wr}.

A black hole is completely described by its quasinormal modes, and the detection of a single quasinormal mode is sufficient to completely determine the spacetime of astrophysical black holes.
The frequency of oscillation and damping time of a single quasinormal determine the mass and the spin of the remnant black hole.
The fundamental quadrupolar mode, which is the dominant mode, was already detected in 22 events~\cite{LIGOScientific:2016lio, LIGOScientific:2020ufj, LIGOScientific:2020tif, LIGOScientific:2021sio}.
The frequency and damping time of the fundamental quadrupolar mode was measured at times later than the peak of amplitude, and the results were compared with the estimate performed using full inspiral–merger–ringdown waveforms.
This kind of measurement provides a consistency check of the black hole spacetime and the theory of General Relativity.

The confident detection of a subdominant quasinormal mode would provide enough information to perform tests of the black hole spacetime using only information from the ringdown.
The frequency of oscillation and damping time of a secondary quasinormal mode also determine the mass and the spin of the remnant black hole, which can be compared with the values obtained from the primary mode.
In analogy to the standard electromagnetic spectroscopy, the multimode analysis of the quasinormal modes is known as \emph{black hole spectroscopy}~\cite{Dreyer:2003bv}.

In this thesis I present the results of the work I developed during my doctoral studies.
In Chapter~\ref{ch:gravitational_waves} I give an introduction for quasinormal modes and black hole spectroscopy with binary black hole systems.
In Chapter~\ref{ch:gw-detection} I present the tools and methodologies needed for gravitational wave data analysis.
In Chapter~\ref{ch:ringdown-qnm} I discuss our analysis of quasinormal modes using numerical simulations~\cite{Ota:2019bzl}.
I present the prospects for detecting two or more quasinormal mode with current and future gravitational wave detectors~\cite{Ota:2021ypb} in Chapter~\ref{ch:spectroscopy-horizon}.
Finally, I present my conclusions in chapter~\ref{ch:conclusions}.
The scripts I wrote to produce the results of this work can be found in~\cite{github}.

\begin{center}
\myclearpage
\par\end{center}


\chapter{Gravitational waves and quasinormal modes}
\label{ch:gravitational_waves}

To date, the only detected sources of gravitational waves (GW) are compact binary coalescences~\cite{LIGOScientific:2018mvr,LIGOScientific:2020ibl,LIGOScientific:2021usb,LIGOScientific:2021djp}.
Those are highly dynamical events, and they are only fully described by Numerical Relativity simulations, which are able to solve the non-linear dynamics of the Einstein field equations.
It is remarkable that the remnant black hole (BH) of a binary black hole (BBH) merger emits gravitational waves which are well approximated by solutions obtained in the linear perturbation analysis of a single BH\@.
This result not only highlights the importance of the perturbation analysis as more than just an idealized scenario, but it gives us decades of theoretical knowledge to analyze gravitational waves from astrophysical sources.
Throughout this chapter, unless stated otherwise, we consider the geometrized unit system $c = G = M_\odot = 1$,
where $c$ is the speed of light, $G$ is the gravitational constant and $M_\odot$ is the mass of the Sun.

\section{Black hole perturbation theory}

A BH is a region of spacetime from which nothing, not even light, can escape.
The event horizon of a BH represents the boundary where everything can only move towards the singularity of the BH\@.
More precisely, there are two regions, the interior and the exterior, and the exterior observers are causally disconnected with the interior.
Mathematically speaking, null geodesics inside the event horizon never reach the future null infinity, that is, events inside the event horizon cannot be the casual past of the null infinity.
Therefore, for classical BHs, all the mass-energy information is lost inside the event horizon.
The first BH solution was found by Karl Schwarzschild when he derived the spherical vacuum solution.
His solution represents an uncharged non-rotating BH, which is only characterized by its mass $M$.
The Schwarzschild metric has the form~\cite{Schwarzschild:1916uq}
\begin{equation}
	ds^2 = - f dt^2 + f^{-1} dr^2 + r^2 d\Omega^2, \qquad f = 1 - \frac{2M}{r}.
\end{equation}

The singularity at the event horizon $r = 2M$ is a coordinate singularity, which becomes regular when a different coordinate system is used.
The tortoise coordinate $r^*$, defined as $dr/dr^* = f$, is a coordinate transformation used to regularize the singularity at the event horizon.
The singularity at $r = 0$ is the point where the spacetime curvature becomes infinite.
BHs were long thought to be just a mathematical artifact of the Einstein field equations, and due to Birkhoff's theorem the Schwarzschild solution was mostly considered as a solution for the spacetime outside spherical objects.

BH linear perturbation theory was first studied in 1957 by Regge and Wheeler~\cite{Regge:1957td}, when they analyzed axial gravitational perturbations in the Schwarzschild spacetime in order to investigate the time stability of the geometry.
The stability of a spacetime metric is necessary to guarantee the viability of the solution, and those analyses where performed well before BHs and event horizons were fully understood.
In the 70s, these perturbations were studied in more detail and the gravitational waves from perturbed BH were discovered.
Zerilli~\cite{Zerilli:1970se,Zerilli:1970wzz} analyzed general perturbations of the Schwarzschild geometry and derived wave equations and radiation emitted by test particles.
Vishveshwara~\cite{Vishveshwara:1970zz} studied the evolution of the perturbations and discovered that, at some late time, the waveform is a damped sinusoidal wave,
which Press~\cite{Press:1971wr} identified as free modes of oscillation of the BH\@.

We are interested in the solutions of the BH perturbation equations that satisfy appropriate boundary conditions:
physically, no information can come from the event horizon, and, therefore no wave leaves the event horizon.
To guarantee that the BH is not continually perturbed, no wave comes from infinity.
A perturbed BH that satisfies these boundary conditions emits gravitational waves which are divided in three stages.
The \emph{transient} is the first stage and depends on the initial perturbation.
The \emph{quasinormal modes} (QNM) stage appears at later times, and they are damped sinusoidal solutions.
Their complex frequencies  $\omega \equiv \omega^r + i\omega^i$ depend solely on the BH properties.
The amplitudes and phases of these modes are determined by the transient, that is, by the initial perturbation.
In the last stage the oscillation ceases with a \emph{power-law tail}~\cite{Price:1971fb,Gundlach:1993tp,Price:1994pm}.

Moreover, BH metrics that are asymptotically flat imply that the solutions must be asymptotically plane waves, that is, the waveform is proportional to $e^{i\omega(t - r_*)}$, when $r_*\to + \infty$, where $r_*$ is the tortoise coordinate~\cite{Nollert:1999ji}.
For classical BHs, nothing leaves the event horizon, and only ingoing waves are present, that is, the wave is proportional to $e^{i\omega(t + r_*)}$ when $r_*\to - \infty$.
All these conditions result in the quasinormal mode frequencies $\omega$, and the linear stability of the BH solution implies that $\omega^i > 0$.
The damping caused by the imaginary part of the frequency is physically expected.
Unlike physical problems involving perturbation whose solutions are normal modes of oscillation,
the boundary conditions we imposed imply that the system must be dissipative,
as the system is open in the BH horizon and at infinity.

The Schwarzschild BH is the simplest BH solution, as its only parameter is the mass of the BH\@.
But even astrophysical BHs cannot be arbitrarily complex and no equation of state is needed to describe one.
The \emph{no-hair} theorems imply that the stationary and asymptotically flat BH solutions of General Relativity are fully described by three parameters: mass, spin and charge~\cite{Israel:1967wq, Israel:1967za, Carter:1971zc, Cardoso:2016ryw}.
The name of the theorem originates from the phrase ``black holes have no hair'', which was coined by Jacob Bekenstein but famously used by John Wheeler.
In this phrase, ``hair'' represents all the information about matter and energy that is lost in the event horizon.
Although charged BHs are stable solutions of the Einstein field equations, described by the Reissner–Nordström and the Kerr-Newman metrics, astrophysical BHs are not expected to be strongly charged.
Any possible charge obtained by a BH will be readily neutralized by its surrounding environment, as astrophysical BHs are not expected to be completely isolated in vacuum.
Therefore, we can consider that astrophysical BHs are fully determined by their mass and spin, and they are described by the Kerr geometry~\cite{Misner:1974qy}.

BH solutions assume that the BH has always existed and is completely surrounded by vacuum.
When we say that astrophysical BHs are Kerr BHs we are making some approximations.
The universe is of course not empty and not even asymptotically flat, as it is undergoing an accelerated expansion.
The approximations are, however, extremely good in the vicinity of a BH and also for GWs, as the perturbation (or merger) does not happen in a cosmological scale (although the redshift of the propagated wave should be considered).
BHs are also one of the possible final stages of the stellar evolution process.
If the mass of a star in its late stage is large enough (but smaller than approximately 50 solar masses), the collapse into a BH is unavoidable.
From stellar evolution theory (and observations!), BHs are expected to be real, and they should be described by the Kerr metric.

The Teukolsky equation~\cite{Teukolsky:1972my} describes linear perturbations on the Kerr metric.
The solution of the Teukolsky equation is obtained by separation of variables, where the angular part is described by the spin-weighted spheroidal harmonics with spin-weight parameter $-2$~\cite{Teukolsky:1973ha, Berti:2005gp}, ${}_{-2}S_{\ell m}(a \omega, \iota, \beta)$,
which depends on the BH spin parameter $a_f \in [0, M_f]$, where $M_f$ is the BH mass\footnote{We use the $f$ subscript here because the remnant BH of a BBH merger can be described by the Kerr metric.}, on the complex frequencies $\omega$,
and on the inclination and azimuth angles $(\iota, \beta)$ of the binary relative to the observer.
When $a_f = 0$, the Kerr BH becomes a Schwarzschild BH, and the spin-weighted \emph{spheroidal} harmonics become the spin-weighted \emph{spherical} harmonics, ${}_{-2}Y_{\ell m}$~\cite{Berti:2005gp}.
The radial part of the perturbation describes the decaying oscillating wave solution.

For a distant observer, the ringdown waveform written as the sum of the polarizations $+$ and $\times$ basis (see Appendix~\ref{ch:linearized_gravity}) of the quasinormal modes is given by
\begin{equation}
    h_+ + i h_{\times} = \frac{M_f}{r} \sum_{\ell m n} A_{\ell m n} e^{i[\omega_{\ell m n}(t - t_i) - \phi_{\ell m n}]} {}_{-2}S_{\ell m}(a \omega_{\ell m n}, \iota, \beta), \qquad t > t_i,
    \label{eq:qnm_polarizations}
\end{equation}
where $A_{\ell m n}$ and $\phi_{\ell m n}$ are, respectively, the amplitude and phase of the mode $(\ell,m,n)$,
$\omega_{\ell m n} \equiv \omega_{\ell m n}^r + i\omega_{\ell m n}^i$ are the corresponding complex quasinormal modes frequencies,
$t_i$ is the initial time of the ringdown and $r$ is the distance between the observer and the source.
The modes with $n=0$ are called \emph{fundamental modes} and the modes with $n > 0$ are the \emph{overtones}.
We define the \emph{harmonic mode} $(\ell, m)$ as the radial solution of the corresponding angular harmonic, given by
\begin{equation}
    h_{\ell m} \equiv \sum_n h_{\ell m n} \equiv \sum_n A_{\ell m n} e^{i[\omega_{\ell m n}(t - t_i) - \phi_{\ell m n}]}.
    \label{eq:qnm_harmonic}
\end{equation}

The complex frequencies $\omega_{\ell m n}$ depend only on the BH mass $M_f$ and spin $a_f$, and they are independent of the initial perturbation.
The quasinormal mode spectrum is the solution of the eigenvalue problem associated with the Teukolsky equation.
A literature review of several methods used to find the quasinormal modes spectrum can be found in~\cite{Konoplya:2011qq}.
The BH mass and the spin can be computed from the frequencies using the fits obtained by~\cite{Berti:2009kk}, which can be inverted to give
\begin{equation}
	M_f = \left[f_1 + f_2\left(\frac{\omega^r/(2\omega^i) - q_1}{q_2}\right)^{\frac{f_3}{q_3}} \right]/\omega^r,
	\quad
	a = M_f\left[1 -  \left(\frac{\omega^r/(2\omega^i) - q_1}{q_2}\right)^{\frac{1}{q_3}} \right],
	\label{eq:mass_spin_qnm}
\end{equation}
where the values of $f_i$ and $q_i$ for each mode $(\ell, m, n)$ are available in~\cite{Berti-ringdown}.
The original form of the above equations can be used to compute the frequencies $\omega_{\ell m n}$ as a function of the BH mass and spin.

\section{Binary black hole merger}
\label{sec:binary-black-hole-merger}

BBH systems are the most common source of gravitational wave detections~\cite{LIGOScientific:2018mvr, LIGOScientific:2020ibl, LIGOScientific:2021usb, LIGOScientific:2021djp}.
The emission of gravitational waves implies that the orbit shrinks until the BHs eventually merge into a single BH\@.
Close to the merger time the remnant BH is highly distorted and the system evolution is not linear.
At late times, in the ringdown, the remnant BH emits gravitational waves that are very well approximated by QNMs~\cite{Buonanno:2006ui,Seidel_2004}.

When the BHs are widely separated, they can be treated as point particles and their orbits are quasi-Kleperian, which
effectively means that their orbital velocity is much smaller than the speed of light~\cite{poisson-gravity}.
In this regime, the dynamical evolution can be treated analytically using Newtonian mechanics with General Relativity corrections.
An expansion can be done in the orbital velocity parameter over the speed of light through the post-Newtonian approximation~\cite{maggiore-vol1,poisson-gravity, Blanchet:2013haa},
which expresses orders of deviations from Newton's law of universal gravitation.

Due to the emission of gravitational waves, the system loses energy.
The loss of energy results in a decrease of all of the orbital parameters: the orbital period, the semi-major axis and the eccentricity~\cite{Peters:1964zz}.
This implies that the orbit of the binary decreases over time, until the BHs merge into a single BH\@.
The decrease of eccentricity implies that the orbit will most likely (but not necessarily) become nearly circular before the merger.
As the orbital period decreases, the binary frequency increases, and so does the emitted GW frequency, as for quasi-circular binaries the GW frequency is twice the orbital frequency~\cite{maggiore-vol1, poisson-gravity}.
The amplitude of the waveform is inversely proportional to the square of semi-major axis, which implies that the amplitude increases over time.

\begin{figure}[htb!]
	\centering
	\includegraphics[width=0.95\linewidth]{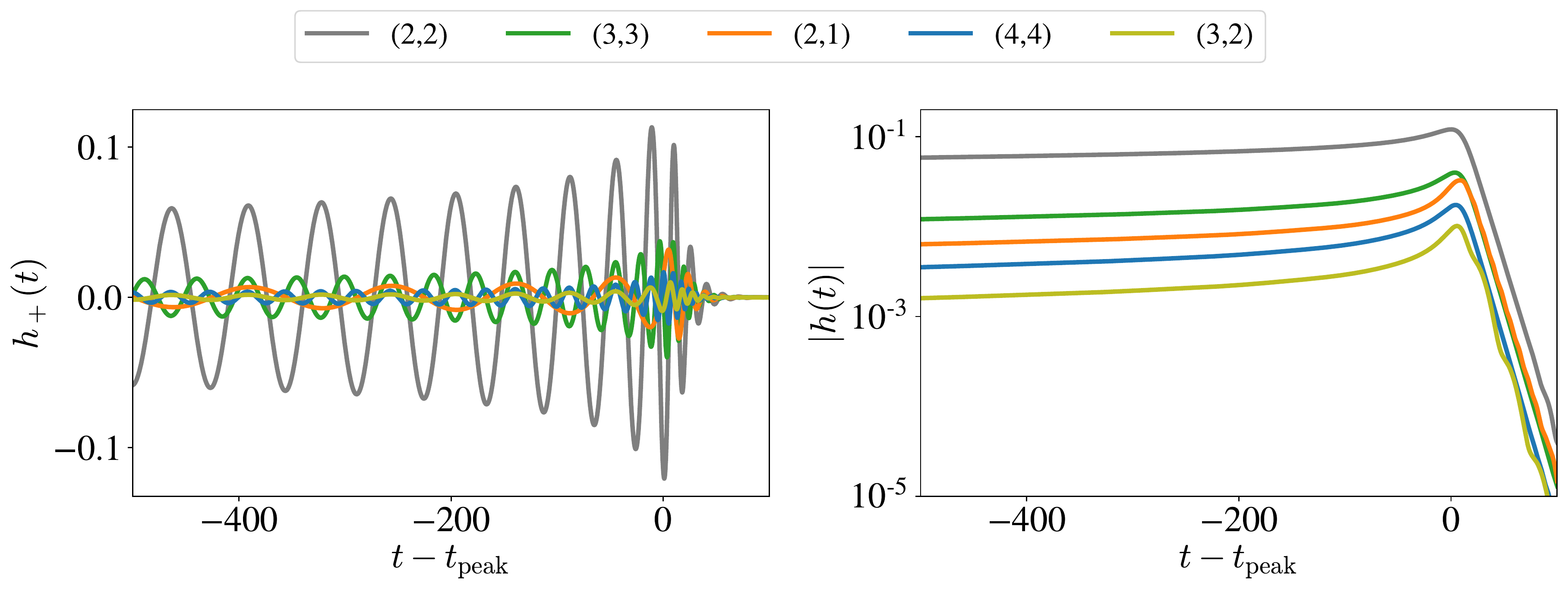}
	\caption{Waveform of the $+$ polarization (left) and the amplitude of the wave (right) as a function of time of
	the binary black hole simulation SXS:BBH:1107,
		which corresponds to a a nonspinning BBH system with initial mass ratio  $q \equiv m_1/m_2 = 10$, for the modes
		$(2,2)$, $(3,3)$, $(4,4)$, $(2,1)$ and $(3,2)$.
	Before the peak of amplitude at $t_{\textrm{peak} }$, the amplitude and the frequency of the modes increase, and there is an exponential decay some time after the peak, which corresponds to the QNM decay.
	The mode mixing can be seen in the $(3,2)$ decay, as there is a wobble in the decay.
	}
	\label{fig:NR_h}
\end{figure}

Therefore, in the inspiral phase, the amplitude and the frequency of the GW waveform will increase over time.
After the approximation is no longer valid, the frequency and amplitude continue to increase, but they are not determined analytically.
The waveform amplitude peaks near the merger time and starts decreasing, and, at a later time,
the waveform is well described by QNMs of the remnant BH~\cite{Buonanno:2006ui}.
This process is depicted in Figures~\ref{fig:waveform_gw150914} and~\ref{fig:NR_h}.

The QNM frequencies are fully characterized by the remnant BH parameters, but their amplitudes and phases depend on the initial perturbation.
In the BBH ringdown these parameters are determined by the initial conditions of the binary system.
Numerical Relativity is needed to solve the evolution of the BBH when the post-Newtonian approximation is no longer valid.
Therefore, the determination of QNMs parameters of a BBH system is dependent on numerical simulations.

The time evolution of the merger process is a Cauchy problem,
where the time evolution of the gravitational field is associated with an initial value problem.
In the Einstein field equations time and space are a part of the same entity: spacetime.
A reformulation of the Einstein's equations that splits time from space is needed for the definition of an initial value problem.
A very commonly used reformulation is called \emph{3+1 formalism}, and the spacetime metric is written as~\cite{baumgarte_shapiro_2010, alcubierre2008introduction}
\[
ds^2 = -\alpha^2dt^2 + \gamma_{ij}(dx^i + \beta^i dt)(dx^j + \beta^j dt),
\]
where $\gamma_{ij}$ is the metric of the 3-dimensional spacial hypersurface,
$\alpha$ is the \emph{lapse function}, which represents the time evolution,
and $\beta^i$ is the \emph{shift vector}, which determines the hyperspace evolution as a function of time.

The numerical evolution of the Einstein field equations is a challenging problem and it is an active area of research.
In this work we analyze numerical simulations from the Simulating eXtreme Spacetimes (SXS)~\cite{SXS-catalog, Boyle:2019kee} project.

In NR simulations, the waveforms are decomposed in terms of the spin-weighted spherical harmonics, ${}_{-2}Y_{\ell m}$.
This represents a problem in the analysis of the ringdown, as the Kerr QNMs are decomposed in terms of the spin-weighted spheroidal harmonics, ${}_{-2}S_{\ell m }$.
The expansion of ${}_{-2}S_{\ell m }$ in terms of the of spherical harmonics includes the ${}_{-2}Y_{\ell m }$ with the same $m$ and different $\ell$ (see equation~(3.7) of~\cite{Press:1973zz}).
This mode mixing is relevant only for $|m| \neq \ell$  and the strongest mode for which the mixing is relevant is the $(\ell, m) =(3,2)$ mode~\cite{Kelly:2012nd,Berti:2014fga}.
However, the $(3,2)$ mode has a very small amplitude and we do not consider it in our analysis.
We only consider the modes $(2,2)$, $(3,3)$, $(4,4)$ and $(2,1)$, which are the strongest modes~\cite{Cotesta:2018fcv}.
Therefore, ${}_{-2}Y_{\ell m }$ is taken here as a good approximation of the ${}_{-2}S_{\ell m }$.
Figure~\ref{fig:NR_h} shows on the left one example of the polarization $+$ of a NR simulation of a nonspinning BBH system with initial mass ratio  $q \equiv m_1/m_2 = 10$, and the amplitude of the waveform is on the right.
The $(2,2)$, $(3,3)$, $(4,4)$, $(2,1)$ and $(3,2)$ modes are shown, and we can see a clear dominance of the $(2,2)$, and the $(3,2)$ mode is the weakest mode (see also Figure~1 of~\cite{Cotesta:2018fcv}).
We can also see in the amplitude plot that, at some time after the peak of amplitude $t_{\textrm{peak} }$, the modes decay exponentially
(linearly in the log-scale), which is compatible with the QNM solution.
However, there is a wobble in the $(3,2)$ mode decay, which is related to the mode mixing.

The remnant properties can be computed from NR simulations through the analysis of the apparent horizon~\cite{Boyle:2019kee}.
Using equation~\eqref{eq:mass_spin_qnm}, the QNM frequencies can be obtained from the computed mass and spin.
These frequencies are compatible with frequencies obtained by fitting equation~\eqref{eq:qnm_harmonic} to the given waveform, as we show in chapter~\ref{ch:ringdown-qnm}.

The decomposition of the waveform in harmonic modes is an important feature of the GW,
as it helps determine better the characteristics of the source.
Detecting one harmonic mode, usually the dominant $(\ell, m) = (2,2)$, is enough to determine some source properties.
However, there can be degenerencies in the binary parameters.
For example, precession and eccentricity may have similar effects on the waveform~\cite{Romero-Shaw:2020thy}.
Thus, the determination of the parameters will be better when more modes are detected.

Furthermore, although all the detections (GW and electromagnetic) are compatible with the existence of BHs,
there could still be a doubt about whether the detected objects are really BHs or some exotic object mimicking a BH\@.
The detection of several modes of the BH waveform increases the evidence of a BH detection.
This test can be done in the whole waveform, as we can see from Figure~\ref{fig:NR_h}.

The deviations in the ringdown are going to be deviations in the QNMs parameters.
The mass and spin of the remnant BH can be determined from the complex frequency of a single quasinormal mode $(\ell,m,n)$,
using equation~\eqref{eq:mass_spin_qnm}.
The remnant BH properties are also determined by the initial condition of the binary, given by the whole waveform evolution.
Therefore, a consistency test can be done by comparing the values estimated by the whole waveform and by the ringdown alone.
This test was performed in some detections where the dominant QNM $(2,2,0)$ was detected, and no deviation from GR was found~\cite{LIGOScientific:2016lio, LIGOScientific:2020tif, LIGOScientific:2020ufj, LIGOScientific:2021sio}.

However, tests using QNMs are not all dependent on the whole waveform.
Once the mass and spin is determined by the dominant QNM, the consistency test can be done with another QNM,
which should have a complex frequency compatible with the determined mass and spin.
Therefore, the no-hair theorem can be tested by detecting two or more QNMs.
In direct analogy to the electromagnetic spectroscopy, the multi-mode analysis of the ringdown is called \emph{black hole spectroscopy}~\cite{Dreyer:2003bv}.
The QNMs are always in the ringdown, and their amplitude always decays exponentially with time, which make their detection very challenging.

\section{Quasinormal modes in the binary black hole ringdown}
\label{sec:qnm-bbh}

\begin{figure}[htb!]
	\centering
	\includegraphics[width=1\linewidth]{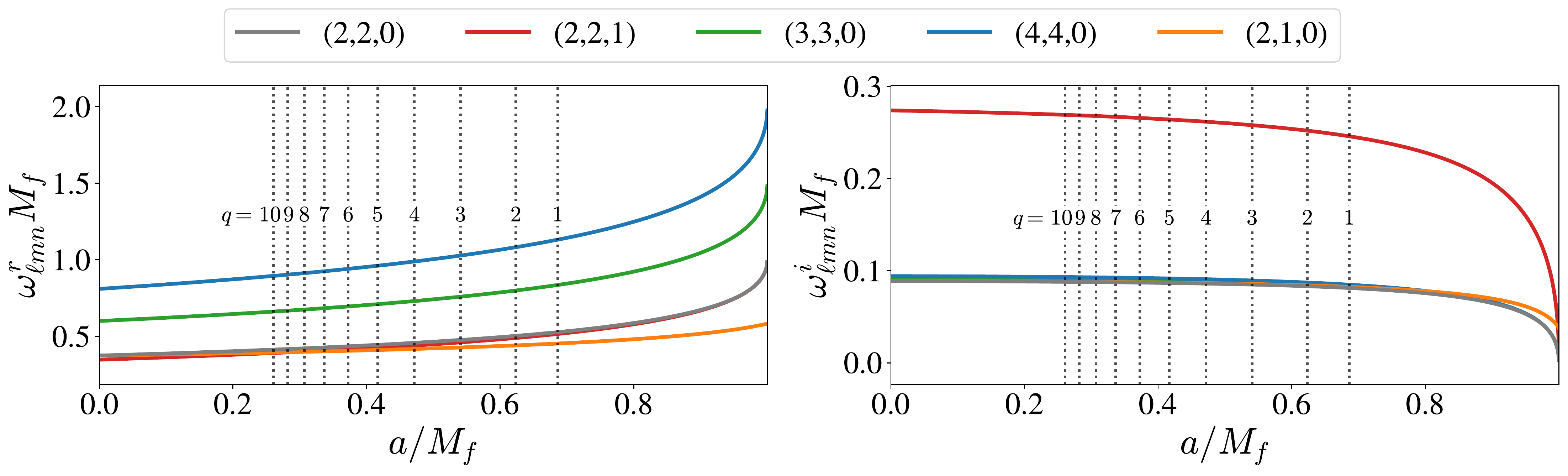}
	\caption{Real (left) and imaginary (right) parts of the dimensionless frequencies for the QNMs modes  $(2,2,0)$, $(2,2,1)$, $(2,1,0)$, $(3,3,0)$ and $(4,4,0)$ as a function of the dimensionless BH spin $a/M_f$.
	The values for the plots were obtained with linear perturbation theory by~\cite{Berti:2009kk, Berti-ringdown}.
	The vertical dotted lines are the same for both plots and indicate the final dimensionless spin of the remnant BH of nonspinning circular binaries with mass ratios $q\equiv m_1/m_2$ ranging from 1 (highest spin) to 10 (lowest spin).}
	\label{fig:qnm_omegas_a}
\end{figure}

An understanding of the properties of QNMs in the ringdown of BBHs is very important when we are analyzing the detections.
Figure~\ref{fig:qnm_omegas_a} shows the real (left) and imaginary (right) parts
of the dimensionless frequencies $M_f \omega_{\ell m n}$ of the modes
$(2, 2, 0)$, $(2, 2, 1)$, $(2, 1, 0)$, $(3, 3, 0)$ and $(4, 4, 0)$ as a function of the
dimensionless spin $a/M_f$.
These values were obtained with BH perturbation theory and are available in~\cite{Berti-ringdown}.
For the modes showed in the plots, the real part of the frequency increases and the imaginary part decreases as the spin increases.

The frequencies of oscillation and damping times are defined in terms of the complex frequencies:
\begin{equation}
	f_{\ell m n} \equiv \frac{\omega^r_{\ell m n}}{2\pi}, \qquad \tau_{\ell m n} \equiv \frac{1}{\omega^i_{\ell m n}}.
	\label{eq:freq_tau_omegas}
\end{equation}
Therefore, Figure~\ref{fig:qnm_omegas_a} shows that fast spinning BHs will have higher frequencies of oscillation and larger damping times.
It is important to stress that these trends are not valid for all $(\ell, m, n)$ modes, as can be seen in Figure 5 of~\cite{Berti:2005ys}.

There are some general relations between the QNMs frequencies and damping times which are independent of the spin of the BH\@.
From the left plot of Figure~\ref{fig:qnm_omegas_a}, we can see that modes with higher $\ell$ have higher frequencies of oscillation.
The right plot shows that fundamental modes have very similar damping times, but the overtone has a much smaller damping time.
In fact, the overtone number is defined such that higher overtones decay faster, and the fundamental mode with $n=0$ is the longest lived tone of a harmonic, that is, $\omega^i_{\ell m (n+1)} > \omega^i_{\ell m n}$ for $n\geq 0$.

The properties described above are valid for any perturbed Kerr BH.
The amplitudes and phases of the QNMs depend on the BBH merger process, and the initial parameters of the BBH determine the dimensionless spin of the remnant.
The initial parameters of a BBH are the initial masses $m_1$ and $m_2$, the initial spins $\mathbf{S}_1$ and $\mathbf{S}_2$ and the orbital eccentricity $e$.
These are also known as \emph{intrinsic} parameters, which are independent of the observer.
For simplicity, we consider here only nonspinning circular binaries, that is $\mathbf{S}_1 = \mathbf{S}_2 = e = 0$.
Although this may seem like an idealized scenario, most LVC detections are compatible with it~\cite{LIGOScientific:2018mvr, LIGOScientific:2020ibl, LIGOScientific:2021usb, LIGOScientific:2021djp}.
We also notice that highly spinning or highly eccentric binaries are very hard to simulate, and there are not many NR simulations available for these cases.
With this assumption, the remnant BH properties will depend only on the mass ratio $q = m_1/m_2 > 1$.\footnote{Notice that, in this chapter, we are only considering intrinsic parameters and treating the problem with dimensionless quantities.
The detected parameters (in S.I. units) depend on the total mass of the binary and the source position relative to the detector, as explained in Chapter~\ref{ch:gw-detection}.}

Larger mass ratio binaries result in BHs with lower spins.
An intuitive way to understand why is thinking about one BH being perturbed by a second one.
When the second BH is much smaller, it will spiral into the larger one without strongly disturbing it.
In the point particle limit, the larger BH will not start spinning at all, and it will remain a Schwarzschild BH\@.
As the smaller BH gets larger, it effectively spins up the larger BH\@.
Therefore, the remnant BH of the equal mass case has the highest spin possible for nonspinning circular binaries.\footnote{This intuitive description is not valid for head-on collision.}
The vertical dotted lines of Figure~\ref{fig:qnm_omegas_a} indicate the final dimensionless spin of BBH with mass ratios ranging from 1 to 10.

The amplitude of the modes also depends on the BBH properties.
The quadrupolar mode $(\ell, m) = (2,2)$ has the largest amplitude.
As the fundamental mode is the longest lived mode of a harmonic, we call the mode $(2,2,0)$ the \emph{dominant mode}.
The other harmonics have amplitude significantly smaller than the $(2,2)$ mode amplitude (see Figures~\ref{fig:NR_h} and~\ref{fig:mass_ratio_fits}),
and we call modes with $(\ell,m)\neq (2,2)$ \emph{higher harmonics}.
Symmetries in the binary make the amplitude of the higher harmonics smaller.
For example, a nonspinning circular binary with mass ratio $q=1$ does not excite the harmonics with odd $m$.
Here we just consider nonspinning circular binaries, but a study on how the initial parameters of the binary impact the amplitude of the higher harmonics can be found in~\cite{Cotesta:2018fcv}.

The amplitudes of the higher overtones of a harmonic mode ($n > 0$) are harder to determine,
as these modes are not separated in the numerical relativity simulations.
The overtones decay much faster than the fundamental mode, and their relative amplitude depends on the considered time.
The initial time when the non-linearities of the merger can be neglected is an open question.
We address these problems and propose a way to confidently compute the amplitude of the first overtone in Chapter~\ref{ch:ringdown-qnm}.
\begin{center}
\myclearpage
\par\end{center}


\chapter{Gravitational wave detection}
\label{ch:gw-detection}

Gravitational wave amplitudes are very small, and the detection of gravitational radiation is very challenging.
Nevertheless, dozens of compact binary mergers have been detected by the laser interferometers that are currently operating.
The detected wave is buried in the noise of the detector, therefore, the characterization of the noise is an essential part of gravitational wave analyses.

\section{Detector frame}

The physical quantities of equation~\eqref{eq:qnm_polarizations} are ``measured'' in the source frame.
As GW detectors can detect objects at a cosmological distance,
the cosmological redshift $z$, which is caused by the expansion of the universe, must be taken into account.
In the detector frame, the redshift is accounted for the masses $M \to (1+z) M$,
 frequencies  $f_{\ell mn} \to f_{\ell mn}/(1+z)$, damping times $\tau_{\ell mn} \to \tau_{\ell mn}(1+z)$~\cite{maggiore-vol1}
and distance, by substituting $r$ with the luminosity distance $D_L = \sqrt{\mathfrak{L}/(4\pi\mathcal{F}})$~\cite{Hogg:1999ad},
where $\mathfrak{L}$ is the luminosity and $\mathcal{F}$ is the energy flux of the source.
Whenever needed, we consider the cosmological parameters obtained by the Planck mission~\cite{Planck:2018vyg}.

\section{Detector response: antenna response pattern}
\label{sec:antenna-response-pattern}

The detected waveform $h(t)$ is the projection of the wave onto the detector: $h(t) = D^{ij}h_{ij}(t)$,
where $D^{ij}$ is a constant tensor which depends on the geometry of the detector~\cite{maggiore-vol1, Sathyaprakash:2009xs}.
The detector antenna response patterns are defined as $F_{+,\times}(\theta, \phi) = D^{ij}e^{(+,\times)}_{ij}$,
where $e^{(+,\times)}_{ij}$ are the tensorial basis of the $+$ and $\times$ polarizations defined in equation~\eqref{eq:polarization_tensor}
and $(\theta, \phi)$ determine the orientation of the source, which defines the propagation direction vector $\mathbf{n}\equiv \mathbf{e}_z$.
Thus, the detected wave is given by
\begin{equation}
	h(t) = F_+(\theta, \phi)h_+(t) + F_{\times} (\theta, \phi)h_{\times} (t).
	\label{eq:h_detector_two_angs}
\end{equation}

\begin{figure}[htb!]
  \centering
  \includegraphics[width=0.45\linewidth]{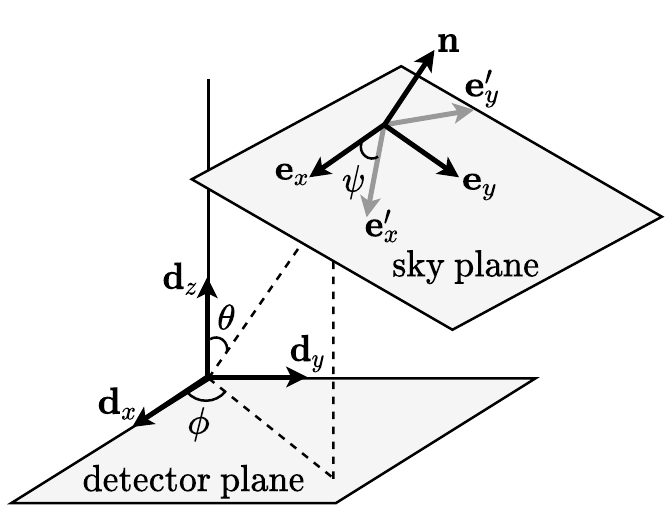}
  \includegraphics[width=0.45\linewidth]{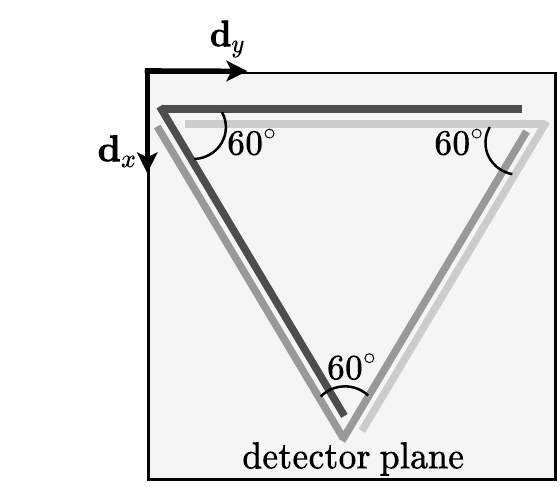}
  \caption{\emph{Left:} relative orientation of the detector and sky frames.
  \emph{Right:} schematic configuration of a triangular interferometer detector, like ET and LISA. These detectors are formed by six V-shaped interferometers two at each corner (one for low frequencies and the other for high frequencies), with arms with a $60^\circ$ opening angle.}
  \label{fig:angles_detector}
\end{figure}

The above equation assumes that the TT gauge basis vectors $\mathbf{e}_{x,y}$ are contained in the plane formed by
the propagation vector $\mathbf{n}$ and the detector frame basis vectors $\mathbf{d}_{x,y}$ (one plane for each direction).
It is usually suitable to choose the vectors $\mathbf{d}_{x,y}$ that correspond to the arm directions.
Although this choice of basis is simple, it is not ideal when there are two or more detectors in operation.
It is more convenient to choose a more general basis $(\mathbf{e}_x^\prime, \mathbf{e}_y^\prime)$, which is a rotation of the basis $(\mathbf{e}_x, \mathbf{e}_y)$ by an angle $\psi$:
\begin{equation*}
    \mathbf{e}^\prime_x = \cos\psi\mathbf{e}_x - \sin\psi\mathbf{e}_y,
    \quad \mathbf{e}^\prime_y = \sin\psi\mathbf{e}_x + \cos\psi\mathbf{e}_y.
\end{equation*}
Figure~\ref{fig:angles_detector} shows on the left the geometric relation between the detector frame basis and the sky plane,
which is the plane orthogonal to the wave propagation.

The rotated polarization tensors are
\begin{align*}
    \mathbf{e}_{+}^\prime \equiv&
    \mathbf{e}_{x}^\prime\otimes\mathbf{e}_{x}^\prime - \mathbf{e}_{y}^\prime\otimes \mathbf{e}_{y}^\prime
    = \cos(2\psi)\mathbf{e}_{+} + \sin(2\psi)\mathbf{e}_{\times},
    \\
    \mathbf{e}_{\times}^\prime \equiv &
    \mathbf{e}_{x}^\prime\otimes\mathbf{e}_{y}^\prime + \mathbf{e}_{y}^\prime\otimes \mathbf{e}_{x}^\prime
    = -\sin(2\psi)\mathbf{e}_{+} + \cos(2\psi)\mathbf{e}_{\times}.
\end{align*}
The antenna patterns  $F_{+,\times}(\theta, \phi)$ and waveform polarizations $h_{+,\times}$ are transformed to the new basis according to the transformations above.
By substituting $F_{+,\times}(\theta, \phi)$ and $h_{+,\times}$ with the transformed $F_{+,\times}^\prime(\theta, \phi)$ and $h_{+,\times}^\prime$, equation~\eqref{eq:h_detector_two_angs} remains unchanged.
However, to consider the general situation, the antenna pattern is parametrized by the polarization angle $\psi$ while the wave polarization is defined in the generic basis $(\mathbf{e}_x^\prime, \mathbf{e}_y^\prime)$.
Dropping the primes, the most general form of the detected wave is given by
\begin{equation}
	h(t) = F_+(\theta, \phi, \psi)h_+(t) + F_{\times} (\theta, \phi, \psi)h_{\times}(t).
	\label{eq:h_detector_all_angs}
\end{equation}

For an interferometer, the detector tensor is given by~\cite{maggiore-vol1, Sathyaprakash:2009xs}
\begin{equation}
    \mathbf{D} = \frac{1}{2}(\mathbf{d}_x\otimes\mathbf{d}_x - \mathbf{d}_y\otimes\mathbf{d}_y),
    \label{eq:detector-tensor}
\end{equation}
where $\mathbf{d}_{x,y}$ are in the direction of the detector's arms.
To obtain the antenna pattern, we just need to write the detector vectors in terms of the polarization vectors.
For an L-shaped detector, that is, interferometers with perpendicular arms, such as LIGO and CE,
$(\mathbf{d}_x, \mathbf{d}_y, \mathbf{d}_z)$ are obtained from $(\mathbf{e}_x^\prime, \mathbf{e}_y^\prime, \mathbf{e}_z^\prime)$
by a rotation around the $\mathbf{e}_y^\prime$ vector by the angle $\theta$
followed by a rotation around the $\mathbf{e}_z^\prime$ vector by the angle $\phi$.
This results in

\begin{subequations}
	\begin{align}
		\label{eq:antenna_plus}
		F_+(\theta, \phi, \psi) = \frac{1}{2}(1 + \cos^2(\theta))\cos(2\phi)\cos(2\psi) - \cos(\theta)\sin(2\phi)\sin(2\psi), \\
		F_{\times}(\theta, \phi, \psi) = \frac{1}{2}(1 + \cos^2(\theta))\cos(2\phi)\sin(2\psi) + \cos(\theta)\sin(2\phi)\cos(2\psi).
		\label{eq:antenna_cross}
	\end{align}
	\label{eq:antenna_pattern}
\end{subequations}

ET is a triangular detector, which is formed by a set of six V-shaped interferometers (the arm opening angle is $60^{\circ}$), as depicted in Figure~\ref{fig:angles_detector}, on the right.
The detector tensors for each interferometer are equivalent to equation~\eqref{eq:detector-tensor}, but with $\mathbf{d}_{x,y}$ being the arms of each interferometer.
The antenna pattern of a single V-shaped interferometer is a factor $\sin60^{\circ}$ smaller than the antenna pattern of an L-shaped interferometer~\cite{Regimbau:2012ir}.
Moreover, all three pairs interferometers are the same, except for a $120^{\circ}$ rotation in the $\phi$ angle.
Although LISA is a triangular detector, its antenna pattern does not have an analytical form, as its arms will be millions of meters long~\cite{Prince:2002hp,Robson:2018ifk}.

\begin{figure}[!h]
	\centering
	\includegraphics[width=0.32\linewidth]{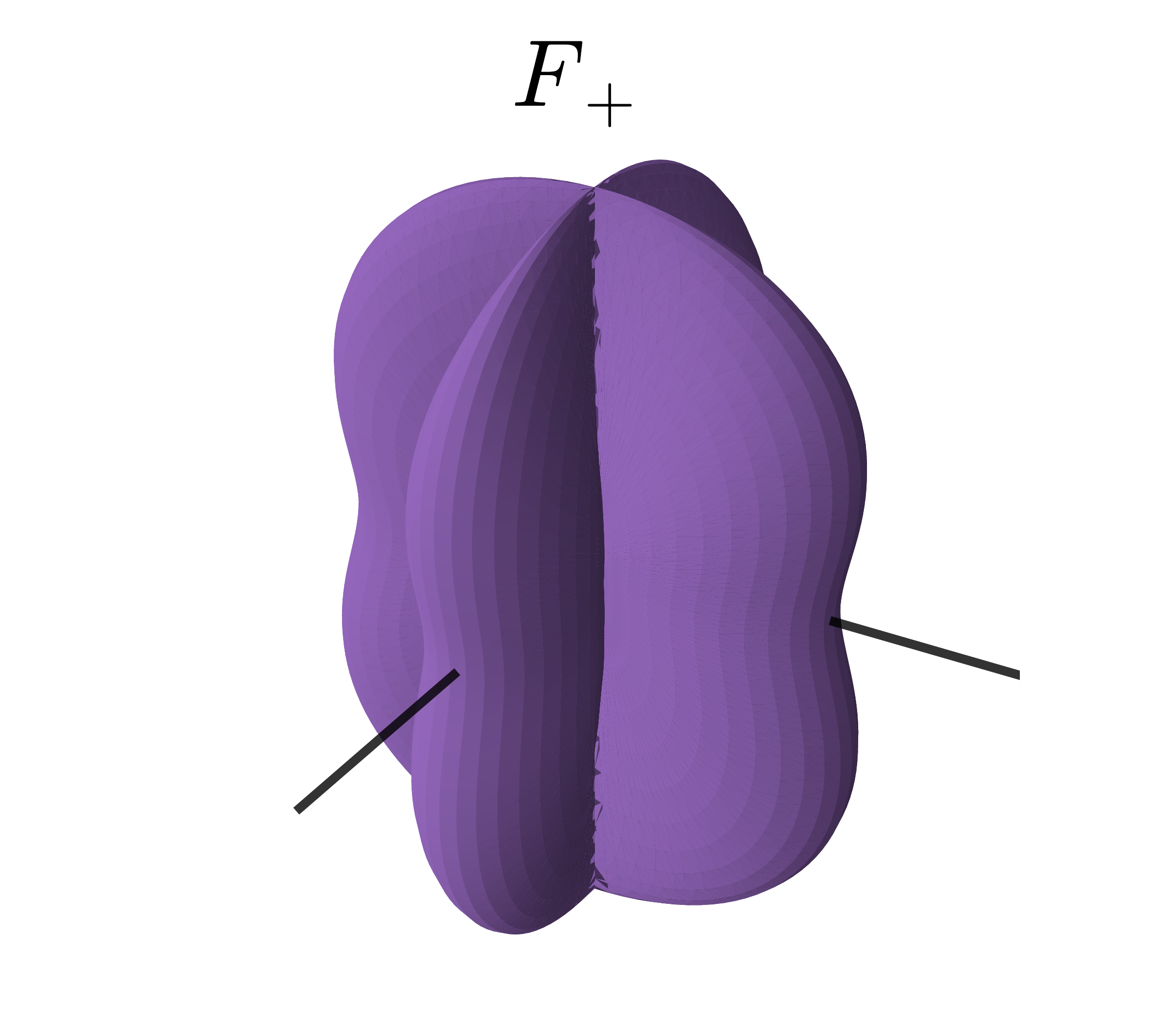}
	\includegraphics[width=0.32\linewidth]{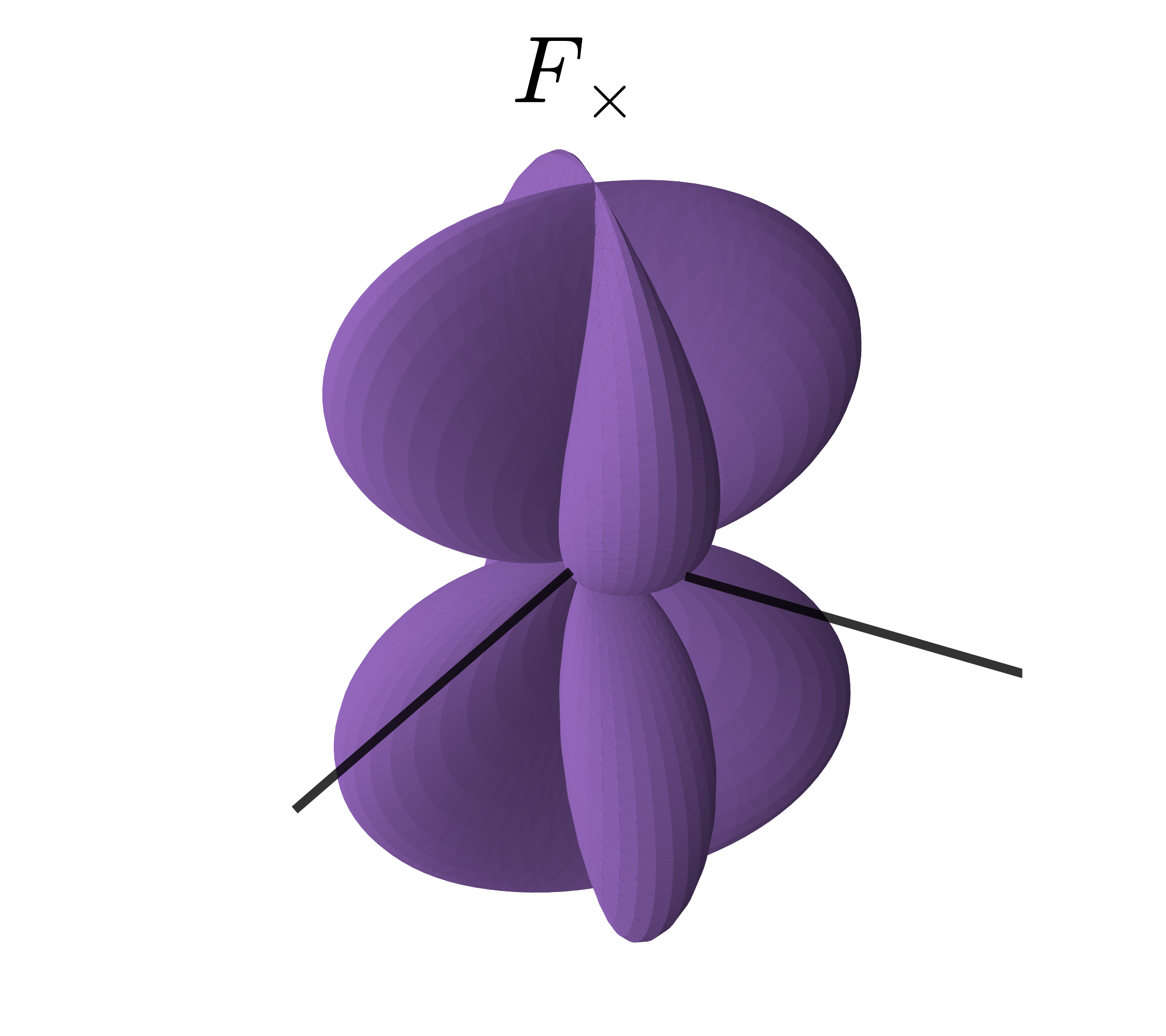}
	\includegraphics[width=0.32\linewidth]{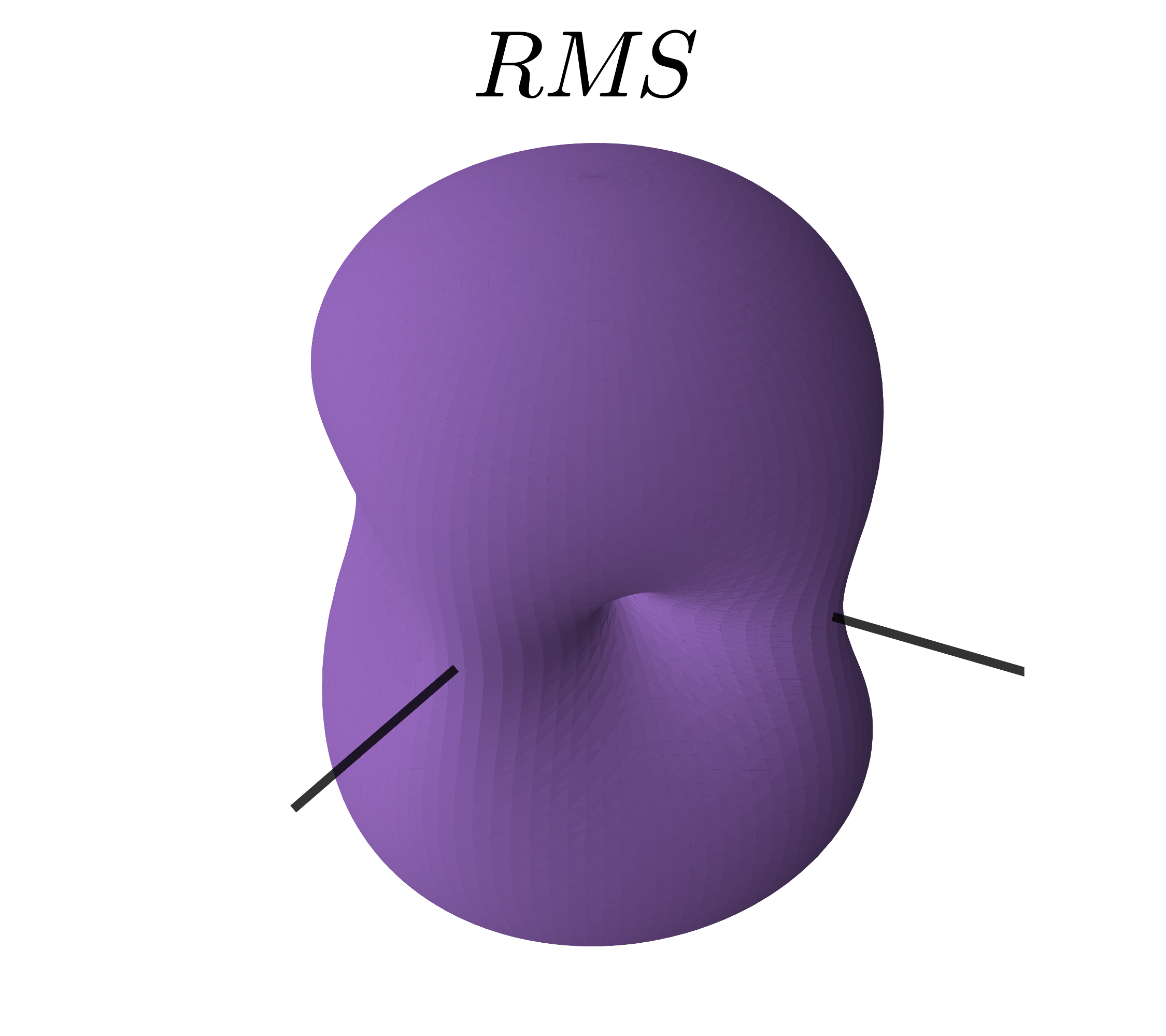}
	\caption{Antenna response pattern for L-shaped GW detectors, such as LIGO and CE, for the $+$ polarization (left), the $\times$ (center) polarization and the root-mean-square of both responses (right).
	The black lines indicates the direction of the detector's arms.
	}
	\label{fig:antenna_pattern}
\end{figure}

Figure~\ref{fig:antenna_pattern} shows the antenna response pattern for an L-shaped detector, given by equations~\eqref{eq:antenna_pattern}.
The root-mean-square (RMS) of the antenna polarization is also shown and the direction of the detector's arms is shown as black lines.
Analysing these plots we can see that the performance of the detector is highly dependent on the GW propagation direction.
The optimal direction is a wave coming in the direction orthogonal to the plane of the detector, as both polarizations will have maximum response on the detector.
A GW propagating parallel to the plane will be the hardest to detect, and there is even a blind spot in the direction that forms a $45^{\circ}$ angle with the detectors arms, which is the ``hole'' in the RMS plot.
Due to the triangular configuration, ET and LISA will not have blind spots.

Having two or more detectors is a strong factor for confirming the detection of a gravitational wave.
But having multiple detectors also avoids missing events coming from the ``wrong'' direction.
The more widely-distributed the detectors are located around the globe, the higher is the chance to spot a GW\@.
One interesting example is the first binary neutron star (BNS) detection GW170817~\cite{LIGOScientific:2018cki}.
This detection happened just three days after GW170814~\cite{LIGOScientific:2017ycc}, the first signal observed by a network of three laser interferometers, the two LIGOs and Virgo.
The signal of the BNS was clearly visible in both LIGO detectors, but only weakly detected Virgo, which was significant to constrain the sky position of the source, as the direction of the GW was near the Virgo's blind spot.
This constrained localization was fully compatible with the electromagnetic counterpart detection (see Figure 1 of~\cite{LIGOScientific:2017ync}).

In this work we consider the angular average of the detector response $\langle ({F^2_{+,\times}})^{1/2}\rangle$,
where $\langle X \rangle  \equiv (1/4\pi^2) \int_0^\pi d\psi \int_0^{2\pi} d\phi \int_0^\pi X \sin\theta d\theta$.
For the L-shaped detectors (LIGO and CE) $\langle ({F^2_{+,\times}})^{1/2}\rangle = 1/\sqrt{5}$,
and for ET (the triangular set of three V-shaped interferometers) $\langle ({F^2_{+,\times}})^{1/2}\rangle = 3/(2\sqrt{5})$~\cite{Regimbau:2012ir}.
The LISA detector does not have an analytical form~\cite{Robson:2018ifk}, and the values for $\langle ({F^2_{+,\times}})^{1/2}\rangle$ are contained in the noise spectral density (see next section) that we use in our analysis.

\section{Detector sensitivity: noise}

The detector sensitivity is characterized by its noise, which may limit which source can or cannot be detected.
The detector noise $n(t)$ is the output of the detector $d(t)$ in the absence of a gravitational wave signal, $d(t) = n(t)$.
As the GW signals are so small, the sources of noise are manifold.
Some examples are earthquakes, thermal noise and quantum noise.

In a theoretical approach, it is common to make some assumptions to simplify the noise characterization.
The first assumption is stationary noise, that is, the noise is independent of time.
This assumption is definitely not valid for long-duration signals.
However, the ringdown decays exponentially, and therefore this approximation is valid for our purposes.
There is also the assumption that the noise is Gaussian.
Although this seems like an idealized assumption, this is generally a good approximation for LVC~\cite{LIGOScientific:2019hgc}.

For Gaussian noise, the joint probability distribution is given by
\begin{equation}
	p(n) = \frac{1}{\det(2\pi C(t_i,t_j))^{\frac{1}{2}}}\exp\left[ -\frac{1}{2}\sum_{ij}(n_i - \mu)(n_j - \mu)C(t_i,t_j)^{-1}_{ij}  \right],
	\label{eq:prob_noise}
\end{equation}
where $\mu$ is the mean of the noise and $C(t_i,t_j)$ is the sample covariance matrix.
For stationary noise, the covariance matrix depends only on the time lag $\delta t = t_i - t_j$, that is,
the stationary noise is characterized by the correlation function~\cite{maggiore-vol1,Sathyaprakash:2009xs,Grishchuk:2000gh, Moore:2014lga}
\[C(\delta t) =\lim_{T\to \infty} \frac{1}{T}\int_{-T/2}^{T/2} n(t)n(t + \delta t)dt.\]
The advantage of stationary noise is clear on the Fourier domain because the covariance matrix is diagonal:
$C_{ij} = \delta_{ij} S_n(f_i)$,
where $\delta_{ij}$ is the Kronecker delta and $S_n(f)$ is the power spectral density (PSD),
which is the Fourier transform of the correlation function.

The PSD has dimension of time, but as it is defined in the Fourier domain, it is commonly written in the dimension $\mathrm{Hz}^{-1}$.
The PSD fully characterizes a Gaussian stationary noise.
But because the detectors measure amplitudes, the amplitude spectral density (ASD), defined as the square root of the PSD, is most commonly used in the literature.

\begin{figure}[!h]
	\centering
	\includegraphics[width=\linewidth]{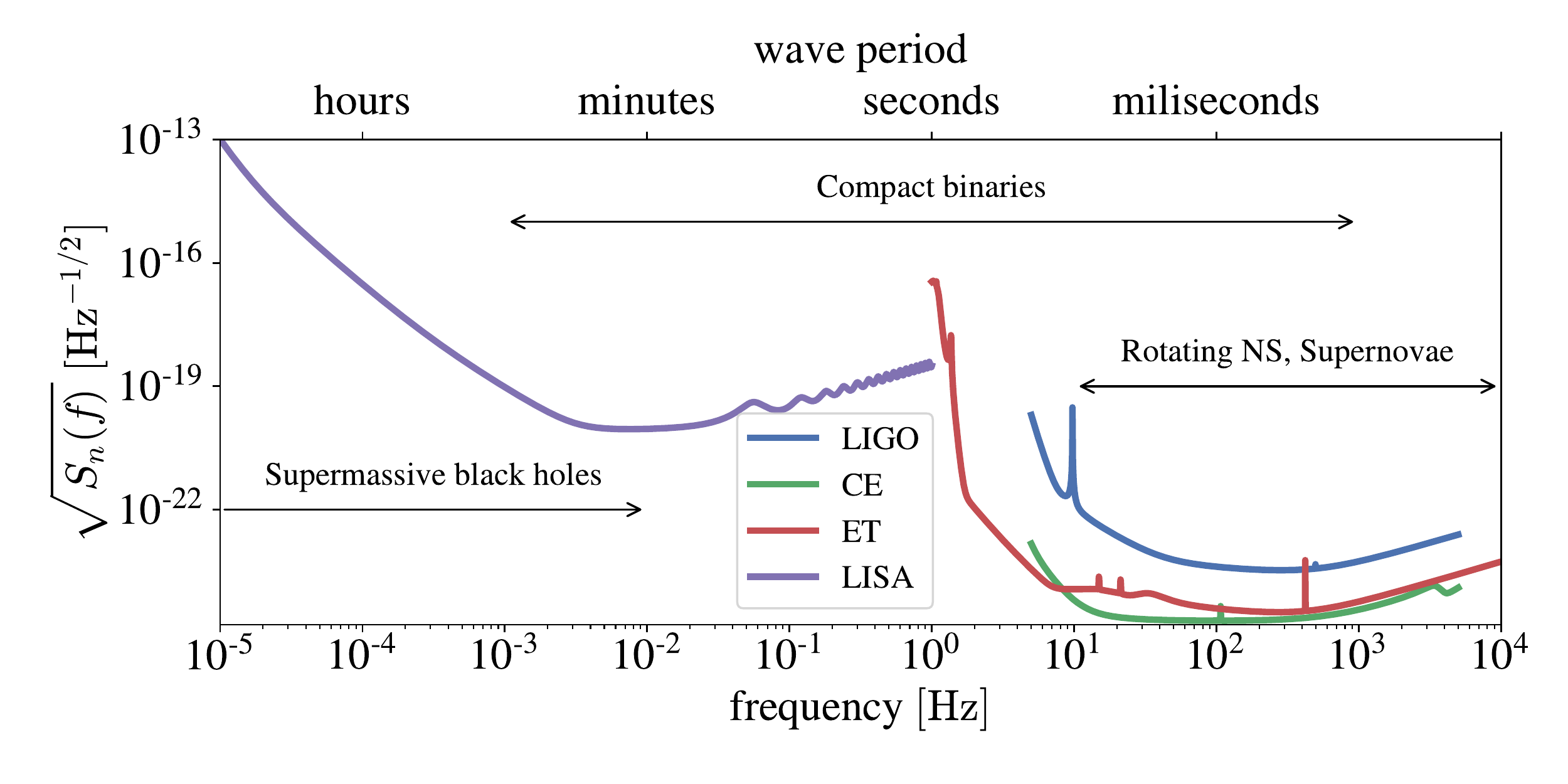}
	\caption{Amplitude spectral density $\sqrt{S_n(f)}$ as a function of the frequency of LIGO (blue), CE (green), ET (red) e LISA (purple).
	The wave period is shown in the upper $x$ axis.
	The gravitational wave sources depend on the frequency bands, the arrows show the approximate frequencies for supermassive black holes, compact binaries, rotating neutron stars and supernovae.
	}
	\label{fig:detectors_Sn}
\end{figure}

Figure~\ref{fig:detectors_Sn} shows the ASD of LIGO at the design sensitivity~\cite{Design-sensitivity}, CE~\cite{CE-psd}, ET~\cite{ET-psd} and LISA\footnote{We are thankful to Quentin Baghi for providing the LISA noise spectral density curve.}.
The performance of each detector is characterized by its PSD,
and they are all limited in low and high frequencies.
The ground based detectors (LIGO, CE and ET) are sensitive to similar frequencies.
The low frequency noise of these detectors is dominated by seismic noise,
which is motion in the interferometer mirrors caused by seismic vibrations, winds, earthquakes, or any ground movement.
The motion produces changes in the density of the ground, which change the gravitational attraction of the ground on the detector.
ET will be more sensitive than LIGO and CE to lower frequencies because it will be build underground,
which reduces the seismic noise.

Another important noise source is quantum noise.
Due to the Heisenberg uncertainty principle, there is a trade-off between shot noise, which is related to photon counting and is a high frequency noise, and radiation pressure noise, which is caused by random motion of the mirrors and is a low frequency noise.
By increasing the laser power, the shot noise will decrease, but the radiation pressure noise increases.
The quantum noise can be reduced by making the interferometer longer and this is one reason why the 3G detectors have lower ADS than LIGO's.

There are several other sources of noise~\cite{Abbott:2016xvh,aLIGO:2020wna}, and the improvement of the detectors is an instrumental challenge.
For example, the Japanese detector KAGRA is sometimes considered a 2.5G detector, as it is the first detector that uses cryogenic technology in the mirrors.

To avoid the seismic noise limitations, LISA will consist of three spacecrafts that form an equilateral triangular detector orbiting around the Sun.
The arms will be 2.5 million kilometers long, which results in a wide low frequency band.
The frequency sensitive band of the detector determines the detectable sources.
LISA is the only detector that will be able to detect supermassive binaries and compact objects captured by supermassive black holes, because the frequency of the wave is inversely proportional to the total mass.
LISA will also detect the inspiral of less massive binaries.

At later times, these binaries may be detected by the ground based detectors~\cite{Sesana:2017vsj}.
These detectors can detect the inspiral, merger and ringdown of stellar mass binaries.
BNSs, rotating neutron stars and supernovae will also be in the ground based detector frequency band.

\section{Detecting gravitational waves: signal}
When a gravitational wave signal $h(t)$ passes through the detector, the output $d(t)$ will be the signal buried in the noise, that is,
\begin{equation}
	d(t) = n(t) + h(t).
\end{equation}
To detect the GW signal we must be able to filter it from the noise.

When the waveform $h(t)$ is known, \emph{matched filtering} is an efficient way to search for GWs.
The matched filtering process consists in maximizing the \emph{signal-to-noise ratio} (SNR) $\rho$, which is defined in terms of the filtered signal $s = \int d(t)K(t) dt$, where $K(t)$ is a filter function.
The SNR is the ratio between the expected value of $s$ when the GW signal is present and the expected value of $s$ when the signal is absent.
The filter $K(t)$ that maximizes the SNR is the Wiener filter~\cite{maggiore-vol1, Sathyaprakash:2009xs, Moore:2014lga}, given by $K(t) \propto \tilde{h}(f)/S_n(f)$, where the tilda denotes the Fourier transform.
Therefore, the SNR is given by
\begin{equation}
	\rho^2 = \langle h | h \rangle = 4 \int_0^\infty\frac{|\tilde{h}(f)|^2}{S_n(f)} df,
	\label{eq:snr}
\end{equation}
where we introduced the noise-weighted inner product
\begin{equation}
	\langle h_1 | h_2 \rangle  \equiv
	2 \int^\infty_0 \frac{\tilde{h}_1^*(f)\tilde{h}_2(f) + \tilde{h}_1(f)\tilde{h}_2^*(f)}{S_n(f)} df.
	\label{eq:inner_product}
\end{equation}

The SNR depends on $h(t)$, which represents the signal we are trying to extract from the data.
As in principle we do not know which signal is in the data, the SNR must be computed for all possible templates we have.
The SNR is then computed in the output of the detector and, when the output matches with the template, the SNR is high and the GW is spotted.

This process is only possible if we have a template for the waveform, which is the case for compact binary mergers.
However, matched filtering cannot be applied to unknown GW signals.
One way to identify unknown signals is by searching for an excess of power in the output~\cite{Sathyaprakash:2009xs}.
This is a suboptimal method and is less efficient than matched filtering.
Although matched filtering is a very efficient way for detecting a signal, it is not the most efficient method to estimate the parameters of the source.

Bayesian inference is an accurate technique for parameter estimation, and it also gives us statistical significance for our data.
In the Bayesian framework we compute the probability of a hypothesis given the data.
The hypothesis is the waveform model $\mathcal{M}$ we choose.
As the model may depend on several parameters, the probability distribution of each parameter determines the probability distribution of the model given the data.
As we are looking for GWs from astrophysical sources, the probability distribution of the parameters can be analyzed to make statements about the source.

Bayes' Theorem states that the posterior probability $p(\vartheta| d, \mathcal{M})$ of the parameters $\vartheta$ from the model $\mathcal{M}$ given the data $d$  is given by
\begin{equation}
	p(\vartheta|d,{\cal M}) = \frac{\mathcal{L}(d|\vartheta,{\cal M})\pi(\vartheta,{\cal M})}{{\cal Z_{\cal M}}},
	\label{eq:posterior}
\end{equation}
where $\pi(\vartheta, {\cal M})$ is the prior distribution for the parameters $\vartheta$, which encodes the previous knowledge about the parameters of the model,
$\mathcal{L}(d|\vartheta, {\cal M})$ is the likelihood of the data $d$ given the model $\mathcal{M}$ with the parameters $\vartheta$ and
${\cal Z_{\cal M}}$ is the evidence of ${\cal M}$, which is a normalization factor, and is given by
\begin{equation}
    \mathcal{Z}_\mathcal{M} = \int \mathcal{L}(d|\vartheta;\mathcal{M})\pi(\vartheta;\mathcal{M}) d\vartheta.
    \label{eq:evidence}
\end{equation}

If we assume a noise model, we can use it to compute the likelihood for the data.
For the detector output with a GW signal, the noise can be written as $n(t) = d(t) - h(t)$.
Assuming a stationary Gaussian noise, the probability distribution of equation~\eqref{eq:prob_noise} can be written in terms of the output data and the waveform model $h(\vartheta,{\cal M})$, resulting in the likelihood~\cite{maggiore-vol1}
\begin{equation}
	\mathcal{L}(s|\vartheta,{\cal M}) \propto e^{-\frac{1}{2}\langle d-h(\vartheta,{\cal M}) | d-h(\vartheta,{\cal M})\rangle}.
	\label{eq:likelihood}
\end{equation}

To compute the posterior probability distribution, the likelihood must be evaluated in the whole parameter space.
A brute force computation may work for simple models but is usually unfeasible, as the computational resources are usually limited.
A common approach to solve this problem is using Markov chain Monte Carlo (MCMC) methods.
These methods consist in choosing arbitrary points in the parameter space that are distant from each other.
Starting at these points, the ``walkers'' move around the parameter space randomly following an algorithm that accepts movements in to regions with high probabilities.
After a chosen number of steps, the walkers' final positions result in a sample for the probability distribution.
The resulting distribution is more precise when more steps are included.
The MCMC methods are really good for estimating the parameters, but they do not compute the evidence.

The nested sampling methods are designed to compute the evidence, and, as a by-product, they also compute samples for the posterior.
These methods consist in a reparameterization of the likelihood such that the evidence becomes a one-dimensional integral,
by considering the function $X$ defined as $dX = \pi(\vartheta; \mathcal{M})d\vartheta$.
Then, a chosen number of ``live'' points are sampled and their likelihood are computed.
A new candidate point is drawn and, if its likelihood is greater than the likelihood of the live point $p_{\textrm{lowest}}$ with the lowest likelihood $\mathcal{L}_{\textrm{lowest}}$, the new point becomes a live point and $p_{\textrm{lowest}}$ is discarded, but stored as a ``dead'' point for the likelihood integral computation.
If the likelihood of the candidate is smaller than $\mathcal{L}_{\textrm{lowest}}$, $p_{\textrm{lowest}}$ is kept as a live point and the candidate point becomes a dead point.
This process is repeated with $\mathcal{L}_{\textrm{lowest} }$ being updated with the lowest likelihood of the live points, until a chosen termination criterion.
The resulting samples are going to be sparse in low likelihood regions and dense in high likelihood regions.
As they sample the whole space, these methods do not require much tuning to find multimodal distributions, unlike MCMC methods that depend highly on the initial position of the walkers.
In this work we will use the nested sampling library \textsc{PyMultiNest}~\cite{Feroz:2008xx, Feroz:2013hea, Buchner:2014nha}.

Parameter estimation using MCMC or nested sampling methods may be faster than the brute force approach, but in some circumstances an approximated result with faster computation times is more convenient.
For a high SNR signal in a stationary Gaussian noise, if the posterior distributions for the parameters are unimodal, the probability distribution can be approximated as a Gaussian centered in the real values for the parameters $\vartheta_{\textrm{real} }$.
Thus, the estimated parameter values will be very close to the real value, $\vartheta = \vartheta_{\textrm{real} } + \delta\vartheta$, as a first order approximation.
The posterior probability then takes the form  $p(\vartheta|d,{\cal M}) \propto \exp \left[-(1/2)\Gamma_{ab}\vartheta^a\vartheta^b \right]$, where $\vartheta^{a,b}$ represents a single parameter, $\Gamma_{ab}$ is the \emph{Fisher matrix}
\begin{equation}
	\Gamma_{ab} \equiv \left(\frac{\partial h}{\partial \vartheta^a}, \frac{\partial h}{\partial \vartheta^b}\right)
	\label{eq:fisher_matrix}
\end{equation}
and the Einstein summation convention notation was used for $a$ and $b$.

The variance of the parameter $\vartheta^a$ is $\sigma_{\vartheta^a} = \sqrt{(\Gamma^{-1})^{aa}}$, where $\Gamma^{-1}$ is the inverse of the Fisher matrix and the repeated indices denote the element of the diagonal~\cite{Finn:1992wt,maggiore-vol1}.
Therefore, for high SNR signals, which are definitely expected for 3G detectors, we can use the Fisher matrix approximation to determine the statistical uncertainties of the parameters.

The methods described above are used to determine the parameters of a model $\mathcal{M}$, assuming the model correctly describes the data.
A good fit of a given model does not guarantee that the model is correct.
Before analysing the output as an astrophysical events, some thresholds are set in the SNR of the signal to avoid false alarms generated by the random noise mimicking events.
The higher the SNR the smaller is the probability that the observed signal is due to noise.
For example, the first detection GW150914 has a network (considers both LIGO detector) SNR $\rho = 23.6$, which is larger than any fake event found in 20300 years of noise only data~\cite{LIGOScientific:2016aoc}.

Another way to determine whether the noise is mimicking an event is doing a hypothesis test.
The Bayes model comparison quantifies the support for a model $\mathcal{M}_A$ over another model $\mathcal{M}_B$.
This is done using Bayes factors, defined as
\begin{equation}
	\mathcal{B}^A_B = \frac{\mathcal{Z}_A}{\mathcal{Z}_B},
	\label{eq:bayes-factor}
\end{equation}
where $\mathcal{Z}_{X}$ is the evidence of the model ${\cal M}_X$, given by equation~\eqref{eq:evidence}.
A model is strongly favored over another when the absolute value of $\ln \mathcal{B}^A_B$ is large; a commonly used value in GW analysis is $\left|\ln\mathcal{B}^A_B\right| = 8$~\cite{Thrane:2018qnx}, that is, the evidence of one model is approximately 3000 greater than the evidence of the other model.

One common test is considering the detection hypothesis as opposed to the noise-only (null) hypothesis.
One example is the detection of the dominant QNM in GW150914.
At 3ms after the peak of amplitude, the Bayes factor of a waveform model $h = n + h_{220}$, where $h_{220}$ is the fundamental quadrupolar QNM and $n$ is the noise, over the noise-only model $h = n$ is $\ln\mathcal{B} \sim 32$~\cite{LIGOScientific:2016lio}.
This gives a high statistical significance for the QNM in the ringdown.

The Bayes factor is also useful to compare different assumptions of our models.
Although an infinite number of QNMs is expected to be in the BBH waveform, a statistical evidence is necessary to confirm that the fitted parameters of all QNMs are actually physical QNMs parameters and not just overfitting.
Bayes factors penalize complex models.
That is, if a very complex model does not significantly increase the evidence compared with a simpler model, the Bayes factor will ``select'' the simpler model.
Therefore, one can use Bayes factors to compare models containing different numbers of modes in the waveform and attach statistical significance to those models, as we show in Chapter~\ref{ch:spectroscopy-horizon}.

\begin{center}
\myclearpage
\par\end{center}


\chapter{Ringdown: quasinormal modes characterization}
\label{ch:ringdown-qnm}
We saw in Section~\ref{sec:qnm-bbh} that the detection of two or more QNMs allow us to perform black hole spectroscopy and test the no-hair theorem.
The dominant $(2,2,0)$ mode was already observed in the ringdown of GW detections~\cite{LIGOScientific:2016lio, LIGOScientific:2020ufj, LIGOScientific:2020tif, LIGOScientific:2021sio}, but a confident detection of a secondary mode is needed to perform black hole spectroscopy.
Before trying to access the detectability of subdominant modes, it is important to have the correct model that describes the QNMs in the ringdown of a BBH\@.
Although it is clear from simulations that the fundamental QNM is compatible with the ringdown~\cite{Seidel_2004, Buonanno:2006ui, Cotesta:2018fcv}, there are some open questions concerning whether and when the non-linear behaviour of the merger can be neglected.

The initial time of the QNMs is still unknown.
There are several works that assess the initial time of the ringdown~\cite{Dorband:2006gg, Berti:2007fi, Thrane:2017lqn, Carullo:2018sfu}, but most of them consider only the dominant mode in the analysis.
Equation~\eqref{eq:qnm_harmonic} tells us that each harmonic mode contains a sum of an infinite number of overtones.
As the overtones decay very fast, most BH spectroscopy analysis neglect their influence~\cite{Cotesta:2018fcv,Kamaretsos:2011um, Kelly:2012nd,Shi:2019hqa, Thrane:2017lqn, Maselli:2019mjd, Baibhav:2018rfk, Baibhav:2020tma}.

Nevertheless, this approximation may be overly simplified, as the importance of overtones has been known for decades~\cite{Leaver:1986gd, Stark:1985da}.
In the context of BBHs, in~\cite{Buonanno:2006ui} the authors show that the properties of the remnant BH can be obtained at early initial times when the overtones are considered in the ringdown modeling: the more overtones that are considered, the earlier the initial time.
It was shown by~\cite{London:2014cma} that the addition of overtones increases the SNR of the signal.
Moreover, it was pointed out~\cite{Baibhav:2017jhs} that the inclusion of overtones decreases the errors in the determination of the BH parameters.
More recently, following the previous results, in~\cite{Giesler:2019uxc} the authors suggested that the linear regime of the ringdown can start as early as the time of the peak of the amplitude of the waveform, when seven overtones are taken into account.
Following the increased interest in BH spectroscopy, the contribution of overtones was further studied~\cite{Bhagwat:2019dtm, Ota:2019bzl, Forteza:2020cve, Mourier:2020mwa, Okounkova:2020vwu, Bustillo:2020buq, Finch:2021iip, Forteza:2021wfq}.

In 2020, Jaramillo et al.~\cite{Jaramillo:2020tuu,Jaramillo:2021tmt} studied the stability of QNMs and found instability of the overtones under small-scale perturbations in the potential.
In a following analysis~\cite{Cheung:2021bol}, Cheung et al.\ found that even the fundamental mode can be destabilized under generic perturbations.
These results have already been noticed by Aguirregabiria and Vishveshwara~\cite{Aguirregabiria:1996zy, Vishveshwara:1996jgz} in 1996, but a detailed analysis was not presented.
The effect of small perturbations in the BBH ringdown and the potential detectability of this effect are not known and further studies are needed.
Therefore, this effect is not considered in our analyses.

The detectability of QNMs is highly dependent on the model considered, as the most robust statistical methods are model dependent.
The number of overtones considered in the model will affect the initial time and the amplitudes of the model.
Each overtone adds at least two (amplitude and phase) free parameters to the model, and such complex models are easily susceptible to overfitting.
Therefore, our goal in this chapter is to assess the contribution of a single overtone of the quadrupolar mode  in the ringdown and compare its contribution with the most relevant fundamental higher harmonics.

\section{Fundamental quasinormal mode}
\label{sec:fundamental-quasinormal-mode}
The first step in our modelling process is to determine when the ringdown of a BBH is compatible with the fundamental QNM\@.
As we stated earlier, the contributions of the overtones should not be neglected in the ringdown model, but they indeed can be neglected at late times, as the overtones decay much faster than the fundamental mode.
As the QNMs have fixed decay time and frequency of oscillation, the results can readily be extended to earlier times, when the overtones are relevant.

In this and the following section we will use the numerical simulation  SXS:BBH:0305, which is a simulation consistent with the first GW detection GW150914~\cite{LIGOScientific:2016aoc}.
It is the simulation of a circular BBH merger with mass ratio $q \equiv m_1/m_2 = 1.2$, initial dimensionless spins $\vec{a}_1 = (0,0,0.33)M$ and $\vec{a}_2 = (0,0,-0.44)M$, where $M = m_1 + m_2$ is the total mass of the binary.
The mass and spin of the remnant extracted from the apparent horizon are $M_f^{\mathrm{NR}} = 0.9520 M$ and $a^{\mathrm{NR}} = 0.6589 M$, respectively.
The decay time, computed from the mass and spin, of the fundamental mode and the first overtone are $\tau_{220} = 1/\omega^i_{220} = 11.8M$ and $\tau_{221} = 1/\omega^i_{221} = 3.9M$, respectively (see Table~\ref{tab:freq_n0_n1}), that is, the first overtone decays approximately three times faster than the fundamental mode.

When the waveform $h_{\ell m}$ of equation~\eqref{eq:qnm_harmonic} is equivalent to the fundamental mode, that is,
$h_{\ell m} = h_{\ell m 0}$, the time derivative $\dot{\theta}_{\ell m}$ of the complex phase, defined as
\begin{equation}
	\theta_{\ell m} \equiv \arctan\left[\frac{\Im(h_{\ell m})}{\Re(h_{\ell m})}\right],
	\label{eq:complex-phase}
\end{equation}
is equal to the real part of the complex frequency $\omega^r_{\ell m 0}$.

Assuming that all the modes are excited simultaneously (or at close times), there is a time $t_{n=0}$ at which all the overtone contributions can be neglected, that is, at this time the overtones have decayed $A_{\ell m n}(t \geq t_{n=0}) \to 0$, for $n>0$, and the derivative of the complex phase will be fully characterized by the real frequency of the fundamental mode  $\dot{\theta}_{\ell m}(t\geq t_{n=0}) \to  \omega^r_{\ell m 0}$.
At times earlier than $t_{n=0}$, the overtones are relevant and $\dot{\theta}_{\ell m}$  is not constant.

\begin{figure}[htb!]
	\centering
	\includegraphics[width = 0.49\linewidth]{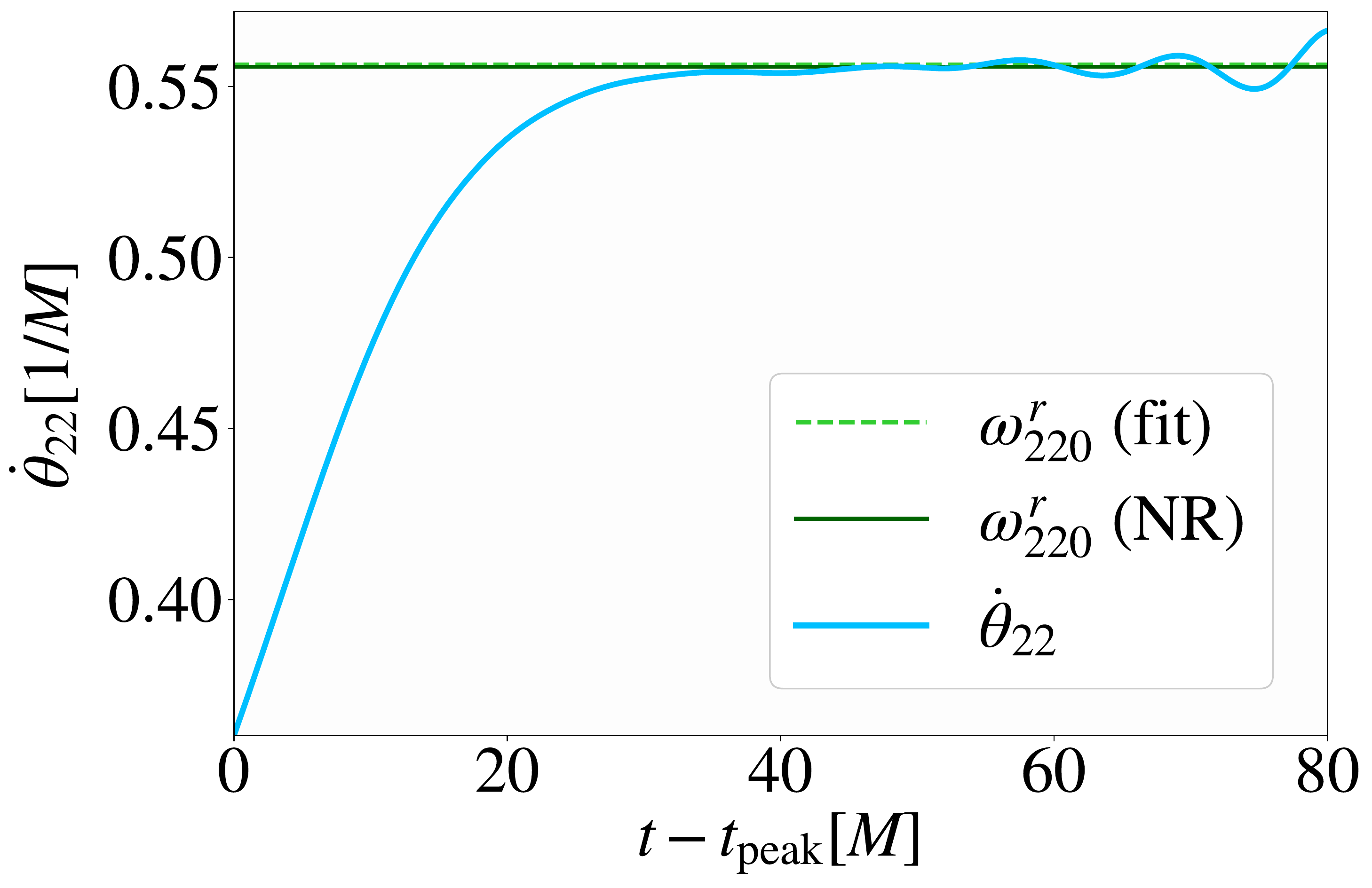}
	\includegraphics[width = 0.49\linewidth]{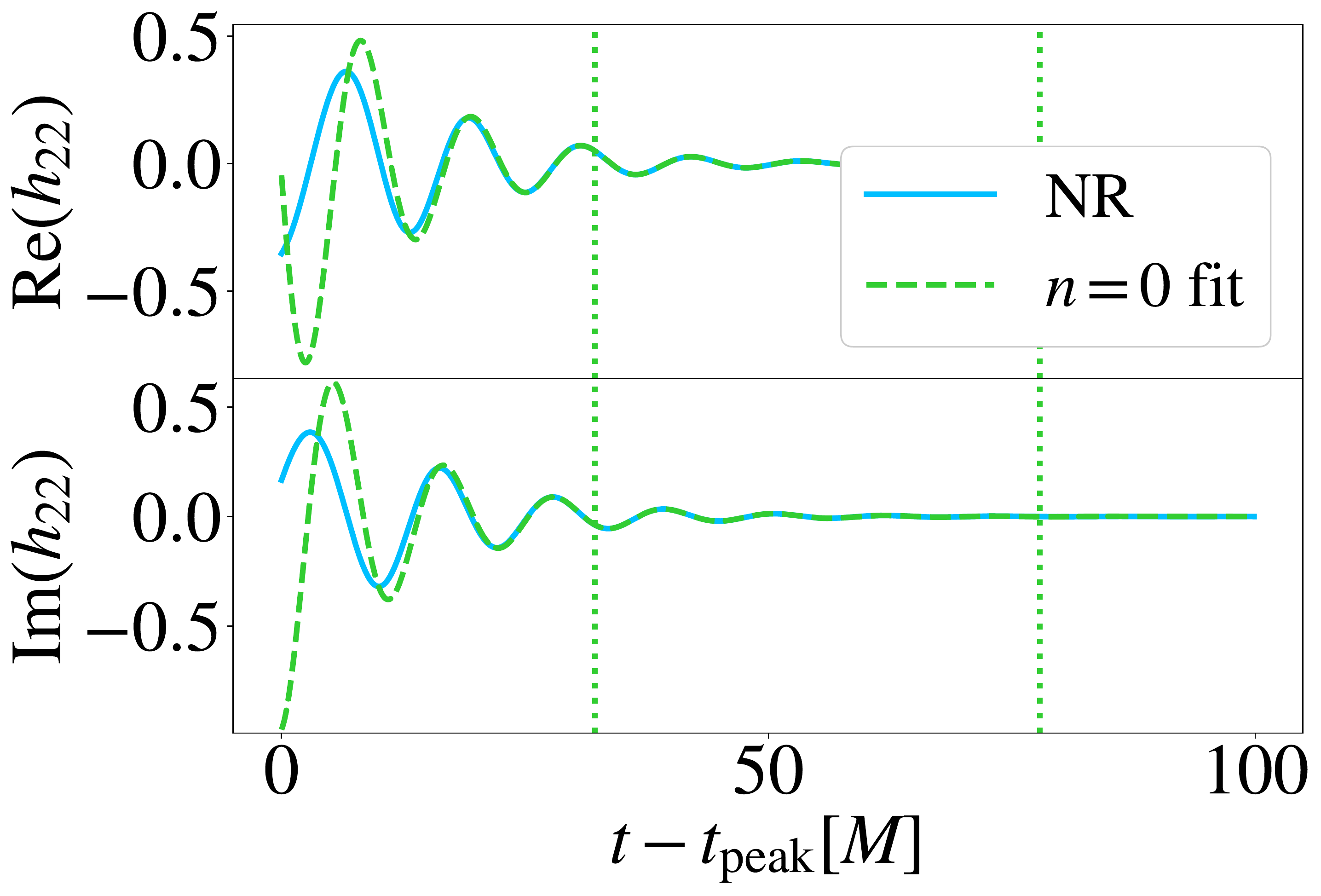}
	\caption[width = \textwidth]{\emph{Left:} time derivative of the complex phase $\theta_{22}$ of the ringdown of the BBH simulation SXS:BBH:0305 as a function of time.
	The horizontal lines indicate the real frequency of the fundamental mode $\omega_{220}^r$ fitted to the waveform on the right plot (dashed) and obtained from the mass and spin computed from the apparent horizon (solid).
	The oscillations at  $t-t_\mathrm{peak} > 75 M$ are late time numerical errors.
	\emph{Right:} real and imaginary parts ($+$ and $\times$ polarizations) of the simulated waveform $h_{22}$ (solid) and the fitted fundamental mode  waveform $h_{220}$ (dashed). The vertical lines indicate the initial and final time of the interval considered in the fit.}
	\label{fig:fit_n0}
\end{figure}

Figure~\ref{fig:fit_n0} on the left shows $\dot{\theta}_{22}(t \geq t_{\mathrm{peak}})$ for the simulation SXS:BBH:0305, where $t_{\mathrm{peak}}$ is the time of the peak of the amplitude of the $(2,2)$ waveform.
We can see that $\dot{\theta}_{22}$ is approximately constant in the interval $30 M \lesssim t-t_\mathrm{peak} \lesssim 75 M$ (with relative variations smaller than 1\%).
In this interval contributions of the overtones and the non-linear behaviour of the merger have already damped and the numerical errors of the simulation at later times can still be neglected.
Therefore, the waveform can be well described by just the fundamental mode.

In this interval, we fit to the simulation the waveform of the fundamental mode
\[
h_{220} = A_{0} e^{-\omega_{220}^i t}\left[\cos(\omega_{220}^r t -\phi_0) + i \sin(\omega_{220}^r t -\phi_0)\right]
\]
where $A_0$, $\phi_0$, $\omega_{220}^r$ and $\omega_{220}^i$ are all fitting parameters.
Figure~\ref{fig:fit_n0} on the right shows the simulated and fitted waveforms.
The fit is very good at the considered interval, but it is clear that the fundamental mode alone does not describe the waveform at times close to the peak of amplitude.
The fitted frequencies are $M\omega_{220}^r = 0.5549$ and $M\omega_{220}^i = 0.0848$, and the BH mass and spin computed from these frequencies
using equations~\eqref{eq:mass_spin_qnm} are $M_f = 0.9553 M$ and $a_f = 0.6632 M$ respectively.
These values differ by 0.3\% and 0.7\% from the values for the mass $M_f^{\mathrm{NR}}$ and the spin $a^{\mathrm{NR}}$ obtained from the apparent horizon of the simulation.
Thus, in the time interval defined by the derivative of the complex phase, the fundamental mode describes well the waveform.

The dashed line in the  $\dot{\theta}_{22}$ plot (left) represents the fitted value for $\omega_{220}^r$, and it is just 0.2\% different from the real frequency value obtained from $M_f^{\mathrm{NR}}$ and $a^{\mathrm{NR}}$, which is shown as a solid line in the plot.
It is clear that $\dot{\theta}_{22}$ reaches  $\omega_{220}^r$ after $t_{n=0} = 30 M$.
Consequently, the assumption that the overtones can be neglected at late times is valid.

\begin{figure}[htb!]
	\centering
	\includegraphics[width = 0.32\linewidth]{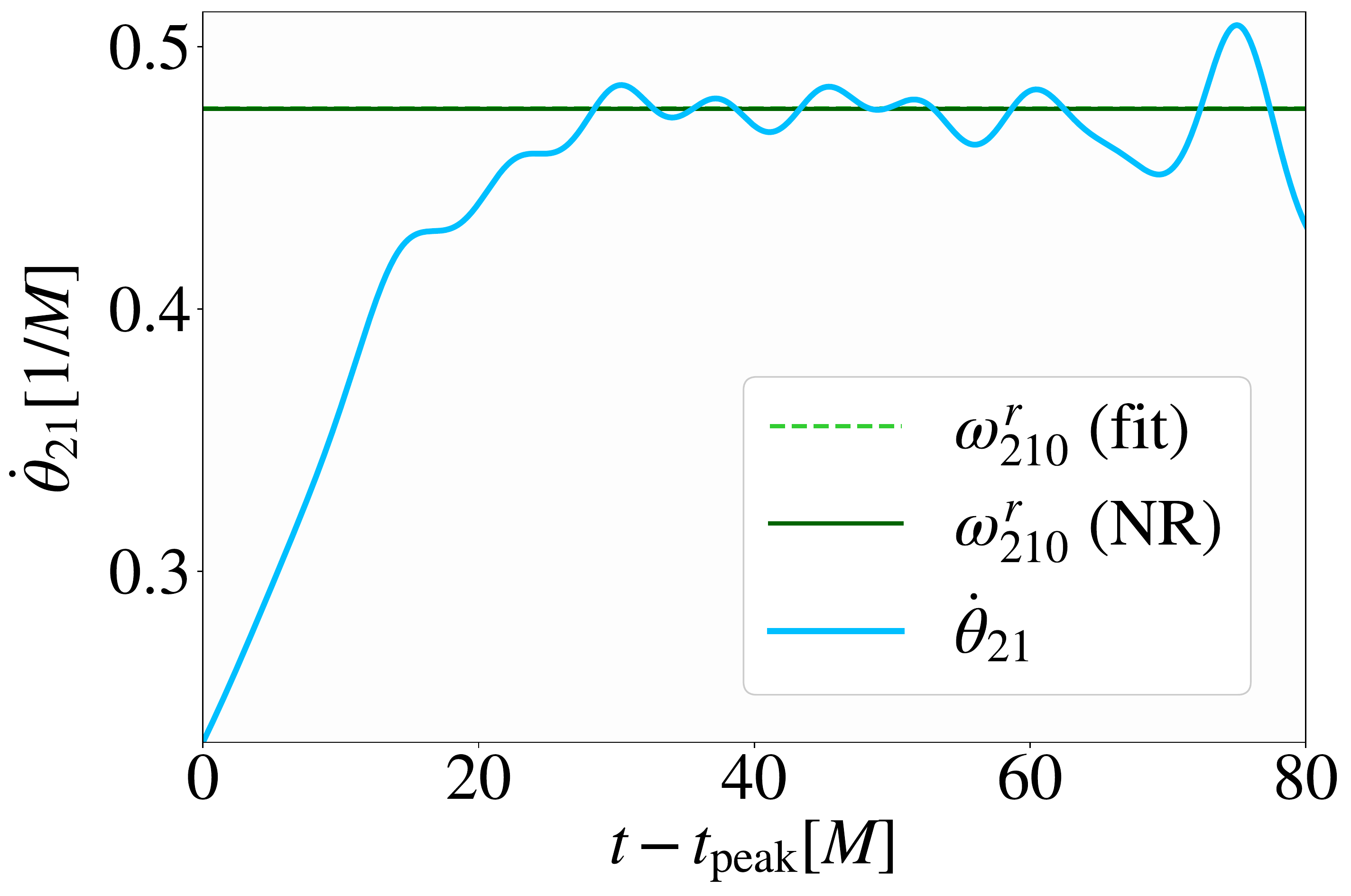}
	\includegraphics[width = 0.32\linewidth]{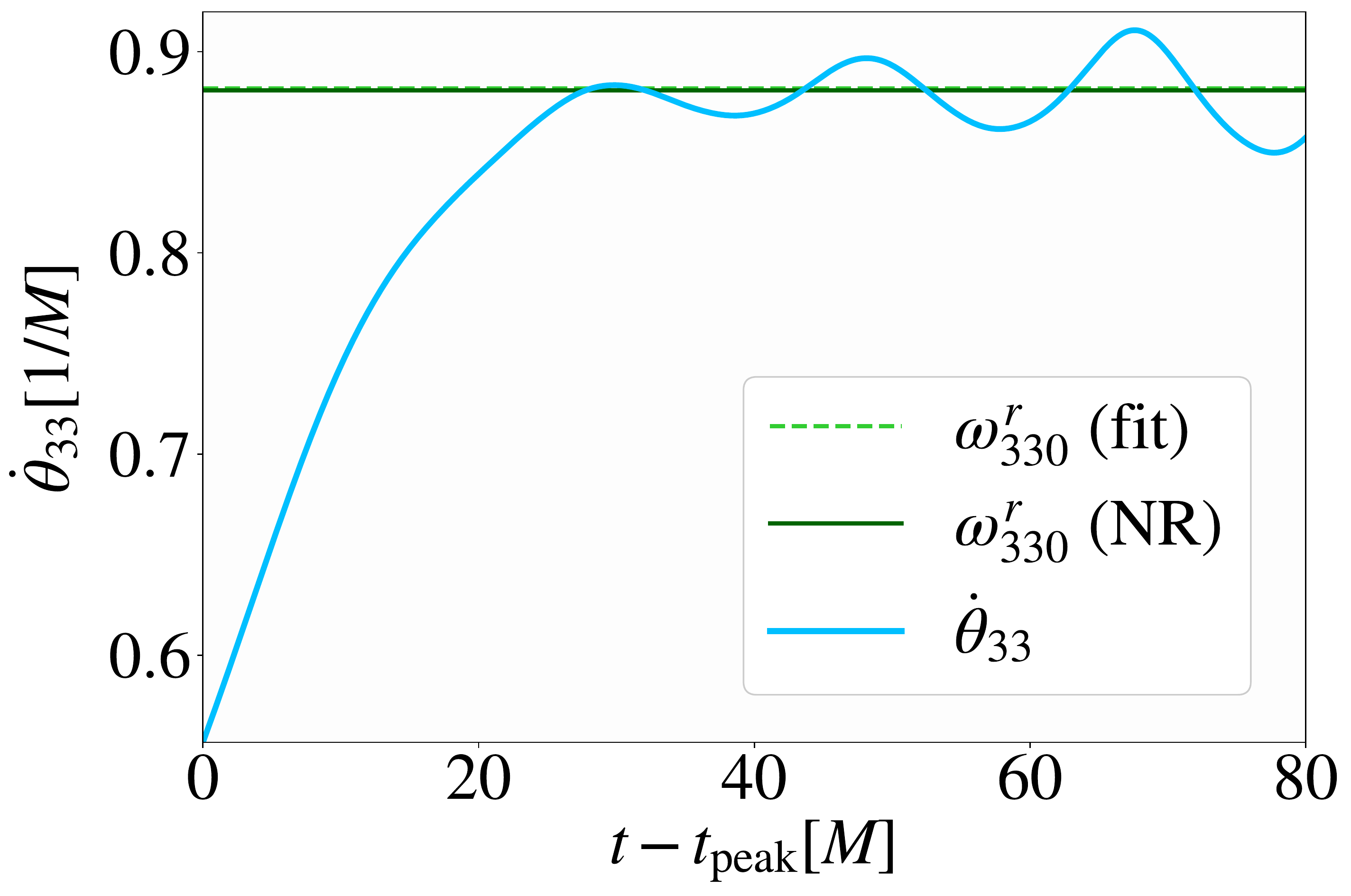}
	\includegraphics[width = 0.32\linewidth]{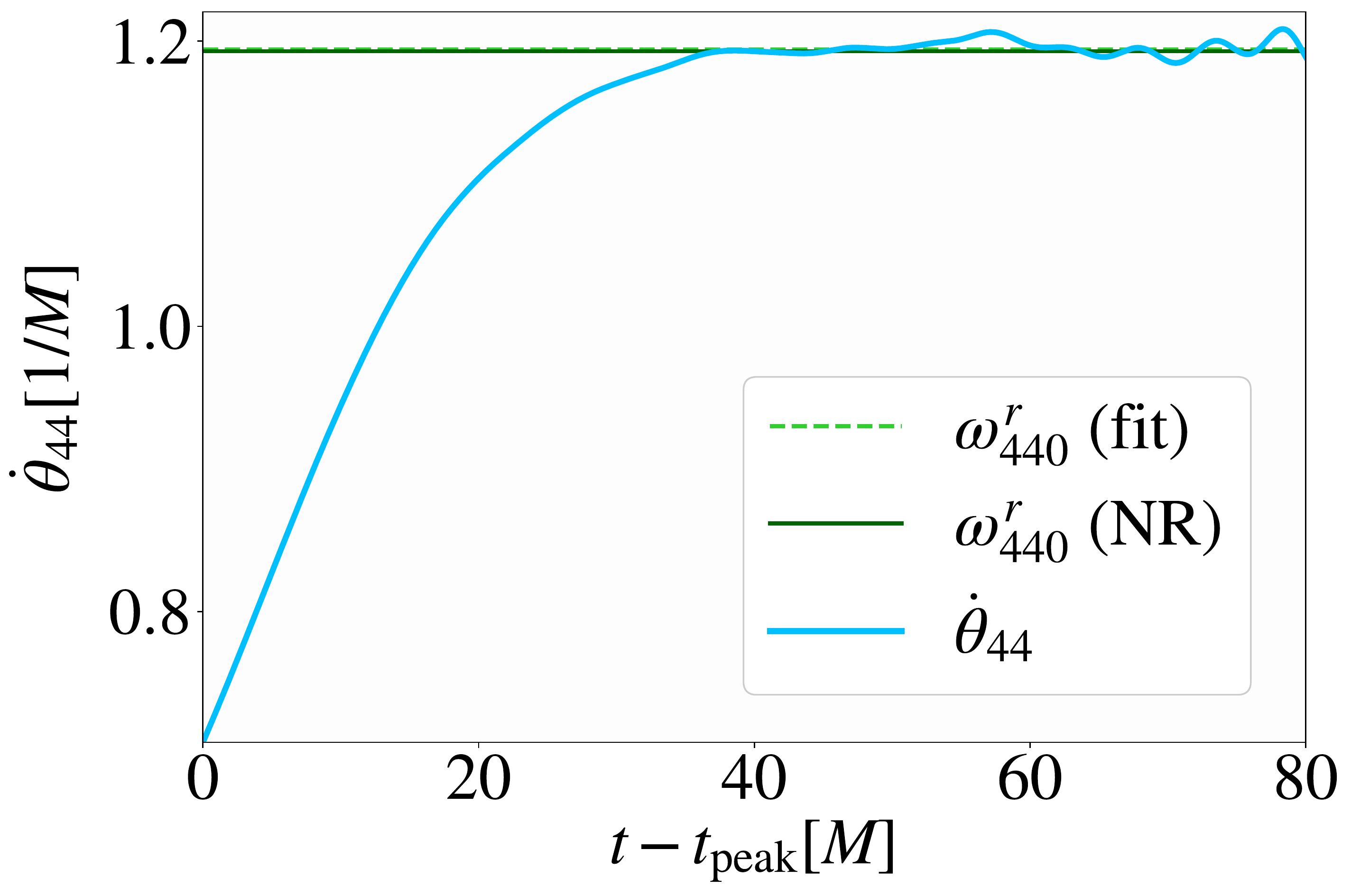}
	\includegraphics[width = 0.32\linewidth]{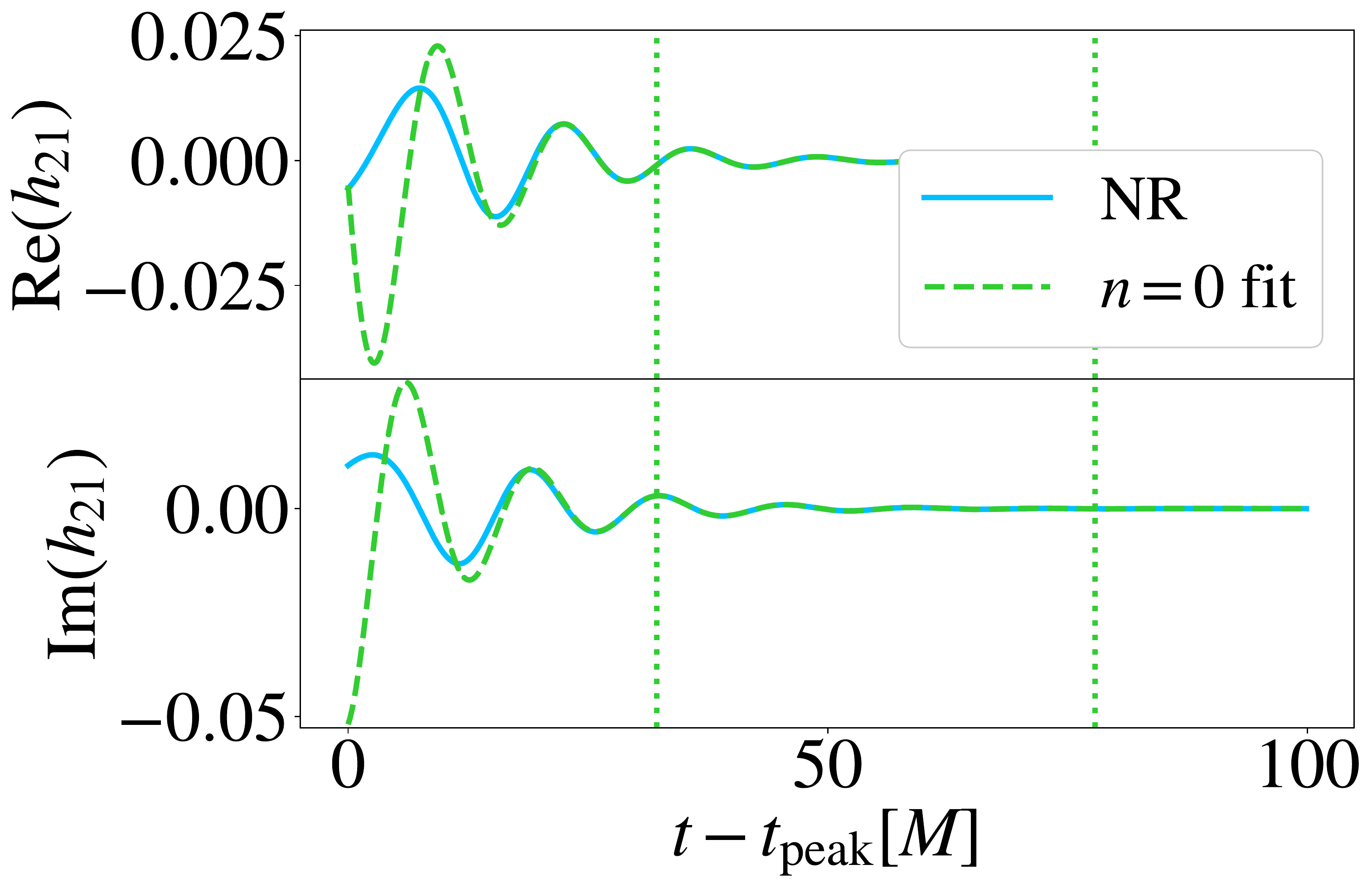}
	\includegraphics[width = 0.32\linewidth]{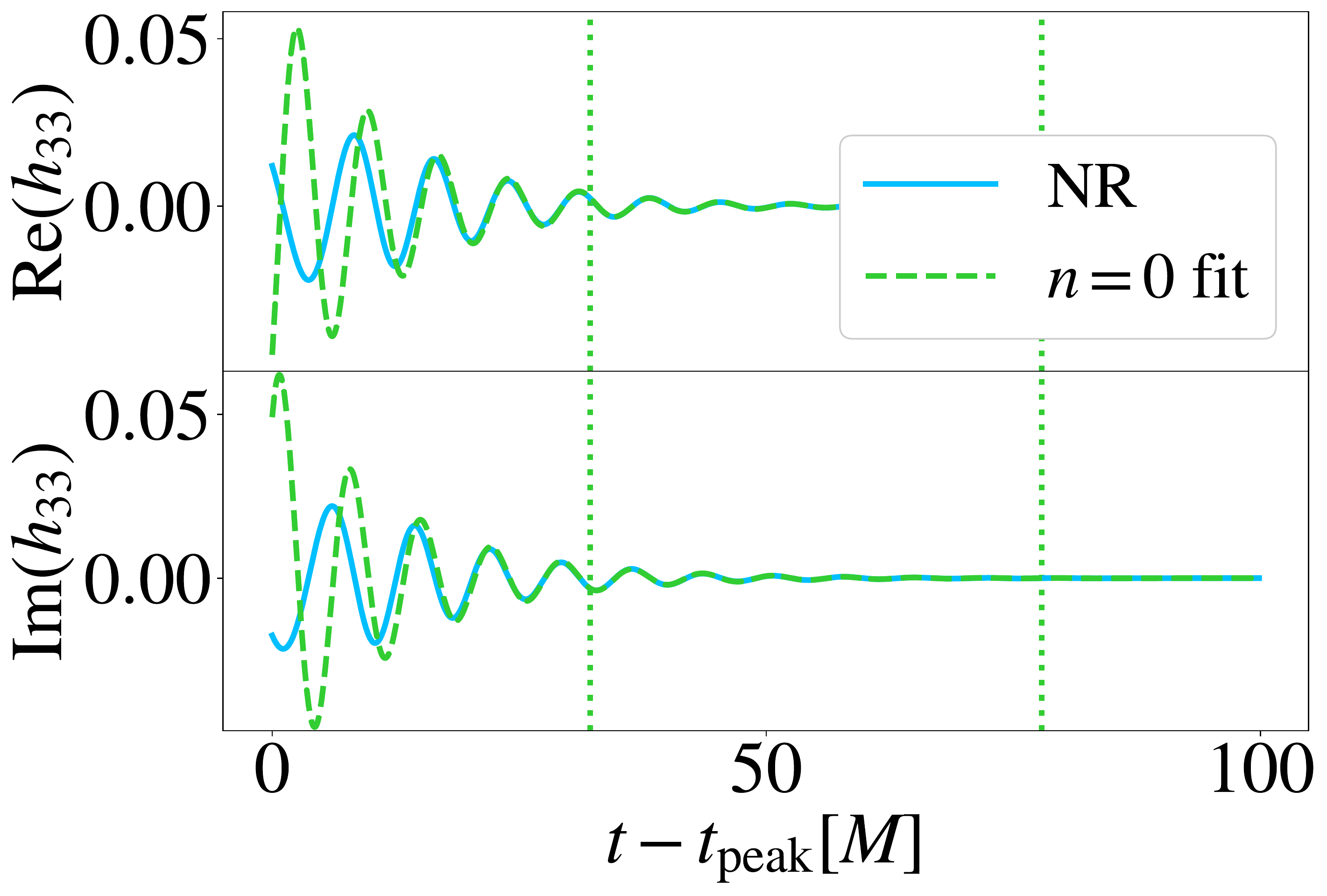}
	\includegraphics[width = 0.32\linewidth]{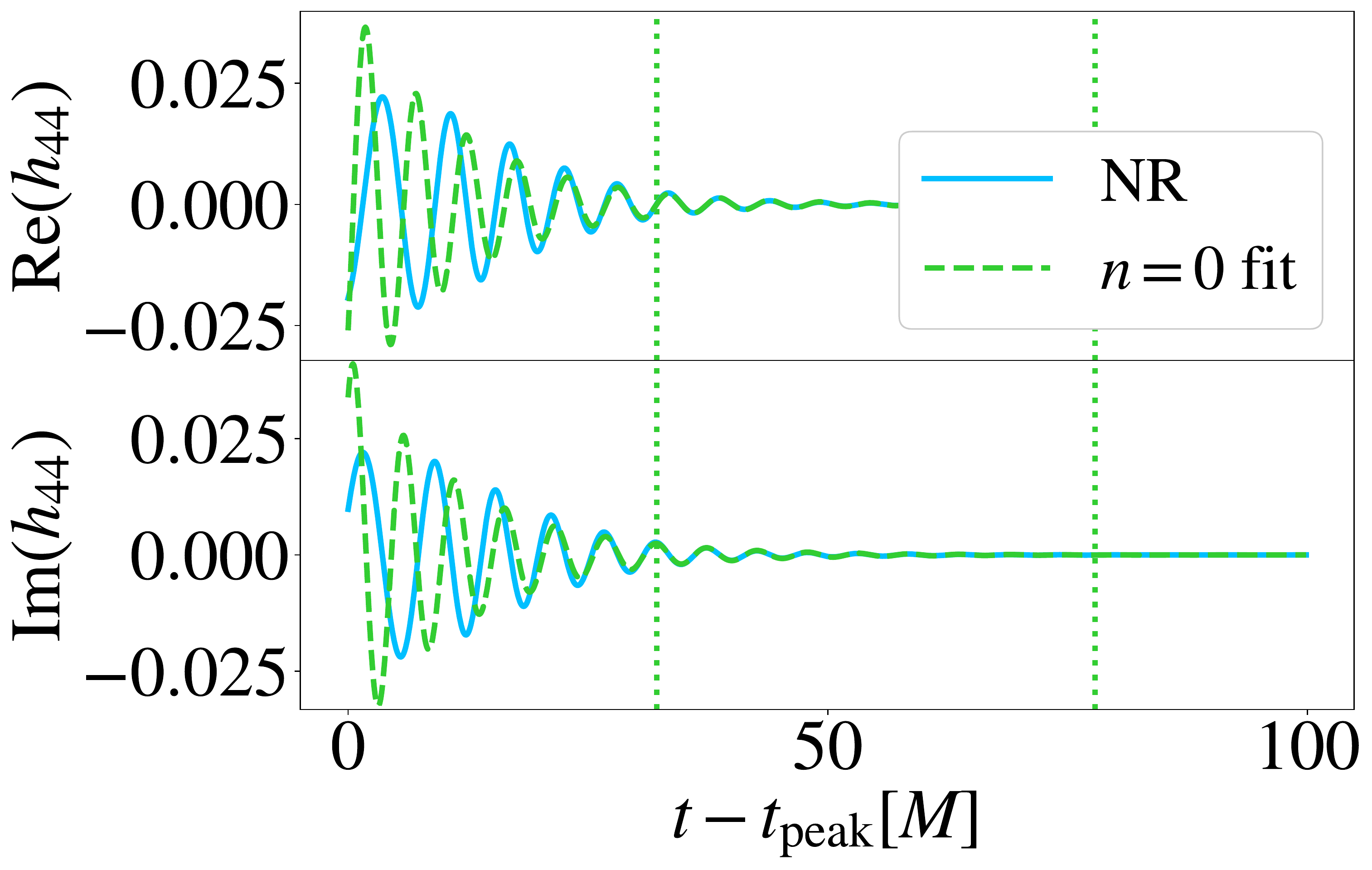}
	\caption{Same as Figure~\ref{fig:fit_n0}, but for the harmonic modes $(2,1)$, $(3,3)$ and $(4,4)$.}
	\label{fig:fit-n0-harmonics}
\end{figure}

The same analysis can be done for the higher harmonics.
Figure~\ref{fig:fit-n0-harmonics} shows the results for the harmonic modes $(2,1)$, $(3,3)$ and $(4,4)$.
The real frequencies $\omega_{210}^r$, $\omega_{330}^r$ and $\omega_{440}^r$ fitted to the waveform are in agreement with the correspondent values obtained from $M_f^{\mathrm{NR}}$ and $a^{\mathrm{NR}}$, within a 0.2\% difference.
Similar to the quadrupolar mode case, $\dot{\theta}_{\ell m}$ rises from lower values towards $\omega_{\ell m 0}^r$, when the overtones and non-linear behavior have been damped, and stays approximately constant until the numerical errors dominates the signal.
The oscillations in $\dot{\theta}_{\ell m}$ can be associated with numerical errors.

\section{First overtone}
\label{sec:overtones}

It is clear from Figures~\ref{fig:fit_n0} and~\ref{fig:fit-n0-harmonics} that the waveform immediately after the peak of amplitude is not fully described by the fundamental quasinormal mode, which would be true if  $\dot{\theta}_{\ell m}(t \geq t_{\mathrm{peak}}) = \omega_{\ell m 0}^r$.
The waveform near the peak differs from the fundamental mode because of the contribution of overtones or the non-linear behaviour caused by the merger.
In~\cite{Giesler:2019uxc}, by including seven overtones in their model, the authors suggest that, after the peak of amplitude, the waveform is fully described by quasinormal modes, which would imply that the non-linear behaviour should end before the peak of amplitude.
More recently, an extensive analysis using many NR simulations and with several overtones found substantial instabilities on the amplitudes of overtones with $n>1$, suggesting that the fits with a large number of overtones are non-physical~\cite{Forteza:2021wfq}.

Moreover, the confident identification of a secondary mode in the ringdown of the detected signals is still being debated~\cite{LIGOScientific:2020ufj, LIGOScientific:2020tif,LIGOScientific:2021sio,Carullo:2019flw, Isi:2019aib, Capano:2021etf, Cotesta:2022pci} (see also Chapter~\ref{ch:spectroscopy-horizon}).
Therefore, in our work we consider the contribution of just one overtone, which avoids non-physical fits to the early time non-linear behaviour of the waveform which could occur in a model with too many free parameters.
As in the previous section, we consider a model containing the fundamental mode and the first overtone as an approximation for sufficient late times when the other overtones and possible non-linear behaviour can be neglected.
In this analysis, we do not consider the frequencies as free parameters, as it was already determined by the fundamental mode analysis that the quasinormal mode is compatible with the ringdown, and the values for these frequencies are given by Table~\ref{tab:freq_n0_n1}.
With the fitted model, we determine the time interval at which the waveform is well described by the fundamental mode and the first overtone.
We stress that this will not necessarily give the initial time at which the linear regime is valid, as the higher overtones ($n \geq 2$) are not taken into account.

\begin{table}[htb!]
\centering
\caption{Complex frequencies for the fundamental mode and first overtone of the quadrupolar mode $(\ell,m) = (2,2)$ for a remnant BH with mass $M_f = 0.9553M$ and dimensionless spin $a = 0.6632M$.}
\begin{tabular*}{\textwidth}{l@{\extracolsep{\fill}} l l}
 \hline\hline
$n$ & $M$$\omega^r_{22n}$ & $M$$\omega^i_{22n}$ \\ \hline
0 & 0.5549 & 0.0848 \\
1 & 0.5427 & 0.2564 \\  \hline\hline
\end{tabular*}
\label{tab:freq_n0_n1}
\end{table}

To compute the initial time $t_0$ of this interval, we use two fitting functions, whose final results will be compared to guarantee consistency.
The choice of two different fitting function avoids overfitting and misleading interpretation of the results.
First we fit to the numerical waveform $h_{22}$ the 4-parameter function
\begin{align}
    h_{22}(t) &=
    A_{220} e^{-\omega_{220}^i t}\left[\cos(\omega_{220}^r t -\phi_{220}) + i \sin(\omega_{220}^r t -\phi_{220})\right] \nonumber\\
    &+ A_{221} e^{-\omega_{221}^i t}\left[\cos(\omega_{221}^r t -\phi_{221}) + i \sin(\omega_{221}^r t -\phi_{221})\right],
    \label{eq:h_fund_overtone}
\end{align}
where $\omega^{(r,i)}_{22n}$ are given by Table~\ref{tab:freq_n0_n1}, and the amplitudes $A_{22n}$ and phases $\phi_{22n}$, for each mode $(n = 0,1)$, are free fitting parameters.
The second fit is done to the complex phase $\dot{\theta}_{22}$ using the 2-parameter function
\begin{align}
    \dot{\theta}_{22}(t) &= \left\{ \omega_{220}^r + R^2 e^{2(\omega_{220}^i - \omega_{221}^i)t} \omega_{221}^r + R e^{(\omega_{220}^i - \omega_{221}^i)t}\right.\nonumber\\
    &\times\left.\left[\left(\omega_{220}^r + \omega_{221}^r\right)\cos((\omega_{220}^r-\omega_{221}^r)t -\phi) +\, (\omega_{221}^i - \omega_{220}^i)\sin((\omega_{220}^r-\omega_{221}^r)t -\phi)\right]\vphantom{ R^2 e^{2(\omega_{220}^i - \omega_{221}^i)t}}\right\}\nonumber\\
    &\times\left[2 R e^{(\omega_{220}^i - \omega_{221}^i)t}\cos((\omega_{220}^r-\omega_{221}^r)t -\phi) + R^2 e^{2 (\omega_{220}^i- \omega_{221}^i) t} + 1 \right]^{-1},
    \label{eq:complex-phase-overtone}
\end{align}
where the amplitude ratio $R \equiv A_{221}/A_{220}$ and the phase difference $\phi \equiv \phi_{220} - \phi_{221}$ are free fitting parameters.
This is just the analytical form of the derivative of equation~\eqref{eq:complex-phase} when $h_{\ell m}$ is substituted by equation~\eqref{eq:h_fund_overtone}.
As $\omega_{221}^i > \omega_{220}^i$, $e^{(\omega_{220}^i - \omega_{221}^i)t} \to 0$ and $\dot{\theta}_{22} \to \omega_{220}^r$ when $t\to \infty$.

The initial time $t_0$ is \emph{not} a free fitting parameter.
We will select $t_0$ by minimizing the mismatch $\mathcal{M}$ between the simulated data $f_{\mathrm{NR}}$ and the fitted function $f_{\mathrm{fit}}$, where $f_{\mathrm{NR, fit}}$ represent the waveform $h_{22}$ or the derivative of the complex phase $\dot{\theta}_{22}$.
The mismatch is defined as
\begin{equation}
\mathcal{M} = 1 - \frac{\langle f_{\mathrm{NR}}, f_{\mathrm{fit}} \rangle}{\sqrt{\langle f_{\mathrm{NR}}, f_{\mathrm{NR}} \rangle\langle f_{\mathrm{fit}}, f_{\mathrm{fit}} \rangle}}.
\label{eq:mismatch}
\end{equation}

Following~\cite{8marcel-grossman}, the inner product can be defined in the usual form:
\begin{equation}
	\langle f_1, f_2 \rangle_{\rm standard} \equiv \abs{\int_{t_0} f_1^* f_2 dt},
	\label{eq:inner1}
\end{equation}
where the star denotes the complex conjugate.
The above inner product presents a problem when it is used to compute the energy of each quasinormal mode, as the sum of the energy of each mode would be different from the total energy of the wave.
This happens because QNMs are not orthogonal and complete with respect to the product defined above (or any other).
To avoid this problem, Nollert~\cite{8marcel-grossman,Nollert:1999ji} suggested the following inner product
\begin{equation}
	\langle f_1, f_2 \rangle_{\rm energy} \equiv \abs{\int_{t_0} (\dot{f}_1)^* \dot{f_2} dt},
	\label{eq:inner2}
\end{equation}
where the dot denotes time derivatives.
To compute the mismatch, we use both inner products defined above.

We consider the mismatch as a function of $t_0$ and the fits are performed starting at each $t_0$ and ending at the final time obtained in section~\ref{sec:fundamental-quasinormal-mode}, $t_\mathrm{final} = 75M$.
This procedure is similar to the one used in~\cite{Giesler:2019uxc}, but there are many other proposed methods for the determination of the initial time~\cite{Thrane:2017lqn,Dorband:2006gg,Berti:2007fi}, for instance, by minimizing the residuals of the fits.

\begin{figure}[htb!]
	\centering
	\includegraphics[width = 0.7\linewidth]{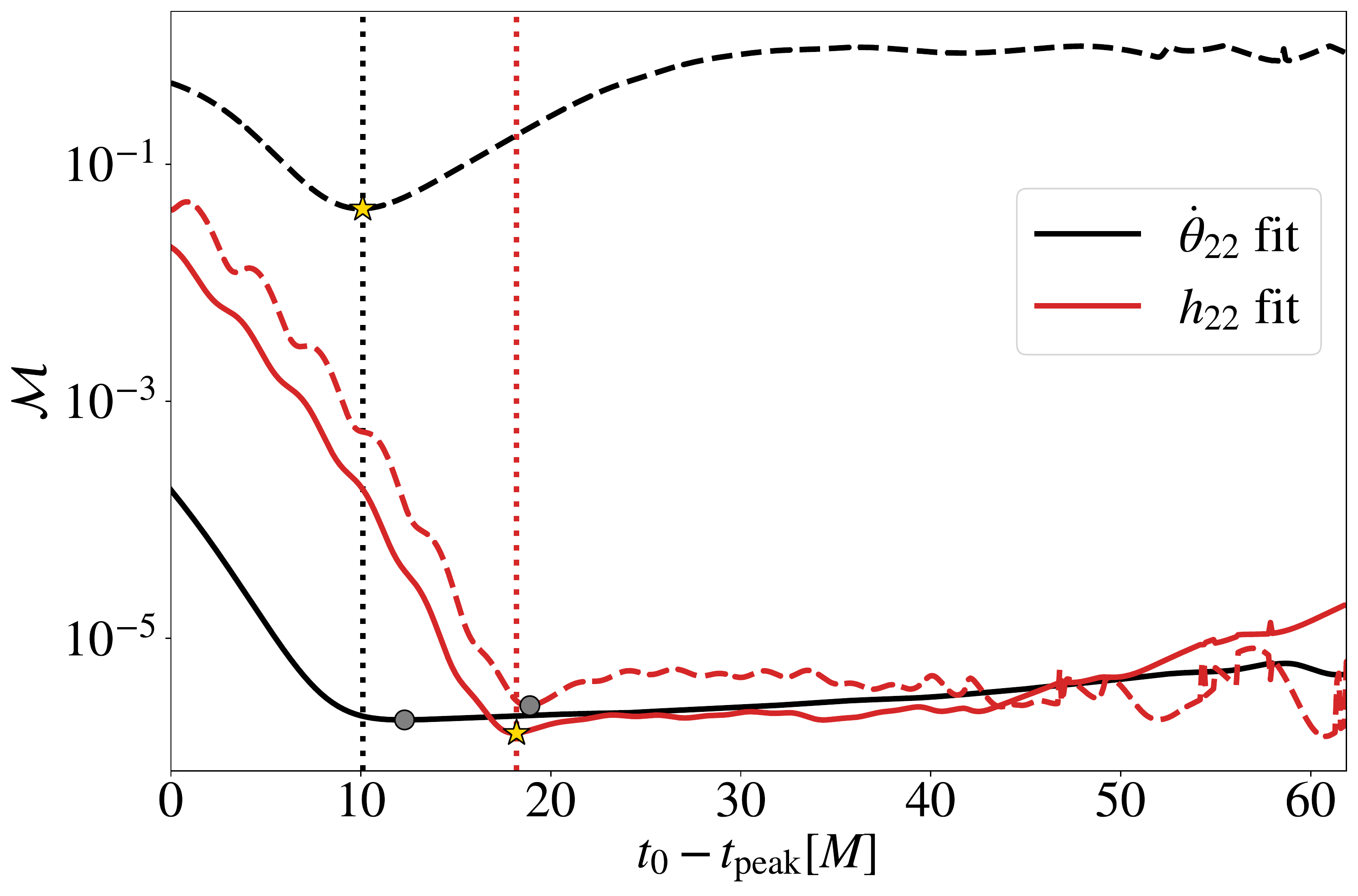}
	\caption{Mismatch~\eqref{eq:mismatch} between the numerical simulation SXS:BBH:0305 and the ringdown models with two modes (fundamental mode plus first overtone) for the quadrupolar mode as a function of time.
	The mismatch is computed using the waveform $h_{22}$~\eqref{eq:h_fund_overtone} (red) and the time derivative of the complex phase $\dot{\theta}_{22}$~\eqref{eq:complex-phase} (black).
	In the solid curves the inner product is computed using equation~\eqref{eq:inner1} and in the dashed curves the inner product is given by~\eqref{eq:inner2}.
	The scatter points indicate the chosen best initial time $t_0$ for each case.
	The gray circles are not considered for the analysis and the yellow stars indicate the methods chosen:  $\dot{\theta}_{22}$ (method I) and $h_{22}$ (method II).}
	\label{fig:mismatch}
\end{figure}

Figure~\ref{fig:mismatch} shows the mismatch between the numerical simulation SXS:BBH:0305 and the two-mode waveform $h_{22}$ (red), given by equation~\eqref{eq:h_fund_overtone}, and the time derivative of the complex phase (black), given by equation~\eqref{eq:complex-phase-overtone}.
The solid lines indicate that the inner product was computed by equation~\eqref{eq:inner1} and in the dashed lines the inner product is given by equation~\eqref{eq:inner2}.
We choose the time $t_0$ as the first minimum of the mismatch $\mathcal{M}$,
ignoring the local minima when the mismatch value is oscillating while it decreases.
The yellow stars and gray circles show the $t_0$ chosen in each case.

We can see that the time $t_0$ depends on the method considered, that is, it depends on the fitting function the chosen inner product.
The dashed black line has a clear minimum at  $t_0 - t_\mathrm{peak} = 10 M$ (indicated by a yellow star).
This method also has the largest mismatch values, which is expected as the denominator of equation~\eqref{eq:mismatch} is close to zero, because $\ddot{\theta}_{22} \to 0$ for $\dot{\theta}_{22} \to \omega^r_{220}$.
In the solid black line, the mismatch stops decreasing and the curve starts flattening at approximately the same time as the minimum of the black dashed line, but the actual first minimum is a bit later (marked in gray circle).
The flattening of the curve at earlier time already indicates that the mismatch is already close to the minimum value.
The red curves have a very similar behaviour, with a minimum close to $t_0 - t_\mathrm{peak} = 15.6M$.
However, the dashed curve presents oscillations while decreasing and the several local minima make the determination of $t_0$ (gray circle) less precise.

The behaviour of these curves is typical in all the simulations we have analyzed (nonspinning circular binaries, see Section~\ref{sec:overtone_mass_ratio}).
Thus, we choose two methods to determine $t_0$, one for each fitting function:
\begin{enumerate}
	\item[I] the global minimum of the mismatch of $\dot{\theta}_{22}$ using the inner product given by equation~\eqref{eq:inner2};
	\item[II] the \emph{first} minimum of the mismatch of $h_{22}$ using the inner product given by equation~\eqref{eq:inner1}.
\end{enumerate}
The times $t_0$ chosen with these methods are marked with yellow stars in Figure~\ref{fig:mismatch}.

\begin{figure}[htb!]
	\centering
	\includegraphics[width=0.9\linewidth]{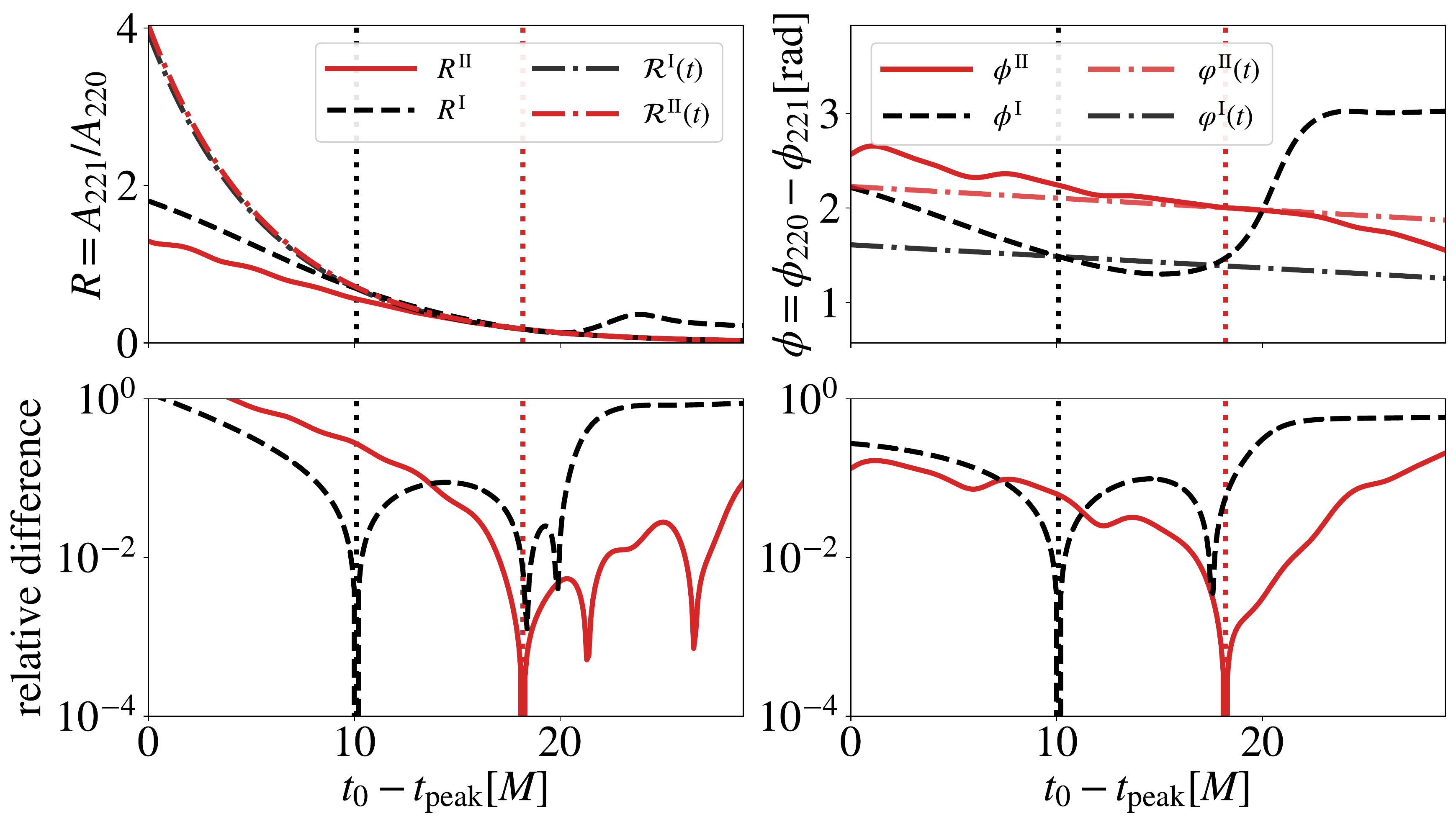}
	\caption{\emph{Top:} amplitude ratio between the first overtone and the fundamental mode $R^{\alpha} = A_{221}/A_{220}$ (left) and phase differences  $\phi^\alpha = \phi_{220} - \phi_{221}$ (right) as a function of time, where $\alpha$ indicated methods I and II defined in Figure~\ref{fig:mismatch} and in the text. The dotted curves are the expected values for the amplitude ratios  $\mathcal{R}^{\alpha}(t)$, given by equation~\eqref{eq:R_t}, and the phase difference  $\varphi^\alpha(t)$, defined in equation~\eqref{eq:phi_t}.
	\emph{Bottom:} relative difference between the expected values and the fitted values.}
	\label{fig:fit_pars}
\end{figure}

Given that $t_0$ is the initial time from which the waveform is well described by the sum of the fundamental mode and the overtone, we have to check whether the QNM parameter values obtained at $t_0$ are compatible with the values obtained when the fits start at later times.
Figure~\ref{fig:fit_pars} on the top shows the fitted amplitude ratio $R$ (left) and phase difference $\phi$ (right) for the methods I (dashed black) and II (solid red) as a function of the initial fitting time $t_0$.
The dotted vertical lines indicate the best $t_0$ chosen for methods I and II (see Figure~\ref{fig:mismatch}).
Both $R$ and $\phi$ decrease with time, as expected (see below).
But we can see that in method I $R$ and $\phi$ increase at late times (which is equivalent of $\ddot{\theta}_{22} \to 0$).
A similar behaviour is present for $\phi$ obtained with method II at later times (not shown in the plot).
These increases happen because after the overtone has decayed, the fitting procedure cannot determine well its parameters.

To compare the values obtained by starting the fitting at different initial times $t_0$ with the expected values,
we find the values at the chosen $t_{0}^{\alpha}$, where  $\alpha = {\mathrm{I, II} }$, as functions of time.
While the linear regime is valid, the amplitude ratio is given by
\begin{equation}
	\mathcal{R}^{\alpha}(t) = R^{\alpha}(t_0^{\alpha})e^{(\omega_{221}^i - \omega_{220}^i) (t_0^{\alpha} - (t-t_\mathrm{peak}))},
	\label{eq:R_t}
\end{equation}
where $R^\alpha(t_0^\alpha)$ is the amplitude ratio $R^\alpha$ fitted at the initial time $t_0^\alpha$, for each method $\alpha$.
Likewise, the phase difference as a function of time is given by
\begin{equation}
	\varphi^\alpha (t) = \phi^\alpha(t_0^\alpha) - (\omega^r_{220} - \omega^r_{221})\left((t - t_\mathrm{peak}) - t_0^\alpha\right).
	\label{eq:phi_t}
\end{equation}

The plots on the top of Figure~\ref{fig:fit_pars} show the expected values for the amplitude ratio $\mathcal{R}^{\alpha}$ and the phase difference $\varphi^{\alpha}$ as dash-dotted curves.
The relative difference between the fitted and the expected values are shown in the plots on the bottom.
We can see that methods I and II are compatible in the determination of the amplitude ratio R, as the expected curves are approximately the same.
The difference in $\phi$ is because the phase difference is more sensitive to the method, especially because the overtone decays very fast.
The amplitude ratio compatibility indicates that both methods can estimate well the contribution of the overtone, and the low mismatch is not due to overfitting.
The large difference between the expected and the fitted values near $t_{\mathrm{peak}}$ is due to non-linear behaviour of the waveform or significant contribution of higher overtones.
The large differences at late times are caused by the exponential decay of $R$.

\begin{figure}[htb!]
	\centering
	\includegraphics[width=0.9\linewidth]{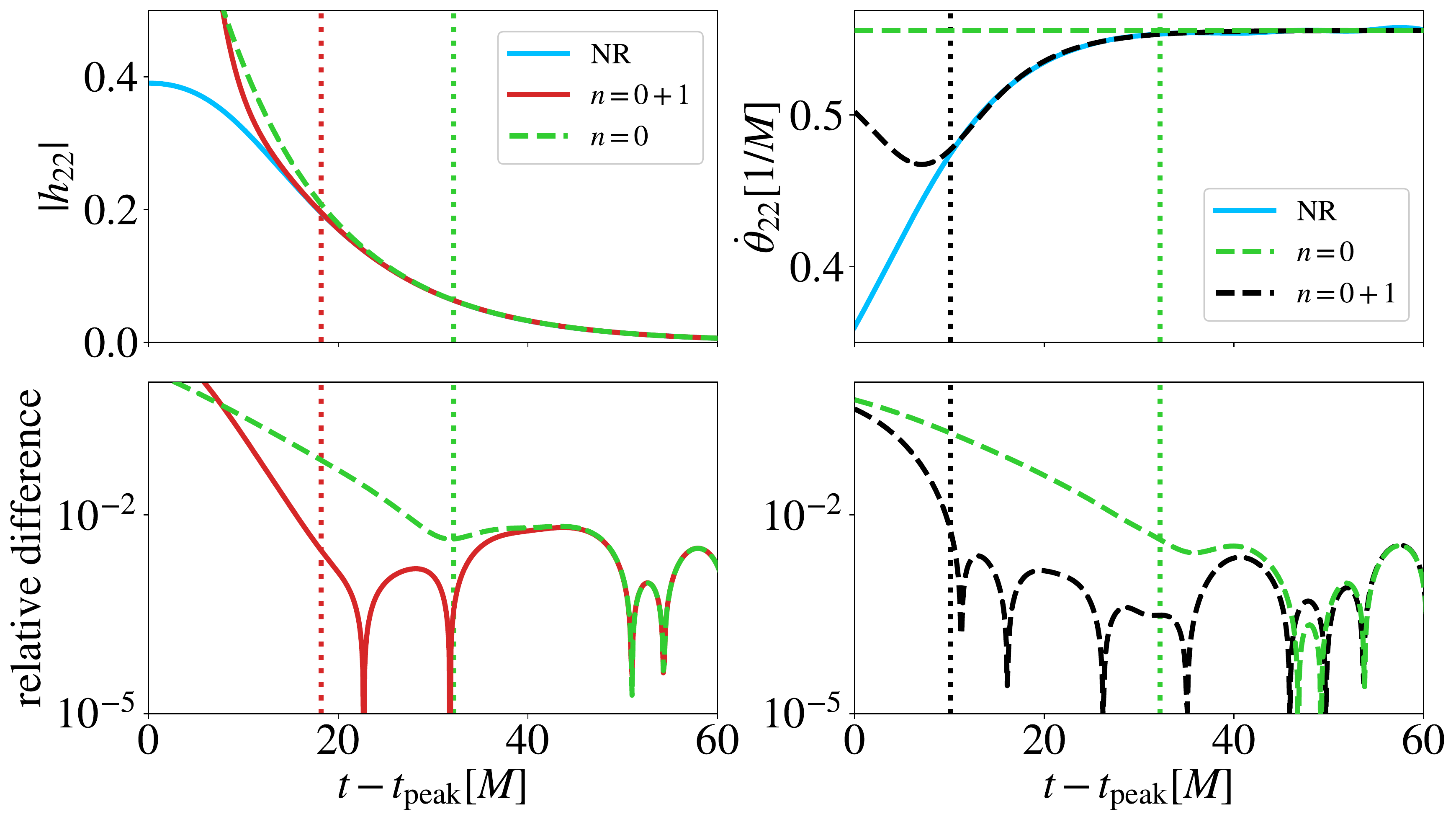}
	\caption{\emph{Top:} waveform amplitude $|h_{22}|$ (left) and the time derivative of the complex phase $\dot{\theta}_{22}$ (right) as a function of time, for the numerical simulation SXS:BBH:0305 (blue), the best fit considering just the fundamental mode $(2,2,0)$ (green) and the best fit considering the contribution of the first overtone $(2,2,0)+(2,2,1)$ (red and black), for the method I (left) ant the method II (right).
	\emph{Bottom:} relative difference between the numerical simulation and the best fits as a function of time.
	The vertical dotted lines in all the plots indicate the initial time consider for each fit, the same as Figures~\ref{fig:fit_n0} and~\ref{fig:mismatch}.}
	\label{fig:fits_h_dtheta}
\end{figure}

The top plots of Figure~\ref{fig:fits_h_dtheta} show the comparison between the simulated waveform SXS:BBH:0301 (blue) and the best fits for the fundamental mode (green) and the fundamental mode + first overtone, for method I (black) and II (red).
The initial times ($t_{n = 0}$, $t_0^{\rm I}$ e $t_0^{\rm II}$) for the fits are indicated by the vertical dotted lines.
In the bottom plots we show the relative difference between the simulation and the fits.
The fundamental mode describes very well the waveform at late times, but the overtone is needed at early times.
Near the peak the models are not good, and the model with two modes is always better or comparable (at times greater than $t_{n=0}$) to the single mode model.

We have seen that the overtone has a significant contribution to the ringdown waveform model, with an amplitude ratio of $R = A_{221}/A_{220} = 0.66$ at $t_0^{\rm I} = t_\mathrm{peak} + 10M$.
For the simulation considered in this section, the amplitude ratios of the fundamental higher harmonics at $t_\mathrm{peak} + 10M$ are  $A_{210}/A_{220} = 0.05$, $A_{330}/A_{220} = 0.07$ e $A_{440}/A_{220} = 0.04$, that is, the overtone has an amplitude almost ten times larger than the amplitude of the most relevant fundamental higher harmonic.

Although the importance of overtones was already known for decades~\cite{Leaver:1986gd, Stark:1985da} and a recent work~\cite{Giesler:2019uxc} suggested that the non-linear dynamics has damped before the peak of amplitude, its contribution to the BBH waveform is still not a closed topic~\cite{Bhagwat:2019dtm, Ota:2019bzl, Forteza:2020cve, Mourier:2020mwa, Okounkova:2020vwu, Bustillo:2020buq, Finch:2021iip, Forteza:2021wfq, Baibhav:2017jhs, Ota:2021ypb}.
We proposed a method that double checks the results to confidently model the overtone and avoids overfitting.
The choice of using just one overtone is conservative but safe, as non-physical behaviours of many parameters are ruled out.

\section{Mass ratio dependence}
\label{sec:overtone_mass_ratio}
As the amplitude and phases of the QNMs depend on the initial conditions, their values will not be constant when different BBH systems are considered.
As most LVKC detections are compatible with nonspinning circular binaries~\cite{LIGOScientific:2018mvr, LIGOScientific:2020ibl, LIGOScientific:2021usb, LIGOScientific:2021djp}, we will restrict our analysis to this kind of system.
In order to assess the dependence on the binary parameters, the results of the previous sections were reproduced for nineteen numerical simulations with mass ratio $q \equiv m_1/m_2 \geq 1$ ranging from 1 to 10 (for circular nonspinning binaries, the other parameters are equal to zero).

\begin{figure}[htb!]
	\centering
	\includegraphics[width=0.9\linewidth]{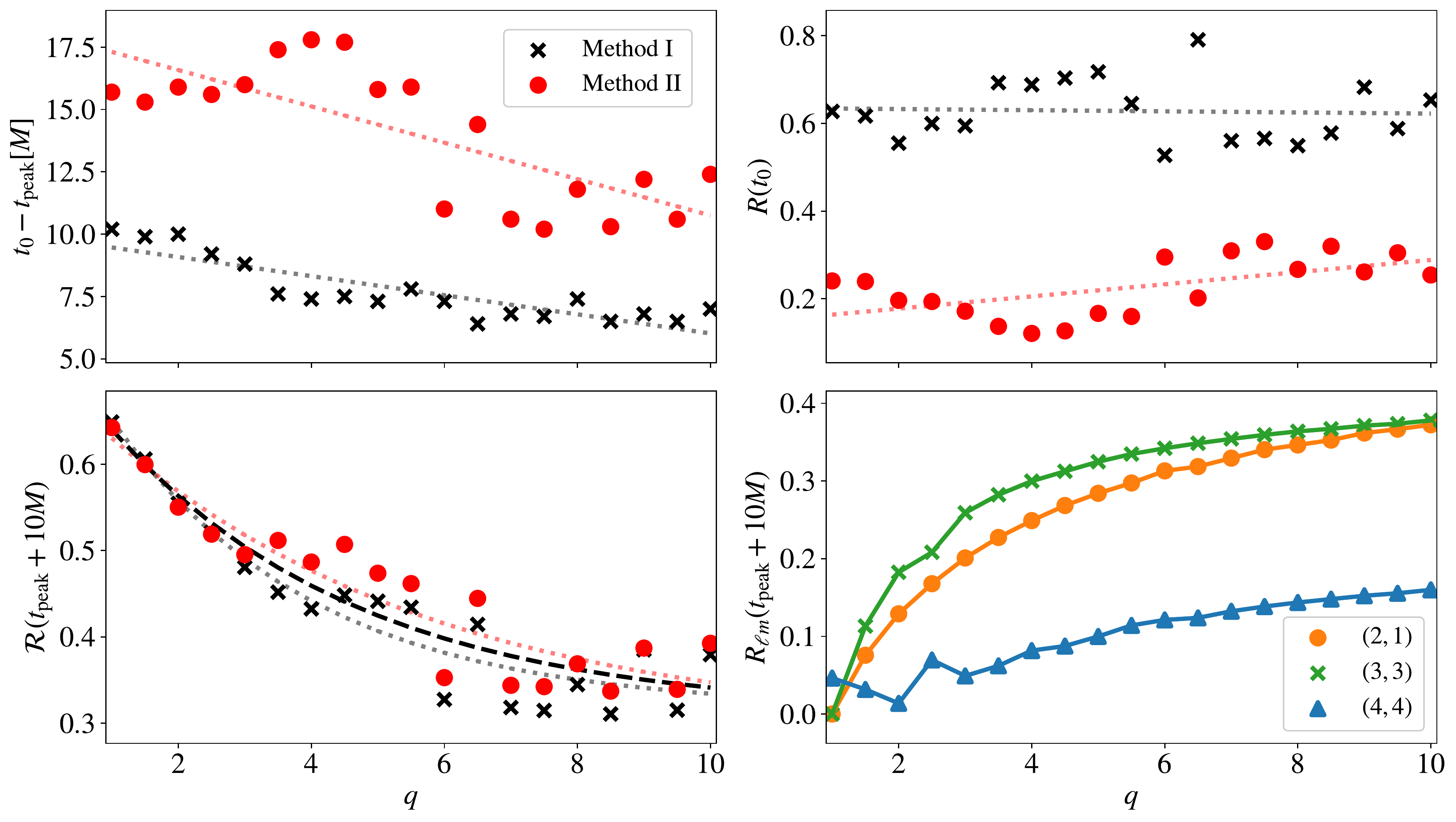}
	\caption{{\textit{Top:} } best fitting initial time $t_0$ (left) and amplitude ratio $R = A_{221}/A_{220}$ at $t_0$ (right) as a function of the mass ratio $q$.
	{\textit{Bottom:} } expected amplitude ratio $\mathcal{R}(t)$, defined in equation~\eqref{eq:R_t} (left) and the ratio $R_{\ell m} = A_{\ell m 0}/A_{220}$ between the fundamental higher harmonics $(\ell, m, n) = (2, 1, 0)$, $(3, 3, 0)$ and $(4, 4, 0)$ and the quadrupolar fundametnal mode $(2, 2, 0)$ (right), both evaluated at a reference time $t = t_{\mathrm{peak}} + 10M$, as a function of the mass ratio $q$.
		The dashed black curve on the left plot shows the best fit of an exponential decay function, considering both methods I and II.}
	\label{fig:mass_ratio_fits}
\end{figure}

The top left of Figure~\ref{fig:mass_ratio_fits} shows the time $t_0$ as a function of the mass ratio $q$.
We can see that the waveform can be well described by the fundamental mode and the first overtone closer to the peak of amplitude for larger mass ratio binaries.
As the mass ratio increases, the non-linear dynamics of the merger and the contribution of higher overtones become less relevant and the linear perturbation regime gets closer to the merger time.

In the top right plot of Figure~\ref{fig:mass_ratio_fits} we show the amplitude ratio $R$ at the initial time $t_0$ as a function of the mass ratio $q$.
The difference seen between the methods I and II is due to the dependence on the time $t_0$, which is smaller for the method I, as we can see in the top left plot.
Nonetheless, the consistency between the methods is seen when we look at the expected amplitude ratio $\mathcal{R}$, given by equation~\eqref{eq:R_t}, at the same reference time.
The expected amplitude ratio at the reference time  $t = t_\mathrm{peak} + 10 M$ is shown in the bottom left plot.
The dashed black curve indicated the best fit for an exponential decay, which takes into account both methods I and II:
\begin{equation}
	\mathcal{R}(q, t_\mathrm{peak} + 10M) = 0.4 e^{-0.3q} + 0.3.
	\label{eq:R_exp_fit}
\end{equation}
The asymptotic amplitude ratio for large mass ratios, $\mathcal{R} \to 0.3$, is compatible with the point particle limit~\cite{Cardoso:amplitude}.\footnote{We thank professor Vitor Cardoso for pointing this out.}

Finally, the bottom right plot of Figure~\ref{fig:mass_ratio_fits} shows the ratio $R_{\ell m} = A_{\ell m 0}/A_{220}$ between the amplitude of the fundamental mode of the higher harmonics $(\ell, m, 0) = (2, 1, 0)$, $(3, 3, 0)$ e $(4, 4, 0)$ and the quadrupolar fundamental mode $(2, 2, 0)$ at $t = t_{\mathrm{peak}} + 10M$.
Comparing with the bottom left plot, we can see that the first overtone $(2,2,1)$ has a larger amplitude ratio $\mathcal{R}$ than all the fundamental higher harmonics for $q \lesssim 5$.
For larger mass ratios, the overtone has an amplitude ratio comparable with the modes $(2,1,0)$ and $(3,3,0)$.
It is important to stress that these comparisons are valid at the reference time $t = t_{\mathrm{peak}} + 10M$, as the amplitude ratio of the overtone is highly dependent on the time considered and the amplitude ratio of the higher harmonics is almost independent of the time, as the decay times of the fundamental modes are very similar and the overtone decays much faster (see Figure~\ref{fig:qnm_omegas_a}).

We have seen that the amplitude of the overtone is greater than or equivalent to the amplitude of the most relevant higher harmonics.
This may indicate that the overtone will be easier to detect than the higher harmonics.
However, the first overtone damps much faster than the higher harmonics (see Figure~\ref{fig:qnm_omegas_a}), which makes its detection more challenging than the fundamental mode.
The detectability of these modes will be discussed in the following chapter.

\begin{center}
\myclearpage
\par\end{center}


\chapter{Black hole spectroscopy horizons}
\label{ch:spectroscopy-horizon}

A confident detection of the dominant $(2,2,0)$ mode was achieved in the first gravitational wave event, GW150914~\cite{LIGOScientific:2016lio}, and this mode has been detected in other events whose ringdown is visible~\cite{LIGOScientific:2020ufj, LIGOScientific:2020tif, LIGOScientific:2021sio}.
In 2019, based on the overtone analysis from NR simulations~\cite{Giesler:2019uxc}, Isi et al.~\cite{Isi:2019aib} reanalysed GW150914 to look for overtone contribution in the detected events.
They started the analysis at the peak of amplitude and showed that, when overtones are considered in the models, the estimated remnant mass and spin are compatible with the inspiral-merger-ringdown (IMR) analysis, and the single-mode analysis is not.
However, the Bayes factor favoring a two-mode model over the single-mode model is smaller than 5~\cite{LIGOScientific:2020tif}.
The lack of statistical evidence is justified by a proposed detectability criterion based on the support for a non-zero amplitude for the overtone.
In 2022, Cotesta et al.~\cite{Cotesta:2022pci} reanalysed these claims and found that the noise can induce an artificial amplitude for the overtone.

Given the uncertainties caused by the low statistical evidence for subdominant modes, the claims for detection of subdominant modes in the observed events~\cite{Isi:2019aib, Capano:2021etf} are still under debate~\cite{Carullo:2019flw, LIGOScientific:2020tif, LIGOScientific:2020ufj, LIGOScientific:2021sio, Cotesta:2022pci}.
In this chapter we analyse the necessary conditions to confidently perform BH spectroscopy.
We compute the black hole spectroscopy horizons, the maximum distance up to which two modes can be detected and distinguished in the ringdown of a BBH event.
We compute the horizons using two conditions.
First, we use the Rayleigh criterion~\cite{Berti:2005ys}, which is a resolvability condition and guarantees the detection of a subdominant mode with high statistical evidence.
In the second approach we use a minimum threshold for the Bayes factor $\ln\mathcal{B} > 8$, which gives high support for a two-mode (or more) model over a single-mode model.

\section{International System of Units}

In the previous chapters, we worked with code units, that is, the complex frequencies $\omega_{\ell m n}$ were given in terms of the BH mass and $c = G = M_{\odot} = 1$.
To evaluate the detectability of the modes, we will work in the International System of units.
The first important thing is to set the mass $M_f$ of the detected BH in kilograms, as the frequencies and amplitude of the QNMs depend on it.
Then the correct factors of $c$ and $G$ must be taken into account.
To transform the simulation time, which is given in units of total mass, to seconds, we must multiply it by the factor $M_f G M_{\odot}/c^{3} = 4.93\times10^{-6} \mathrm{s}$.
The frequency in Hertz is obtained with the inverse factor.
The amplitude of equation~\eqref{eq:qnm_polarizations} should be multiplied by the factor $G/c^{2}$ to guarantee that the strain $h$ is dimensionless.

The oscillation frequencies and damping times of the QNMs are obtained from the complex frequencies using equation~\eqref{eq:freq_tau_omegas}.
The factor $M_f G M_{\odot}/c^{3}$ indicates that the damping time will be larger (the modes will take longer to decay) for larger masses and the frequencies of oscillation will be smaller (the period of the wave will be longer).
In Figure~\ref{fig:detectors_Sn} we see that each detector is sensitive to a certain frequency band, that is, the detectability of the QNMs is highly dependent on the mass, as we see below.

\section{Detectability measure}
\label{sec:detectability-metric}

The SNR~\eqref{eq:snr} is a standard quantity used to determine the detectability of gravitational wave sources, and it was used in many BH spectroscopy works~\cite{Berti:2005ys, Baibhav:2017jhs, Baibhav:2018rfk, Baibhav:2020tma, Bhagwat:2019dtm, Forteza:2020cve}.
A typical threshold used in GW analyses is $\rho \geq 8$.
For the ringdown this threshold may or may not be high enough depending on how the SNR is defined.
As we saw in Chapter~\ref{ch:gw-detection}, the SNR is defined in terms of the template we choose.
For the ringdown the best template should contain the contribution of all QNMs, given by equation~\eqref{eq:qnm_polarizations}.
In this case, $\rho_{\mathrm{ringdown}} \geq 8$ is enough to assert the detectability of the dominant mode $(2,2,0)$~\cite{LIGOScientific:2016lio}, but it is not high enough to confidently detect a subdominant mode~\cite{Carullo:2019flw, LIGOScientific:2020tif, LIGOScientific:2021sio, LIGOScientific:2020ufj, Cotesta:2022pci}.

A threshold for SNR $\rho_{\ell mn}$ of a single mode $(\ell, m, n)$ may give a more accurate estimate for the detectability of each mode, and it was used in some works~\cite{Berti:2005ys, Baibhav:2020tma, Baibhav:2018rfk} (see also Figure~\ref{fig:Q_rayleigh_snr} on the right).
However, although the SNR of a single QNM is easily defined, its value depends on the parameters of the QNM, which cannot be estimated independently of other QNMs.
Therefore, in practice the single mode SNR is more difficult to compute than the ringdown SNR\@.
Moreover, the ringdown SNR gives more general information about the signal strength, as we show in the example below.

\begin{figure}[htb!]
	\centering
	\includegraphics[width = 0.98\linewidth]{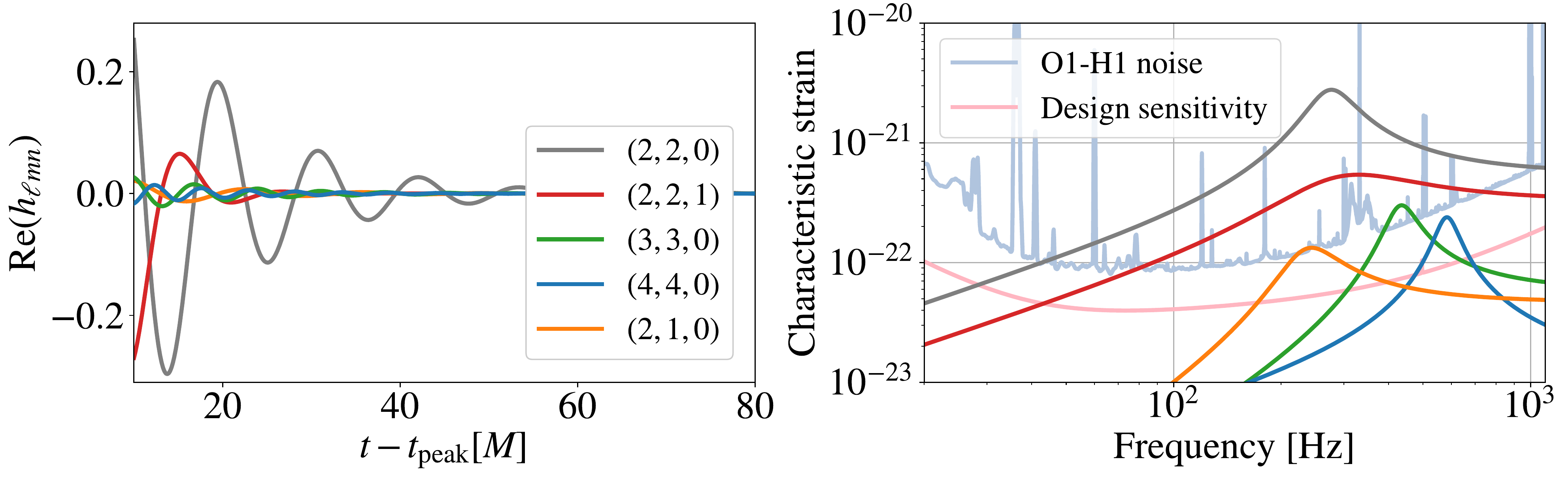}
	\caption{\emph{Left:} real part of the fitted waveforms for the $(2,2,0)$ (gray), $(2,2,1)$ (red), $(3,3,0)$ (green), $(4,4,0)$ (blue) and $(2,1,0)$ (orange) modes of the ringdown of the BBH simulation SXS:BBH:0305 (compatible with GW150914).
	At early times the amplitude of the $(2,2,1)$ is much higher than the amplitude of the higher harmonics.
	\emph{Right:} characteristic strain~\cite{Moore:2014lga} of each QNM, rescaled with the GW150914 values, $M_f =  63 M_\odot$ and $D_L = 440 \text{Mpc}$, and for the noise of Advanced LIGO Hanford near GW150914 (light blue) and Advanced LIGO at design sensitivity (light pink).}
	\label{fig:modes_strain}
\end{figure}

Figure~\ref{fig:modes_strain} on the left shows the strain of the $+$ polarization (real part) of the modes $(2,2,0)$, $(2,2,0)$, $(3,3,0)$, $(4,4,0)$ and $(2,1,0)$ of the simulation SXS:BBH:0305, with parameters computed as described in Chapter~\ref{ch:ringdown-qnm} and using method~II.
As we saw before, the amplitudes of the higher harmonics are much smaller than the first overtone amplitude, but the overtone decays very fast.
To compute the SNRs, the waveforms are scaled with the GW150914 values~\cite{LIGOScientific:2016aoc}, that is,  $M_f =  63 M_\odot$ and $D_L = 440 \text{Mpc}$.
For simplicity we consider just the $+$ polarization in the computation of the SNRs, which is equivalent to consider the optimal direction of the source with $F_{+} = 1$ and $F_{\cross} = 0$.
Figure~\ref{fig:modes_strain} on the right shows the characteristic strain~\cite{Moore:2014lga} of each mode, the LIGO Hanford strain near the time of the GW150914 event and the LIGO noise curve at the design sensitivity.
The position of the signal curves relative to the noise curves gives us an estimate of the SNR of each mode.

Using the LIGO Hanford noise data~\cite{GW150914-data}, we compute the SNR $\rho_{\ell mn}$ for each mode, using the best fits for each mode and equation~\eqref{eq:snr}.
We find $\rho_{220} = 12.19$, $\rho_{221} = 3.42$, $\rho_{330} = 0.74$, $\rho_{440} = 0.40$ and $\rho_{210} = 0.69$.
Therefore, even though the first overtone decays three times faster than the higher harmonics, its SNR is at least 4.7 times greater than the higher harmonics SNRs.
However, when we look for the ringdown SNR $\rho_{220, \ell m n}$ that considers the contribution of the dominant mode and only one subdominant mode $(\ell, m,n)$, we see different trends.
We have $\rho_{220,221} = 9.32$, $\rho_{220,330} = 12.46$, $\rho_{220,440} = 12.24$ and $\rho_{220,210} = 12.46$, which indicates that the overtone decreases the ringdown SNR\@.
This is expected, as we see in the top left of Figure~\ref{fig:fits_h_dtheta} the dominant mode alone does not describe the waveform very well at $t = t_{\mathrm{peak}} + 10M$, and has a higher amplitude than the numerical simulation.
The inclusion of the overtone corrects this excess of amplitude, which lowers the two-mode SNR\@.
The ringdown SNR, considering all the above modes, is $\rho_{\mathrm{ringdown}} = 9.66$.
This example exposes some problems in the interpretation of a single-mode SNR value, without looking at the whole problem.

Furthermore, interpreting the comparison of the SNRs of different sources is considerably challenging.
For example, for nonspinning circular BBH mergers with a fixed total mass and orbital frequency, binaries with lower mass ratios have a greater gravitational wave luminosity.
Thus, two BBH systems with the same mass and distance will have different SNRs depending on their mass ratios.
In~\cite{Ota:2019bzl}, we found that the minimum SNR to satisfy the Rayleigh criterion (more about the Rayleigh criterion in Section~\ref{sec:horizon-rayleigh-criterion}) is approximately constant for mass ratio ranging from 1 to 10 for the overtone, and for mass ratio ranging 5 to 10 for the $(3,3,0)$ mode (see also Figure 8 of~\cite{Forteza:2020cve}).
These results may lead to the conclusion that the Rayleigh criterion poses the same challenge for different mass ratios, which is not true because the amplitude of the GW waveform is much smaller for high mass ratio binaries.

Lastly, different detectors also result in different SNRs for the same source.
With these considerations and motivated by the ringdown horizons computed by~\cite{Baibhav:2018rfk}, we use the source distance as a more universal metric to assess the detectability of the modes.
This metric solves the problems stated above.
Moreover, quoting directly the distance helps with the identification of other quantities of interest, such as the event rate, which depends on the volume considered.
In the following sections we present the \emph{black hole spectroscopy horizons}, which measures the maximum distance, averaged over sky location and source inclination, of a BBH merger event, as a function of the remnant mass and binary mass ratio, up to which two or more QNMs are detectable in the GW ringdown.
With the BH spectroscopy horizons the comparison of QNM detectability between different detectors and across several orders of magnitude in mass range is straightforward.

Besides the QNMs parameters, the detectability of the modes depends on the sky location and source inclination.
The sky location dependence is decoded in the antenna response patterns, as described in section~\ref{sec:antenna-response-pattern}.
An optimal source location would be one such that $F_+ = 1$ and $F_{\cross} = 0$~\cite{maggiore-vol1}, that is,
the source comes from a direction orthogonal to the detector's plane.
Such optimal location requires luck, and we are interested in assessing the detectability of subdominant modes for general detections.
Therefore, we consider the angular average $\langle ({F^2_{+,\times}})^{1/2}\rangle$, given for each detector in section~\ref{sec:antenna-response-pattern}.

The binary inclination is a very relevant factor in the mode detection.
The harmonic modes~\eqref{eq:qnm_harmonic} are defined in relation to the spheroidal harmonics, which are functions of the inclination angle $\iota$ and the azimuthal angle $\phi$~\cite{Boyle:2019kee,Berti:2005gp}.
Therefore, disregarding the mode mixing between spherical and spheroidal harmonics (see Section~\ref{sec:binary-black-hole-merger}), the effective amplitude of each mode should be  $|_{-2}Y_{\ell m}(\iota,\varphi)A_{\ell mn}|$, which is independent of the azimuthal angle $\varphi$.

\begin{figure}[htb!]
	\centering
	\includegraphics[width = 0.7\linewidth]{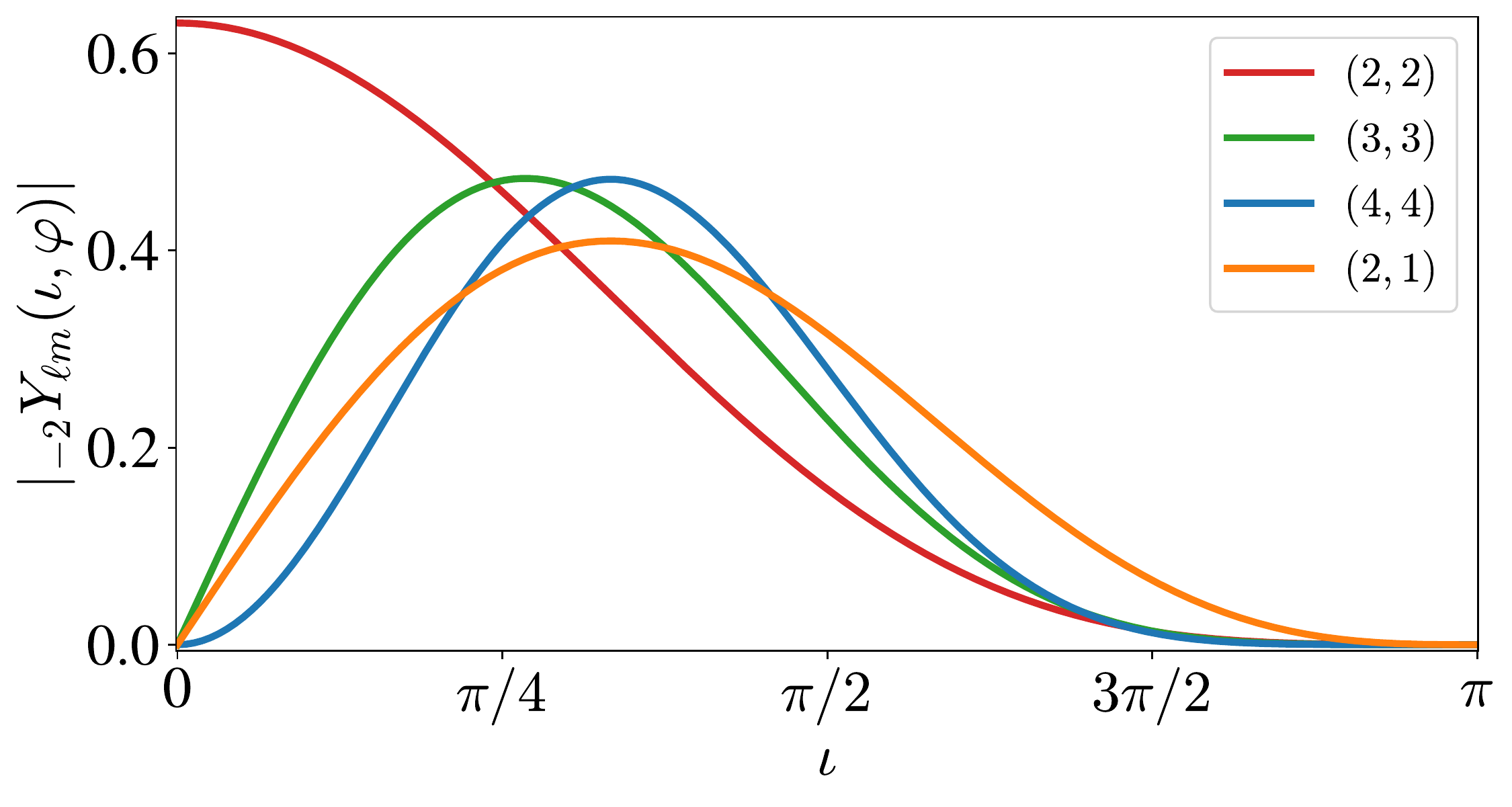}
	\caption{Absolute value of the spin weighted spherical harmonic $_{-2}Y_{\ell m}$, for $(\ell, m)$ equal to $(2,2)$ (red), $(3,3)$ (green), $(4,4)$ (blue) and $(2,1)$ (orange).
	For face-on binaries ($\iota = 0$), the $_{-2}Y_{22}$ is the only non-zero spherical harmonic.
	For edge-on orientation ($\iota = \pi/2$), the higher harmonics have larger contribution than $(2,2)$.}
	\label{fig:spherical_harmonics}
\end{figure}

Figure~\ref{fig:spherical_harmonics} shows the absolute value of the spin-weighted spherical harmonic $_{-2}Y_{\ell m}$, for the $(2,2)$, $(3,3)$, $(4,4)$ and $(2,1)$ harmonics, as a function of the inclination angle $\iota$.
For face-on orientation ($\iota = 0$), the only harmonic with non-zero contribution is the $(2,2)$ harmonic.
For edge-on events ($\iota = \pi/2$) the higher harmonics $(3,3)$, $(4,4)$ and $(2,1)$ have larger contribution than the $(2,2)$ harmonic.
On average, the harmonics have similar contributions.
As we are working with QNMs, we consider the full angular average of the spin-weighted \emph{spheroidal} harmonics $\langle _{-2}S_{\ell m}\rangle = 1/\sqrt{4\pi}$, which does not favor one mode over another.

However, face-on events are much stronger than edge-on events, that is, they can be detected at larger distances.
Identical events with the same SNR are detected out to a distance (see section 7.7.2 of~\cite{maggiore-vol1})
\[d_\mathrm{sight} \propto \left|\frac{1+\cos^2\iota}{2}\right|,\]
which is two times larger if they are detected face-on rather than with edge-on inclination.
Therefore, the expected number of detections as a function of the binary inclination angle is

\[
N(\iota) = \left(\frac{\textrm{events rate}}{\rm volume}\right) \times \textrm{(observation time)} \times V(\iota),
\]
where $V(\iota) \propto d^3_{\mathrm{sight} }$ is the observed volume.
This gives an observational preference for the fundamental quadrupolar mode and its overtone.
For example, on average the amplitude ratio between the $(2,2)$ and the $(3,3)$ harmonics is boosted by a factor
$\langle|N(\iota)Y_{22}(\iota,\varphi)|\rangle/\langle|N(\iota)Y_{33}(\iota,\varphi)|\rangle \sim 1.5$.

With this considerations in mind we stress that the BH spectroscopy horizons computed below are angle averaged over sky location and source inclination, and are \emph{not} the best prospects for detection.
Notice that in the LIGO literature, a ``range'' refers to the angular average, while a ``horizon'' is defined as the maximum distance obtained with optimal values.
Thus, our horizon definition is equivalent to the range definition used by LIGO\@.

\section{Rayleigh criterion}
\label{sec:horizon-rayleigh-criterion}
The Rayleigh criterion was originally proposed to distinguish two bright stars very near each other in the sky.
The adaptation for QNMs was proposed by~\cite{Berti:2005ys} to distinguish two modes in the ringdown.
The adapted criterion sets conditions for both frequencies $f_{\ell m n}$ and damping times $\tau_{\ell m n}$.
For two modes $(\ell,m,n) \ne (\ell',m',n')$, the Rayleigh criterion is
\begin{subequations}
	\begin{align}
		\label{eq:rayleigh_f}
		|f_{\ell mn} - f_{\ell'm'n'}|\equiv\Delta f_{\ell mn,\ell'm'n'}  > {\mathrm{max}}(\sigma_{f_{\ell mn}},\sigma_{f_{\ell' m'n'}}),\\
		|\tau_{\ell mn} - \tau_{\ell'm'n'}|\equiv\Delta \tau_{\ell mn,\ell'm'n'} > {\mathrm{max}}(\sigma_{\tau_{\ell mn}},\sigma_{\tau_{\ell' m'n'}}),
		\label{eq:rayleigh_tau}
	\end{align}
	\label{eq:rayleigh_both}
\end{subequations}
where $\sigma_{f_{\ell mn}}$ and $\sigma_{\tau_{\ell mn}}$ are uncertainties in the frequencies and damping times, that is, the observations are reported as $f_{\ell mn} \pm \sigma_{f_{\ell mn}}$ and $\tau_{\ell mn} \pm \sigma_{\tau_{\ell mn}}$.

As the fundamental quadrupolar mode is the dominant mode, we consider $(\ell,m,n) = (2,2,0)$ and $(\ell', m', n')$ one of the most relevant subdominant modes: $(2,2,1)$, $(3,3,0)$, $(4,4,0)$ or $(2,1,0)$.
When \emph{both} inequalities~\eqref{eq:rayleigh_both} are satisfied, both pairs of frequencies and damping times are independently determined.
Thus, the observed values can be used to compute two pairs for the mass and spin of the remnant black hole~\eqref{eq:mass_spin_qnm}, which can be compared to test the no-hair theorem (see section~\ref{sec:testing-the-no-hair-theorem}).

\begin{figure}[htb!]
	\centering
	\includegraphics[width=1\linewidth]{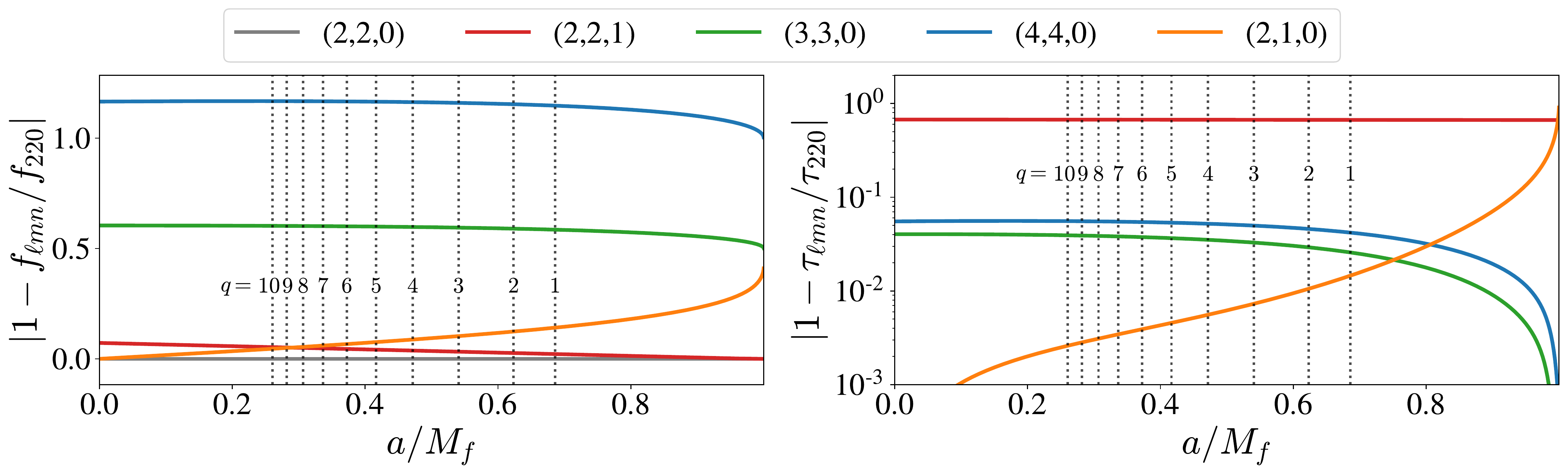}
	\caption{Relative differences of frequencies (left) and damping times (right) between the dominant QNM mode $(2, 2, 0)$ and subdominant modes $(2,2,1)$ (red), $(3,3,0)$ (green), $(4,4,0)$ (blue) and $(2,1,0)$ (orange) as a function of the BH dimensionless spin $a/M_f$.
	The values for the plots were obtained with linear perturbation theory by~\cite{Berti:2009kk, Berti-ringdown}.
	The vertical dotted lines are the same for all the plots and indicate the final dimensionless spin of the remnant BH of nonspinning circular binaries with mass ratios $q$ raging from 1 (highest spin) to 10 (lowest spin).}
	\label{fig:qnm_df_dtau_a}
\end{figure}

Figure~\ref{fig:qnm_df_dtau_a} shows the relative difference of frequencies (left) and damping times (right) between the dominant QNM mode $(2, 2, 0)$ and subdominant modes $(2,2,1)$, $(3,3,0)$, $(4,4,0)$ and $(2,1,0)$ as a function of the BH dimensionless spin $a/M_f$.
This is just a transformation of Figure~\ref{fig:qnm_omegas_a} using equations~\eqref{eq:freq_tau_omegas}.
In the left plot we can see that  $\Delta f_{220,\ell m n}$ is greater for modes with greater $\ell$ and modes with $\ell = 2$ have frequencies of oscillation very similar to the frequency of the $(2,2,0)$ mode.
On the right we see that the damping times of the fundamental modes $(\ell,m,0)$ are very similar, and the overtone has a considerably different damping time.

From Figure~\ref{fig:qnm_df_dtau_a} we can conclude that the Rayleigh criterion~\eqref{eq:rayleigh_both} poses different challenges for different modes and systems.
The differences $\Delta f_{220,\ell m n}$ and $\Delta \tau_{220,\ell m n}$ are smaller for larger spins (smaller mass ratios), except for the $(2,1,0)$ mode, whose differences grow with the spin.
Moreover, modes with $\ell \neq 2$ will easily satisfy the frequency condition~\eqref{eq:rayleigh_f}, whereas the damping time~\eqref{eq:rayleigh_tau} is easily resolved for the overtone.
As the $(2,2,0)$ and $(2,1,0)$ modes have very similar frequencies and damping times, both conditions~\eqref{eq:rayleigh_both} are very restrictive for the $(2,1,0)$ mode.

We consider the uncertainties $\sigma_{f_{\ell mn}}$ and $\sigma_{\tau_{\ell mn}}$ given by the Fisher matrix formalism~\eqref{eq:fisher_matrix}.
The parameters considered are the QNMs intrinsic parameters
\[\vartheta^a = \{f_{220},\tau_{220}, A_{220}, \phi_{220}, f_{\ell m n},\tau_{\ell m n }, R, \phi_{\ell m n}\},\]
where $R = A_{\ell m n}/A_{220}$ is the amplitude ratio between the subdominant mode $(\ell,m,n) \neq (2,2,0)$ and the dominant mode $(2,2,0)$.

\begin{table}[htb!]
	\centering
	\caption{Amplitudes, phases and complex frequencies of the QNMs for the ringdown of nonspinning circular binaries with mass ratios $q=1.5$ (SXS:BBH:0593) and $q=10$ (SXS:BBH:1107). The amplitudes and phases are scaled at the reference time $t = t_{\mathrm{peak}} + 10 M$.}
	\begin{tabular*}{\textwidth}{c|@{\extracolsep{\fill}} c c c c } \hline\hline
		$(\ell,m,n)$ & $A_{\ell mn}$ & $\phi_{\ell mn}$ [rad]& $\omega^r_{\ell mn} MG/c^3$ & $\omega^i_{\ell mn} MG/c^3$ \\ \hline\hline
		& \multicolumn{4}{c}{$q = 1.5$} \\ \hline
		$(2,2,0)$ & 0.40 & 0.41 & 0.517 & 0.082 \\
		$(2,2,1)$ & 0.28 & 4.59 & 0.505 & 0.248 \\
		$(3,3,0)$ & 0.05 & 6.16 & 0.821 & 0.084 \\
		$(4,4,0)$ & 0.01 & 5.46 & 1.112 & 0.086 \\
		$(2,1,0)$ & 0.03 & 5.26 & 0.448 & 0.083 \\ \hline
		&\multicolumn{4}{c}{$q = 10$} \\ \hline
		$(2,2,0)$ & 0.14 & 3.56 & 0.412 & 0.088 \\
		$(2,2,1)$ & 0.07 & 1.17 & 0.390 & 0.269 \\
		$(3,3,0)$ & 0.05 & 3.32 & 0.661 & 0.092 \\
		$(4,4,0)$ & 0.02 & 3.32 & 0.894 & 0.093 \\
		$(2,1,0)$ & 0.05 & 5.57 & 0.394 & 0.088 \\  \hline\hline
	\end{tabular*}
	\label{tab:qnm_pars}
\end{table}

\begin{figure}[!htb]
	\centering
	\includegraphics[width=0.95\linewidth]{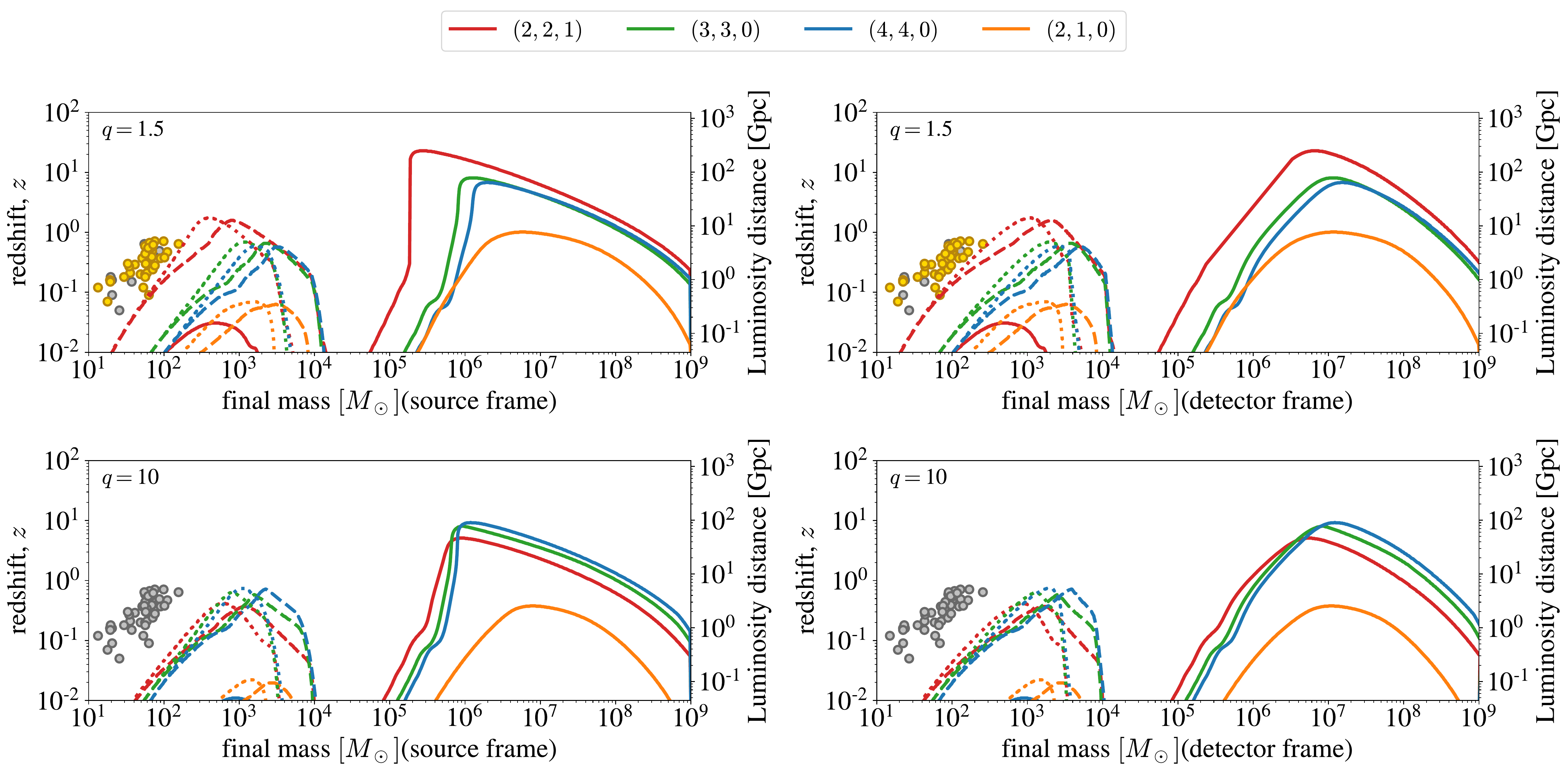}
	\caption{BH spectroscopy horizons using the Rayleigh criterion~\eqref{eq:rayleigh_both} and imposing the detectability condition $\rho_{\ell m n} > 8$ for nonspinning circular binaries with mass ratio $q = 1.5$ (top) and $q = 10$ (bottom), in the source frame (left) and the detector frame (right), for the subdominant modes $(2,2,1)$ (red), $(3,3,0)$ (green), $(4,4,0)$ (blue) and $(2,1,0)$ (orange).
	The solid curves are the LIGO horizons (for $M<10^4 M_{\odot}$) and the LISA horizons (for $M>10^4 M_{\odot}$), the dotted curves are the CE horizons and the dashed curves are the ET horizons.
	The circles indicate the confirmed BBH detections until the O3a run, the yellow (gray) circles are the detections (not) compatible with the systems considered.
	A signal with $(2,2,0) + (\ell,m,n)$ satisfies the Rayleigh criterion if the event is inside the horizon curves.}
	\label{fig:rayleigh_horizons}
\end{figure}

\begin{figure}[!htb]
	\centering
	\includegraphics[width=1\linewidth]{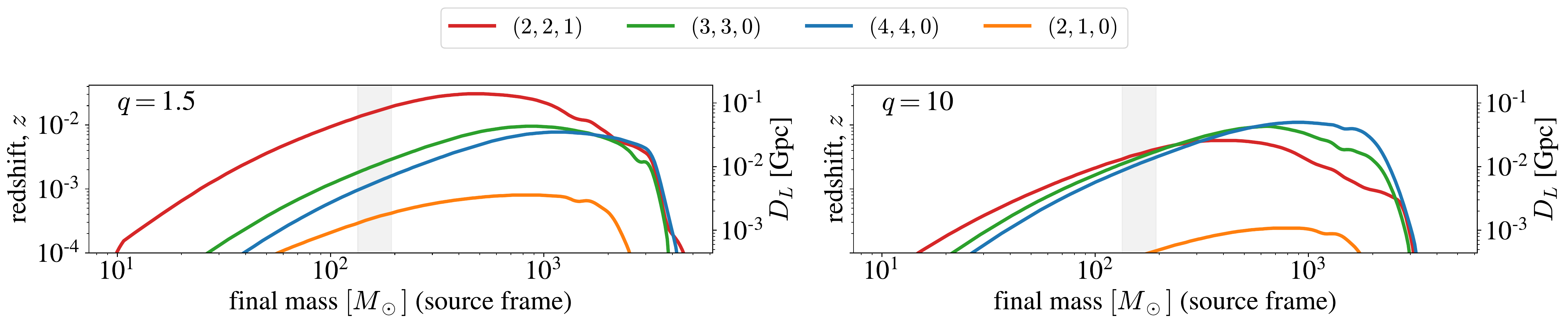}
	\caption{Zoom-in on the LIGO curves from Figure~\ref{fig:rayleigh_horizons}.
	The gray vertical band indicates the estimated mass of the most massive event GW190521.
	For known BBH sources (masses smaller than GW190521), the overtone has larger Rayleigh horizons for small and large mass ratios.}
	\label{fig:rayleigh_horizons_LIGO}
\end{figure}

Figure~\ref{fig:rayleigh_horizons} shows the BH spectroscopy horizons using the Rayleigh criterion, that is, the maximum distance at which the dominant mode $(2,2,0)$ and a subdominant mode $(\ell,m,n)$ satisfy the Rayleigh conditions~\eqref{eq:rayleigh_both}.
We also impose the detectability condition $\rho_{\ell m n} > 8$, which is equivalent to the ``ringdown horizons'' of~\cite{Baibhav:2018rfk}, at the reference time $t = t_{\mathrm{peak}} + 10 M$.
We consider the horizons for low mass ratio $q = 1.5$ (top) and high mass ratio $q=10$ (bottom), with QNM values given in Table~\ref{tab:qnm_pars}.
We do not consider the equal mass case because, in this case, modes with odd $m$ are not excited (see Figure~\ref{fig:mass_ratio_fits}).
The solid curves indicate the horizons for LIGO (for masses $M_f< 10^4$) and LISA (for masses $M_f>10^4$), the dotted curves are the horizons for CE and the dashed curves are the ET horizons.
The circles indicate the BBH observed by the LVKC up to the O3a observing run~\cite{LIGOScientific:2018mvr, LIGOScientific:2020ibl, LIGOScientific:2021usb}.
Events inside the region delimited by the curves satisfy the Rayleigh criterion, and events outside the curve do not.
The horizons for LIGO are also shown separately in Figure~\ref{fig:rayleigh_horizons_LIGO}.

We can see that all the curves have similar trends, that is, the way the horizons depend on the mass does not change when different detectors are considered.
The horizons are small for lower masses because the amplitude of the gravitational wave is proportional to the mass.
As the mass increases the horizon gets larger, until it reaches a peak and starts to decrease.
This is because the frequency of oscillation is inversely proportional to the mass, and the SNR decreases as the frequency approaches the detectors' low-frequency sensitivity limit (see Figure~\ref{fig:detectors_Sn}).

For LIGO, the horizon of the overtone is 10 to 80 times larger than the higher harmonic horizon for masses smaller than the mass at the peak of the overtone horizon.
For larger masses, modes with $\ell > 2$ have larger horizons, which is expected, as these modes have higher frequencies (see Table~\ref{tab:qnm_pars}).
The $(2,1,0)$ mode will always have the smallest horizon, as its frequency and damping time are very similar to the dominant mode's values.

There are two unique behaviours of the LISA curves ($M>10^4$), which distinguish them from the curves relevant for the ground-based detectors ($M\lesssim10^4$).
First, for low mass ratio ($q=1.5$) the overtone has the largest horizon for all masses.
This is because the LISA noise curve does not abruptly increase for low frequencies (see Figure~\ref{fig:detectors_Sn}), which is the case for the ground-based detectors due to the seismic noise.
Second, the LISA horizons increase vertically in the source frame (left plots).
This is a counterintuitive behaviour, as there seems to be two maximum distances up to which the Rayleigh criterion is satisfied.
This happens because LISA can detect systems at cosmological distances, and there is an interplay between the cosmological redshift and the colored noise.
Black holes with low mass emit GW with higher frequencies, and they should have high amplitudes (low distances) to compensate the high frequency noise.
As the BH gets further the amplitude decreases and the signal is buried in the noise, but for distances far enough, the cosmological redshift decreases the detected frequency and the signal can be detected as the frequencies lie on the detector's most sensitive frequency band.
In the detector frame the final masses are also scaled with the cosmological redshift, and these trends are not observed.

As GW190521~\cite{LIGOScientific:2020ufj} is the most massive stellar mass BH detected to date, the Rayleigh criterion predicts that, for detections compatible with the already observed BBHs, the first overtone will be easier to detect than any higher harmonic, even for high mass ratio binaries (but see Section~\ref{sec:bayes-horizons}).
For larger masses the harmonics should be easier to detect, however the existence of more massive stellar mass BHs is still uncertain.
The LISA sources are even more uncertain, and the population of supermassive binaries is still undetermined.
Notwithstanding, LISA will be extremely sensitive and will probably be able to distinguish more than one secondary mode of a detected BBH ringdown.

\section{Bayesian inference}
\label{sec:bayes-horizons}

The Rayleigh criterion guarantees that the frequencies and damping times of a pair of QNMs are distinguishable with the precision of equations~\eqref{eq:rayleigh_both}.
As an alternative to the Rayleigh criterion, model selection can be used to confirm the presence of subdominant modes in the ringdown signal.

Let us consider a ringdown signal $d(t) = d_2(t)$, which is described by the noise $n(t)$, the dominant mode $(2,2,0)$ and a subdominant mode $(\ell, m, n)$, that is,
\begin{equation}
	d_2 = n + h_{220} + h_{\ell mn},
	\label{eq:data_2modes}
\end{equation}
where we dropped the time dependence to simplify the notation, $h_{220}$ and $h_{\ell mn}$ are the injected QNMs with the parameters given by Table~\ref{tab:qnm_pars} and $n$ is the injected noise.
We will work in the frequency domain, that is, the signals and noise will be injected in the frequency domain, and the noise is generated from a Gaussian distribution with zero mean and $\sigma = \sqrt{S_n(f)/ df}/2$, where $S_n(f)$ is the noise spectral density of the considered detector and $df$ is the frequency step.
To compute the BH spectroscopy horizon, we set a threshold on the Bayes factor~\eqref{eq:bayes-factor}.
As we saw in Chapter~\ref{ch:gw-detection}, the Bayes factor depends on the data, the models and on the prior distributions $\pi(\vartheta, {\cal M})$.

For the two-mode analysis, we consider the following models
\begin{itemize}
    \item $\mathcal{M}_1$, a \emph{single-mode} model with 4 parameters $\vartheta_{1} = \{ A, \phi_{220}, f_{220}, \tau_{220}\}$,
    \item $\mathcal{M}_2$, a \emph{two-mode} model with 8 parameters $\vartheta_{2} = \{A$, $\phi_{220}$, $f_{220}$, $\tau_{220}$, $R_{\ell mn}$, $\phi_{\ell m n}$, $f_{\ell m n}$, $\tau_{\ell m n} \}$,
\end{itemize}
where $R_{\ell m n} = A_{\ell m n}/A$ is the ratio between the amplitude of the $(\ell,m,n)$ subdominant mode and the amplitude of the dominant mode $(2,2,0)$.
The global amplitude parameter $A$ is proportional to $M_f/D_L$ (see equation~\eqref{eq:qnm_polarizations}), and all the parameters are defined in the detector frame.

The prior distribution is the initial information we have about the model parameters.
As the spectroscopic analysis looks for small contributions in an already detected signal, we can consider priors informed by the mass and redshift (distance) of the remnant BH\@.
We assume that the analysis of the IMR determined the mass and redshift with 50\% uncertainties, in the 90\% credible interval, that is, $[M_{\mathrm{min} },M_{\mathrm{max} }] = [0.5M_f,1.5M_f]$ and $[z_{\mathrm{min} },z_{\mathrm{max} }] = [0.5z,1.5z]$, where $M_f$ and $z$ are the injected remnant mass and redshift, respectively.
This assumption is not restrictive in a ringdown analysis, as all the events used in LIGO ringdown analysis have parameters estimated with precision inside this interval~\cite{LIGOScientific:2016lio,LIGOScientific:2020tif,LIGOScientific:2021sio}.

The QNM parameters also depend on the spin $a$.
Nonetheless, the dependence is very small (except for the amplitude ratio $R_{\ell m n}$) and we will fix $a$ to the actual value in each case.
Therefore, the prior distributions considered are strongly informed by the Theory of General Relativity, as following:

The prior distributions for the dominant mode $(2,2,0)$, valid for both models  ${\cal M}_1$ and ${\cal M}_2$, are
\begin{itemize}
\item $\pi(A;\mathcal{M}_{(1,2)})$ is log-uniform in the interval $[A_{\mathrm{min} },A_{\mathrm{max} }]$, where $A_{\mathrm{min}} = A(M_{\mathrm{min}},z_{\mathrm{max}},a)/10$ and $A_{\mathrm{max}} = 10\times A(M_{\mathrm{max}},z_{\mathrm{min}},a)$,
\item $\pi(\phi_{220};\mathcal{M}_{(1,2)})$ is uniform in the interval $[0,2\pi]$,
\item $\pi(f_{220};\mathcal{M}_{(1,2)})$ is log-uniform in the interval $[f^{\mathrm{min}}_{220},f^{\mathrm{max}}_{220}]$, where $f^{\mathrm{min}}_{220} = f_{220}(M_{\mathrm{max}},z_{\mathrm{max}},a)$ and $f^{\mathrm{max}}_{220} = f_{220}(M_{\mathrm{min}},z_{\mathrm{min}},a)$,
\item $\pi(\tau_{220};\mathcal{M}_{(1,2)})$ is uniform in the interval $[\tau^{\mathrm{min}}_{220},\tau^{\mathrm{max}}_{220}]$, where $\tau^{\mathrm{min}}_{220} = \tau_{220}(M_{\mathrm{min}},z_{\mathrm{min}},a)$ and $\tau^{\mathrm{max}}_{220} = \tau_{220}(M_{\mathrm{max}},z_{\mathrm{max}},a)$,
\end{itemize}
and the prior distributions for the subdominant mode $(\ell,m,n)$ in model ${\cal M}_2$ are
\begin{itemize}
\item $\pi(R_{\ell mn};\mathcal{M}_{2})$ is uniform in the interval $[0,0.9]$,
\item $\pi(\phi_{\ell m n};\mathcal{M}_{2})$ is uniform in the interval $[0,2\pi]$,
\item $\pi(f_{\ell mn};\mathcal{M}_{2})$ and $\pi(\tau_{\ell mn};\mathcal{M}_{2})$ are defined similarly to $\pi(f_{220};\mathcal{M}_{(1,2)})$ and $\pi(\tau_{220};\mathcal{M}_{(1,2)})$, but considering the subdominant mode $(\ell,m,n)$ and the model $\mathcal{M}_2$.
\end{itemize}

The log-uniform priors in the global amplitude and frequencies are scale-invariant priors.
As frequencies are inversely proportional to the mass, a uniform prior would result in a preference for smaller masses.

To check if the models ${\cal M}_1$ and ${\cal M}_2$ can describe well the signal $d_2 = n + h_{220} + h_{\ell m n}$, we compute the deviations between the injected parameters $\vartheta^\alpha_{\mathrm{inj}}$ and the posterior probability distribution $\vartheta^\alpha$ of each model.
We calculate the deviations in units of ``$\sigma$'', associated with the normal distribution, obtained from the quantile function $Q(x)$ of the integral
\begin{equation}
x = \int_{-\infty}^{\vartheta_{\mathrm{inj}}^\alpha}p(\vartheta^{\alpha}|d_2,\mathcal{M}_{(1,2)})\ d\vartheta^{\alpha}.
\label{eq:x_gaussian}
\end{equation}
\begin{figure}[!htb]
	\centering
	\includegraphics[width=1\linewidth]{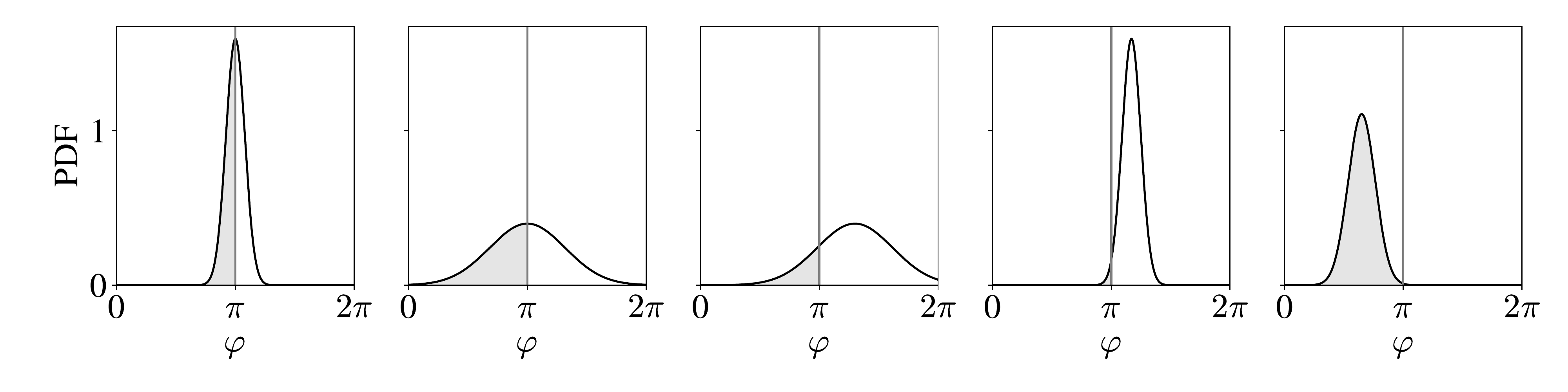}
	\caption{Illustrative examples of the posterior distributions of a fictional phase parameter $\varphi\in [0, 2\pi]$.
	The vertical lines indicate the injected fictional value $\varphi_{\mathrm{inj}} = \pi$ and the gray region represents the integration region of equation~\eqref{eq:x_gaussian}.
	The first two panels on the left are examples in which the deviation between the injected and the estimated parameter is zero, as the peak of the distribution is centered in the injected parameter.
	The third, fourth and fifth panels represent  $-1\sigma$, $-2\sigma$ and $3\sigma$ deviations, respectively.}
	\label{fig:gaussian_densities}
\end{figure}

Figure~\ref{fig:gaussian_densities} shows some illustrative examples for posterior probability distributions of a fictional phase parameter $\varphi \in [0, 2\pi]$, with the injected value  $\varphi_{\mathrm{inj}} = \pi$, indicated by the vertical lines.
On the first two panels on the left the posterior distribution is centered at $\pi$ and, in both cases, $x = 0.5$ and $Q(x) = 0$, but the uncertainty of the second panel is larger.
In the third panel the peak is slightly larger than $\pi$ and the integral value is $x \approx 0.159$ and the deviation (in units of $\sigma$) is $Q(x) \approx -1$.
In the fourth panel the peak has the same peak value as the third panel, but the uncertainty is much smaller.
In this case $x \approx 0.02$ and the deviation is $Q(x)\approx -2$.
In the last panel the peak is significantly smaller than $\pi$ and the uncertainty of the distribution is relatively small, in this case  $x\approx0.999$ and $Q(x) \approx 3$.

\begin{figure}[!htb]
	\centering
	\includegraphics[width=.8\linewidth]{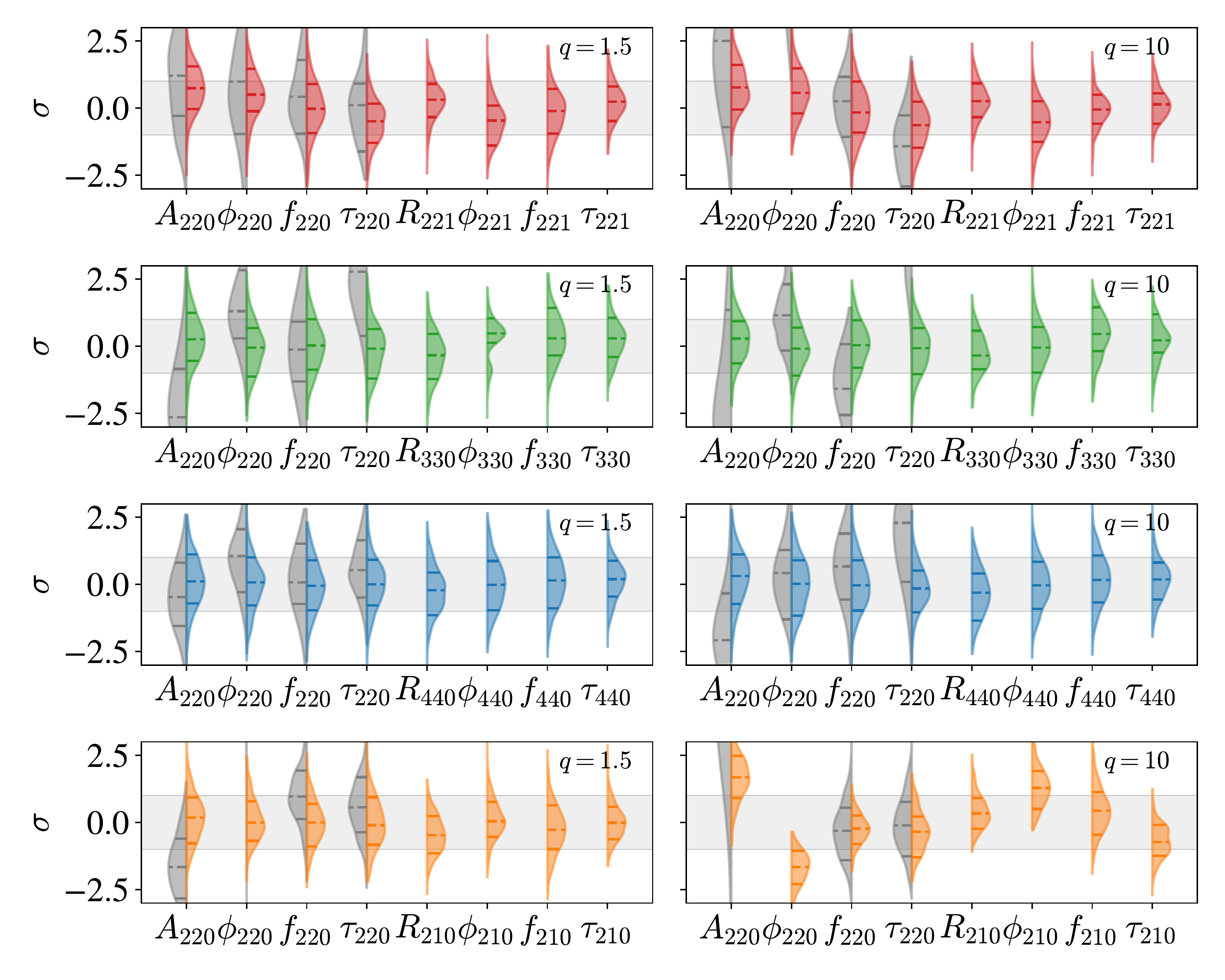}
	\caption{Violin plots of the distributions of the deviations between the injected values $\vartheta^a_{\mathrm{inj}}$ and the posterior probability distribution $\vartheta^\alpha$, for signals $d_2 = n + h_{220} + h_{\ell mn}$, and models $\mathcal{M}_1$ (gray) and $\mathcal{M}_2$ (colored), for nonspinning circular binaries with mass ratios $q = 1.5$ (left) and $q = 10$ (right).
	The gray band indicates the $\pm 1\sigma$ interval.
	The model that neglects the subdominant mode $\mathcal{M}_1$ produces results that are more biased and have larger deviations.}
	\label{fig:desvios_violin}
\end{figure}

Figure~\ref{fig:desvios_violin} shows as violin plots the distribution of the deviations between injected and estimated parameters for the models $\mathcal{M}_1$ (gray left-sided violins) and $\mathcal{M}_2$ (colored right-sided violins) for signals $d_2 = n + h_{220} + h_{\ell mn}$.
For each subdominant mode $(\ell, m, n)$, we generated 500 signals $d_2$ with independent noise realizations, for 50 masses and 10 redshifts (for each mass), distributed uniformly around the Bayes factor spectroscopy horizon for each subdominant mode (see Figure~\ref{fig:bayes_horizons_2modes}).
We can see that the correct model $\mathcal{M}_2$ can estimate the parameters with a small bias, as the distributions are centered near zero.
$\mathcal{M}_1$ produces larger bias because it neglects part of the signal, thus there are larger deviations.
Notwithstanding, these deviations are not significantly large, as half of the injected events considered in this sample are outside the BH spectroscopy horizon, where the subdominant mode can be neglected.
Thus, both models predict the correct parameters, but $\mathcal{M}_1$ has significant bias when the subdominant mode signal cannot be neglected.

To compute the BH spectroscopy horizons we will use the Bayes factors $\mathcal{B}_1^2$, which quantifies the support of the model $\mathcal{M}_2$ over the model $\mathcal{M}_1$.
We take the detectability criterion $\ln \mathcal{B}_1^2 > 8$, that is, the evidence $\mathcal{Z}_2$ of the two-mode model must be approximately 3000 times larger than the evidence $\mathcal{Z}_1$ of the single-mode model.
\begin{figure}[!htb]
	\centering
	\includegraphics[width=.48\linewidth]{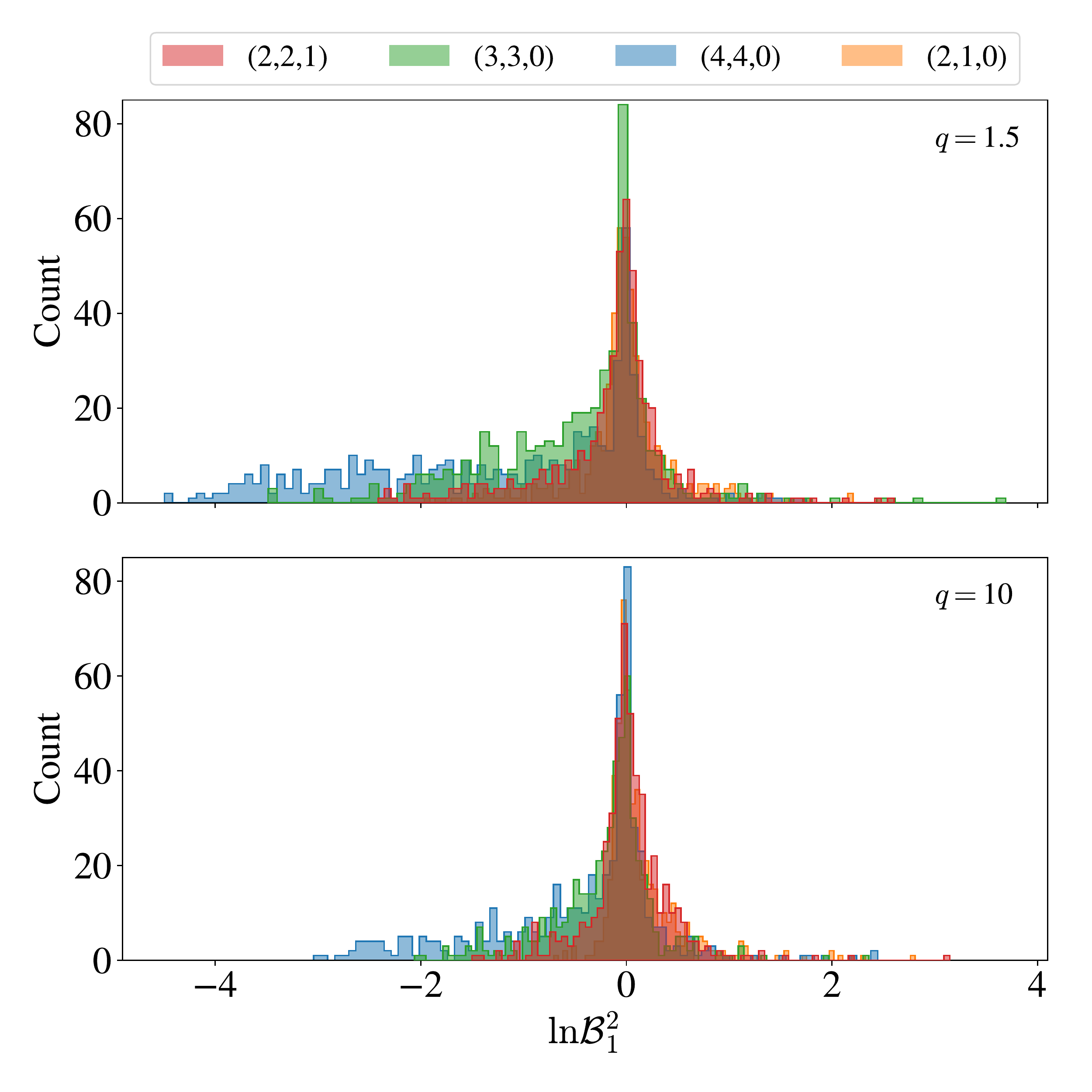}
	\includegraphics[width=.48\linewidth]{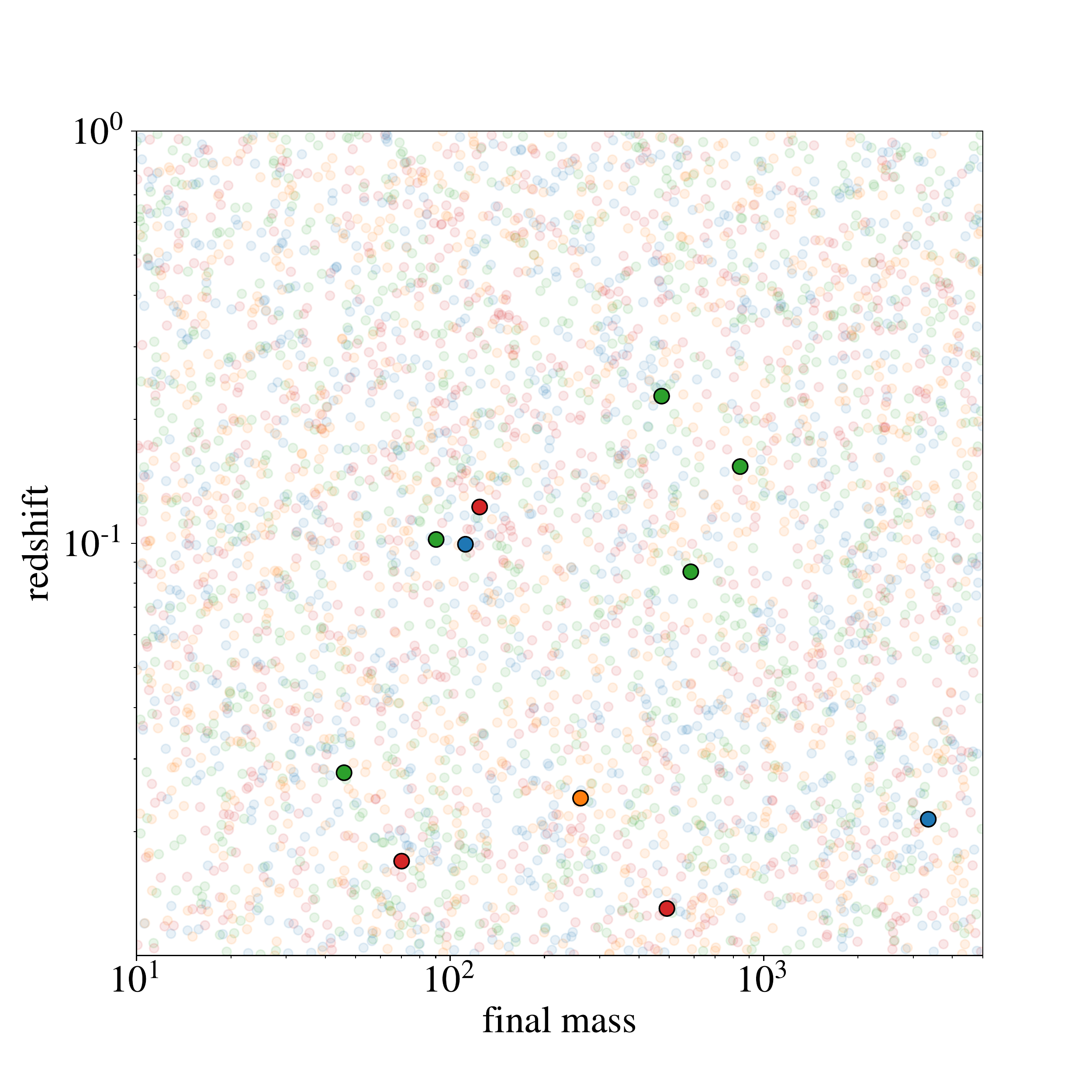}
	\caption{\emph{Left:} histogram of the natural logarithm of the Bayes factor $\mathcal{B}_1^2$, of the two-mode model $\mathcal{M}_2$ over the single-mode model $\mathcal{M}_1$, for 2000 signals containing just the dominant mode $d_1 = n + h_{220}$, for the binary mass ratios $q = 1.5$ (top) and $q = 10$ (bottom).
	We considered 2000 signals with $M_f \in [1, 5\times 10^3] M_{\odot}$ and $z\in [10^{-2}, 1]$, 500 for each prior for the secondary mode in the $\mathcal{M}_2$ model:  $(2,2,1)$ (red), $(3,3,0)$ (green), $(4,4,0)$ (blue) e $(2,1,0)$ (orange).
	For most of the signals, the correct model $\mathcal{M}_1$ is favored ($\ln\mathcal{B}^2_1 \lesssim 0$), but there are some cases that the incorrect model is slightly favored $\mathcal{M}_2$ ($2 \lesssim \ln\mathcal{B}^2_1 \lesssim 4$).
	Our threshold ($\ln \mathcal{B}_1^2>8$) for detectability is more than 2 times larger than the largest $\ln \mathcal{B}_1^2$ found in these injections.
	\emph{Right:} mass and redshift of the injected signals. The outliers found  ($2 \lesssim \ln\mathcal{B}^2_1 \lesssim 4$) are highlighted and there is no visible pattern for the outliers.
	This indicates that the support for the incorrect model $\mathcal{M}_2$ is due to the noise injection.}
	\label{fig:hist_logB}
\end{figure}

To check whether this threshold is large enough we compute the Bayes factor $\mathcal{B}_1^2$ for signals that do not contain the secondary mode, that is, the signal is described as $d_1 = n + h_{220}$.
For these signals, any support for the two-mode model $\mathcal{M}_2$ is caused by the noise.
We computed the Bayes factor $\mathcal{B}_1^2$ of 2000 signals $d_1$ with $M_f \in [1, 5\times 10^3] M_{\odot}$ and $z\in [10^{-2}, 1]$, 500 for each prior for the secondary mode in the $\mathcal{M}_2$ model.
These results are shown in Figure~\ref{fig:hist_logB} on the left.
As expected, most events favor the correct mode $\mathcal{M}_1$, with $\ln {\cal B}^2_1 \lesssim 0$.
None of the injected events had $\ln {\cal B}^2_1 > 4$, however 11 events (0.6\% of all cases) have $2 \lesssim \ln\mathcal{B}^2_1 \lesssim 4$ ($10 \lesssim {\cal B}^2_1 \lesssim 55$), which are values considered as ``strong'' evidence according to the Kass-Raftery scale~\cite{Kass:1995loi}.
Figure~\ref{fig:hist_logB} shows on the right the mass and redshift of all injected events, and the 11 outliers are highlighted.
We cannot identify any pattern for the outliers, which indicates that the high Bayes factor is caused by the noise injection.
This result shows that such generic scales, such as the  Kass-Raftery scale, are not appropriate for different types of data (or noise) and they should be avoided when a more detailed analysis is possible.

\begin{figure}[!htb]
	\centering
	\includegraphics[width=.75\linewidth]{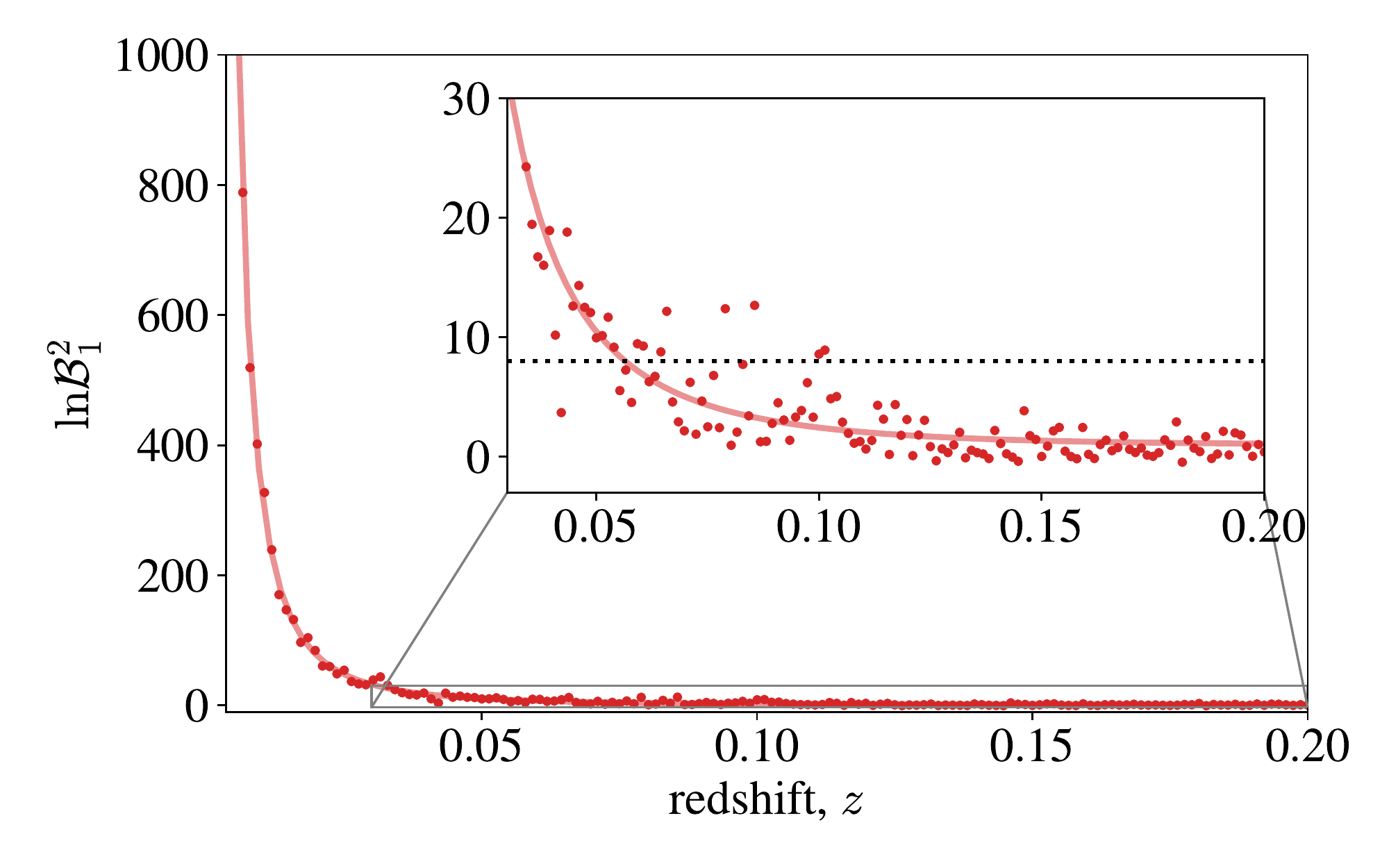}
	\caption{Bayes factor between a two-mode model $\mathcal{M}_2$, considering the overtone $(2,2,1)$ as the secondary mode, and a single-mode model $\mathcal{M}_1$ as a function of the redshift.
	In this example we generated 150 injected signals $d_2 = n + h_{220} + h_{221}$ with the fixed final mass $M_f = 156.4 M_\odot$, mass ratio $q=1.5$ and LIGO noise realizations.
	The solid curve is a Laurent polynomial $c_1 + c_2z^{-1} + c_3 z^{-2}$ fitted to the points.
	The redshift where $\ln {\cal B}^2_1 = 8$ is determined using the fitted curve.}
	\label{fig:bayes_redshift_fit}
\end{figure}

\begin{figure}[!htb]
	\centering
	\includegraphics[width=.49\linewidth]{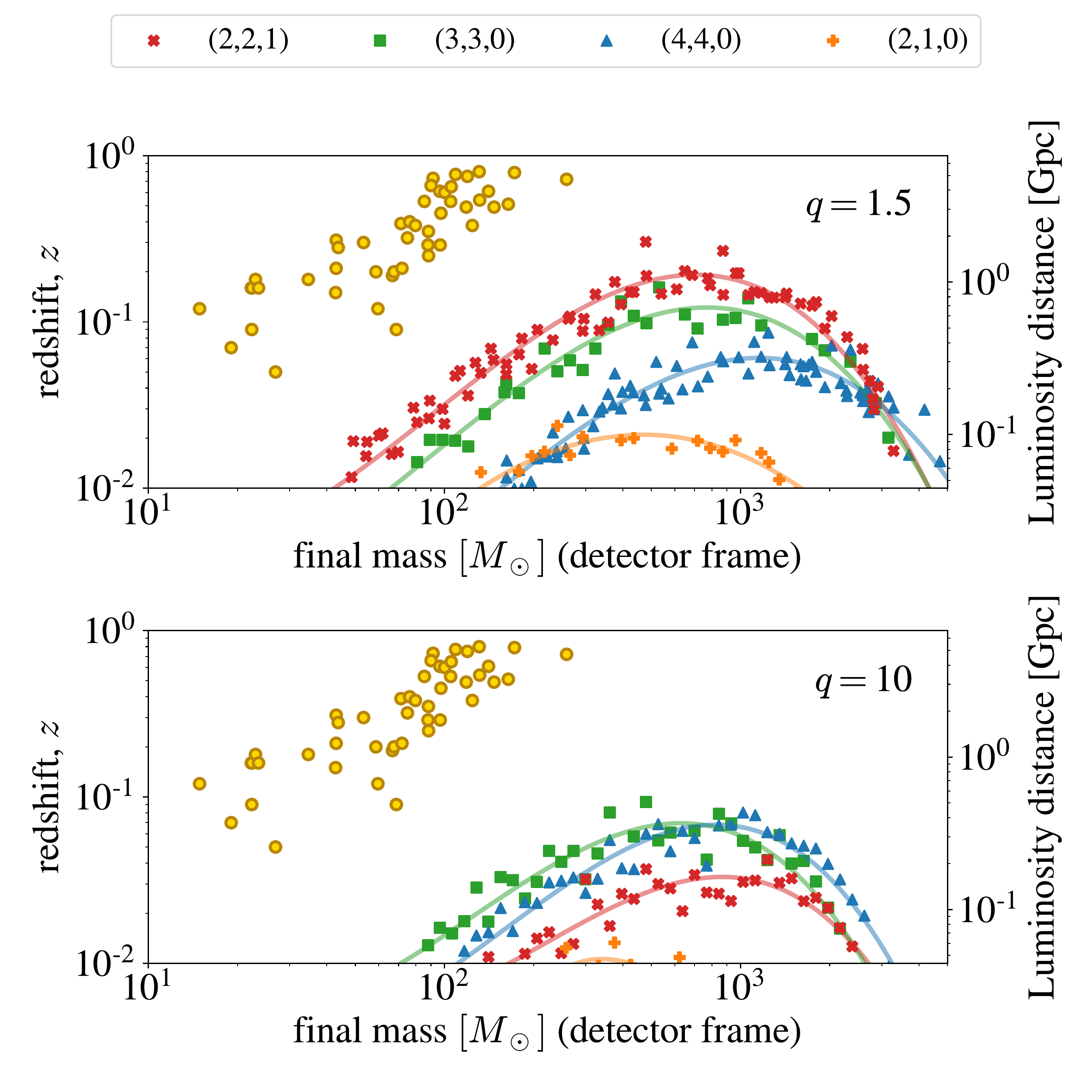}
	\includegraphics[width=.49\linewidth]{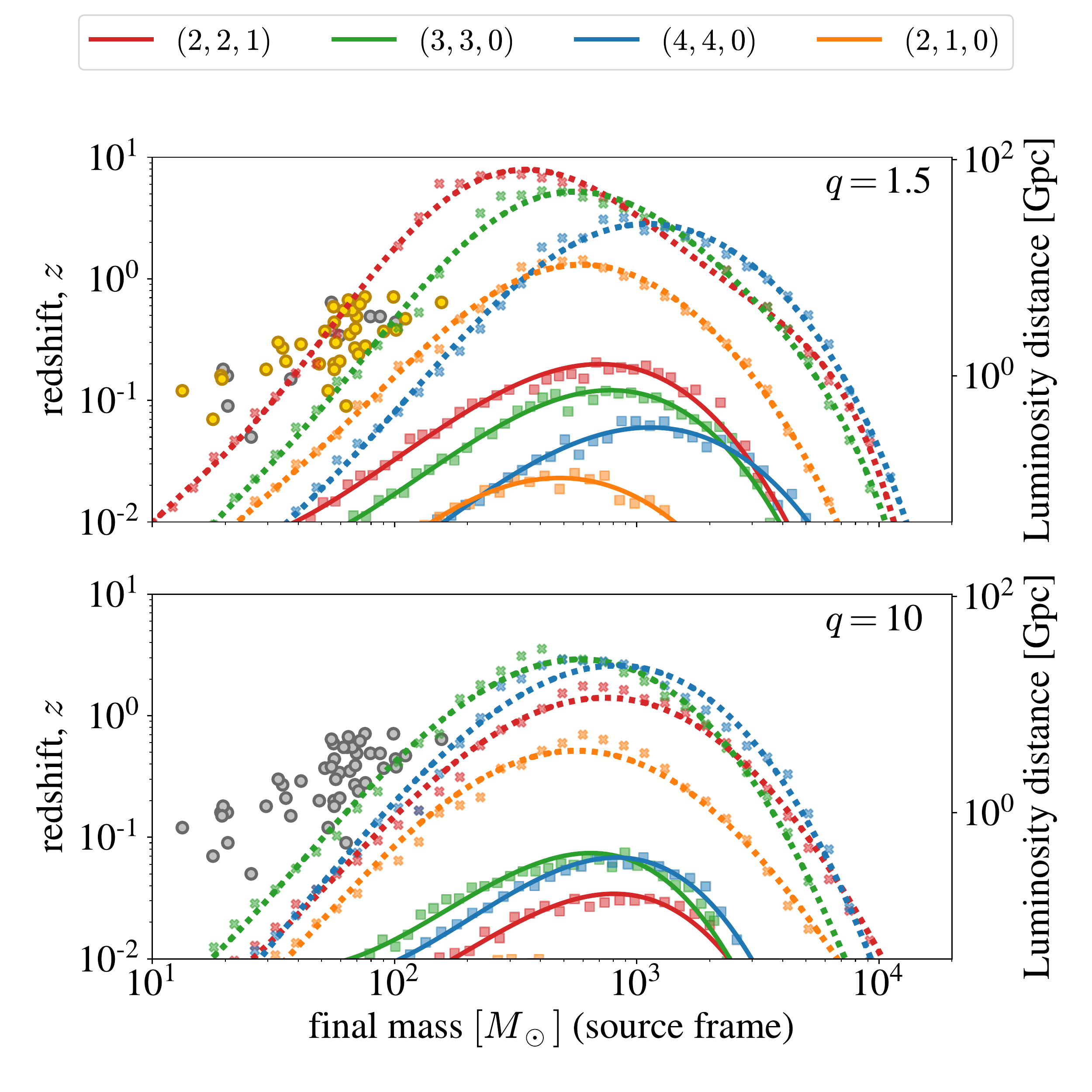}
	\caption{BH spectroscopy horizons for signals $d_2 = n + h_{220} + h_{\ell mn}$, with the Bayes factor threshold $\ln {\cal B}^2_1 > 8$. \emph{Left}: rough estimate of the BH spectroscopy horizons for the LIGO detector, before the refinement as described in the text.
	\emph{Right}: refined BH spectroscopy horizons for LIGO (squared and solid lines) and CE (crosses and dashed lines).
	We can see that after the refinement the points are more uniformly distributed with low scattering in the redshift.
	In both plots we considered the secondary modes $(2,2,1)$ (red), $(3,3,0)$ (green), $(4,4,0)$ (blue) and $(2,1,0)$ (orange), and nonspinning circular binaries with mass ratios $q = 1.5$ (top) and $q = 10$ (bottom).
	The yellow (gray) circles are the detected events (not) compatible with the NR simulation considered.}
	\label{fig:bayes_horizons_2modes}
\end{figure}

The computational cost required to compute Bayes factors is extremely high, and the time needed to find the root $\ln\mathcal{B}_{1}^{2} - 8 = 0$
with independent noise realizations can be extremely long if we use standard root-finding algorithms.
Nonetheless, for a fixed mass the Bayes factor is monotonic as a function of the redshift, with a dependence similar to a Laurent polynomial  $c_1 + c_2z^{-1} + c_3 z^{-2}$.
Figure~\ref{fig:bayes_redshift_fit} shows an example of 150 injected signals $d_2 = n = h_{220} + h_{221}$ of events with fixed mass and mass ratio.
The monotonic behaviour is clear, but there is a large scatter for close Bayes factor values (see the inset plot).
This scatter is caused by different noise realizations, and the curve would be much smoother if all the events had the same noise.
We choose the redshift where  $\ln {\cal B}^2_1 = 8$ by fitting a Laurent polynomial  $c_1 + c_2z^{-1} + c_3 z^{-2}$ to the points and finding its root (solid line).

To obtain the spectroscopy horizon we first compute many events with different redshifts distributed in a logarithm scale, in the intervals of Figure~\ref{fig:rayleigh_horizons}, for a fixed mass and find the root with the Laurent polynomial as described above.
As the redshifts are very spread out, this give us an rough estimation for the Bayes factor BH horizons, shown in Figure~\ref{fig:bayes_horizons_2modes} on the left.
Then, we refine the results by computing 10 (30) events distributed within $\pm50\%$ around the first estimated horizon, for the LIGO (CE) detector.
This refinement step results in BH horizon points with small scattering, as shown in Figure~\ref{fig:bayes_horizons_2modes} on the right.
To obtain the horizon curves we fit polynomials to the obtained points, whose coefficients are found in Tables~\ref{tab:fits_2_modes} (LIGO) and~\ref{tab:fits_2_modes_ce} (CE).

\begin{table}[!htb]
	\centering
	\caption{Fitting coefficients for the BH spectroscopy horizons for the LIGO detector shown in Figure~\ref{fig:bayes_horizons_2modes} on the right, where we considered the threshold $\ln \mathcal{B}_1^2 = 8$.
	The polynomials were fitted in the logarithm scale: $\ln z^{\rm spec, B}_{\ell mn} = a_0 + a_1 \ln M_f + a_2 (\ln M_f)^2 + a_3 (\ln M_f)^3$, where $z^{\rm spec, B}_{\ell mn}$ is the redshift at the horizon distance and $M_f$ is the mass of the remnant BH.
	The fits are only valid in the interval shown in Figure~\ref{fig:bayes_horizons_2modes}.}
	\begin{tabular*}{\textwidth}{c |@{\extracolsep{\fill}} c c c c}  \hline\hline
		coefficient & $(2,2,1)$ & $(3,3,0)$ & $(4,4,0)$ & $(2,1,0)$ \\ \hline\hline
		& \multicolumn{4}{c}{$q = 1.5$}\\ \hline
		$a_0$	&	-0.6685	&	-0.7298	&	-0.5887	&	-0.0022	\\
		$a_1$	&	4.0504	&	4.6084	&	3.8802	&	-1.3164	\\
		$a_2$	&	-6.8045	&	-8.3415	&	-7.1914	&	7.0597	\\
		$a_3$	&	1.2680	&	2.3177	&	1.3209	&	-11.071	\\\hline
		 & \multicolumn{4}{c}{$q = 10$}\\ \hline
		$a_0$	&	-0.8621	&	-0.9526	&	-1.1328	&	0	\\
		$a_1$	&	5.7614	&	6.1710	&	7.8589	&	0	\\
		$a_2$	&	-11.651	&	-12.138	&	-16.930	&	0	\\
		$a_3$	&	4.8845	&	5.3732	&	9.4642	&	0	\\ \hline\hline
	\end{tabular*}
	\label{tab:fits_2_modes}
\end{table}
\begin{table}[!htb]
\caption{Same as Table~\ref{tab:fits_2_modes}, but for the CE detector and higher degree polynomials.}
	\begin{tabular*}{\textwidth}{c |@{\extracolsep{\fill}} c c c c }  \hline\hline
		coefficient & $(2,2,1)$ & $(3,3,0)$ & $(4,4,0)$ & $(2,1,0)$  \\ \hline\hline
		& \multicolumn{4}{c|}{$q = 1.5$}\\ \hline
		$a_0$	&	-0.2331 	&	-0.2081 	&	-0.1342		&	-0.1491 	\\
		$a_1$	&	3.3135 		&	3.1694 		&	2.1851		&	 2.3180 	\\
		$a_2$	&	-18.5618 	&	-19.3023 	&	-14.4192	&	-14.5241 	\\
		$a_3$	&	51.7728 	&	59.5351 	&	48.7304		&	46.3409 	\\
		$a_4$	&	-75.6249 	&	-98.0044 	&	-88.7223	&	-79.3446 	\\
		$a_5$	&	57.1219 	&	84.0752 	&	84.6849		&	71.1739		\\
		$a_6$	&	-19.7984 	&	-32.0157 	&	-36.0500	&	-28.7501 	\\ \hline
		& \multicolumn{4}{c}{$q = 10$}\\ \hline
		$a_0$	&	0.1135  	&	0.0647  	&	-0.0107		&	0.1703	\\
		$a_1$	&	-1.5090  	&	-1.0693  	&	-0.3054		&	-1.9806	\\
		$a_2$	&	5.8306  	&	4.1535  	&	1.4965		&	7.0495	\\
		$a_3$	&	-6.9124  	&	-3.9086  	&	0.1276		&	-8.0004	\\
		$a_4$	&	-0.0702  	&	-1.6608  	&	-4.3371		&	-0.1510	\\ \hline\hline
	\end{tabular*}
\label{tab:fits_2_modes_ce}
\end{table}

For $q = 1.5$ and masses smaller than $\sim 2000 M_{\odot}$ ($\sim 800 M_{\odot}$) for LIGO (CE), the overtone $(2,2,1)$ has the largest horizon, followed by the $(3,3,0)$, $(2,1,0)$ and $(4,4,0)$ horizons, respectively.
For larger masses the $(4,4,0)$ has the largest horizon, which is expected, as the $(4,4,0)$ mode has the highest frequency.
For the $q = 10$ case, the higher harmonics $(3,3,0)$ and $(4,4,0)$ have horizons larger than the overtone's for all masses.
In this case, the $(2,1,0)$ has consistently the smallest horizon.

The BH spectroscopy horizons depend on the amplitude, frequency of oscillation and damping time of each QNM mode.
For $q = 1.5$ the overtone $(2,2,1)$ has an amplitude at least 5 times higher than the amplitude of the fundamental higher harmonics (see Table~\ref{tab:qnm_pars}), which results in a larger horizon.
However, for $q = 10$ the amplitude of the overtone is still larger than the amplitude of the harmonics, but the $(3,3,0)$ and the $(4,4,0)$ modes have larger horizons.
This indicates that the difference between the frequency of the dominant $(2,2,0)$ mode and the frequency of the subdominant mode is more relevant than the difference in the damping times, as the $(3,3,0)$ and $(4,4,0)$ have damping times very similar to the $(2,2,0)$ damping time but very distinct frequencies, whilst the overtone and the dominant mode have very distinct damping times and very similar frequencies (see also Figure~\ref{fig:qnm_df_dtau_a}).
Furthermore, the peak of the $(4,4,0)$ horizon is approximately at the same distance as the $(3,3,0)$ peak (but for larger mass), even though the amplitude of the $(3,3,0)$ mode is more than twice the amplitude of the $(4,4,0)$.
This again indicates that the high frequency of the $(4,4,0)$ mode facilitates its detection, as the frequency remains in the detector's most sensitive band for more massive BHs, which results in a higher SNR for the mode.

The $(2,1,0)$ mode has consistently the smallest horizon for the high mass ratio case ($q = 10$).
Although its amplitude is larger than the $(4,4,0)$ amplitude and similar to the $(3,3,0)$ and $(2,2,1)$ amplitudes.
The smaller horizon is due to the fact that \emph{both} frequency and damping time of the $(2,1,0)$ mode are extremely similar to the $(2,2,0)$ values,
which makes it very difficult to distinguish a two-mode signal from a single-mode signal with amplitude $\sim A_{220} + A_{210}$.

These results indicate the importance of \emph{both} $\Delta f_{220, \ell mn}$ and $\Delta \tau_{220, \ell mn}$ used in the Rayleigh criterion~\eqref{eq:rayleigh_both}.
The qualitative behaviour of the Bayes factor horizons is very similar to the Rayleigh horizon (Figure~\ref{fig:rayleigh_horizons}), although the distance of the horizon and the order of the modes is different.
The difference between both methods is discussed in Section~\ref{sec:comparison-methods}.

Considering the rates for BBH mergers with primary masses $45 M_{\odot} < M_1 < 100 M_{\odot}$ estimated as $0.70^{+0.65}_{-0.35}\ {\rm Gpc}^{-3}\ {\rm year}^{-1}$~\cite{LIGOScientific:2020kqk}, we compute the comoving volume~\cite{Hogg:1999ad} of the $(2,2,1)$ mode horizon for $q = 1.5$ and final mass $M = 156.3 M_{\odot}$ (similar to GW190521).
The event rates for Bayes factor BH spectroscopy for LIGO at design sensitivity is  $0.03 - 0.10\ {\mathrm{year}}^{-1}$ and for CE is $(0.6 - 2.3) \times 10^3\ {\mathrm{year}}^{-1}$.

\section{Multimode black hole spectroscopy horizons}
\label{sec:multimode-bh-spectroscopy-horizons}

The results presented above are valid when the tertiary and all the subsequent modes can be neglected.
But from Figure~\ref{fig:bayes_horizons_2modes} we can see that some of different mode horizons are very close, and this approximation may not be valid.
To address a more realistic case, we consider a signal containing the five most relevant modes~\cite{Cotesta:2018fcv}, that is,
\begin{equation}
    d_5 = n + h_{220} + h_{221} + h_{330} + h_{440} + h_{210}.
    \label{eq:data-5modes}
\end{equation}

As we are analysing a signal with many modes we will consider the following models:
\begin{itemize}
    \item $\mathcal{M}_{1}$, same \emph{single-mode} model defined before,
    \item $\mathcal{M}_{2}$, a \emph{two-mode} model, with an \emph{unspecified} secondary mode $(\ell_1,m_1,n_1)$ and 8 parameters $\vartheta_{2} = \{A$, $\phi_{220}$, $f_{220}$, $\tau_{220}$, $R_{\ell_1 m_1n_1}$, $\phi_{\ell_1 m_1 n_1}$, $f_{\ell_1 m_1 n_1}$, $\tau_{\ell_1 m_1 n_1} \}$,
    \item $\mathcal{M}_{3}$, a \emph{three-mode} model, with two \emph{unspecified} subdominant modes $(\ell_1,m_1,n_1)$ and $(\ell_2,m_2,n_2)$  and 12 parameters $\vartheta_{3} = \{A$, $\phi_{220}$, $f_{220}$, $\tau_{220}$, $R_{\ell_1 m_1 n_1}$, $\phi_{\ell_1 m_1 n_1}$, $f_{\ell_1 m_1 n_1}$, $\tau_{\ell_1 m_1 n_1}$, $R_{\ell_2 m_2 n_2}$, $\phi_{\ell_2 m_2 n_2}$, $f_{\ell_2 m_2 n_2}$, $\tau_{\ell_2 m_2 n_2}\}$.
\end{itemize}
The two-mode model $\mathcal{M}_2$ has the same number of parameters as the one used before, but now we consider much broader priors, which are not limited to a specified secondary mode.
Following the same prescription described before, the prior distributions for the secondary mode $(\ell_1,m_1,n_1)$  will be in an interval that allows the values for the most relevant subdominant modes $(2,2,1)$, $(3,3,0)$, $(4,4,0)$ and $(2,1,0)$.
The same prescription is considered for the secondary and tertiary mode of model $\mathcal{M}_3$.

\begin{figure}[!htb]
	\centering
	\includegraphics[width=.49\linewidth]{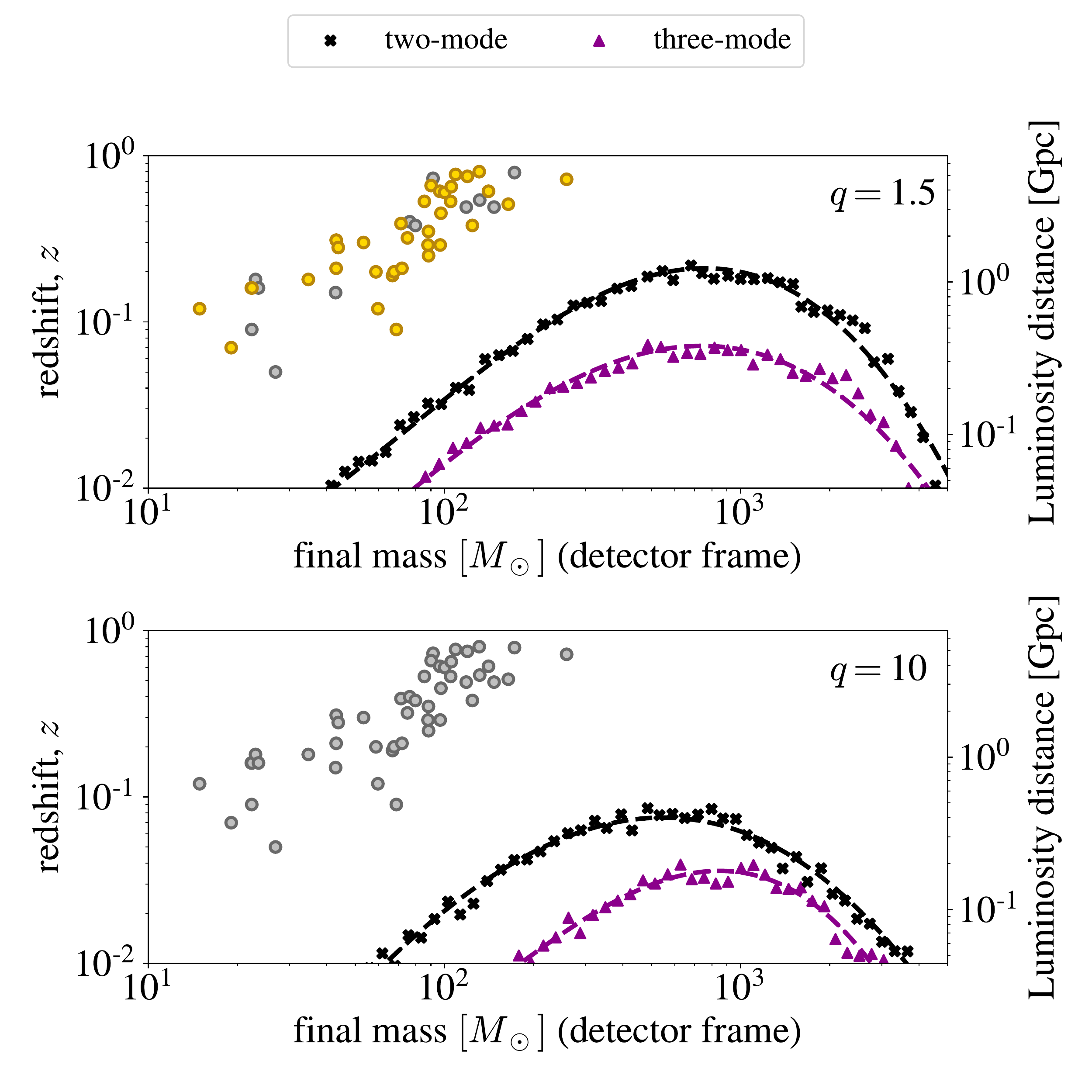}
	\includegraphics[width=.49\linewidth]{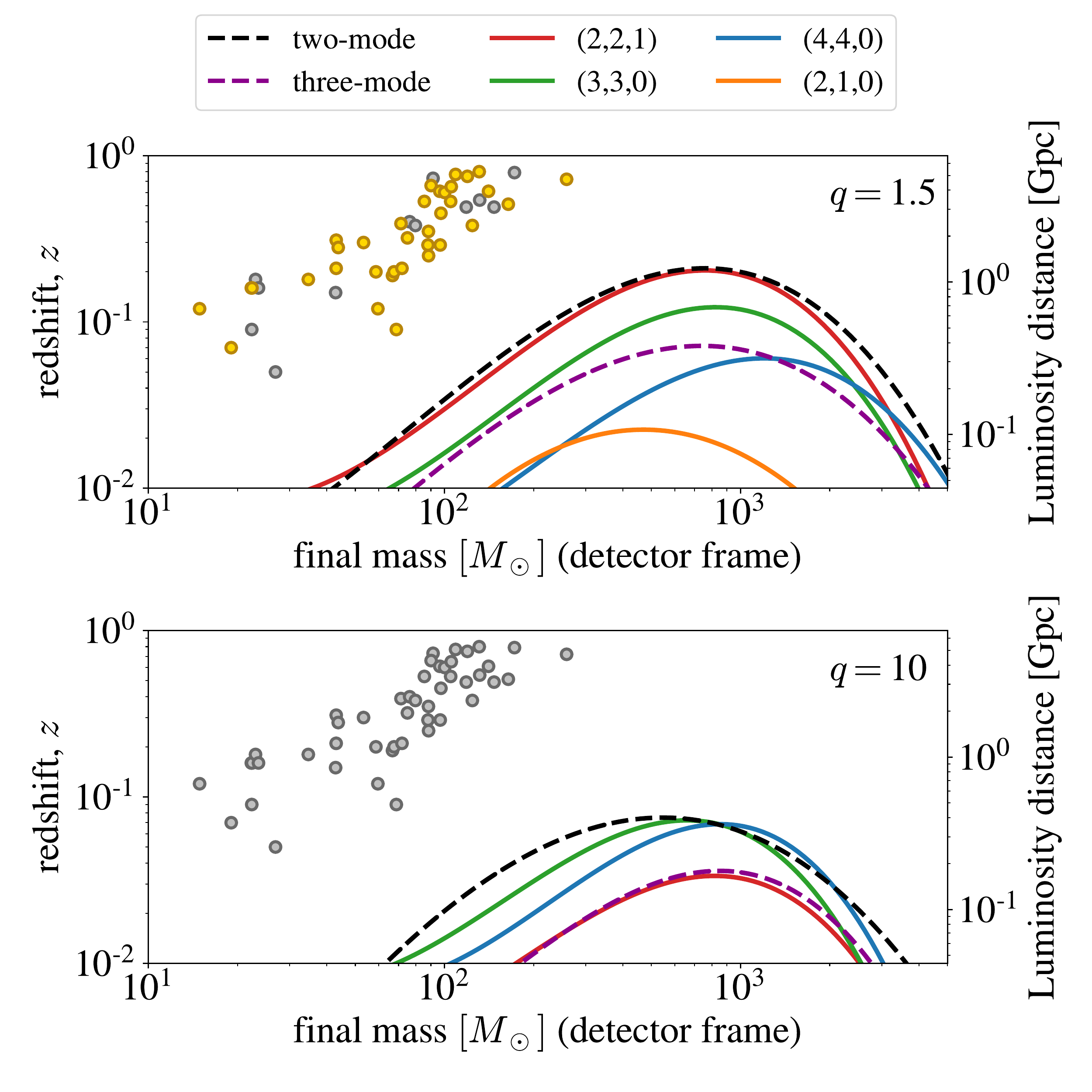}
	\caption{\emph{Left}: LIGO BH spectroscopy horizons for signals $d_5 = n + h_{220} + h_{221} + h_{330} + h_{440} + h_{210}$, with the threshold $\ln {\cal B}^n_{n-1} > 8$, for $n=2$ (black) and $n=3$ (purple).
	\emph{Right}: comparison between the \emph{unspecified} horizons (dashed) and the specified two-mode horizons (solid).
	The unspecified horizon is compatible with the largest specified two-mode horizon and the three-mode horizon is more restrictive.}
	\label{fig:horizon_bayes_multi}
\end{figure}

Figure~\ref{fig:horizon_bayes_multi} on the left shows the unspecified horizons for LIGO, using the threshold $\ln \mathcal{B}^n_{n-1} > 8$,
where the Bayes factor is between the models $\mathcal{M}_n$ and $\mathcal{M}_{n-1}$, for  $n = 3$ and $n = 2$ modes.
As expected, the two-mode horizon (black) is larger than the three-mode horizon (purple).
Moreover, the horizons for $q = 1.5$ are larger than the horizons for $q = 10$, which is also expected because, for the same total mass, nonspinning BBHs with more asymmetric initial masses emit less energy in the form of GW~\cite{Kamaretsos:2011um, Cotesta:2018fcv, London:2014cma}, resulting in modes with smaller amplitudes (see Table~\ref{tab:qnm_pars}).
The dashed curves are fitted polynomials with coefficients given in Table~\ref{tab:fits_n_modes}.

\begin{table}[!htb]
\caption{Fitting coefficients for the BH spectroscopy horizons for the LIGO detector shown in Figure~\ref{fig:horizon_bayes_multi} on the left, where we considered the threshold $\ln \mathcal{B}_{n-1}^n = 8$, where $n = \{2,3\}$ represents the number of unspecified subdominant modes.
	The polynomials were fitted on the logarithm scale: $\ln z^{\textrm{spec, B} }_{\ell mn} = a_0 + a_1 \ln M_f + a_2 (\ln M_f)^2 + a_3 (\ln M_f)^3$, where $z^{\textrm{spec, B} }_{\ell mn}$ is the redshift at the horizon distance and $M_f$ is the mass of the remnant BH.
	The fits are only valid in the interval shown in Figure~\ref{fig:horizon_bayes_multi}.}
\begin{tabular*}{\textwidth}{c |@{\extracolsep{\fill}}c c | c c} \hline\hline
coefficient & 2 modes & 3 modes  & 2 modes & 3 modes \\ \hline\hline
& \multicolumn{2}{c|}{$q = 1.5$}&\multicolumn{2}{c}{$q = 10$}\\ \hline
$a_0$ & -0.4238	 & -0.3022 	 & -0.1121	 & -0.7089	 \\
$a_1$ & 2.2212	 & 1.3808 	 & -0.2280	 & 4.5011	 \\
$a_2$ & -2.3470	 & -0.4939 	 & 3.7164	 & -8.1516	 \\
$a_3$ & -2.2267	 & -3.9584 	 & -7.2915	 &  1.6289	 \\\hline\hline
\end{tabular*}
\label{tab:fits_n_modes}
\end{table}

Figure~\ref{fig:horizon_bayes_multi} on the right shows a comparison between the unspecified two- and three-mode horizons of the signal $d_5$ and the specified horizons obtained from signals $d_2$.
For $q = 1.5$ the unspecified two-mode horizon is compatible with the $(2,2,1)$ horizon, and the three-mode horizon is more restrictive than the second largest specified horizon.
For $q = 10$ the two-mode unspecified horizon is compatible with both the $(3,3,0)$ and the $(4,4,0)$ horizons, which have approximately the same size.
The three-mode horizon is more restrictive and seems to be compatible with the $(2,2,1)$ horizon, but this is a misleading conclusion as we see below.

\begin{figure}[!htb]
	\centering
	\includegraphics[width=.49\linewidth]{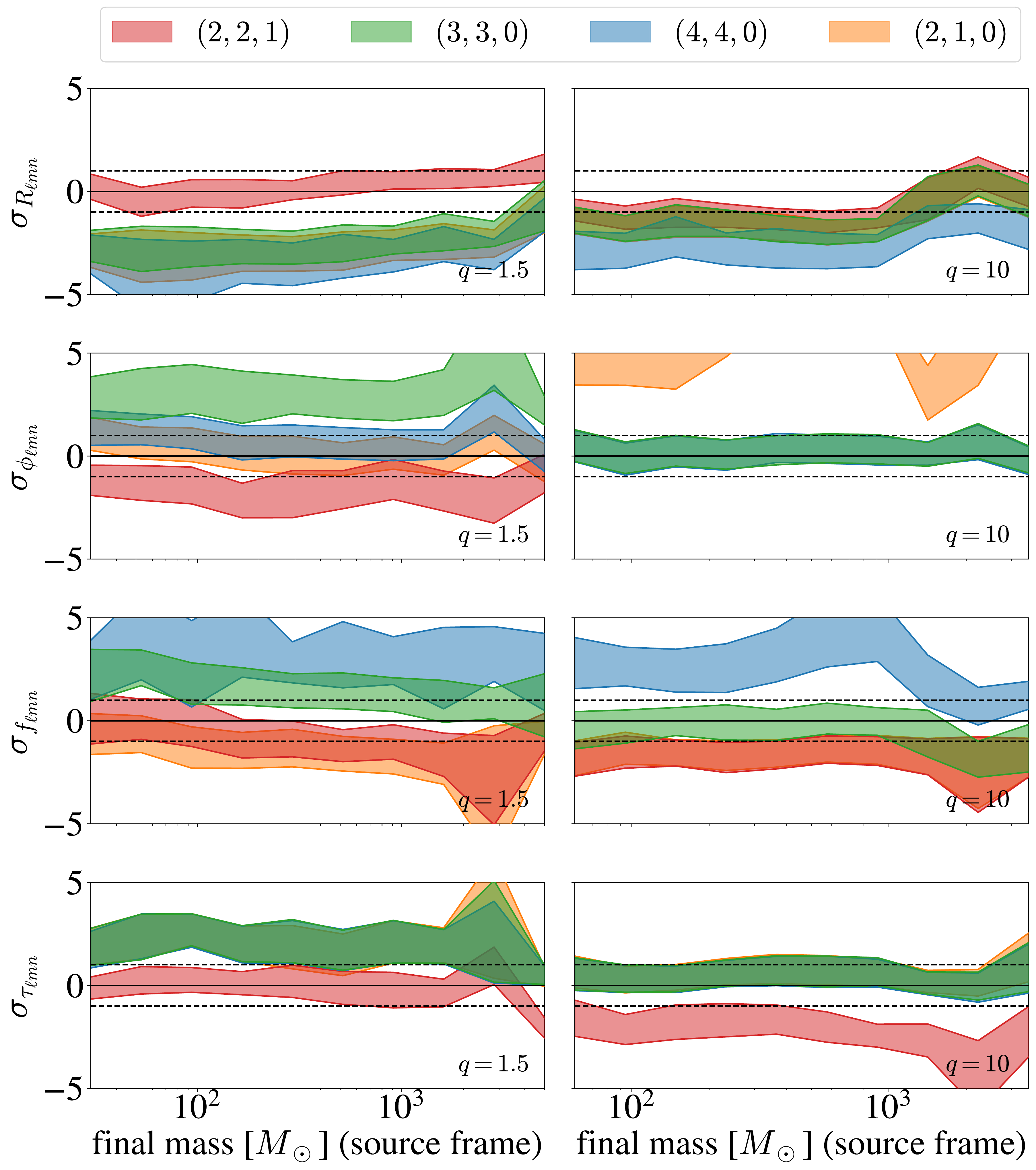}
	\includegraphics[width=.49\linewidth]{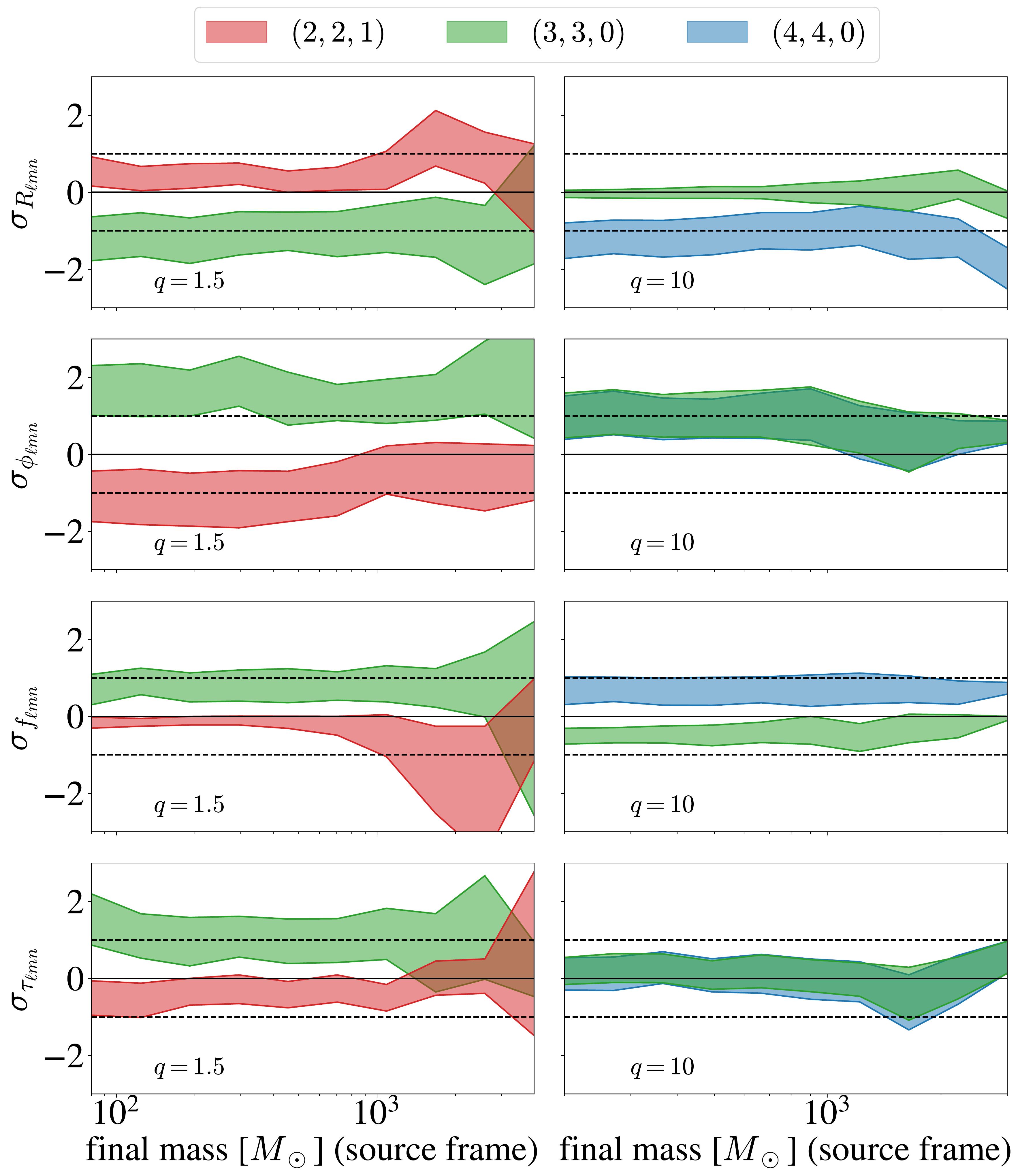}
	\caption{\emph{Left}: deviation between the posterior probability of the parameters $\vartheta^a$ of the model $\mathcal{M}_2$ and the injected parameters $\vartheta^a_{\mathrm{inj}}$, for each subdominant mode, for the signal $d_5 = n + h_{220} + h_{221} + h_{330} + h_{440} + h_{210}$ and LIGO detector.
	The colored bands indicate the $\pm 1\sigma$ credible interval of the deviation distribution.
	For $q = 1.5$ the secondary mode is the $(2,2,1)$, and for $q = 10$ the secondary is the $(3,3,0)$.
	\emph{Right}: secondary and tertiary modes identified for the $\mathcal{M}_3$ model.
	For $q = 1.5$ the tertiary mode is the $(3,3,0)$, and for $q = 10$ the $(4,4,0)$ is the tertiary mode.}
	\label{fig:desvios}
\end{figure}

To determine which are the secondary and tertiary modes in the two-mode and three-mode horizons we compute the deviation between the estimated parameters and the injected parameters (see equation~\eqref{eq:x_gaussian} and associated discussion).
Figure~\ref{fig:desvios} shows the deviations computed for events \emph{at} the horizon distances for the two-mode model $\mathcal{M}_2$ (left)
 and the three-mode model $\mathcal{M}_3$ (right).
We considered the mass interval of Figure~\ref{fig:horizon_bayes_multi} and selected 10 masses uniformly distributed on the logarithm scale.
For each mass we generated 100 events with individual noise realizations.
The colored bands show the $\pm 1\sigma$ credible interval for the deviation distribution.

For $q = 1.5$ the $(2,2,1)$ mode is the secondary mode in almost the entire mass range.
For masses larger than $\sim 10^3 M_\odot$ the secondary mode is compatible with the $(3,3,0)$ mode, which has higher frequency and stays longer in the detector's most sensitive band.
This transition can be clearly seen in the deviation of the amplitude ratios  $R_{\ell m n}$ and the damping times $\tau_{\ell m n}$.
For $q = 10$ the secondary mode is the $(3,3,0)$ and the tertiary mode is the $(4,4,0)$.
These results are compatible with the horizons of Figure~\ref{fig:horizon_bayes_multi}.
The larger offsets in the deviations of the tertiary mode (right) is due to the non-negligible contribution of the other subdominant modes.

\section{Comparison between different methods}
\label{sec:comparison-methods}

The Rayleigh criterion~\eqref{eq:rayleigh_both} and the threshold in the Bayes factor result in different horizons.
The Rayleigh criterion guarantees the resolvability of two modes, while the threshold in the Bayes factor guarantees a high statistical support for a two-mode model over a single-mode model.
There is no reason to expect the same result for these two criteria.
In these section we analyze the differences between both methods, and we also analyze alternatives to the Rayleigh criterion proposed in the literature.

\subsection{Rayleigh vs.\ Bayes}
By comparing Figures~\ref{fig:rayleigh_horizons} and~\ref{fig:bayes_horizons_2modes}, we can see that the BH spectroscopy horizons obtained using the Rayleigh criterion are consistently smaller than the horizons obtained using the Bayes factor threshold.
Two factors could influence this results: the Fisher matrix approximation may not be valid at large distances and/or the Rayleigh criterion may be inherently too restrictive.

\begin{figure*}[!htb]
    \centering
    \includegraphics[width = 1.0\linewidth]{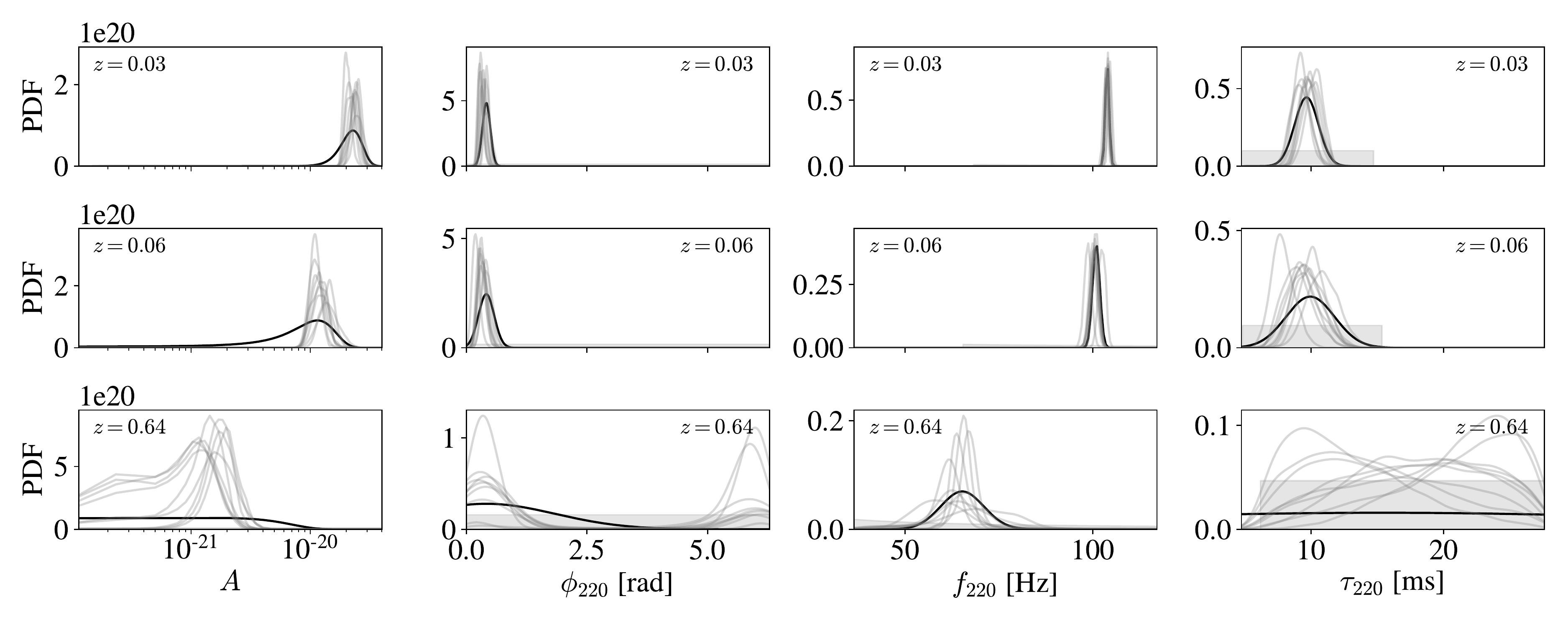}
    \includegraphics[width = 1.0\linewidth]{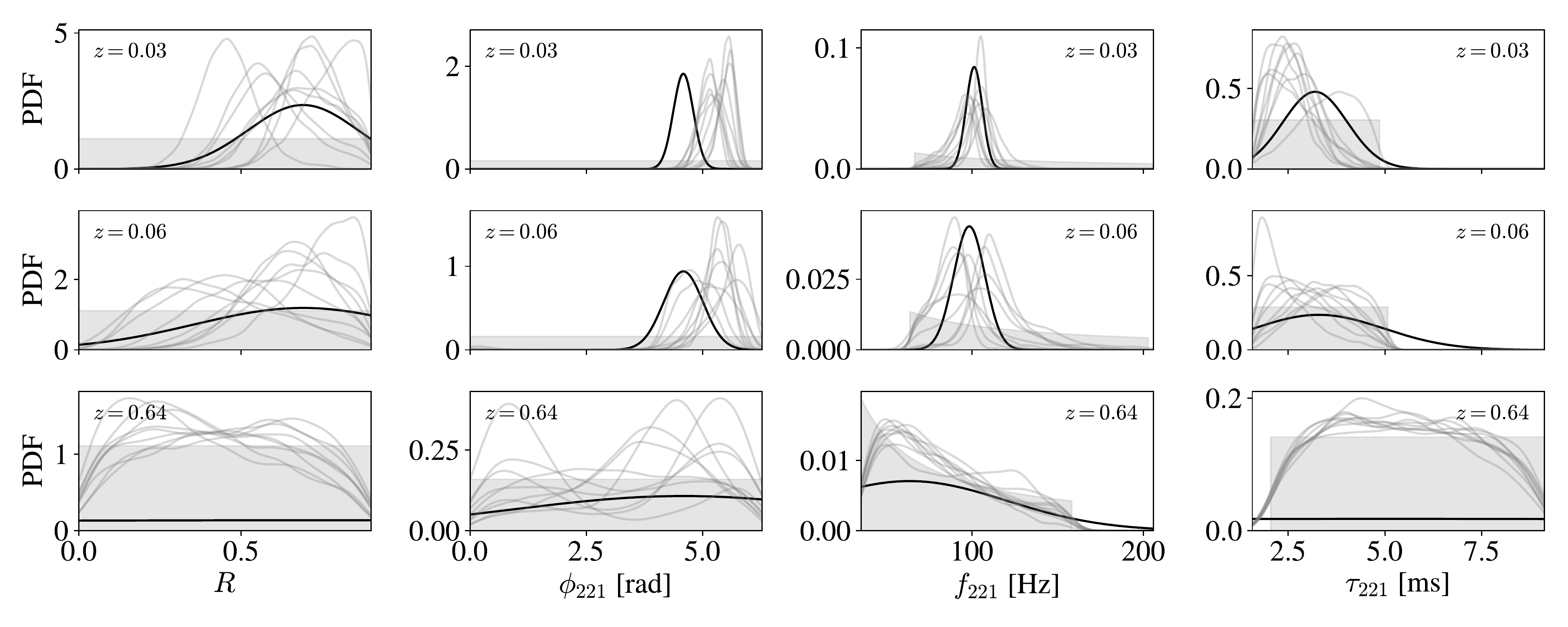}
    \caption{Posterior probability distributions for the eight ringdown parameters for a signal $d_2 = n + h_{220} + h_{221}$.
	The remnant mass is fixed $M_f = 156.3 M_{\odot}$ and the mass ratio is $q = 1.5$ (similar to GW190521).
	We considered three distances: a distance \emph{inside} the $(2,2,1)$ Bayes horizons (first and fourth columns), the distance \emph{of} the horizon $z^{\textrm{spec, B} }_{221}$ (second and fifth columns) and a distance \emph{outside} the horizon (third and last columns).
	The gray region indicates the prior distributions, the gray curves are posterior distributions for 10 simulated signals (different noise realizations) and the black curves are the Gaussian distributions constructed with the uncertainties estimated by the Fisher matrix formalism.
	The injected parameters correspond to the peak of the Gaussian distributions.}
    \label{fig:fisher_parameters_gw190521}
\end{figure*}

The Fisher matrix formalism is only valid for high SNRs~\cite{Vallisneri:2007ev}, and using it for low SNR signals can result in wrong conclusions.
Figure~\ref{fig:fisher_parameters_gw190521} shows a comparison between the posterior probability distributions, obtained using Bayesian inference, and the estimated Fisher matrix uncertainties (in the form of a Gaussian).
We considered the remnant mass and mass ratio compatible with the event GW190521, the most massive event detected so far,  $M_f = 153.6 M_{\odot}$ and $q=1.5$.
We generated 10 events with distinct noise realizations for signals $d_2 = n + h_{220} + h_{221}$ and three distances:
a distance \emph{inside} the Bayes factor BH spectroscopy horizon of the $(2,2,1)$ mode,
the distance \emph{at} the horizon $z^{\textrm{spec, B} }_{221}$ and a distance \emph{outside} the horizon, compatible with the distance of GW190521.

For the cases \emph{inside} the horizon, the Fisher uncertainties (black Gaussian distributions) are compatible with the posterior probability distributions, but for $A$, $R$ and $\tau_{221}$ the Fisher matrix already predicts larger uncertainties.
For the events \emph{at} the horizon distance the Fisher uncertainties for $R$ and $\tau_{221}$ are considerably larger.
For the cases \emph{outside} the horizon all the uncertainties estimated by the Fisher matrix formalism are larger than the posterior distribution.
We notice that the posterior distributions of the $(2,2,1)$ parameters are equivalent to the prior, which is expected as outside the horizon the secondary mode is negligible.

As expected, the Fisher matrix gives good estimates when the SNR is high (the distance is small).
The estimate at the distance inside the horizon $z = 0.03$ is good, but note that this distance is already outside the Rayleigh horizon.
This indicates that the limits of the Rayleigh horizons are not caused by an incorrect use of the Fisher matrix formalism, but the Rayleigh criterion is inherently too restrictive.

\begin{figure*}[!htb]
    \centering
    \includegraphics[width = 0.7\linewidth]{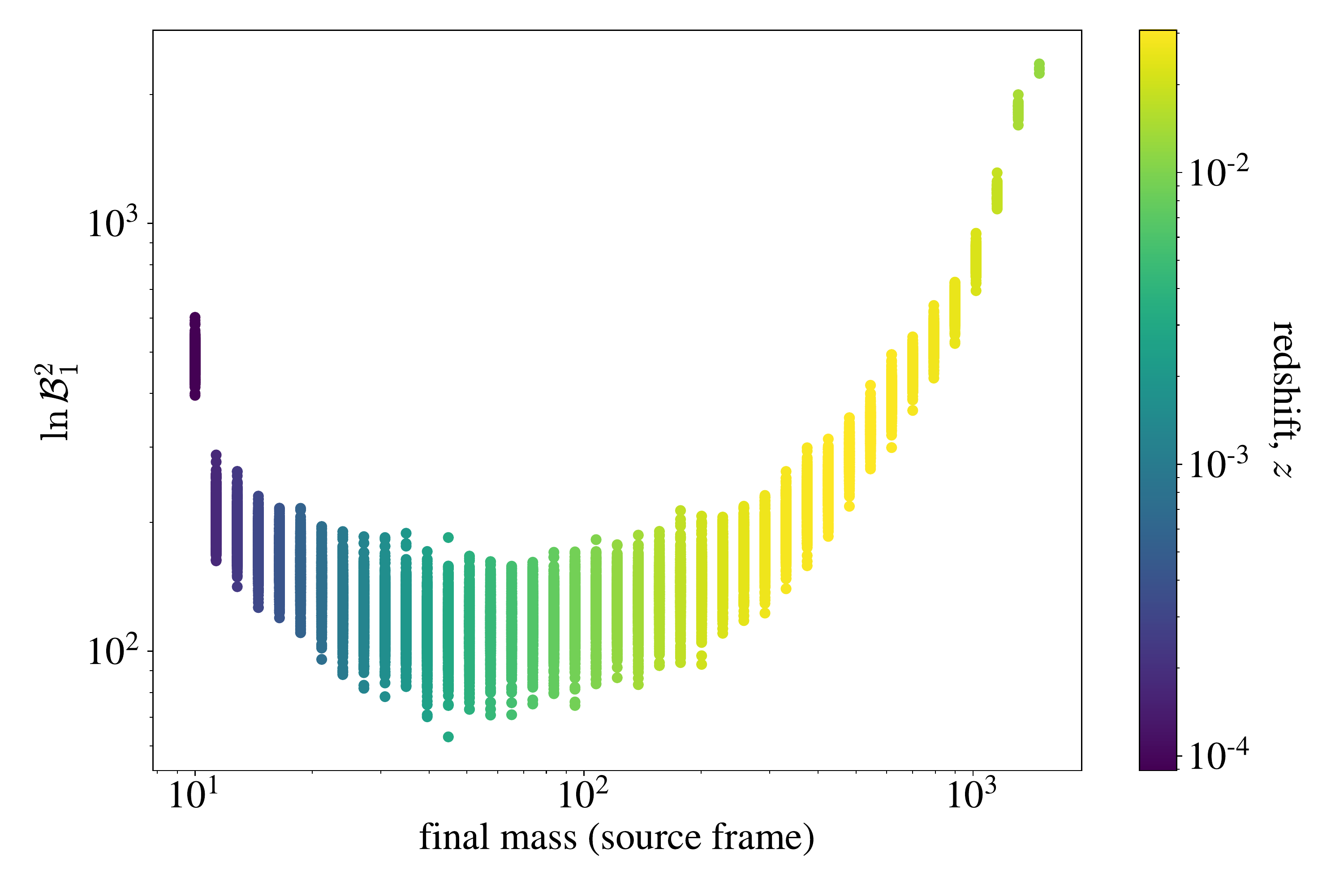}
    \caption{Bayes factors $\ln\mathcal{B}_1^2$ as a function of remnant mass for the overtone at the corresponding Rayleigh horizons for LIGO and $q = 1.5$.
	We considered 50 log-distributed masses and 500 injections with independent noise realizations for each mass.
	We found large Bayes factor at the Rayleigh horizon, which indicates a high statistical evidence.}
    \label{fig:bayes_factor}
\end{figure*}

In Figure~\ref{fig:bayes_factor} we show the Bayes factors $\ln\mathcal{B}_1^2$ for the overtone at the corresponding Rayleigh horizons for LIGO and $q = 1.5$ (Figure~\ref{fig:rayleigh_horizons_LIGO}).
We considered 50 log-distributed masses and 500 injections with independent noise realizations for each mass.
We found $\ln \mathcal{B}_1^2 \gtrsim 60$ ($\mathcal{B}_1^2 \gtrsim 10^{26}$), which are very high Bayes factors.
Furthermore, the Bayes factor is not constant with the mass; it has a minimum value near $M_f = 45 M_{\odot}$ and monotonically increases for larger and lower masses.
For higher harmonics the Bayes factors are even larger at the Rayleigh horizons, as the Rayleigh criterion is very restrictive for these modes due to the damping time condition.
This indicates that when the Rayleigh criterion is satisfied, there will be no doubt whether the secondary (and even a subsequent) mode is detectable in the signal.

\subsection{Variations of the resolvability criterion}
\label{subsec:variations-of-the-resolvability-criterion}
We have seen that the Rayleigh criterion is too restrictive compared to the Bayes factor threshold.
When the Rayleigh criterion was first introduced~\cite{Berti:2005ys}, two conditions were analyzed.
The ``critical'' SNR $\rho_{\mathrm{crit}}$ needed to resolve \emph{either} frequency~\eqref{eq:rayleigh_f} or damping time~\eqref{eq:rayleigh_tau} and the larger SNR $\rho_{\mathrm{both}}$ needed to resolve \emph{both} conditions.
They found that the minimum SNR $\rho_{\mathrm{crit}}$ needed to resolve one condition was one or two orders of magnitude smaller than the minimum SNR $\rho_{\mathrm{both}}$ needed to resolve both conditions.

\begin{figure}[!htb]
	\centering
	\includegraphics[width = 0.48\linewidth]{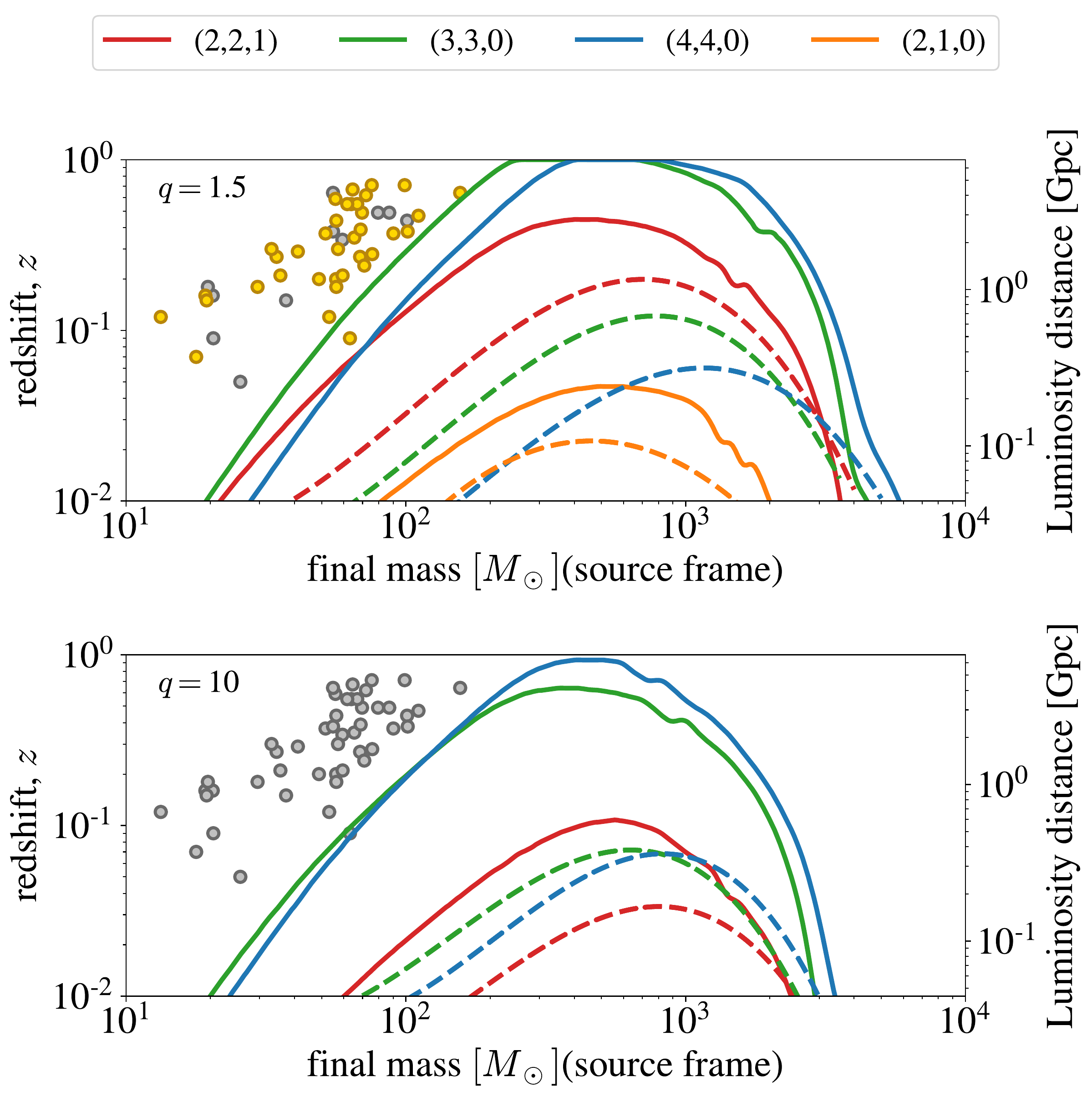}
	\includegraphics[width = 0.48\linewidth]{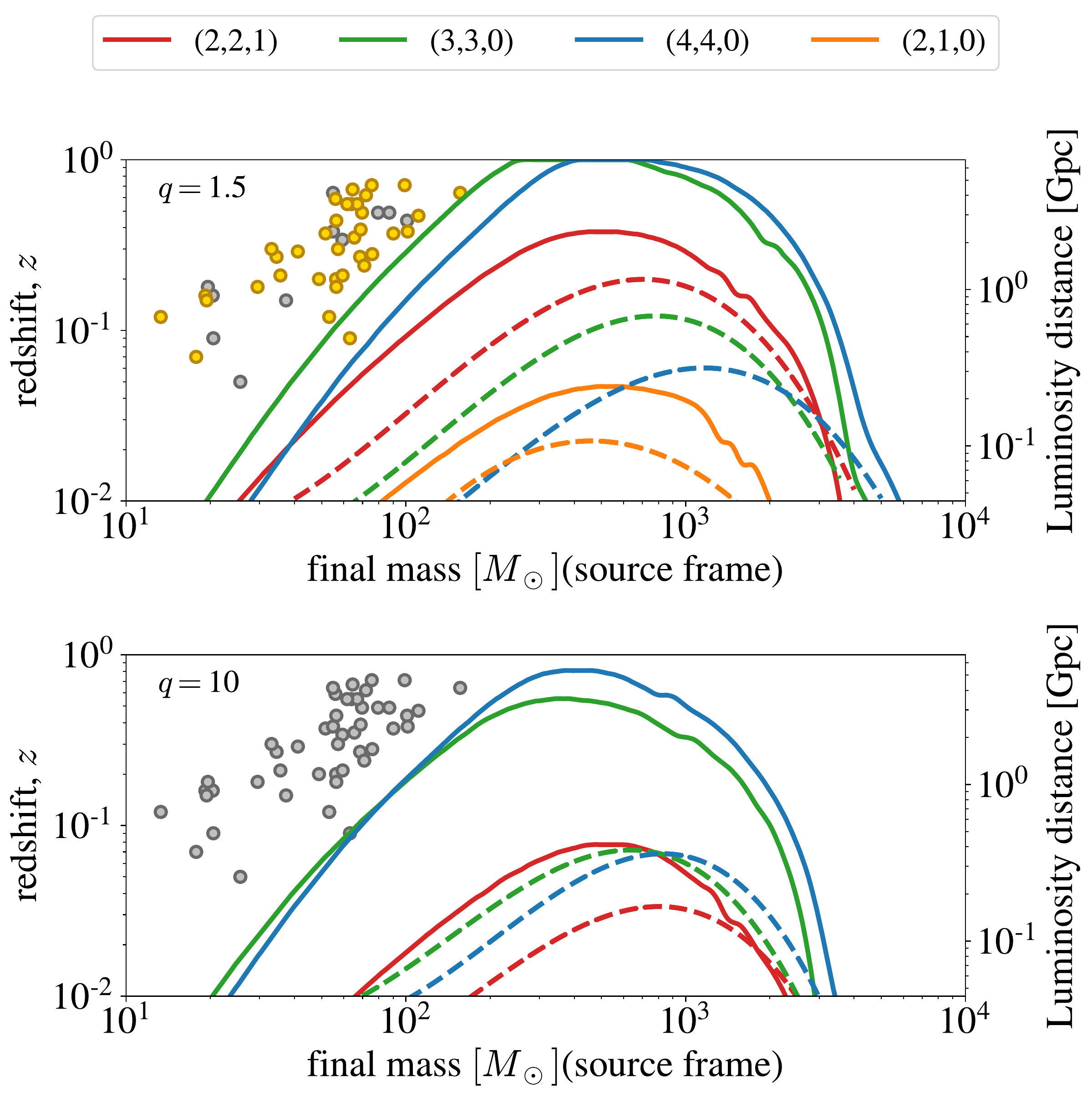}
	\caption{LIGO BH spectroscopy horizons  (solid lines) obtained by requiring only one of Rayleigh conditions~\eqref{eq:rayleigh_f} or~\eqref{eq:rayleigh_tau} to be satisfied (left) and obtained with condition~\eqref{eq:resolvability_posterior} (right). The errors are estimated with a Fisher matrix analysis. The dashed curves show the LIGO BH spectroscopy horizons obtained with a Bayes factor threshold $\ln \mathcal{B} > 8$ (same as in Figure~\ref{fig:bayes_horizons_2modes}). Circles show detections from GWTC-1 and GWTC-2; yellow circles are compatible with the mass ratio in each case, whereas gray circles are not. The one-condition Rayleigh horizons are much less restrictive and favor subdominant modes with $\ell \neq 2$.}
	\label{fig:rayleigh_one}
\end{figure}

In section~\ref{sec:horizon-rayleigh-criterion} we required that both conditions should be satisfied, which would be equivalent to the $\rho_{\mathrm{both}}$ analysis.
Using just one condition is not restrictive enough, and this condition alone is not enough to assess the detectability of subdominant modes.
The left panel of Figure~\ref{fig:rayleigh_one} shows the LIGO BH spectroscopy horizons by requiring the resolvability of the frequency \emph{or} the damping time.
As it requires just the easiest conditions, these horizons are much larger than the horizons that require both conditions, shown in Figure~\ref{fig:rayleigh_horizons_LIGO}.
Requiring just one condition favors the higher harmonics with $\ell \neq 2$, which can easily resolve the frequencies.
The Fisher matrix formalism does not estimate the uncertainties well at these distances, and the incorrect estimate for the damping time of the subdominant mode is larger the incorrect estimate for frequency of the subdominant mode (see last two panels of Figure~\ref{fig:fisher_parameters_gw190521}).
This reduces the $(2,2,1)$ horizons, as this mode can resolve the damping time condition more easily than the frequency condition.
Some recent works have also confirmed that requiring a single condition to be satisfied is not very restrictive~\cite{Forteza:2020cve,Bhagwat:2016ntk}.

Due to the high restriction of requiring both conditions to be satisfied, Isi and Farr~\cite{Isi:2021iql} argued that this is not necessary to distinguish two modes in the ringdown.
As an alternative, they suggested a condition needed to distinguish the 2-dimensional posteriors of the frequency and damping times,
\begin{equation}
	\frac{(f_{\ell m n} - f_{\ell' m'n'})^2}{\sigma^2_{f_{\ell m n}} + \sigma^2_{f_{\ell' m' n'}}} + \frac{(\tau_{\ell m n} - \tau_{\ell' m'n'})^2}{\sigma^2_{\tau_{\ell m n}} + \sigma^2_{\tau_{\ell' m' n'}}} \gtrsim 1.
	\label{eq:resolvability_posterior}
\end{equation}
Figure~\ref{fig:rayleigh_one} shows on the right panel the LIGO BH spectroscopy horizons obtained using equation~\eqref{eq:resolvability_posterior} and the Fisher matrix formalism.
We can see that these horizons are very close to the Rayleigh horizons that require just one criterion.
This is expected, as the easiest condition to be satisfied in the Rayleigh criterion will be the dominant term in equation~\eqref{eq:resolvability_posterior}, whereas the other difference will be almost negligible.

We notice that for both conditions above, the event GW150914 is inside the $(3,3,0)$ horizon, but no evidence was found for this mode in the detected event~\cite{Carullo:2019flw, LIGOScientific:2020tif}.
Therefore, we conclude that requiring just a single Rayleigh condition or the condition given by equation~\eqref{eq:resolvability_posterior} is not enough to assess detectability of the modes.

An alternative to the very restrictive damping time condition was proposed by~\cite{Forteza:2020cve}.
They consider the  the quality factor $Q_{\ell m n } = \pi f_{\ell m n }\tau_{\ell m n }$ instead of the damping time, replacing equation~\eqref{eq:rayleigh_tau} by
\begin{equation}
	|Q_{\ell mn} - Q_{\ell' m'n'}|  > {\rm max}(\sigma_{Q_{\ell mn}},\sigma_{Q_{\ell' m'n'}}).
	\label{eq:Q_rayleigh}
\end{equation}
Note that the damping times are replaced by the quality factors in the Fisher matrix, which results in different estimates for the uncertainties of all parameters, including the frequencies.

\begin{figure}[!htb]
	\centering
	\includegraphics[width = 0.48\linewidth]{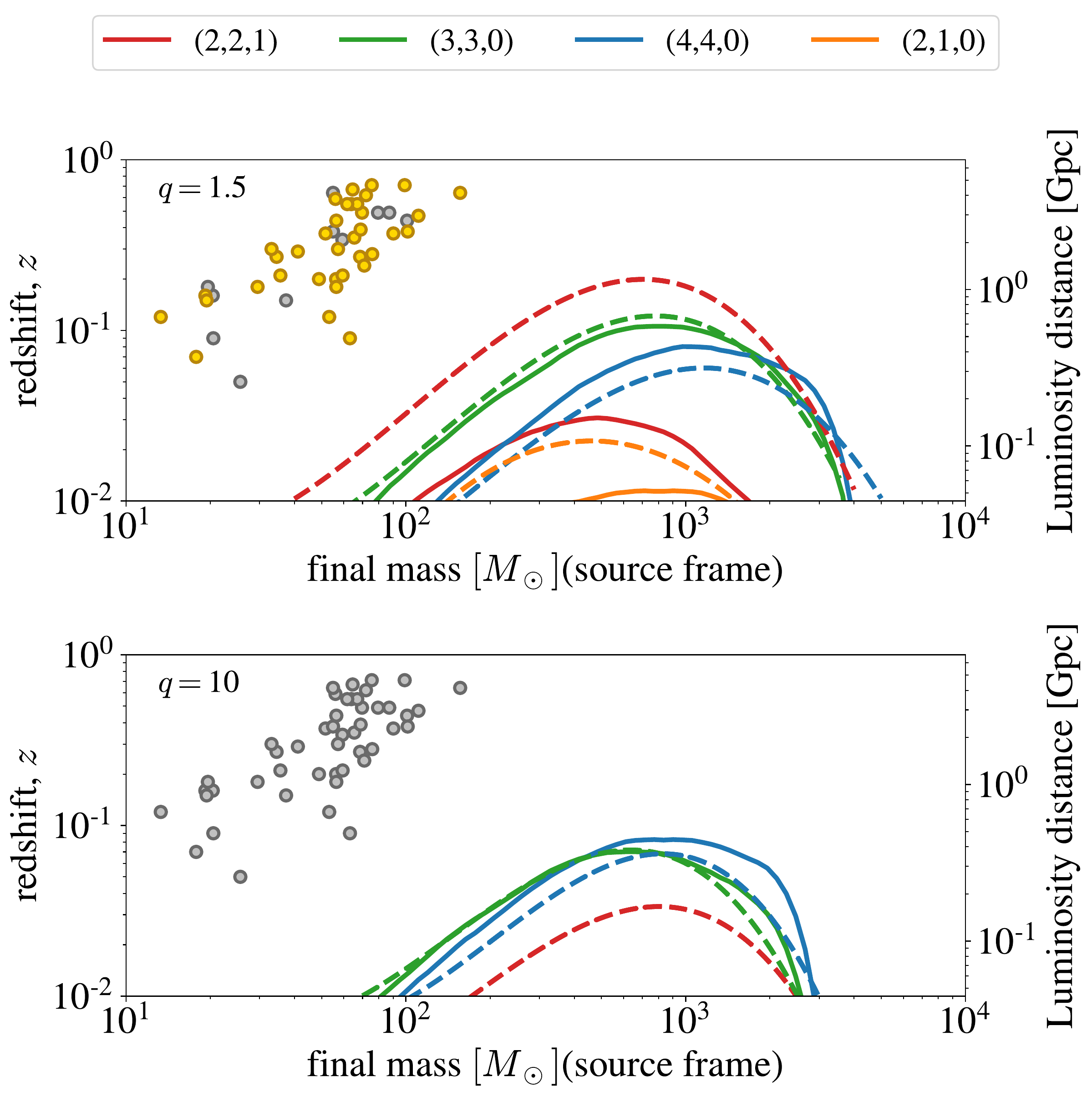}
	\includegraphics[width = 0.48\linewidth]{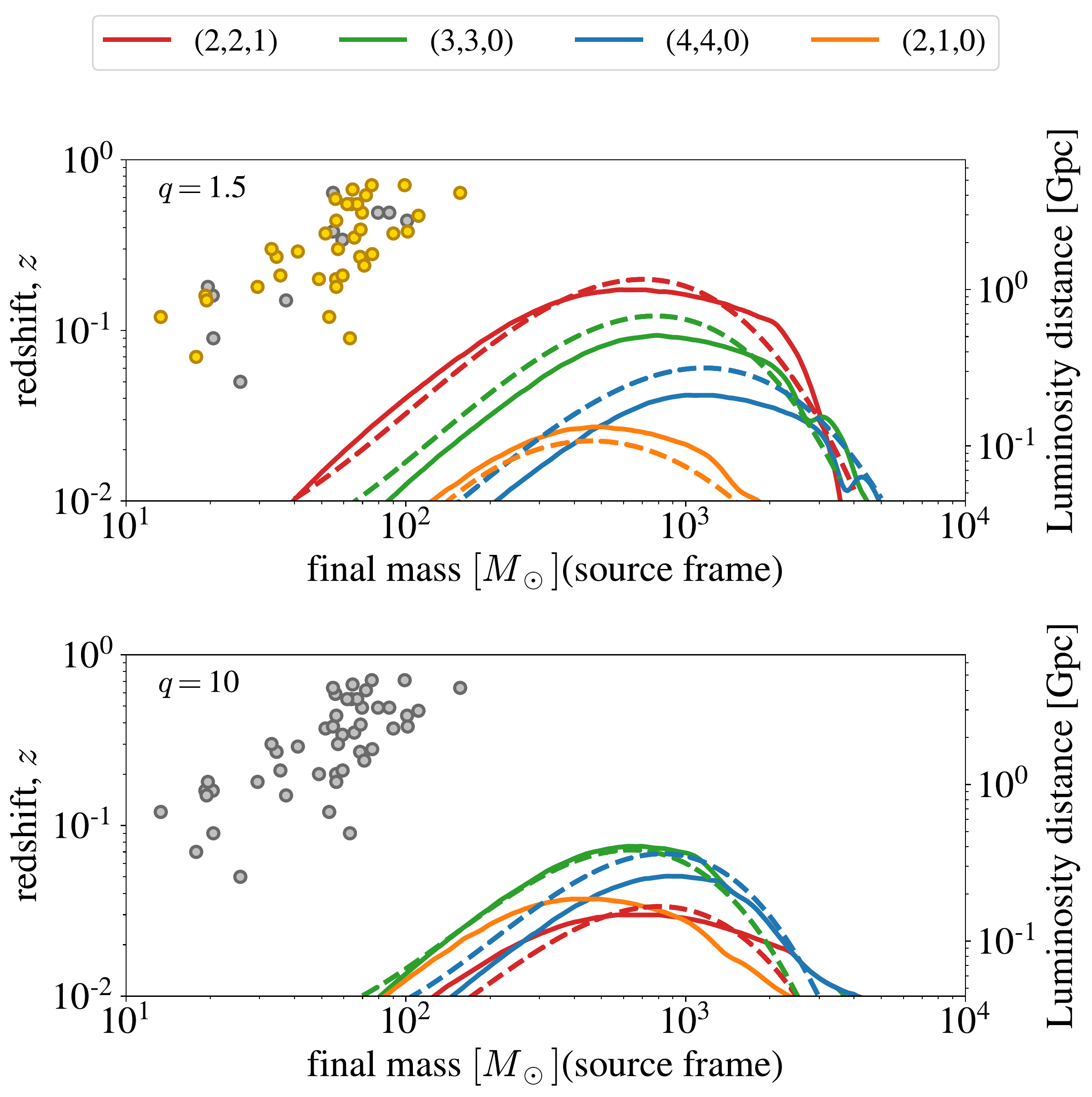}
	\caption{\emph{Left}: same as Figure \ref{fig:rayleigh_one}, but using the quality factors $Q_{\ell m n } = \pi f_{\ell m n }\tau_{\ell m n }$ instead of the damping times in the Rayleigh conditions~~\eqref{eq:rayleigh_both} (solid lines).
	The horizons obtained for modes with $\ell \neq 2$ are very close to the horizons obtained with the Bayes factor threshold (dashed lines), but the horizons of the modes with $\ell = 2$ are more restrictive.
	\emph{Right}: LIGO SNR horizons obtained by requiring $\rho_{\ell m n} > 8$ for each subdominant mode $(\ell, m,n)$. The SNR horizons are close to the Bayes factor threshold, with the exception of the $(2,1,0)$ mode for $q=10$.}
	\label{fig:Q_rayleigh_snr}
\end{figure}

Figure~\ref{fig:Q_rayleigh_snr} shows on the left panel the LIGO BH spectroscopy horizons obtained by requiring the resolvability of the frequencies~\eqref{eq:rayleigh_f} \emph{and} the quality factors~\eqref{eq:Q_rayleigh}.
This condition greatly favors the harmonics, but has almost no effect on the overtone horizon.
This is expected, as the quality factor condition is much easier to satisfy than the damping time condition when the frequency condition is easily satisfied.
Therefore the $(2,2,1)$ and $(2,1,0)$ modes still have very restrictive horizons, but the $(3,3,0)$ and $(4,4,0)$ horizons are very close to the Bayes factor horizons.

The different prescriptions presented above are variations of the resolvability criterion of two modes in the ringdown.
We have seen that these variations depend on the modes considered, and they may not give better results than the standard Rayleigh criterion, as they are either not restrictive at all or favor modes with higher frequencies.
As we explained in Section~\ref{sec:detectability-metric}, the SNR is a standard metric for detectability, and it indeed gives us useful information, when the correct template is used.
When we computed the Rayleigh criterion, we imposed a detectability condition of the SNR $\rho_{\ell m n} > 8$, which was not considered in the criteria above as it is more restrictive than those criteria.
For the Rayleigh criterion with both conditions, the SNR criterion is superfluous, as the Rayleigh criteria is much more restrictive.

In Figure~\ref{fig:Q_rayleigh_snr} on the right we show the ringdown horizons, where  $\rho_{\ell m n} = 8$, which are much more restrictive than the horizons shown in Figure~\ref{fig:rayleigh_one}.
We notice that the SNR condition is very close to the Bayes factor horizons.
But this condition is not sufficient to assess detectability, as the $(2,1,0)$ SNR horizon for $q = 10$ is very large, as the single-mode SNR is independent other modes.
However, we consider the SNR condition as a necessary condition for detectability.
In~\cite{Baibhav:2018rfk} the ringdown horizons using the SNR conditions were first computed, and our horizons are smaller than theirs due to the difference in the amplitudes of the modes, as we take the start time to be $t = t_{\mathrm{peak} } + 10 M$ instead of the peak of amplitude.

There are some other criteria proposed in the literature which we did not explore.
In~\cite{Isi:2021iql}, the authors argue that it is necessary that the amplitude posterior distribution of the subdominant mode excludes zero in the 90\% credible interval, without the need of more restrictive statistical measurements, like the Bayes factor.
They found that this condition holds for the GW150914 event (and simulations similar to the event), and this analysis was also confirmed by~\cite{Bustillo:2020buq}.
More recently,~\cite{Cotesta:2022pci} reevaluated this kind of analysis and showed that this result can be artificial evidence induced by the noise.
A similar condition was proposed by~\cite{Forteza:2020cve}, where they imposed that the mode amplitude ratio $R$ should exclude zero at the 1$\sigma$ level ($\sigma_R < R$).
They also proposed  a ``measurability'' criterion, which checks for relative accuracy for the estimated QNM parameters.

\section{Testing the no-hair theorem}
\label{sec:testing-the-no-hair-theorem}

In the previous sections we computed the BH spectroscopy horizon, which is the maximum distance of a BBH event, averaged over the sky location and the binary inclination, up to which a secondary mode can be detected in the signal.
The Rayleigh criterion results in very restrictive horizons, at which the Bayes factors are extremely large and the precision in the determination of the QNMs parameters are given by equations~\eqref{eq:rayleigh_both}.
The Bayes factor horizons are larger, and they do not guarantee a very precise parameter estimation, but they give statistical evidence for the presence of a secondary mode.
We called the horizons of both methods BH \emph{spectroscopy} horizons, which implies that at these horizons BH spectroscopy should be achieved with high confidence, and the difference in the horizon distance between the two methods is reflected in the precision of the test of the no-hair theorem.

With the detection of two QNMs, two pairs of mass and spin can be computed using equations~\eqref{eq:mass_spin_qnm}.
That is, given an injected signal $d_2 = n + h_{220} + h_{\ell m n}$, we can estimate the mass and the spin given by each QNM in the model.
For this analysis we use Bayesian inference to estimate the model parameters, and we consider the following models
\begin{itemize}
    \item $\mathcal{M}_1$, a \emph{single-mode} model with 4 parameters $\vartheta_{1} = \{ A_{220}, \phi_{220}, M_{f{220}}, a_{220}\}$,
    \item $\mathcal{M}_2$, a \emph{two-mode} model with 8 parameters $\vartheta_{2} = \{A_{220}$, $\phi_{220}$, $M_{f220}$, $a_{220}$, $A_{221}$, $\phi_{221}$, $M_{f 221}$, $a_{221} \}$,
\end{itemize}
where $A_{(220,221)}$ are the amplitudes, $\phi_{(220, 221)}$ are the phases, $M_{f (220, 221)}$ are the remnant masses and $a_{(220, 221)}$ are the remnant spins.
The redshift (distance) is fixed to the injected value and the amplitudes are scaled by the estimated mass (see equation~\eqref{eq:qnm_polarizations}).
To compute the QNM complex frequencies $\omega_{\ell m n}$ from the mass and spin we do an interpolation in the frequency values shown in Figure~\ref{fig:qnm_omegas_a}.
We do not use equations~\eqref{eq:mass_spin_qnm} because the interpolations are more precise, as equations~\eqref{eq:mass_spin_qnm} are a general fitting formula.

The prior distributions, valid for both models  ${\cal M}_1$ and ${\cal M}_2$ and modes $(2,2,0)$ and $(2,2,1)$, are
\begin{itemize}
\item $\pi(A_{(220,221)};\mathcal{M}_{(1,2)})$ is log-uniform in the interval $[0, 10]$,
\item $\pi(\phi_{(220, 221)};\mathcal{M}_{(1,2)})$ is uniform in the interval $[0,2\pi]$,
\item $\pi(M_{f(220,221)};\mathcal{M}_{(1,2)})$ is uniform in the interval $[M_f/10, 10 M_f]$, where $M_f$ is the remnant mass of the injected signal,
\item $\pi(a_{(220,221)};\mathcal{M}_{(1,2)})$ is uniform in the interval $[0, 0.999]$.
\end{itemize}
\begin{figure}[!htb]
	\centering
	\includegraphics[width = 0.48\linewidth]{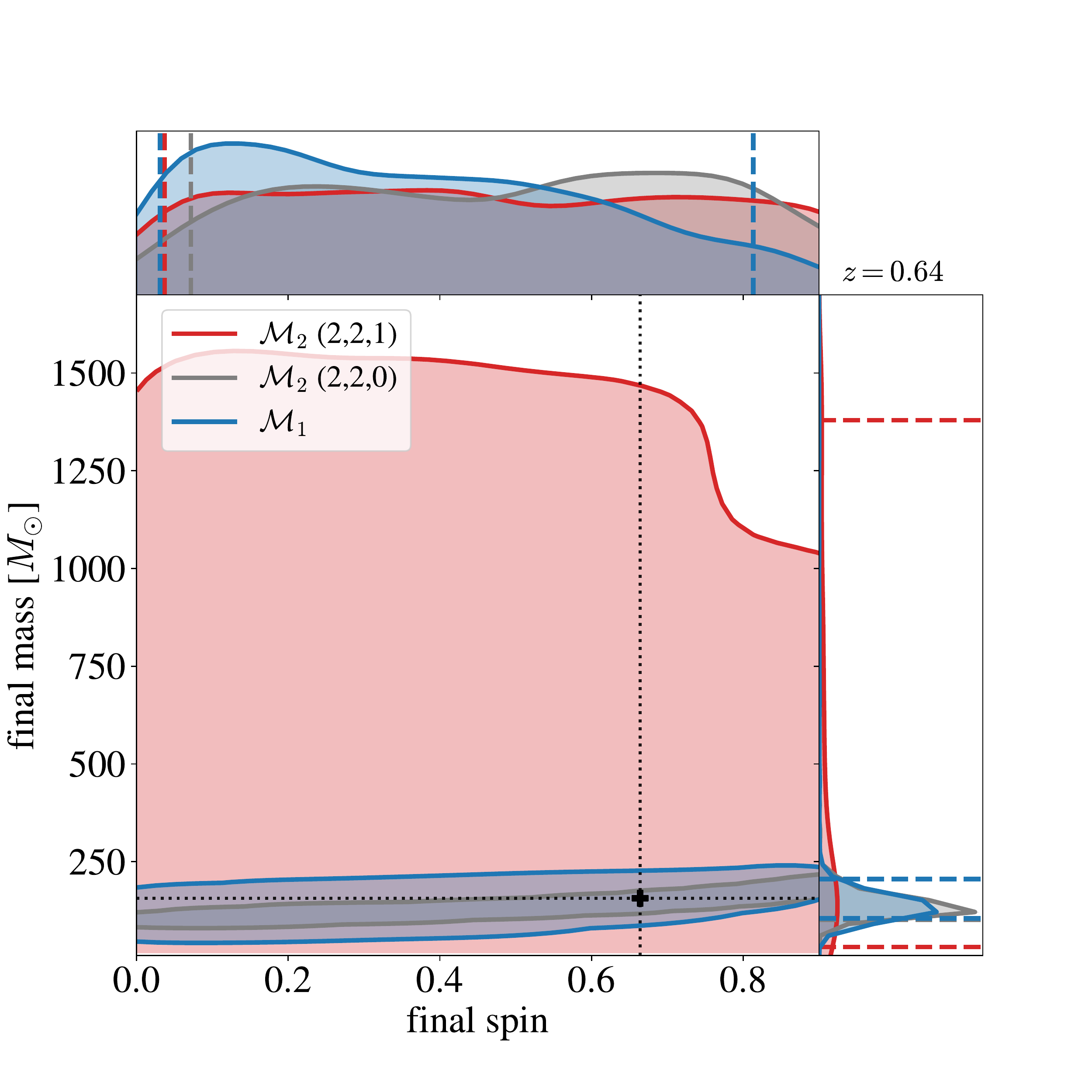}
	\includegraphics[width = 0.48\linewidth]{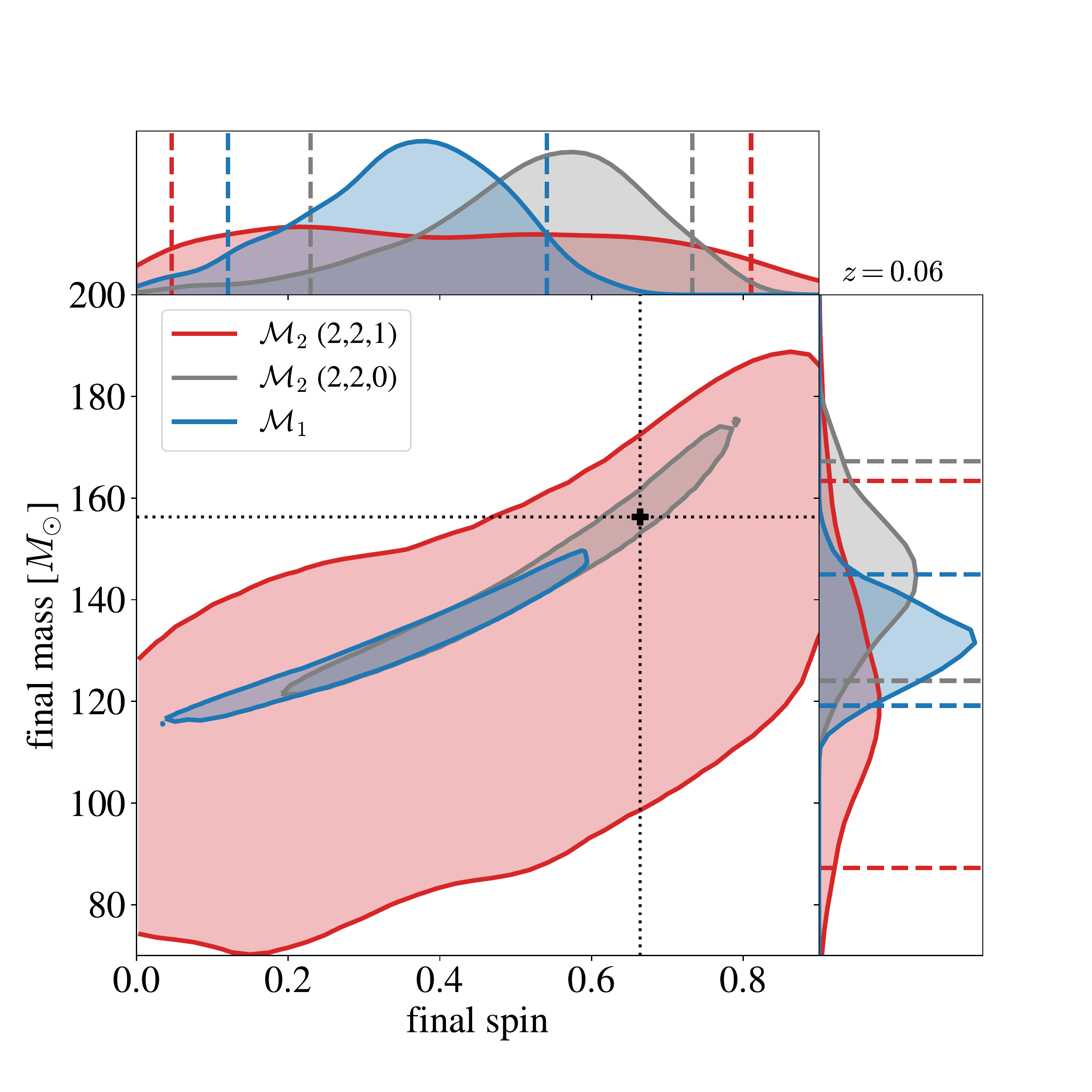}
	\includegraphics[width = 0.48\linewidth]{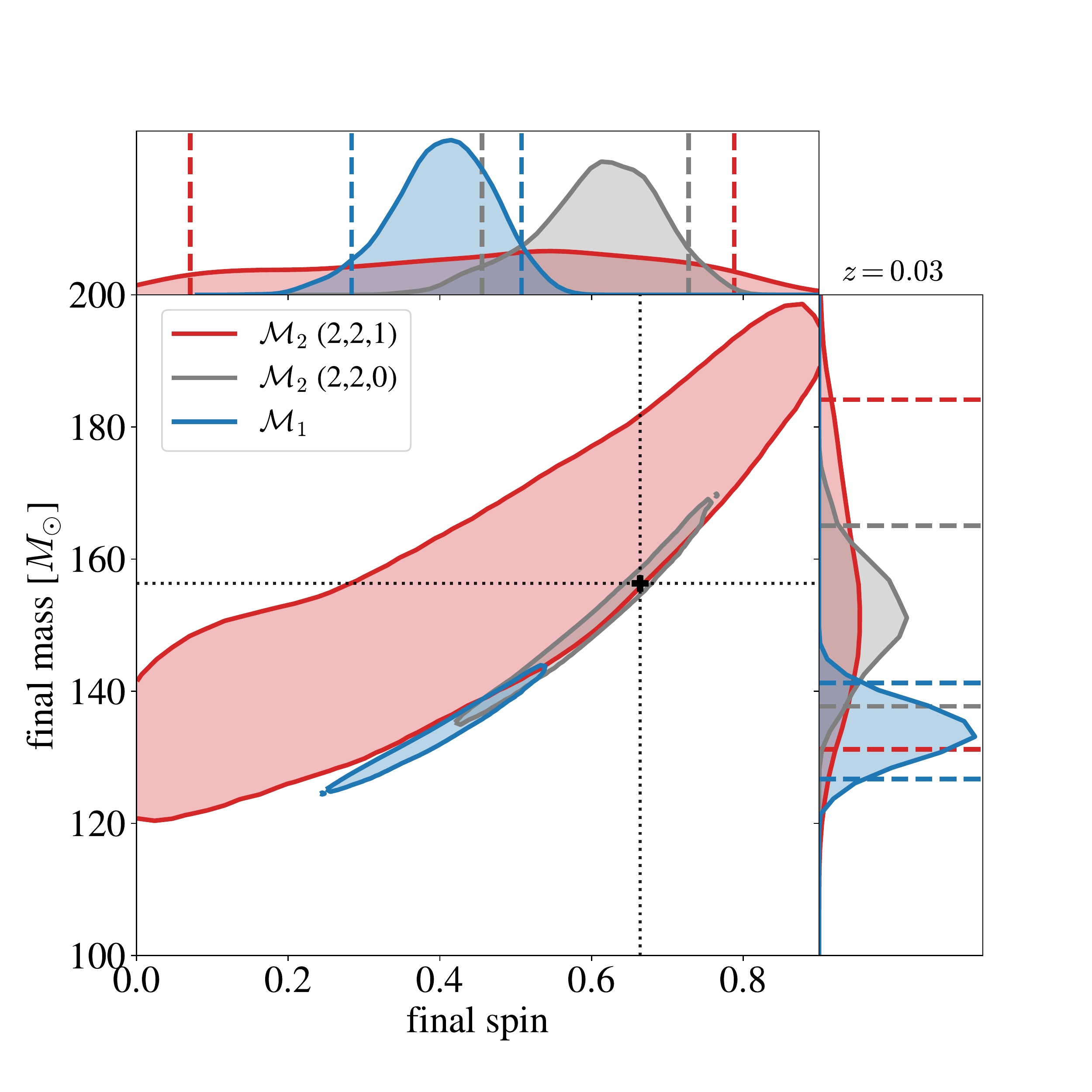}
	\includegraphics[width = 0.48\linewidth]{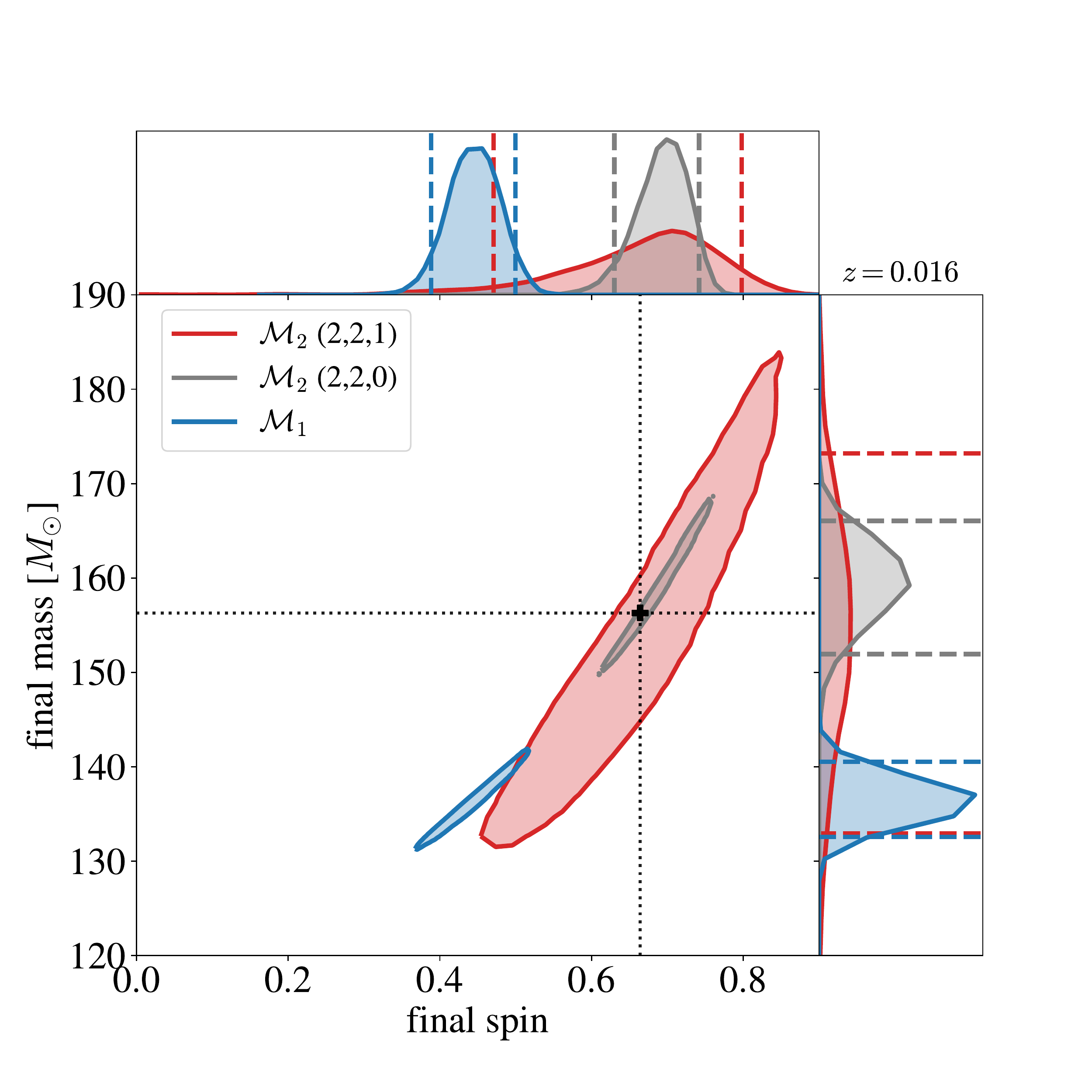}
	\caption{Mass and spin inferred from the ringdown  signal $d_2 = n + h_{220} + h_{221}$, with remnant mass $M_f = 156.3 M_{\odot}$ and the mass ratio $q = 1.5$ (similar to GW190521).
	The black cross and dotted lines indicate the injected parameters.
	The parameters are estimated with the single-mode model $\mathcal{M}_1$ (blue) and with the $(2,2,0)$ mode (gray) and $(2,2,1)$ mode (red) of the two-mode model $\mathcal{M}_2$.
	We considered four distances: $z = 0.64$ a distance \emph{outside} the horizons and compatible with the distance of GW190521 (top left), $z=0.6$ the distance \emph{of} the \emph{Bayes factor} horizon $z^{\textrm{spec, B} }_{221}$ (top right), $z = 0.3$ a distance \emph{between} the Bayes factor horizon and the Rayleigh horizon (bottom left) and $z=0.016$ the distance \emph{of} the \emph{Rayleigh} horizon $z^{\textrm{spec, R} }_{221}$ (bottom right).
	The contours enclose 90\% of the posterior distribution and the dashed lines are the limits to the 90\% credible interval of the 1D posteriors.
}
	\label{fig:mass_spin_nohair}
\end{figure}

Figure~\ref{fig:mass_spin_nohair} shows the mass and spin inferred from the  $d_2 = n + h_{220} + h_{221}$ signal at different distances.
We considered a nonspinning circular BBH similar to GW190521, with mass ratio $q = 1.5$ and remnant mass $M_f = 156.3 M_{\odot}$.
At a distance compatible with GW190521 with redshift $z = 0.64$ (top left), the overtone is extremely uninformative and its posterior contour is very broad.
The contours of the $(2,2,0)$ mode for both the single-mode (blue) and two-mode (gray) models are compatible with each other and consistent with the injected parameters, but the two-mode model contour is smaller.
At the Bayes factor horizon distance $z^{\textrm{spec, B} }_{221} = 0.06$, the uncertainties of all modes decrease.
The overtone still has large uncertainties but the estimated mass uncertainty is approximately within $\pm 50\%$ the injected value.
The dominant mode contours are much more restrictive and the single-mode model estimate is not compatible with the injected values.
This shows that, although the parameter estimation precision of the overtone is not very high, a two-mode model is necessary to correctly estimate the parameters with the dominant mode.
As the distance of the event gets closer, the uncertainties decreases and the two-mode model always estimate the parameters correctly.

The difference between the Bayes factor horizons (top right) and the Rayleigh horizon (bottom right) is in the precision of the test.
Although the Bayes factor threshold does not impose any uncertainty limit for the parameters, it guarantees that at this distance the secondary mode is not negligible.
This is clear in the one-mode model contour, which is incompatible with the injected values, even tough the values estimated with the overtone have large uncertainties.
We also note that the uncertainty of the overtone contour is always considerably larger than the dominant mode contours, which is expected as the overtone is a subdominant weaker mode.

We note that this analysis is different from the mass-spin 2D posteriors presented in LIGO analyses~\cite{LIGOScientific:2020ufj,LIGOScientific:2020tif,LIGOScientific:2021sio}, in which the Kerr model is assumed to be true and a single pair for mass and spin is computed for the models with subdominant modes.
The tests of the no-hair theorem done by LIGO look for deviations in the frequencies and damping times of the subdominant mode.
Our analysis is similar to the analysis presented in Figure 4 of~\cite{Gossan:2011ha}, where the mass and spin of each mode are determined independently of other modes.

\begin{center}
\myclearpage
\par\end{center}


\chapter{Conclusions}
\label{ch:conclusions}

Black hole spectroscopy is the proposal to use two or more quasinormal modes to test the no-hair theorem~\cite{Dreyer:2003bv}.
There is some suggestive evidence for a secondary mode in gravitational wave data~\cite{Isi:2019aib, Capano:2021etf}, but the low statistical evidence for the extra mode still makes the detection situation uncertain~\cite{LIGOScientific:2020ufj, LIGOScientific:2020tif, LIGOScientific:2021sio, Carullo:2019flw, Cotesta:2022pci}.
In this work we proposed a conservative criterion to assess the detectability of subdominant modes, in order to guarantee strong statistical evidence for the presence of such modes in the ringdown.

A detailed computation of the quasinormal mode parameters used in the analysis is an important first step.
In Chapter~\ref{ch:ringdown-qnm} we analyzed numerical relativity simulations from the SXS project~\cite{SXS-catalog, Boyle:2019kee} and studied the contribution of the first overtone of the quadruplar mode and the higher harmonics in the ringdown of nonspinning circular binary black hole mergers~\cite{Ota:2019bzl}.
The main goal was to identify the subdominant mode parameters and compare the relevance of each mode as a function of the mass ratio.

The higher harmonic modes have already been extensively studied before~\cite{Cotesta:2018fcv,Kamaretsos:2011um, Kelly:2012nd,Shi:2019hqa, Thrane:2017lqn, Maselli:2019mjd, Baibhav:2018rfk, Baibhav:2020tma} and we mainly focused on the overtone $(2,2,1)$.
To confidently determine the overtone parameters and avoid any doubt caused by a non-physical fitting, we considered two methods.
We fitted to the waveform $h_{22}$ and to the time derivative of the complex phase $\theta_{22}$ functions containing the contribution of the fundamental mode and the overtone and determined the initial time $t_0$ at which the waveform is well described by the $(2,2,0)$ and the $(2,2,1)$ modes.
For a simulation similar to the first detection, GW150914, we found an initial time compatible with other ringdown analysis~\cite{Thrane:2017lqn, Carullo:2019flw}.
For nonspinning circular binaries we showed that the initial time is smaller for larger mass ratios.

Rescaling the excitation amplitudes at the reference time $t = t_\mathrm{peak} +  10 M$,
we found that the overtone has amplitude greater than all the higher harmonics for mass ratios of 1:1 to 5:1 and has amplitude comparable with the most relevant higher harmonics, $(3,3,0)$ and $(2,1,0)$, for mass ratios of 5:1 to 10:1.
We fitted an exponential curve to the ratio between the overtone amplitude and the fundamental mode amplitude and found the asymptotic value $\mathcal{R} = 0.3$, which is compatible with the point particle limit~\cite{Cardoso:amplitude}.

In Chapter~\ref{ch:spectroscopy-horizon} we used the results obtained in the previous chapter to assess the detectability of the subdominant modes.
We computed the \emph{black hole spectroscopy horizon}, which is the maximum distance of a binary black hole merger event up to which two or more modes can be confidently detected in the signal.
We first computed the horizons using the Rayleigh criterion~\cite{Berti:2005ys}, which assesses whether two quasinormal modes are resolvable in a gravitational wave detection.
We found horizons that are excessively restrictive, but with general trends compatible with the Bayes factor analysis we performed next.

The Bayes factor horizons were obtained by considering the threshold on the Bayes factor $\ln \mathcal{B} > 8$.
We found larger horizons but still smaller distances than the current detections.
This is compatible with the LVKC analyses~\cite{LIGOScientific:2020tif,LIGOScientific:2021sio}, which found a low Bayes factor ($\ln\mathcal{B} < 3$) for any subdominant mode.
The best prospects for the detection of a secondary mode are for low mass ratio binaries.
For binaries with mass ratio 1.5:1 the secondary mode is the overtone and the tertiary mode is the $(3,3,0)$ higher harmonic.
The horizons for high mass ratio binaries are much smaller, as, in this case, less energy is emitted in the form of gravitational waves~\cite{Kamaretsos:2011um, Cotesta:2018fcv, London:2014cma}.
For mass ratio 10:1 the secondary and tertiary modes are the $(3,3,0)$ and $(4,4,0)$ higher harmonics, respectively.

The horizon results show that analyzing just the amplitude of the modes is not enough to determine the detectability.
The $(2,1,0)$ mode has a significant amplitude compared with the $(3,3,0)$ and the $(2,2,1)$ for large mass ratio, but it is very difficult to detect this mode, as it has frequency and damping time very similar to the dominant $(2,2,0)$ mode.

We estimated the mass and the spin of an event at several distances.
We found that when the event is outside the horizon, a single-mode model can estimate the mass and spin correctly and the subdominant mode estimate is highly unconstrained.
At the Bayes factor horizon distance the single-mode model estimate has biased results, and a two-mode model is necessary to correctly estimate the remnant parameters.
The closer the source is the smaller are the uncertainties and the test is more constrained.

However, with the current detectors we will need luck to detect an event as close as the Bayes factor horizon.
The rate of events at the Bayes horizon , for the low mass ratio case, final mass $M = 156.3 M_{\odot}$ and the overtone, is $0.03 - 0.10\ {\mathrm{year}}^{-1}$ for LIGO at the design sensitivity.
For the third generation detector, Cosmic Explorer, the rate is improved to $(0.6 - 2.3) \times 10^3\ {\mathrm{year}}^{-1}$.
For larger masses the horizons are even larger: intermediate mass black holes with masses between $10^2 M_{\odot}$ and $10^3 M_{\odot}$ are the best sources for ground based detectors, but their rates (and existence) are still uncertain.
The existence of supermassive binary black hole mergers is not confirmed, but LISA will be very sensitive and will most probably be able to detect subdominant modes in the ringdown.

A generalization of this work could be the inclusion of spinning and/or non-circular binaries.
Moreover, the network of detectors would increase the spectroscopy horizon, but a factor $\lesssim 2$.
A promising analysis is coherent mode stacking~\cite{Yang:2017zxs,Zimmerman:2019wzo}, which proposes to use multiple detections to increase the signal-to-noise ratio of a subdominant mode.
The method was proposed to analyze higher harmonics, and cannot be easily generalised for overtones, which is the most relevant mode for events similar to the current detections (low mass ratio).
However, the method can be used to compute the number of ringdown events needed to detect higher harmonics with a significant Bayes factor.


\appendix
\chapter{Linearized gravity}
\label{ch:linearized_gravity}
The Einstein field equations are a set of ten independent  second-order differential equations for the metric tensor field $\mathrm{g}_{\mu\nu}$.
The nonlinearity of the field equations makes it considerably hard to solve them analytically.
We can gain some insight on the theory by considering a linear weak field approximation.

We will consider that our spacetime differs from flat spacetime by a small perturbation in such a way that we can decompose our metric tensor into the Minkowski metric plus a small perturbation $h_{\mu\nu}$, i.e.,
\begin{equation}
	\mathrm{g}_{\mu\nu} = \eta_{\mu\nu} + h_{\mu\nu} + \mathcal{O}([h_{\mu\nu}]^2),
    \label{eq:metric_linear_pert}
\end{equation}
where $\eta_{\mu\nu}=\text{diag}(-1,1,1,1)$ is the Minkowski metric and $\abs{h_{\mu\nu}} \ll1$.
As we are only dealing with first order terms in $h_{\mu\nu}$, we only need the Minkowski metric to raise and lower the indices of the perturbation, i.e., $h^{\mu\nu} = \eta^{\mu\rho}\eta^{\nu\sigma}h_{\rho\sigma}$.
We assume that the inverse metric can be written as a sum of the inverse flat metric and some term proportional to the inverse perturbation metric, that is, $\mathrm{g}^{\mu\nu} = \eta^{\mu\nu} + \alpha h^{\mu\nu} + \mathcal{O}([h_{\mu\nu}]^2)$, where $\alpha$ is a constant.
Recalling that $\mathrm{g}_{\mu\rho}\mathrm{g}^{\mu\sigma} = \delta^\sigma_\rho$, we have
\begin{align*}
    \delta^\sigma_\rho =& (\eta_{\mu\rho} + h_{\mu\rho})(\eta^{\mu\sigma} + \alpha h^{\mu\sigma}) +  \mathcal{O}([h_{\mu\nu}]^2) \\
    =& \eta_{\mu\rho}\eta^{\mu\sigma} +\alpha \eta_{\mu\rho} h^{\mu\sigma} + \eta^{\mu\sigma} h_{\mu\rho}+  \mathcal{O}([h_{\mu\nu}]^2).
\end{align*}
The first term in the equality equals $\delta^\sigma_\rho$ and, therefore, the sum of the other terms must be zero, that is,
\begin{align*}
    \alpha \eta_{\mu\rho} h^{\mu\sigma} =& - \eta^{\mu\sigma} h_{\mu\rho}\\
    =&- \eta^{\mu\sigma}(\eta_{\mu\nu}\eta_{\rho\delta}h^{\nu\delta})\\
    =& - \delta^\sigma_\nu\eta_{\rho\delta}h^{\nu\delta} \\
    =& - \eta_{\delta\rho}h^{\delta\sigma},
\end{align*}
where in the last line we considered the symmetry of the flat and perturbation metrics. As $\delta$ is a dummy index\footnote{When a summation is performed over an index we call this index a \textit{dummy index}. If there is no sum performed over one index we call it a \textit{free index}.}, we can relabel it to $\mu$ and we find that $\alpha = -1$.
Therefore, the inverse metric is given by
\begin{equation}
	\mathrm{g}^{\mu\nu} = \eta^{\mu\nu} - h^{\mu\nu} + \mathcal{O}([h_{\mu\nu}]^2).
\end{equation}

We want to find the linearized version of the Einstein field equations.
We begin by determining the Christoffel symbols.
The Minkowski metric in the canonical form has constant terms, therefore any partial derivative is zero. Hence, the Christoffel symbols are
\begin{align}
	\Gamma^\rho_{\mu\nu} =& \frac{1}{2}\mathrm{g}^{\rho\lambda}(\mathrm{g}_{\mu\lambda,\nu} + \mathrm{g}_{\lambda\nu,\mu} - \mathrm{g}_{\mu\nu,\lambda}) + \mathcal{O}([h_{\mu\nu}]^2)\nonumber\\
	=& \frac{1}{2}\eta^{\rho\lambda}(h_{\mu\lambda,\nu} + h_{\lambda\nu,\mu} - h_{\mu\nu,\lambda}) + \mathcal{O}([h_{\mu\nu}]^2),
\end{align}
where the term $h^{\rho\lambda}$ of the inverse metric is of second-order when contracted with the derivatives of $h_{\mu\nu}$.
With it we can compute the Riemann tensor:
\begin{align}
    R^\mu_{\nu\rho\sigma} =& \Gamma^\mu_{\ \sigma\nu,\rho} -\Gamma^\mu_{\ \rho\nu,\sigma} + \Gamma^\mu_{\ \rho\lambda}\Gamma^\lambda_{\ \sigma\nu} - \Gamma^\mu_{\sigma\lambda}\Gamma^\lambda_{\ \rho\nu} \nonumber\\
    =& \frac{1}{2}\eta^{\mu\lambda}(h_{\sigma\lambda,\nu\rho} + h_{\lambda\nu,\sigma\rho} - h_{\sigma\nu,\lambda\rho} - h_{\rho\lambda,\nu\sigma} - h_{\lambda\nu,\rho\sigma} + h_{\rho\nu,\lambda\sigma})\nonumber \\
    =& \frac{1}{2}\eta^{\mu\lambda}(h_{\sigma\lambda,\nu\rho}  - h_{\sigma\nu,\lambda\rho} - h_{\rho\lambda,\nu\sigma}  + h_{\rho\nu,\lambda\sigma}),
    \label{riemann}
\end{align}
where the two last terms of the first line are of second-order in the metric perturbation and in the last line we considered symmetry of the partial derivatives.
Computing the Ricci tensor:
\begin{align}
    R_{\mu\nu} =& \Gamma^\rho_{\mu\nu,\rho} - \Gamma^\rho_{\mu\rho,\nu} + \Gamma^\rho_{\mu\nu}\Gamma^\beta_{\beta\rho} - \Gamma^\rho_{\mu\beta}\Gamma^\beta_{\nu\rho}\nonumber\\
    =& \frac{1}{2}\eta^{\rho\lambda}(h_{\mu\lambda,\nu\rho} + h_{\lambda\nu,\mu\rho} - h_{\mu\nu,\lambda\rho}-h_{\mu\lambda,\rho\nu} - h_{\lambda\rho,\mu\nu} + h_{\mu\rho,\lambda\nu}) + \mathcal{O}([h_{\mu\nu}]^2)\nonumber \\
    =& \frac{1}{2}( h^\rho_{\nu,\mu\rho} - h_{\mu\nu,\rho}^{\ \ \ \ \ \rho} - h^\rho_{\rho,\mu\nu} + h_{\mu\rho,\nu}^{\ \ \ \ \ \rho}) + \mathcal{O}([h_{\mu\nu}]^2)\nonumber\\
    =&  \frac{1}{2}( h_{\nu\rho,\mu}^{\ \ \ \ \ \rho}+ h_{\mu\rho,\nu}^{\ \ \ \ \ \rho} - \Box h_{\mu\nu} - h_{,\mu\nu} ) + \mathcal{O}([h_{\mu\nu}]^2),
    \label{ricci}
\end{align}
where the two last terms of the first line are of second-order in the metric perturbation, in the third line we considered the symmetry of the indices of the metric perturbation and of the partial derivatives, in the fourth line we considered $g^{\rho\sigma}A_{\rho\sigma} = A^\rho_{\ \rho} = A^{\ \rho}_\rho$, for any tensor $A_{\rho\sigma}$, $h = h^\rho_{\ \rho} = \eta^{\rho\sigma}h_{\rho\sigma}$ is the trace of the metric and $\Box = \partial_{\rho}\partial_{\rho}$ is the d'Alembertian operator.
Finally, the Ricci scalar is given by
\begin{align}
    R =& \mathrm{g}^{\mu\nu}R_{\mu\nu} = \frac{\eta^{\mu\nu}}{2}( h_{\nu\rho,\mu}^{\ \ \ \ \ \rho}+ h_{\mu\rho,\nu}^{\ \ \ \ \ \rho} - \Box h_{\mu\nu} - h_{,\mu\nu} ) + \mathcal{O}([h_{\mu\nu}]^2) \nonumber\\
    =& h_{\ \ \ ,\mu\rho}^{\mu\rho} - \Box h+ \mathcal{O}([h_{\mu\nu}]^2).
\end{align}
Inserting in the Einstein field equations, we have
\begin{align}
    8\pi T_{\mu\nu} =& R_{\mu\nu} -\frac{1}{2}\mathrm{g}_{\mu\nu}R =R_{\mu\nu} -\frac{1}{2}\eta_{\mu\nu}R+ \mathcal{O}([h_{\mu\nu}]^2)\nonumber \\
    =& \frac{1}{2}[ h_{\nu\rho,\mu}^{\ \ \ \ \ \rho}+ h_{\mu\rho,\nu}^{\ \ \ \ \ \rho} - \Box h_{\mu\nu} - h_{,\mu\nu}  - \eta_{\mu\nu}( h_{\ \ \ ,\rho\sigma}^{\rho\sigma} - \Box h)] + \mathcal{O}([h_{\mu\nu}]^2),
\end{align}
where $T_{\mu\nu}$ is the energy-momentum tensor.
We can simplify the equation above by writing it in terms of the trace free metric $\bar{h}_{\mu\nu}\equiv h_{\mu\nu} - \frac{1}{2}\eta_{\mu\nu}h$:
\begin{align}
    16\pi T_{\mu\nu}=&  \left(h_{\nu\rho,\mu}^{\ \ \ \ \ \rho} - \frac{1}{2}h_{,\mu\nu}\right)+ \left(h_{\mu\rho,\nu}^{\ \ \ \ \ \rho}- \frac{1}{2}h_{,\mu\nu}\right) \nonumber\\
    & - \left(\Box h_{\mu\nu}- \frac{1}{2}\eta_{\mu\nu}\Box h\right)   - \eta_{\mu\nu}( hv - \frac{1}{2}\Box h) + \mathcal{O}([h_{\mu\nu}]^2)\nonumber\\
    =& \bar{h}_{\nu\rho,\mu}^{\ \ \ \ \ \rho} +\bar{h}_{\mu\rho,\nu}^{\ \ \ \ \ \rho} -\Box\bar{h}_{\mu\nu}-\eta_{\mu\nu}\bar{h}_{\ \ \ ,\rho\sigma}^{\rho\sigma}+ \mathcal{O}([h_{\mu\nu}]^2).
    \label{fieldeq1}
\end{align}

The expression of equation (\ref{fieldeq1}) can be further simplified by considering gauge transformations.
The decomposition (\ref{eq:metric_linear_pert}) of the metric is not unique.
This means that we can find more than one different small perturbation $h_{\mu\nu}$ which leaves measurable physical quantities, like the curvature, unchanged.

Let us consider a infinitesimal coordinate transformation
\begin{equation}
	x^{\prime\mu} = x^\mu + \xi^\mu, \qquad \abs{\xi^\mu}\ll 1,
	\label{gaugetrans}
\end{equation}
where $\xi^\mu$ is an arbitrary vector field such that $\xi_{\mu ,\nu}$ is of the same order of $h_{\mu\nu}$.
In this new coordinate the metric is given by
\begin{align}
	\mathrm{g}^\prime_{\mu\nu} =& \mathrm{g}_{\rho\sigma}\pdv{x^\rho}{x^{\prime\mu}}\pdv{x^\sigma}{x^{\prime\nu}}\nonumber\\
	=& (\eta_{\rho\sigma} + h_{\rho\sigma})(\delta^\rho_\mu - \xi^\rho_{\ ,\mu})(\delta^\sigma_\nu - \xi^\sigma_{\ ,\nu}) \nonumber \\
	=& \eta_{\mu\nu} + h_{\mu\nu} - \xi_{\mu,\nu} - \xi_{\nu,\mu} + \mathcal{O}([h_{\mu\nu}]^2).\nonumber
\end{align}
Therefore, the new perturbation $h_{\mu\nu}^\text{new}$ is related with the old one $h_{\mu\nu}^\text{old}$ by
\begin{equation}
	h_{\mu\nu}^\text{new} = h_{\mu\nu}^\text{old}- \xi_{\mu,\nu} - \xi_{\nu,\mu}.
	\label{gaugelinear}
\end{equation}
We call this a \textit{gauge transformation in linearized theory}.
Now we need to show that the metric perturbations $h_{\mu\nu}^\text{new}$ and $h_{\mu\nu}^\text{old}$ represent physically equivalent spacetimes, that is, our theory is invariant under these transformations.
The curvature tensors are the components of the left side of the Einstein field equations, if they remain unchanged by these coordinate transformations the field equations will also remain unchanged.
Let us compute the changes of the linearized Ricci tensor.
Using equation (\ref{ricci}), we have
\begin{align}
    \delta R_{\mu\nu} =& R_{\mu\nu}^\text{new} - R_{\mu\nu}^\text{old} \nonumber \\
    =& \frac{1}{2}[ (h_{\nu\rho}- \xi_{\nu,\rho} - \xi_{\rho,\nu})_{,\mu}^{\ \ \rho}+ (h_{\mu\rho}- \xi_{\mu,\rho} - \xi_{\rho,\mu})_{,\nu}^{\ \ \rho} - \Box(h_{\mu\nu} - \xi_{\mu,\nu} - \xi_{\nu,\mu}) \nonumber\\
    &- \eta^{\rho\sigma}(h_{\rho\sigma} - \xi_{\rho,\sigma} - \xi_{\sigma,\rho})_{,\mu\nu} ] - \frac{1}{2}( h_{\nu\rho,\mu}^{\ \ \ \ \ \rho}+ h_{\mu\rho,\nu}^{\ \ \ \ \ \rho} - \Box h_{\mu\nu} - h_{,\mu\nu} )\nonumber\\
    =& \frac{1}{2}(-\xi_{\nu,\rho\mu}^{\ \ \ \ \ \rho} -\xi_{\rho,\nu\mu}^{\ \ \ \ \ \rho}-\xi_{\mu,\rho\nu}^{\ \ \ \ \ \rho}-\xi_{\rho,\mu\nu}^{\ \ \ \ \ \rho} + \xi_{\mu,\nu\rho}^{\ \ \ \ \ \rho} + \xi_{\nu,\mu\rho}^{\ \ \ \ \ \rho} +\xi_{\rho,\mu\nu}^{\ \ \ \ \ \rho} + \xi_{\rho,\mu\nu}^{\ \ \ \ \ \rho})\nonumber\\
    =& 0,
\end{align}
where in the last line we considered the symmetry in the partial derivatives.
As the Ricci scalar is just a contraction of the Ricci tensor, it also invariant under these transformations.
Therefore, the Einstein field equations is invariant under gauge transformations of the form (\ref{gaugelinear}).

\section{Lorentz gauge}
One gauge transformation we can consider is the one that turns the field equations (\ref{fieldeq1}) into a wave equation.
This is done by eliminating the terms of equation (\ref{fieldeq1}) that do not have a wave-like form.
We select a gauge transformation such that $(\bar{h}^\text{new})_{\mu\rho,}^{\ \ \ \ \rho} = 0.$
In this case, we have
\begin{align}
    0 =& (\bar{h}^\text{new})_{\mu\rho,}^{\ \ \ \ \rho} - \frac{1}{2}\eta_{\mu\rho}(\bar{h}^\text{new})_,^{\ \rho}\nonumber \\
    =& (h^\text{old})_{\mu\rho,}^{\ \ \ \ \rho}- \xi_{\mu,\rho}^{\ \ \ \ \rho} - \xi_{\rho,\mu}^{\ \ \ \ \rho} - \frac{1}{2}\eta_{\mu\rho}[(h^\text{old})_,^{\ \rho}- \xi_{\alpha,}^{\ \  \alpha \rho} - \xi_{\alpha,}^{\ \  \alpha \rho}]\nonumber \\
    =& (\bar{h}^\text{old})_{\mu\rho,}^{\ \ \ \ \rho}  - \Box\xi_{\mu},
\end{align}
where we considered the symmetry over the partial derivatives and relabeled the dummy index $\alpha$ to $\rho$.
Hence, any metric perturbation $h_{\mu\nu}$ can be put into the Lorentz gauge by making the coordinate transformation (\ref{gaugetrans}), in which $\xi^\mu$ satisfies $ \Box\xi_{\mu} = \bar{h}_{\mu\rho,}^{\ \ \ \ \rho}$.
In the Lorentz gauge the field equations (\ref{fieldeq1}) take the form
\begin{equation}
	\Box \bar{h}_{\mu\nu} = -16\pi T_{\mu\nu},
\end{equation}
which in vacuum reduces to a wave equation
\begin{equation}
	\Box \bar{h}_{\mu\nu} = 0.
\end{equation}
The Lorentz gauge indicates that a metric perturbation in flat spacetime propagates in spacetime as gravitational radiation.

\section{Transverse and traceless gauge}
The Lorentz gauge introduced in the last section does not determine the perturbation $h_{\mu\nu}$ uniquely, because there is remaining gauge freedom we can choose to impose to the perturbation which leaves the Lorentz gauge unchanged.
One gauge we can use is the Transverse and Traceless (TT) gauge, in which the metric perturbation is purely spatial, traceless and transverse (Lorentz gauge), i.e.,
\begin{equation}
	h^\text{TT}_{\mu 0} = 0, \qquad h^\text{TT} = \eta^{\mu\nu}h^\text{TT}_{\mu\nu} = 0, \qquad \partial^\mu h^\text{TT}_{\mu\nu}=0.
	\label{ttgauge}
\end{equation}
The traceless condition implies that $\bar{h}^\text{TT}_{\mu\nu}=h^\text{TT}_{\mu\nu}$.
We already found the transverse condition for the vector $\xi^\mu$.
The other two conditions are straightforward.
We have that any perturbation $h_{\mu\nu}$ can be put into the TT gauge by making the coordinate transformation (\ref{gaugelinear}), in which $\xi^\mu$ satisfies
\begin{align*}
     \xi_{\mu,0} - \xi_{0,\mu} = h_{\mu 0}, \qquad 2\xi_{\mu,}^{\ \ \mu} = h, \qquad \Box\xi_\mu = h_{\mu\nu,}^{\ \ \ \ \nu}.
\end{align*}

In vacuum, because of the transverse condition, the TT gauge perturbation equation of motion takes the wave form
\begin{equation}
	\Box h_{\mu\nu}^\text{TT} = 0.
\end{equation}
The simplest solutions of equation above are plane waves, which are given by
\begin{equation}
	h^\text{TT}_{\mu\nu} = \Re[A_{\mu\nu}e^{ik_\alpha x^\alpha}].
	\label{planewave}
\end{equation}
For simplicity, we can work with both real and imaginary parts of the solution and take the real part in the end.
Because of the gauge conditions (\ref{ttgauge}), the amplitude is purely spatial and traceless, and because of the symmetry in $h^\text{TT}_{\mu\nu}$, the amplitude is symmetric:
\begin{equation}
	A_{\mu 0} = 0,\qquad\qquad \eta^{\mu\nu}A_{\mu\nu} = 0, \qquad\qquad A_{\mu\nu} = A_{\nu\mu}.
\end{equation}
Using equation (\ref{planewave}), we can find a condition for the wave vector $k_\alpha$:
\begin{align*}
   0=& \eta^{\rho\sigma}h^{\text{TT}}_{\mu\nu,\rho\sigma} =  \eta^{\rho\sigma}\partial_{\rho}\partial_{\sigma} (A_{\mu\nu}e^{ik_\alpha x^\alpha}) \\
   =&\eta^{\rho\sigma}(ik_{\rho})(ik_{\sigma})A_{\mu\nu}e^{ik_{\alpha} x^{\alpha}}\\
   =& - k_{\rho} k^{\rho} h^{\text{TT}}_{\mu\nu,\rho\sigma}.
\end{align*}
Since we are not interested in the trivial solution of $h^\text{TT}_{\mu\nu,\rho\sigma}$, we have $k_\rho k^\rho = 0$, i.e., the wave vector is a null vector.
This implies that gravitational waves propagate at the speed of light.

The transverse condition name is justified by
$$
0 = \partial^\mu h^\text{TT}_{\mu\nu} = ik^\mu A_{\mu\nu}e^{ik_\alpha x^\alpha} \quad \to \quad k^\mu A_{\mu\nu} = 0,
$$
that is, the wave vector is orthogonal to the amplitude.

We can get a better insight if we choose a reference frame such that the wave is traveling in the $x^3 = z$ direction.
In that case, $k^\mu \to(\omega,0,0,\omega)$, where $\omega$ is the frequency of the wave and $k^3 = \omega$ because $k^\mu$ is a null vector.
By the transverse condition, we have
$$0= k^\mu A_{\mu\nu} = \omega(A_{0\nu} + A_{3\nu}) = \omega A_{3\nu} \quad \to A_{3\nu} = 0,$$
where we considered the purely spatial condition $A_{0\nu}=0$.
Therefore, the only nonzero components of the amplitude are $A_{11}$, $A_{12}$, $A_{21}$ and $A_{22}$.
The traceless condition implies $A_{22} = -A_{11}$ and the symmetry implies $A_{12} = A_{21}$.
Denoting $A_{11} = h_{+}$ and $A_{12} = h_{\times}$, for the $+$ and $\times$ polarizations, we have
\begin{equation}
	A_{\mu\nu} = \mqty(0&0&0&0 \\ 0 & h_{+} & h_{\times} &0 \\0 & h_{\times} & - h_{+} &0 \\ 0&0&0&0).
	\label{eq:amplitude_h_tt}
\end{equation}
Hence, a plane wave traveling in the $x^3 = z$ direction is fully characterized by $h_{+}$ and $h_{\times}$, i.e., the wave has two degrees of freedom.
Therefore, the basis polarization tensors are defined as
\begin{equation}
    \mathbf{e}_+ \equiv \mathbf{e}_x\otimes\mathbf{e}_x - \mathbf{e}_y\otimes \mathbf{e}_y, \qquad
    \mathbf{e}_{\times} \equiv \mathbf{e}_x\otimes\mathbf{e}_y + \mathbf{e}_y\otimes \mathbf{e}_x,
    \label{eq:polarization_tensor}
\end{equation}
where $(\mathbf{e}_x, \mathbf{e}_y, \mathbf{e}_z)$ is an orthonormal spatial basis in which the metric takes the form of equation~\eqref{eq:amplitude_h_tt}.
Thus, the metric perturbation can be written as
\begin{equation}
    \mathbf{h} = h_+\mathbf{e}_+ + h_{\times}\mathbf{e}_{\times}.
    \label{eq:h_polarization_tensors}
\end{equation}

%

\bibliographystyle{apsrev4-1}

\bibliography{main}

\begin{thebibliography}{130}%
\makeatletter
\providecommand \@ifxundefined [1]{%
 \@ifx{#1\undefined}
}%
\providecommand \@ifnum [1]{%
 \ifnum #1\expandafter \@firstoftwo
 \else \expandafter \@secondoftwo
 \fi
}%
\providecommand \@ifx [1]{%
 \ifx #1\expandafter \@firstoftwo
 \else \expandafter \@secondoftwo
 \fi
}%
\providecommand \natexlab [1]{#1}%
\providecommand \enquote  [1]{``#1''}%
\providecommand \bibnamefont  [1]{#1}%
\providecommand \bibfnamefont [1]{#1}%
\providecommand \citenamefont [1]{#1}%
\providecommand \href@noop [0]{\@secondoftwo}%
\providecommand \href [0]{\begingroup \@sanitize@url \@href}%
\providecommand \@href[1]{\@@startlink{#1}\@@href}%
\providecommand \@@href[1]{\endgroup#1\@@endlink}%
\providecommand \@sanitize@url [0]{\catcode `\\12\catcode `\$12\catcode
  `\&12\catcode `\#12\catcode `\^12\catcode `\_12\catcode `\%12\relax}%
\providecommand \@@startlink[1]{}%
\providecommand \@@endlink[0]{}%
\providecommand \url  [0]{\begingroup\@sanitize@url \@url }%
\providecommand \@url [1]{\endgroup\@href {#1}{\urlprefix }}%
\providecommand \urlprefix  [0]{URL }%
\providecommand \Eprint [0]{\href }%
\providecommand \doibase [0]{http://dx.doi.org/}%
\providecommand \selectlanguage [0]{\@gobble}%
\providecommand \bibinfo  [0]{\@secondoftwo}%
\providecommand \bibfield  [0]{\@secondoftwo}%
\providecommand \translation [1]{[#1]}%
\providecommand \BibitemOpen [0]{}%
\providecommand \bibitemStop [0]{}%
\providecommand \bibitemNoStop [0]{.\EOS\space}%
\providecommand \EOS [0]{\spacefactor3000\relax}%
\providecommand \BibitemShut  [1]{\csname bibitem#1\endcsname}%
\let\auto@bib@innerbib\@empty
\bibitem [{\citenamefont {Abbott}\ \emph
  {et~al.}(2016{\natexlab{a}})\citenamefont {Abbott} \emph
  {et~al.}}]{LIGOScientific:2016aoc}%
  \BibitemOpen
  \bibfield  {author} {\bibinfo {author} {\bibfnamefont {B.~P.}\ \bibnamefont
  {Abbott}} \emph {et~al.} (\bibinfo {collaboration} {LIGO Scientific,
  Virgo}),\ }\href {\doibase 10.1103/PhysRevLett.116.061102} {\bibfield
  {journal} {\bibinfo  {journal} {Phys. Rev. Lett.}\ }\textbf {\bibinfo
  {volume} {116}},\ \bibinfo {pages} {061102} (\bibinfo {year}
  {2016}{\natexlab{a}})},\ \Eprint {http://arxiv.org/abs/1602.03837}
  {arXiv:1602.03837 [gr-qc]} \BibitemShut {NoStop}%
\bibitem [{\citenamefont {Taylor}\ and\ \citenamefont
  {Weisberg}(1982)}]{Taylor:1982zz}%
  \BibitemOpen
  \bibfield  {author} {\bibinfo {author} {\bibfnamefont {J.~H.}\ \bibnamefont
  {Taylor}}\ and\ \bibinfo {author} {\bibfnamefont {J.~M.}\ \bibnamefont
  {Weisberg}},\ }\href {\doibase 10.1086/159690} {\bibfield  {journal}
  {\bibinfo  {journal} {Astrophys. J.}\ }\textbf {\bibinfo {volume} {253}},\
  \bibinfo {pages} {908} (\bibinfo {year} {1982})}\BibitemShut {NoStop}%
\bibitem [{\citenamefont {Weisberg}\ and\ \citenamefont
  {Taylor}(2005)}]{Weisberg:2004hi}%
  \BibitemOpen
  \bibfield  {author} {\bibinfo {author} {\bibfnamefont {J.~M.}\ \bibnamefont
  {Weisberg}}\ and\ \bibinfo {author} {\bibfnamefont {J.~H.}\ \bibnamefont
  {Taylor}},\ }\href@noop {} {\bibfield  {journal} {\bibinfo  {journal} {ASP
  Conf. Ser.}\ }\textbf {\bibinfo {volume} {328}},\ \bibinfo {pages} {25}
  (\bibinfo {year} {2005})},\ \Eprint {http://arxiv.org/abs/astro-ph/0407149}
  {arXiv:astro-ph/0407149} \BibitemShut {NoStop}%
\bibitem [{\citenamefont {Abbott}\ \emph
  {et~al.}(2018{\natexlab{a}})\citenamefont {Abbott} \emph
  {et~al.}}]{LIGOScientific:2018cki}%
  \BibitemOpen
  \bibfield  {author} {\bibinfo {author} {\bibfnamefont {B.~P.}\ \bibnamefont
  {Abbott}} \emph {et~al.} (\bibinfo {collaboration} {LIGO Scientific,
  Virgo}),\ }\href {\doibase 10.1103/PhysRevLett.121.161101} {\bibfield
  {journal} {\bibinfo  {journal} {Phys. Rev. Lett.}\ }\textbf {\bibinfo
  {volume} {121}},\ \bibinfo {pages} {161101} (\bibinfo {year}
  {2018}{\natexlab{a}})},\ \Eprint {http://arxiv.org/abs/1805.11581}
  {arXiv:1805.11581 [gr-qc]} \BibitemShut {NoStop}%
\bibitem [{\citenamefont {Maggiore}(2007)}]{maggiore-vol1}%
  \BibitemOpen
  \bibfield  {author} {\bibinfo {author} {\bibfnamefont {M.}~\bibnamefont
  {Maggiore}},\ }\href@noop {} {\emph {\bibinfo {title} {Gravitational Waves:
  Volume 1: Theory and Experiments}}}\ (\bibinfo  {publisher} {Oxford
  University Press},\ \bibinfo {year} {2007})\BibitemShut {NoStop}%
\bibitem [{\citenamefont {Poisson}\ and\ \citenamefont
  {Will}(2014)}]{poisson-gravity}%
  \BibitemOpen
  \bibfield  {author} {\bibinfo {author} {\bibfnamefont {E.}~\bibnamefont
  {Poisson}}\ and\ \bibinfo {author} {\bibfnamefont {C.~M.}\ \bibnamefont
  {Will}},\ }\href@noop {} {\emph {\bibinfo {title} {Gravity: Newtonian,
  Post-Newtonian, Relativistic}}}\ (\bibinfo  {publisher} {Cambridge University
  Press},\ \bibinfo {year} {2014})\BibitemShut {NoStop}%
\bibitem [{\citenamefont {Abbott}\ \emph
  {et~al.}(2021{\natexlab{a}})\citenamefont {Abbott} \emph
  {et~al.}}]{LIGOScientific:2020ibl}%
  \BibitemOpen
  \bibfield  {author} {\bibinfo {author} {\bibfnamefont {R.}~\bibnamefont
  {Abbott}} \emph {et~al.} (\bibinfo {collaboration} {LIGO Scientific,
  Virgo}),\ }\href {\doibase 10.1103/PhysRevX.11.021053} {\bibfield  {journal}
  {\bibinfo  {journal} {Phys. Rev. X}\ }\textbf {\bibinfo {volume} {11}},\
  \bibinfo {pages} {021053} (\bibinfo {year} {2021}{\natexlab{a}})},\ \Eprint
  {http://arxiv.org/abs/2010.14527} {arXiv:2010.14527 [gr-qc]} \BibitemShut
  {NoStop}%
\bibitem [{\citenamefont {Abbott}\ \emph
  {et~al.}(2016{\natexlab{b}})\citenamefont {Abbott} \emph
  {et~al.}}]{Abbott:2016xvh}%
  \BibitemOpen
  \bibfield  {author} {\bibinfo {author} {\bibfnamefont {B.~P.}\ \bibnamefont
  {Abbott}} \emph {et~al.},\ }\href {\doibase 10.1103/PhysRevD.93.112004}
  {\bibfield  {journal} {\bibinfo  {journal} {Phys. Rev. D}\ }\textbf {\bibinfo
  {volume} {93}},\ \bibinfo {pages} {112004} (\bibinfo {year}
  {2016}{\natexlab{b}})},\ \bibinfo {note} {[Addendum: Phys.Rev.D 97, 059901
  (2018)]},\ \Eprint {http://arxiv.org/abs/1604.00439} {arXiv:1604.00439
  [astro-ph.IM]} \BibitemShut {NoStop}%
\bibitem [{\citenamefont {Buikema}\ \emph {et~al.}(2020)\citenamefont {Buikema}
  \emph {et~al.}}]{aLIGO:2020wna}%
  \BibitemOpen
  \bibfield  {author} {\bibinfo {author} {\bibfnamefont {A.}~\bibnamefont
  {Buikema}} \emph {et~al.} (\bibinfo {collaboration} {aLIGO}),\ }\href
  {\doibase 10.1103/PhysRevD.102.062003} {\bibfield  {journal} {\bibinfo
  {journal} {Phys. Rev. D}\ }\textbf {\bibinfo {volume} {102}},\ \bibinfo
  {pages} {062003} (\bibinfo {year} {2020})},\ \Eprint
  {http://arxiv.org/abs/2008.01301} {arXiv:2008.01301 [astro-ph.IM]}
  \BibitemShut {NoStop}%
\bibitem [{\citenamefont {Abbott}\ \emph
  {et~al.}(2018{\natexlab{b}})\citenamefont {Abbott} \emph
  {et~al.}}]{KAGRA:2013rdx}%
  \BibitemOpen
  \bibfield  {author} {\bibinfo {author} {\bibfnamefont {B.~P.}\ \bibnamefont
  {Abbott}} \emph {et~al.} (\bibinfo {collaboration} {KAGRA, LIGO Scientific,
  Virgo, VIRGO}),\ }\href {\doibase 10.1007/s41114-020-00026-9} {\bibfield
  {journal} {\bibinfo  {journal} {Living Rev. Rel.}\ }\textbf {\bibinfo
  {volume} {21}},\ \bibinfo {pages} {3} (\bibinfo {year}
  {2018}{\natexlab{b}})},\ \Eprint {http://arxiv.org/abs/1304.0670}
  {arXiv:1304.0670 [gr-qc]} \BibitemShut {NoStop}%
\bibitem [{\citenamefont {Shoemaker}(2019)}]{Shoemaker:2019bqt}%
  \BibitemOpen
  \bibfield  {author} {\bibinfo {author} {\bibfnamefont {D.}~\bibnamefont
  {Shoemaker}} (\bibinfo {collaboration} {LIGO Scientific}),\ }\href@noop {} {\
   (\bibinfo {year} {2019})},\ \Eprint {http://arxiv.org/abs/1904.03187}
  {arXiv:1904.03187 [gr-qc]} \BibitemShut {NoStop}%
\bibitem [{\citenamefont {Abbott}\ \emph
  {et~al.}(2021{\natexlab{b}})\citenamefont {Abbott} \emph
  {et~al.}}]{LIGOScientific:2021djp}%
  \BibitemOpen
  \bibfield  {author} {\bibinfo {author} {\bibfnamefont {R.}~\bibnamefont
  {Abbott}} \emph {et~al.} (\bibinfo {collaboration} {LIGO Scientific, VIRGO,
  KAGRA}),\ }\href@noop {} {\  (\bibinfo {year} {2021}{\natexlab{b}})},\
  \Eprint {http://arxiv.org/abs/2111.03606} {arXiv:2111.03606 [gr-qc]}
  \BibitemShut {NoStop}%
\bibitem [{\citenamefont {Abbott}\ \emph {et~al.}(2019)\citenamefont {Abbott}
  \emph {et~al.}}]{LIGOScientific:2018mvr}%
  \BibitemOpen
  \bibfield  {author} {\bibinfo {author} {\bibfnamefont {B.~P.}\ \bibnamefont
  {Abbott}} \emph {et~al.} (\bibinfo {collaboration} {LIGO Scientific,
  Virgo}),\ }\href {\doibase 10.1103/PhysRevX.9.031040} {\bibfield  {journal}
  {\bibinfo  {journal} {Phys. Rev. X}\ }\textbf {\bibinfo {volume} {9}},\
  \bibinfo {pages} {031040} (\bibinfo {year} {2019})},\ \Eprint
  {http://arxiv.org/abs/1811.12907} {arXiv:1811.12907 [astro-ph.HE]}
  \BibitemShut {NoStop}%
\bibitem [{\citenamefont {Abbott}\ \emph
  {et~al.}(2021{\natexlab{c}})\citenamefont {Abbott} \emph
  {et~al.}}]{LIGOScientific:2021usb}%
  \BibitemOpen
  \bibfield  {author} {\bibinfo {author} {\bibfnamefont {R.}~\bibnamefont
  {Abbott}} \emph {et~al.} (\bibinfo {collaboration} {LIGO Scientific,
  VIRGO}),\ }\href@noop {} {\  (\bibinfo {year} {2021}{\natexlab{c}})},\
  \Eprint {http://arxiv.org/abs/2108.01045} {arXiv:2108.01045 [gr-qc]}
  \BibitemShut {NoStop}%
\bibitem [{\citenamefont {Collaboration}(2018)}]{LIGOScientific:T1700231}%
  \BibitemOpen
  \bibfield  {author} {\bibinfo {author} {\bibfnamefont {L.~S.}\ \bibnamefont
  {Collaboration}} (\bibinfo {collaboration} {LIGO Scientific Collaboration}),\
  }\href@noop {} {\  (\bibinfo {year} {2018})}\BibitemShut {NoStop}%
\bibitem [{\citenamefont {Punturo}\ \emph {et~al.}(2010)\citenamefont {Punturo}
  \emph {et~al.}}]{Punturo:2010zz}%
  \BibitemOpen
  \bibfield  {author} {\bibinfo {author} {\bibfnamefont {M.}~\bibnamefont
  {Punturo}} \emph {et~al.},\ }\href {\doibase 10.1088/0264-9381/27/19/194002}
  {\bibfield  {journal} {\bibinfo  {journal} {Class. Quant. Grav.}\ }\textbf
  {\bibinfo {volume} {27}},\ \bibinfo {pages} {194002} (\bibinfo {year}
  {2010})}\BibitemShut {NoStop}%
\bibitem [{\citenamefont {Abbott}\ \emph
  {et~al.}(2017{\natexlab{a}})\citenamefont {Abbott} \emph
  {et~al.}}]{LIGOScientific:2016wof}%
  \BibitemOpen
  \bibfield  {author} {\bibinfo {author} {\bibfnamefont {B.~P.}\ \bibnamefont
  {Abbott}} \emph {et~al.} (\bibinfo {collaboration} {LIGO Scientific}),\
  }\href {\doibase 10.1088/1361-6382/aa51f4} {\bibfield  {journal} {\bibinfo
  {journal} {Class. Quant. Grav.}\ }\textbf {\bibinfo {volume} {34}},\ \bibinfo
  {pages} {044001} (\bibinfo {year} {2017}{\natexlab{a}})},\ \Eprint
  {http://arxiv.org/abs/1607.08697} {arXiv:1607.08697 [astro-ph.IM]}
  \BibitemShut {NoStop}%
\bibitem [{\citenamefont {Amaro-Seoane}\ \emph {et~al.}(2017)\citenamefont
  {Amaro-Seoane} \emph {et~al.}}]{amaroseoane2017laser}%
  \BibitemOpen
  \bibfield  {author} {\bibinfo {author} {\bibfnamefont {P.}~\bibnamefont
  {Amaro-Seoane}} \emph {et~al.},\ }\href@noop {} {\enquote {\bibinfo {title}
  {Laser interferometer space antenna},}\ } (\bibinfo {year} {2017}),\ \Eprint
  {http://arxiv.org/abs/1702.00786} {arXiv:1702.00786 [astro-ph.IM]}
  \BibitemShut {NoStop}%
\bibitem [{\citenamefont {Luo}\ \emph {et~al.}(2016)\citenamefont {Luo} \emph
  {et~al.}}]{TianQin:2015yph}%
  \BibitemOpen
  \bibfield  {author} {\bibinfo {author} {\bibfnamefont {J.}~\bibnamefont
  {Luo}} \emph {et~al.} (\bibinfo {collaboration} {TianQin}),\ }\href {\doibase
  10.1088/0264-9381/33/3/035010} {\bibfield  {journal} {\bibinfo  {journal}
  {Class. Quant. Grav.}\ }\textbf {\bibinfo {volume} {33}},\ \bibinfo {pages}
  {035010} (\bibinfo {year} {2016})},\ \Eprint
  {http://arxiv.org/abs/1512.02076} {arXiv:1512.02076 [astro-ph.IM]}
  \BibitemShut {NoStop}%
\bibitem [{\citenamefont {Weber}(1960)}]{Weber:1960zz}%
  \BibitemOpen
  \bibfield  {author} {\bibinfo {author} {\bibfnamefont {J.}~\bibnamefont
  {Weber}},\ }\href {\doibase 10.1103/PhysRev.117.306} {\bibfield  {journal}
  {\bibinfo  {journal} {Phys. Rev.}\ }\textbf {\bibinfo {volume} {117}},\
  \bibinfo {pages} {306} (\bibinfo {year} {1960})}\BibitemShut {NoStop}%
\bibitem [{\citenamefont {Weber}(1969)}]{Weber:1969bz}%
  \BibitemOpen
  \bibfield  {author} {\bibinfo {author} {\bibfnamefont {J.}~\bibnamefont
  {Weber}},\ }\href {\doibase 10.1103/PhysRevLett.22.1320} {\bibfield
  {journal} {\bibinfo  {journal} {Phys. Rev. Lett.}\ }\textbf {\bibinfo
  {volume} {22}},\ \bibinfo {pages} {1320} (\bibinfo {year}
  {1969})}\BibitemShut {NoStop}%
\bibitem [{\citenamefont {Tyson}(1973)}]{Tyson:1973ra}%
  \BibitemOpen
  \bibfield  {author} {\bibinfo {author} {\bibfnamefont {J.~A.}\ \bibnamefont
  {Tyson}},\ }\href {\doibase 10.1103/PhysRevLett.31.326} {\bibfield  {journal}
  {\bibinfo  {journal} {Phys. Rev. Lett.}\ }\textbf {\bibinfo {volume} {31}},\
  \bibinfo {pages} {326} (\bibinfo {year} {1973})}\BibitemShut {NoStop}%
\bibitem [{\citenamefont {Aguiar}\ \emph {et~al.}(2002)\citenamefont {Aguiar}
  \emph {et~al.}}]{Aguiar:2002eq}%
  \BibitemOpen
  \bibfield  {author} {\bibinfo {author} {\bibfnamefont {O.~D.}\ \bibnamefont
  {Aguiar}} \emph {et~al.},\ }\href {\doibase 10.1088/0264-9381/19/7/397}
  {\bibfield  {journal} {\bibinfo  {journal} {Class. Quant. Grav.}\ }\textbf
  {\bibinfo {volume} {19}},\ \bibinfo {pages} {1949} (\bibinfo {year}
  {2002})}\BibitemShut {NoStop}%
\bibitem [{\citenamefont {Aguiar}\ \emph {et~al.}(2005)\citenamefont {Aguiar}
  \emph {et~al.}}]{Aguiar:2005tp}%
  \BibitemOpen
  \bibfield  {author} {\bibinfo {author} {\bibfnamefont {O.~D.}\ \bibnamefont
  {Aguiar}} \emph {et~al.},\ }\href {\doibase 10.1088/0264-9381/22/10/011}
  {\bibfield  {journal} {\bibinfo  {journal} {Class. Quant. Grav.}\ }\textbf
  {\bibinfo {volume} {22}},\ \bibinfo {pages} {S209} (\bibinfo {year}
  {2005})}\BibitemShut {NoStop}%
\bibitem [{\citenamefont {Aguiar}\ \emph {et~al.}(2008)\citenamefont {Aguiar}
  \emph {et~al.}}]{Aguiar:2008zz}%
  \BibitemOpen
  \bibfield  {author} {\bibinfo {author} {\bibfnamefont {O.~D.}\ \bibnamefont
  {Aguiar}} \emph {et~al.},\ }\href {\doibase 10.1088/0264-9381/25/11/114042}
  {\bibfield  {journal} {\bibinfo  {journal} {Class. Quant. Grav.}\ }\textbf
  {\bibinfo {volume} {25}},\ \bibinfo {pages} {114042} (\bibinfo {year}
  {2008})}\BibitemShut {NoStop}%
\bibitem [{\citenamefont {Oliveira}\ and\ \citenamefont
  {Aguiar}(2016)}]{Oliveira:2016gds}%
  \BibitemOpen
  \bibfield  {author} {\bibinfo {author} {\bibfnamefont {N.~F.}\ \bibnamefont
  {Oliveira}}\ and\ \bibinfo {author} {\bibfnamefont {O.~D.}\ \bibnamefont
  {Aguiar}},\ }\href {\doibase 10.1007/s13538-016-0436-1} {\bibfield  {journal}
  {\bibinfo  {journal} {Braz. J. Phys.}\ }\textbf {\bibinfo {volume} {46}},\
  \bibinfo {pages} {596} (\bibinfo {year} {2016})}\BibitemShut {NoStop}%
\bibitem [{\citenamefont {Boyle}\ \emph {et~al.}(2019)\citenamefont {Boyle}
  \emph {et~al.}}]{Boyle:2019kee}%
  \BibitemOpen
  \bibfield  {author} {\bibinfo {author} {\bibfnamefont {M.}~\bibnamefont
  {Boyle}} \emph {et~al.},\ }\href {\doibase 10.1088/1361-6382/ab34e2}
  {\bibfield  {journal} {\bibinfo  {journal} {Class. Quant. Grav.}\ }\textbf
  {\bibinfo {volume} {36}},\ \bibinfo {pages} {195006} (\bibinfo {year}
  {2019})},\ \Eprint {http://arxiv.org/abs/1904.04831} {arXiv:1904.04831
  [gr-qc]} \BibitemShut {NoStop}%
\bibitem [{SXS()}]{SXS-catalog}%
  \BibitemOpen
  \href@noop {} {}\bibinfo {howpublished}
  {\url{https://data.black-holes.org/waveforms/index.html}}\BibitemShut
  {NoStop}%
\bibitem [{\citenamefont {Blanchet}(2014)}]{Blanchet:2013haa}%
  \BibitemOpen
  \bibfield  {author} {\bibinfo {author} {\bibfnamefont {L.}~\bibnamefont
  {Blanchet}},\ }\href {\doibase 10.12942/lrr-2014-2} {\bibfield  {journal}
  {\bibinfo  {journal} {Living Rev. Rel.}\ }\textbf {\bibinfo {volume} {17}},\
  \bibinfo {pages} {2} (\bibinfo {year} {2014})},\ \Eprint
  {http://arxiv.org/abs/1310.1528} {arXiv:1310.1528 [gr-qc]} \BibitemShut
  {NoStop}%
\bibitem [{\citenamefont {Buonanno}\ \emph {et~al.}(2007)\citenamefont
  {Buonanno}, \citenamefont {Cook},\ and\ \citenamefont
  {Pretorius}}]{Buonanno:2006ui}%
  \BibitemOpen
  \bibfield  {author} {\bibinfo {author} {\bibfnamefont {A.}~\bibnamefont
  {Buonanno}}, \bibinfo {author} {\bibfnamefont {G.~B.}\ \bibnamefont {Cook}},
  \ and\ \bibinfo {author} {\bibfnamefont {F.}~\bibnamefont {Pretorius}},\
  }\href {\doibase 10.1103/PhysRevD.75.124018} {\bibfield  {journal} {\bibinfo
  {journal} {Phys. Rev. D}\ }\textbf {\bibinfo {volume} {75}},\ \bibinfo
  {pages} {124018} (\bibinfo {year} {2007})},\ \Eprint
  {http://arxiv.org/abs/gr-qc/0610122} {arXiv:gr-qc/0610122} \BibitemShut
  {NoStop}%
\bibitem [{\citenamefont {Seidel}(2004)}]{Seidel_2004}%
  \BibitemOpen
  \bibfield  {author} {\bibinfo {author} {\bibfnamefont {E.}~\bibnamefont
  {Seidel}},\ }\href {\doibase 10.1088/0264-9381/21/3/021} {\bibfield
  {journal} {\bibinfo  {journal} {Classical and Quantum Gravity}\ }\textbf
  {\bibinfo {volume} {21}},\ \bibinfo {pages} {S339} (\bibinfo {year}
  {2004})}\BibitemShut {NoStop}%
\bibitem [{\citenamefont {Regge}\ and\ \citenamefont
  {Wheeler}(1957)}]{Regge:1957td}%
  \BibitemOpen
  \bibfield  {author} {\bibinfo {author} {\bibfnamefont {T.}~\bibnamefont
  {Regge}}\ and\ \bibinfo {author} {\bibfnamefont {J.~A.}\ \bibnamefont
  {Wheeler}},\ }\href {\doibase 10.1103/PhysRev.108.1063} {\bibfield  {journal}
  {\bibinfo  {journal} {Phys. Rev.}\ }\textbf {\bibinfo {volume} {108}},\
  \bibinfo {pages} {1063} (\bibinfo {year} {1957})}\BibitemShut {NoStop}%
\bibitem [{\citenamefont {Vishveshwara}(1970)}]{Vishveshwara:1970zz}%
  \BibitemOpen
  \bibfield  {author} {\bibinfo {author} {\bibfnamefont {C.~V.}\ \bibnamefont
  {Vishveshwara}},\ }\href {\doibase 10.1038/227936a0} {\bibfield  {journal}
  {\bibinfo  {journal} {Nature}\ }\textbf {\bibinfo {volume} {227}},\ \bibinfo
  {pages} {936} (\bibinfo {year} {1970})}\BibitemShut {NoStop}%
\bibitem [{\citenamefont {Press}(1971)}]{Press:1971wr}%
  \BibitemOpen
  \bibfield  {author} {\bibinfo {author} {\bibfnamefont {W.~H.}\ \bibnamefont
  {Press}},\ }\href {\doibase 10.1086/180849} {\bibfield  {journal} {\bibinfo
  {journal} {Astrophys. J. Lett.}\ }\textbf {\bibinfo {volume} {170}},\
  \bibinfo {pages} {L105} (\bibinfo {year} {1971})}\BibitemShut {NoStop}%
\bibitem [{\citenamefont {Abbott}\ \emph
  {et~al.}(2016{\natexlab{c}})\citenamefont {Abbott} \emph
  {et~al.}}]{LIGOScientific:2016lio}%
  \BibitemOpen
  \bibfield  {author} {\bibinfo {author} {\bibfnamefont {B.~P.}\ \bibnamefont
  {Abbott}} \emph {et~al.} (\bibinfo {collaboration} {LIGO Scientific,
  Virgo}),\ }\href {\doibase 10.1103/PhysRevLett.116.221101} {\bibfield
  {journal} {\bibinfo  {journal} {Phys. Rev. Lett.}\ }\textbf {\bibinfo
  {volume} {116}},\ \bibinfo {pages} {221101} (\bibinfo {year}
  {2016}{\natexlab{c}})},\ \bibinfo {note} {[Erratum: Phys.Rev.Lett. 121,
  129902 (2018)]},\ \Eprint {http://arxiv.org/abs/1602.03841} {arXiv:1602.03841
  [gr-qc]} \BibitemShut {NoStop}%
\bibitem [{\citenamefont {Abbott}\ \emph
  {et~al.}(2020{\natexlab{a}})\citenamefont {Abbott} \emph
  {et~al.}}]{LIGOScientific:2020ufj}%
  \BibitemOpen
  \bibfield  {author} {\bibinfo {author} {\bibfnamefont {R.}~\bibnamefont
  {Abbott}} \emph {et~al.} (\bibinfo {collaboration} {LIGO Scientific,
  Virgo}),\ }\href {\doibase 10.3847/2041-8213/aba493} {\bibfield  {journal}
  {\bibinfo  {journal} {Astrophys. J. Lett.}\ }\textbf {\bibinfo {volume}
  {900}},\ \bibinfo {pages} {L13} (\bibinfo {year} {2020}{\natexlab{a}})},\
  \Eprint {http://arxiv.org/abs/2009.01190} {arXiv:2009.01190 [astro-ph.HE]}
  \BibitemShut {NoStop}%
\bibitem [{\citenamefont {Abbott}\ \emph
  {et~al.}(2021{\natexlab{d}})\citenamefont {Abbott} \emph
  {et~al.}}]{LIGOScientific:2020tif}%
  \BibitemOpen
  \bibfield  {author} {\bibinfo {author} {\bibfnamefont {R.}~\bibnamefont
  {Abbott}} \emph {et~al.} (\bibinfo {collaboration} {LIGO Scientific,
  Virgo}),\ }\href {\doibase 10.1103/PhysRevD.103.122002} {\bibfield  {journal}
  {\bibinfo  {journal} {Phys. Rev. D}\ }\textbf {\bibinfo {volume} {103}},\
  \bibinfo {pages} {122002} (\bibinfo {year} {2021}{\natexlab{d}})},\ \Eprint
  {http://arxiv.org/abs/2010.14529} {arXiv:2010.14529 [gr-qc]} \BibitemShut
  {NoStop}%
\bibitem [{\citenamefont {Abbott}\ \emph
  {et~al.}(2021{\natexlab{e}})\citenamefont {Abbott} \emph
  {et~al.}}]{LIGOScientific:2021sio}%
  \BibitemOpen
  \bibfield  {author} {\bibinfo {author} {\bibfnamefont {R.}~\bibnamefont
  {Abbott}} \emph {et~al.} (\bibinfo {collaboration} {LIGO Scientific, VIRGO,
  KAGRA}),\ }\href@noop {} {\  (\bibinfo {year} {2021}{\natexlab{e}})},\
  \Eprint {http://arxiv.org/abs/2112.06861} {arXiv:2112.06861 [gr-qc]}
  \BibitemShut {NoStop}%
\bibitem [{\citenamefont {Dreyer}\ \emph {et~al.}(2004)\citenamefont {Dreyer},
  \citenamefont {Kelly}, \citenamefont {Krishnan}, \citenamefont {Finn},
  \citenamefont {Garrison},\ and\ \citenamefont
  {Lopez-Aleman}}]{Dreyer:2003bv}%
  \BibitemOpen
  \bibfield  {author} {\bibinfo {author} {\bibfnamefont {O.}~\bibnamefont
  {Dreyer}}, \bibinfo {author} {\bibfnamefont {B.~J.}\ \bibnamefont {Kelly}},
  \bibinfo {author} {\bibfnamefont {B.}~\bibnamefont {Krishnan}}, \bibinfo
  {author} {\bibfnamefont {L.~S.}\ \bibnamefont {Finn}}, \bibinfo {author}
  {\bibfnamefont {D.}~\bibnamefont {Garrison}}, \ and\ \bibinfo {author}
  {\bibfnamefont {R.}~\bibnamefont {Lopez-Aleman}},\ }\href {\doibase
  10.1088/0264-9381/21/4/003} {\bibfield  {journal} {\bibinfo  {journal}
  {Class. Quant. Grav.}\ }\textbf {\bibinfo {volume} {21}},\ \bibinfo {pages}
  {787} (\bibinfo {year} {2004})},\ \Eprint
  {http://arxiv.org/abs/gr-qc/0309007} {arXiv:gr-qc/0309007} \BibitemShut
  {NoStop}%
\bibitem [{\citenamefont {Ota}\ and\ \citenamefont
  {Chirenti}(2020)}]{Ota:2019bzl}%
  \BibitemOpen
  \bibfield  {author} {\bibinfo {author} {\bibfnamefont {I.}~\bibnamefont
  {Ota}}\ and\ \bibinfo {author} {\bibfnamefont {C.}~\bibnamefont {Chirenti}},\
  }\href {\doibase 10.1103/PhysRevD.101.104005} {\bibfield  {journal} {\bibinfo
   {journal} {Phys. Rev. D}\ }\textbf {\bibinfo {volume} {101}},\ \bibinfo
  {pages} {104005} (\bibinfo {year} {2020})},\ \Eprint
  {http://arxiv.org/abs/1911.00440} {arXiv:1911.00440 [gr-qc]} \BibitemShut
  {NoStop}%
\bibitem [{\citenamefont {Ota}\ and\ \citenamefont
  {Chirenti}(2022)}]{Ota:2021ypb}%
  \BibitemOpen
  \bibfield  {author} {\bibinfo {author} {\bibfnamefont {I.}~\bibnamefont
  {Ota}}\ and\ \bibinfo {author} {\bibfnamefont {C.}~\bibnamefont {Chirenti}},\
  }\href {\doibase 10.1103/PhysRevD.105.044015} {\bibfield  {journal} {\bibinfo
   {journal} {Phys. Rev. D}\ }\textbf {\bibinfo {volume} {105}},\ \bibinfo
  {pages} {044015} (\bibinfo {year} {2022})},\ \Eprint
  {http://arxiv.org/abs/2108.01774} {arXiv:2108.01774 [gr-qc]} \BibitemShut
  {NoStop}%
\bibitem [{git()}]{github}%
  \BibitemOpen
  \href@noop {} {}\bibinfo {howpublished}
  {\url{https://github.com/iaraota}}\BibitemShut {NoStop}%
\bibitem [{\citenamefont {Schwarzschild}(1916)}]{Schwarzschild:1916uq}%
  \BibitemOpen
  \bibfield  {author} {\bibinfo {author} {\bibfnamefont {K.}~\bibnamefont
  {Schwarzschild}},\ }\href@noop {} {\bibfield  {journal} {\bibinfo  {journal}
  {Sitzungsber. Preuss. Akad. Wiss. Berlin (Math. Phys. )}\ }\textbf {\bibinfo
  {volume} {1916}},\ \bibinfo {pages} {189} (\bibinfo {year} {1916})},\ \Eprint
  {http://arxiv.org/abs/physics/9905030} {arXiv:physics/9905030} \BibitemShut
  {NoStop}%
\bibitem [{\citenamefont {Zerilli}(1970{\natexlab{a}})}]{Zerilli:1970se}%
  \BibitemOpen
  \bibfield  {author} {\bibinfo {author} {\bibfnamefont {F.~J.}\ \bibnamefont
  {Zerilli}},\ }\href {\doibase 10.1103/PhysRevLett.24.737} {\bibfield
  {journal} {\bibinfo  {journal} {Phys. Rev. Lett.}\ }\textbf {\bibinfo
  {volume} {24}},\ \bibinfo {pages} {737} (\bibinfo {year}
  {1970}{\natexlab{a}})}\BibitemShut {NoStop}%
\bibitem [{\citenamefont {Zerilli}(1970{\natexlab{b}})}]{Zerilli:1970wzz}%
  \BibitemOpen
  \bibfield  {author} {\bibinfo {author} {\bibfnamefont {F.~J.}\ \bibnamefont
  {Zerilli}},\ }\href {\doibase 10.1103/PhysRevD.2.2141} {\bibfield  {journal}
  {\bibinfo  {journal} {Phys. Rev. D}\ }\textbf {\bibinfo {volume} {2}},\
  \bibinfo {pages} {2141} (\bibinfo {year} {1970}{\natexlab{b}})}\BibitemShut
  {NoStop}%
\bibitem [{\citenamefont {Price}(1972)}]{Price:1971fb}%
  \BibitemOpen
  \bibfield  {author} {\bibinfo {author} {\bibfnamefont {R.~H.}\ \bibnamefont
  {Price}},\ }\href {\doibase 10.1103/PhysRevD.5.2419} {\bibfield  {journal}
  {\bibinfo  {journal} {Phys. Rev. D}\ }\textbf {\bibinfo {volume} {5}},\
  \bibinfo {pages} {2419} (\bibinfo {year} {1972})}\BibitemShut {NoStop}%
\bibitem [{\citenamefont {Gundlach}\ \emph {et~al.}(1994)\citenamefont
  {Gundlach}, \citenamefont {Price},\ and\ \citenamefont
  {Pullin}}]{Gundlach:1993tp}%
  \BibitemOpen
  \bibfield  {author} {\bibinfo {author} {\bibfnamefont {C.}~\bibnamefont
  {Gundlach}}, \bibinfo {author} {\bibfnamefont {R.~H.}\ \bibnamefont {Price}},
  \ and\ \bibinfo {author} {\bibfnamefont {J.}~\bibnamefont {Pullin}},\ }\href
  {\doibase 10.1103/PhysRevD.49.883} {\bibfield  {journal} {\bibinfo  {journal}
  {Phys. Rev. D}\ }\textbf {\bibinfo {volume} {49}},\ \bibinfo {pages} {883}
  (\bibinfo {year} {1994})},\ \Eprint {http://arxiv.org/abs/gr-qc/9307009}
  {arXiv:gr-qc/9307009} \BibitemShut {NoStop}%
\bibitem [{\citenamefont {Price}\ and\ \citenamefont
  {Pullin}(1994)}]{Price:1994pm}%
  \BibitemOpen
  \bibfield  {author} {\bibinfo {author} {\bibfnamefont {R.~H.}\ \bibnamefont
  {Price}}\ and\ \bibinfo {author} {\bibfnamefont {J.}~\bibnamefont {Pullin}},\
  }\href {\doibase 10.1103/PhysRevLett.72.3297} {\bibfield  {journal} {\bibinfo
   {journal} {Phys. Rev. Lett.}\ }\textbf {\bibinfo {volume} {72}},\ \bibinfo
  {pages} {3297} (\bibinfo {year} {1994})},\ \Eprint
  {http://arxiv.org/abs/gr-qc/9402039} {arXiv:gr-qc/9402039} \BibitemShut
  {NoStop}%
\bibitem [{\citenamefont {Nollert}(1999{\natexlab{a}})}]{Nollert:1999ji}%
  \BibitemOpen
  \bibfield  {author} {\bibinfo {author} {\bibfnamefont {H.-P.}\ \bibnamefont
  {Nollert}},\ }\href {\doibase 10.1088/0264-9381/16/12/201} {\bibfield
  {journal} {\bibinfo  {journal} {Class. Quant. Grav.}\ }\textbf {\bibinfo
  {volume} {16}},\ \bibinfo {pages} {R159} (\bibinfo {year}
  {1999}{\natexlab{a}})}\BibitemShut {NoStop}%
\bibitem [{\citenamefont {Israel}(1967)}]{Israel:1967wq}%
  \BibitemOpen
  \bibfield  {author} {\bibinfo {author} {\bibfnamefont {W.}~\bibnamefont
  {Israel}},\ }\href {\doibase 10.1103/PhysRev.164.1776} {\bibfield  {journal}
  {\bibinfo  {journal} {Phys. Rev.}\ }\textbf {\bibinfo {volume} {164}},\
  \bibinfo {pages} {1776} (\bibinfo {year} {1967})}\BibitemShut {NoStop}%
\bibitem [{\citenamefont {Israel}(1968)}]{Israel:1967za}%
  \BibitemOpen
  \bibfield  {author} {\bibinfo {author} {\bibfnamefont {W.}~\bibnamefont
  {Israel}},\ }\href {\doibase 10.1007/BF01645859} {\bibfield  {journal}
  {\bibinfo  {journal} {Commun. Math. Phys.}\ }\textbf {\bibinfo {volume}
  {8}},\ \bibinfo {pages} {245} (\bibinfo {year} {1968})}\BibitemShut {NoStop}%
\bibitem [{\citenamefont {Carter}(1971)}]{Carter:1971zc}%
  \BibitemOpen
  \bibfield  {author} {\bibinfo {author} {\bibfnamefont {B.}~\bibnamefont
  {Carter}},\ }\href {\doibase 10.1103/PhysRevLett.26.331} {\bibfield
  {journal} {\bibinfo  {journal} {Phys. Rev. Lett.}\ }\textbf {\bibinfo
  {volume} {26}},\ \bibinfo {pages} {331} (\bibinfo {year} {1971})}\BibitemShut
  {NoStop}%
\bibitem [{\citenamefont {Cardoso}\ and\ \citenamefont
  {Gualtieri}(2016)}]{Cardoso:2016ryw}%
  \BibitemOpen
  \bibfield  {author} {\bibinfo {author} {\bibfnamefont {V.}~\bibnamefont
  {Cardoso}}\ and\ \bibinfo {author} {\bibfnamefont {L.}~\bibnamefont
  {Gualtieri}},\ }\href {\doibase 10.1088/0264-9381/33/17/174001} {\bibfield
  {journal} {\bibinfo  {journal} {Class. Quant. Grav.}\ }\textbf {\bibinfo
  {volume} {33}},\ \bibinfo {pages} {174001} (\bibinfo {year} {2016})},\
  \Eprint {http://arxiv.org/abs/1607.03133} {arXiv:1607.03133 [gr-qc]}
  \BibitemShut {NoStop}%
\bibitem [{\citenamefont {Misner}\ \emph {et~al.}(1973)\citenamefont {Misner},
  \citenamefont {Thorne},\ and\ \citenamefont {Wheeler}}]{Misner:1974qy}%
  \BibitemOpen
  \bibfield  {author} {\bibinfo {author} {\bibfnamefont {C.~W.}\ \bibnamefont
  {Misner}}, \bibinfo {author} {\bibfnamefont {K.}~\bibnamefont {Thorne}}, \
  and\ \bibinfo {author} {\bibfnamefont {J.}~\bibnamefont {Wheeler}},\
  }\href@noop {} {\emph {\bibinfo {title} {{Gravitation}}}}\ (\bibinfo
  {publisher} {W. H. Freeman},\ \bibinfo {address} {San Francisco},\ \bibinfo
  {year} {1973})\BibitemShut {NoStop}%
\bibitem [{\citenamefont {Teukolsky}(1972)}]{Teukolsky:1972my}%
  \BibitemOpen
  \bibfield  {author} {\bibinfo {author} {\bibfnamefont {S.~A.}\ \bibnamefont
  {Teukolsky}},\ }\href {\doibase 10.1103/PhysRevLett.29.1114} {\bibfield
  {journal} {\bibinfo  {journal} {Phys. Rev. Lett.}\ }\textbf {\bibinfo
  {volume} {29}},\ \bibinfo {pages} {1114} (\bibinfo {year}
  {1972})}\BibitemShut {NoStop}%
\bibitem [{\citenamefont {Teukolsky}(1973)}]{Teukolsky:1973ha}%
  \BibitemOpen
  \bibfield  {author} {\bibinfo {author} {\bibfnamefont {S.~A.}\ \bibnamefont
  {Teukolsky}},\ }\href {\doibase 10.1086/152444} {\bibfield  {journal}
  {\bibinfo  {journal} {Astrophys. J.}\ }\textbf {\bibinfo {volume} {185}},\
  \bibinfo {pages} {635} (\bibinfo {year} {1973})}\BibitemShut {NoStop}%
\bibitem [{\citenamefont {Berti}\ \emph
  {et~al.}(2006{\natexlab{a}})\citenamefont {Berti}, \citenamefont {Cardoso},\
  and\ \citenamefont {Casals}}]{Berti:2005gp}%
  \BibitemOpen
  \bibfield  {author} {\bibinfo {author} {\bibfnamefont {E.}~\bibnamefont
  {Berti}}, \bibinfo {author} {\bibfnamefont {V.}~\bibnamefont {Cardoso}}, \
  and\ \bibinfo {author} {\bibfnamefont {M.}~\bibnamefont {Casals}},\ }\href
  {\doibase 10.1103/PhysRevD.73.109902} {\bibfield  {journal} {\bibinfo
  {journal} {Phys. Rev. D}\ }\textbf {\bibinfo {volume} {73}},\ \bibinfo
  {pages} {024013} (\bibinfo {year} {2006}{\natexlab{a}})},\ \bibinfo {note}
  {[Erratum: Phys.Rev.D 73, 109902 (2006)]},\ \Eprint
  {http://arxiv.org/abs/gr-qc/0511111} {arXiv:gr-qc/0511111} \BibitemShut
  {NoStop}%
\bibitem [{\citenamefont {Konoplya}\ and\ \citenamefont
  {Zhidenko}(2011)}]{Konoplya:2011qq}%
  \BibitemOpen
  \bibfield  {author} {\bibinfo {author} {\bibfnamefont {R.~A.}\ \bibnamefont
  {Konoplya}}\ and\ \bibinfo {author} {\bibfnamefont {A.}~\bibnamefont
  {Zhidenko}},\ }\href {\doibase 10.1103/RevModPhys.83.793} {\bibfield
  {journal} {\bibinfo  {journal} {Rev. Mod. Phys.}\ }\textbf {\bibinfo {volume}
  {83}},\ \bibinfo {pages} {793} (\bibinfo {year} {2011})},\ \Eprint
  {http://arxiv.org/abs/1102.4014} {arXiv:1102.4014 [gr-qc]} \BibitemShut
  {NoStop}%
\bibitem [{\citenamefont {Berti}\ \emph {et~al.}(2009)\citenamefont {Berti},
  \citenamefont {Cardoso},\ and\ \citenamefont {Starinets}}]{Berti:2009kk}%
  \BibitemOpen
  \bibfield  {author} {\bibinfo {author} {\bibfnamefont {E.}~\bibnamefont
  {Berti}}, \bibinfo {author} {\bibfnamefont {V.}~\bibnamefont {Cardoso}}, \
  and\ \bibinfo {author} {\bibfnamefont {A.~O.}\ \bibnamefont {Starinets}},\
  }\href {\doibase 10.1088/0264-9381/26/16/163001} {\bibfield  {journal}
  {\bibinfo  {journal} {Class. Quant. Grav.}\ }\textbf {\bibinfo {volume}
  {26}},\ \bibinfo {pages} {163001} (\bibinfo {year} {2009})},\ \Eprint
  {http://arxiv.org/abs/0905.2975} {arXiv:0905.2975 [gr-qc]} \BibitemShut
  {NoStop}%
\bibitem [{Ber()}]{Berti-ringdown}%
  \BibitemOpen
  \href@noop {} {}\bibinfo {howpublished}
  {\url{https://pages.jh.edu/~eberti2/ringdown/}}\BibitemShut {NoStop}%
\bibitem [{\citenamefont {Peters}(1964)}]{Peters:1964zz}%
  \BibitemOpen
  \bibfield  {author} {\bibinfo {author} {\bibfnamefont {P.~C.}\ \bibnamefont
  {Peters}},\ }\href {\doibase 10.1103/PhysRev.136.B1224} {\bibfield  {journal}
  {\bibinfo  {journal} {Phys. Rev.}\ }\textbf {\bibinfo {volume} {136}},\
  \bibinfo {pages} {B1224} (\bibinfo {year} {1964})}\BibitemShut {NoStop}%
\bibitem [{\citenamefont {Baumgarte}\ and\ \citenamefont
  {Shapiro}(2010)}]{baumgarte_shapiro_2010}%
  \BibitemOpen
  \bibfield  {author} {\bibinfo {author} {\bibfnamefont {T.~W.}\ \bibnamefont
  {Baumgarte}}\ and\ \bibinfo {author} {\bibfnamefont {S.~L.}\ \bibnamefont
  {Shapiro}},\ }\href {\doibase 10.1017/CBO9781139193344} {\emph {\bibinfo
  {title} {Numerical Relativity: Solving Einstein's Equations on the
  Computer}}}\ (\bibinfo  {publisher} {Cambridge University Press},\ \bibinfo
  {year} {2010})\BibitemShut {NoStop}%
\bibitem [{\citenamefont {Alcubierre}(2008)}]{alcubierre2008introduction}%
  \BibitemOpen
  \bibfield  {author} {\bibinfo {author} {\bibfnamefont {M.}~\bibnamefont
  {Alcubierre}},\ }\href@noop {} {\emph {\bibinfo {title} {Introduction to 3+1
  Numerical Relativity}}},\ International Series of Monogr\ (\bibinfo
  {publisher} {OUP Oxford},\ \bibinfo {year} {2008})\BibitemShut {NoStop}%
\bibitem [{\citenamefont {Press}\ and\ \citenamefont
  {Teukolsky}(1973)}]{Press:1973zz}%
  \BibitemOpen
  \bibfield  {author} {\bibinfo {author} {\bibfnamefont {W.~H.}\ \bibnamefont
  {Press}}\ and\ \bibinfo {author} {\bibfnamefont {S.~A.}\ \bibnamefont
  {Teukolsky}},\ }\href {\doibase 10.1086/152445} {\bibfield  {journal}
  {\bibinfo  {journal} {Astrophys. J.}\ }\textbf {\bibinfo {volume} {185}},\
  \bibinfo {pages} {649} (\bibinfo {year} {1973})}\BibitemShut {NoStop}%
\bibitem [{\citenamefont {Kelly}\ and\ \citenamefont
  {Baker}(2013)}]{Kelly:2012nd}%
  \BibitemOpen
  \bibfield  {author} {\bibinfo {author} {\bibfnamefont {B.~J.}\ \bibnamefont
  {Kelly}}\ and\ \bibinfo {author} {\bibfnamefont {J.~G.}\ \bibnamefont
  {Baker}},\ }\href {\doibase 10.1103/PhysRevD.87.084004} {\bibfield  {journal}
  {\bibinfo  {journal} {Phys. Rev. D}\ }\textbf {\bibinfo {volume} {87}},\
  \bibinfo {pages} {084004} (\bibinfo {year} {2013})},\ \Eprint
  {http://arxiv.org/abs/1212.5553} {arXiv:1212.5553 [gr-qc]} \BibitemShut
  {NoStop}%
\bibitem [{\citenamefont {Berti}\ and\ \citenamefont
  {Klein}(2014)}]{Berti:2014fga}%
  \BibitemOpen
  \bibfield  {author} {\bibinfo {author} {\bibfnamefont {E.}~\bibnamefont
  {Berti}}\ and\ \bibinfo {author} {\bibfnamefont {A.}~\bibnamefont {Klein}},\
  }\href {\doibase 10.1103/PhysRevD.90.064012} {\bibfield  {journal} {\bibinfo
  {journal} {Phys. Rev. D}\ }\textbf {\bibinfo {volume} {90}},\ \bibinfo
  {pages} {064012} (\bibinfo {year} {2014})},\ \Eprint
  {http://arxiv.org/abs/1408.1860} {arXiv:1408.1860 [gr-qc]} \BibitemShut
  {NoStop}%
\bibitem [{\citenamefont {Cotesta}\ \emph {et~al.}(2018)\citenamefont
  {Cotesta}, \citenamefont {Buonanno}, \citenamefont {Boh\'e}, \citenamefont
  {Taracchini}, \citenamefont {Hinder},\ and\ \citenamefont
  {Ossokine}}]{Cotesta:2018fcv}%
  \BibitemOpen
  \bibfield  {author} {\bibinfo {author} {\bibfnamefont {R.}~\bibnamefont
  {Cotesta}}, \bibinfo {author} {\bibfnamefont {A.}~\bibnamefont {Buonanno}},
  \bibinfo {author} {\bibfnamefont {A.}~\bibnamefont {Boh\'e}}, \bibinfo
  {author} {\bibfnamefont {A.}~\bibnamefont {Taracchini}}, \bibinfo {author}
  {\bibfnamefont {I.}~\bibnamefont {Hinder}}, \ and\ \bibinfo {author}
  {\bibfnamefont {S.}~\bibnamefont {Ossokine}},\ }\href {\doibase
  10.1103/PhysRevD.98.084028} {\bibfield  {journal} {\bibinfo  {journal} {Phys.
  Rev. D}\ }\textbf {\bibinfo {volume} {98}},\ \bibinfo {pages} {084028}
  (\bibinfo {year} {2018})},\ \Eprint {http://arxiv.org/abs/1803.10701}
  {arXiv:1803.10701 [gr-qc]} \BibitemShut {NoStop}%
\bibitem [{\citenamefont {Romero-Shaw}\ \emph {et~al.}(2020)\citenamefont
  {Romero-Shaw}, \citenamefont {Lasky}, \citenamefont {Thrane},\ and\
  \citenamefont {Bustillo}}]{Romero-Shaw:2020thy}%
  \BibitemOpen
  \bibfield  {author} {\bibinfo {author} {\bibfnamefont {I.~M.}\ \bibnamefont
  {Romero-Shaw}}, \bibinfo {author} {\bibfnamefont {P.~D.}\ \bibnamefont
  {Lasky}}, \bibinfo {author} {\bibfnamefont {E.}~\bibnamefont {Thrane}}, \
  and\ \bibinfo {author} {\bibfnamefont {J.~C.}\ \bibnamefont {Bustillo}},\
  }\href {\doibase 10.3847/2041-8213/abbe26} {\bibfield  {journal} {\bibinfo
  {journal} {Astrophys. J. Lett.}\ }\textbf {\bibinfo {volume} {903}},\
  \bibinfo {pages} {L5} (\bibinfo {year} {2020})},\ \Eprint
  {http://arxiv.org/abs/2009.04771} {arXiv:2009.04771 [astro-ph.HE]}
  \BibitemShut {NoStop}%
\bibitem [{\citenamefont {Berti}\ \emph
  {et~al.}(2006{\natexlab{b}})\citenamefont {Berti}, \citenamefont {Cardoso},\
  and\ \citenamefont {Will}}]{Berti:2005ys}%
  \BibitemOpen
  \bibfield  {author} {\bibinfo {author} {\bibfnamefont {E.}~\bibnamefont
  {Berti}}, \bibinfo {author} {\bibfnamefont {V.}~\bibnamefont {Cardoso}}, \
  and\ \bibinfo {author} {\bibfnamefont {C.~M.}\ \bibnamefont {Will}},\ }\href
  {\doibase 10.1103/PhysRevD.73.064030} {\bibfield  {journal} {\bibinfo
  {journal} {Phys. Rev. D}\ }\textbf {\bibinfo {volume} {73}},\ \bibinfo
  {pages} {064030} (\bibinfo {year} {2006}{\natexlab{b}})},\ \Eprint
  {http://arxiv.org/abs/gr-qc/0512160} {arXiv:gr-qc/0512160} \BibitemShut
  {NoStop}%
\bibitem [{\citenamefont {Hogg}(1999)}]{Hogg:1999ad}%
  \BibitemOpen
  \bibfield  {author} {\bibinfo {author} {\bibfnamefont {D.~W.}\ \bibnamefont
  {Hogg}},\ }\href@noop {} {\  (\bibinfo {year} {1999})},\ \Eprint
  {http://arxiv.org/abs/astro-ph/9905116} {arXiv:astro-ph/9905116} \BibitemShut
  {NoStop}%
\bibitem [{\citenamefont {Aghanim}\ \emph {et~al.}(2020)\citenamefont {Aghanim}
  \emph {et~al.}}]{Planck:2018vyg}%
  \BibitemOpen
  \bibfield  {author} {\bibinfo {author} {\bibfnamefont {N.}~\bibnamefont
  {Aghanim}} \emph {et~al.} (\bibinfo {collaboration} {Planck}),\ }\href
  {\doibase 10.1051/0004-6361/201833910} {\bibfield  {journal} {\bibinfo
  {journal} {Astron. Astrophys.}\ }\textbf {\bibinfo {volume} {641}},\ \bibinfo
  {pages} {A6} (\bibinfo {year} {2020})},\ \bibinfo {note} {[Erratum:
  Astron.Astrophys. 652, C4 (2021)]},\ \Eprint
  {http://arxiv.org/abs/1807.06209} {arXiv:1807.06209 [astro-ph.CO]}
  \BibitemShut {NoStop}%
\bibitem [{\citenamefont {Sathyaprakash}\ and\ \citenamefont
  {Schutz}(2009)}]{Sathyaprakash:2009xs}%
  \BibitemOpen
  \bibfield  {author} {\bibinfo {author} {\bibfnamefont {B.~S.}\ \bibnamefont
  {Sathyaprakash}}\ and\ \bibinfo {author} {\bibfnamefont {B.~F.}\ \bibnamefont
  {Schutz}},\ }\href {\doibase 10.12942/lrr-2009-2} {\bibfield  {journal}
  {\bibinfo  {journal} {Living Rev. Rel.}\ }\textbf {\bibinfo {volume} {12}},\
  \bibinfo {pages} {2} (\bibinfo {year} {2009})},\ \Eprint
  {http://arxiv.org/abs/0903.0338} {arXiv:0903.0338 [gr-qc]} \BibitemShut
  {NoStop}%
\bibitem [{\citenamefont {Regimbau}\ \emph {et~al.}(2012)\citenamefont
  {Regimbau} \emph {et~al.}}]{Regimbau:2012ir}%
  \BibitemOpen
  \bibfield  {author} {\bibinfo {author} {\bibfnamefont {T.}~\bibnamefont
  {Regimbau}} \emph {et~al.},\ }\href {\doibase 10.1103/PhysRevD.86.122001}
  {\bibfield  {journal} {\bibinfo  {journal} {Phys. Rev. D}\ }\textbf {\bibinfo
  {volume} {86}},\ \bibinfo {pages} {122001} (\bibinfo {year} {2012})},\
  \Eprint {http://arxiv.org/abs/1201.3563} {arXiv:1201.3563 [gr-qc]}
  \BibitemShut {NoStop}%
\bibitem [{\citenamefont {Prince}\ \emph {et~al.}(2002)\citenamefont {Prince},
  \citenamefont {Tinto}, \citenamefont {Larson},\ and\ \citenamefont
  {Armstrong}}]{Prince:2002hp}%
  \BibitemOpen
  \bibfield  {author} {\bibinfo {author} {\bibfnamefont {T.~A.}\ \bibnamefont
  {Prince}}, \bibinfo {author} {\bibfnamefont {M.}~\bibnamefont {Tinto}},
  \bibinfo {author} {\bibfnamefont {S.~L.}\ \bibnamefont {Larson}}, \ and\
  \bibinfo {author} {\bibfnamefont {J.~W.}\ \bibnamefont {Armstrong}},\ }\href
  {\doibase 10.1103/PhysRevD.66.122002} {\bibfield  {journal} {\bibinfo
  {journal} {Phys. Rev. D}\ }\textbf {\bibinfo {volume} {66}},\ \bibinfo
  {pages} {122002} (\bibinfo {year} {2002})},\ \Eprint
  {http://arxiv.org/abs/gr-qc/0209039} {arXiv:gr-qc/0209039} \BibitemShut
  {NoStop}%
\bibitem [{\citenamefont {Robson}\ \emph {et~al.}(2019)\citenamefont {Robson},
  \citenamefont {Cornish},\ and\ \citenamefont {Liu}}]{Robson:2018ifk}%
  \BibitemOpen
  \bibfield  {author} {\bibinfo {author} {\bibfnamefont {T.}~\bibnamefont
  {Robson}}, \bibinfo {author} {\bibfnamefont {N.~J.}\ \bibnamefont {Cornish}},
  \ and\ \bibinfo {author} {\bibfnamefont {C.}~\bibnamefont {Liu}},\ }\href
  {\doibase 10.1088/1361-6382/ab1101} {\bibfield  {journal} {\bibinfo
  {journal} {Class. Quant. Grav.}\ }\textbf {\bibinfo {volume} {36}},\ \bibinfo
  {pages} {105011} (\bibinfo {year} {2019})},\ \Eprint
  {http://arxiv.org/abs/1803.01944} {arXiv:1803.01944 [astro-ph.HE]}
  \BibitemShut {NoStop}%
\bibitem [{\citenamefont {Abbott}\ \emph
  {et~al.}(2017{\natexlab{b}})\citenamefont {Abbott} \emph
  {et~al.}}]{LIGOScientific:2017ycc}%
  \BibitemOpen
  \bibfield  {author} {\bibinfo {author} {\bibfnamefont {B.~P.}\ \bibnamefont
  {Abbott}} \emph {et~al.} (\bibinfo {collaboration} {LIGO Scientific,
  Virgo}),\ }\href {\doibase 10.1103/PhysRevLett.119.141101} {\bibfield
  {journal} {\bibinfo  {journal} {Phys. Rev. Lett.}\ }\textbf {\bibinfo
  {volume} {119}},\ \bibinfo {pages} {141101} (\bibinfo {year}
  {2017}{\natexlab{b}})},\ \Eprint {http://arxiv.org/abs/1709.09660}
  {arXiv:1709.09660 [gr-qc]} \BibitemShut {NoStop}%
\bibitem [{\citenamefont {Abbott}\ \emph
  {et~al.}(2017{\natexlab{c}})\citenamefont {Abbott} \emph
  {et~al.}}]{LIGOScientific:2017ync}%
  \BibitemOpen
  \bibfield  {author} {\bibinfo {author} {\bibfnamefont {B.~P.}\ \bibnamefont
  {Abbott}} \emph {et~al.} (\bibinfo {collaboration} {LIGO Scientific, Virgo,
  Fermi GBM, INTEGRAL, IceCube, AstroSat Cadmium Zinc Telluride Imager Team,
  IPN, Insight-Hxmt, ANTARES, Swift, AGILE Team, 1M2H Team, Dark Energy Camera
  GW-EM, DES, DLT40, GRAWITA, Fermi-LAT, ATCA, ASKAP, Las Cumbres Observatory
  Group, OzGrav, DWF (Deeper Wider Faster Program), AST3, CAASTRO, VINROUGE,
  MASTER, J-GEM, GROWTH, JAGWAR, CaltechNRAO, TTU-NRAO, NuSTAR, Pan-STARRS,
  MAXI Team, TZAC Consortium, KU, Nordic Optical Telescope, ePESSTO, GROND,
  Texas Tech University, SALT Group, TOROS, BOOTES, MWA, CALET, IKI-GW
  Follow-up, H.E.S.S., LOFAR, LWA, HAWC, Pierre Auger, ALMA, Euro VLBI Team, Pi
  of Sky, Chandra Team at McGill University, DFN, ATLAS Telescopes, High Time
  Resolution Universe Survey, RIMAS, RATIR, SKA South Africa/MeerKAT}),\ }\href
  {\doibase 10.3847/2041-8213/aa91c9} {\bibfield  {journal} {\bibinfo
  {journal} {Astrophys. J. Lett.}\ }\textbf {\bibinfo {volume} {848}},\
  \bibinfo {pages} {L12} (\bibinfo {year} {2017}{\natexlab{c}})},\ \Eprint
  {http://arxiv.org/abs/1710.05833} {arXiv:1710.05833 [astro-ph.HE]}
  \BibitemShut {NoStop}%
\bibitem [{\citenamefont {Abbott}\ \emph
  {et~al.}(2020{\natexlab{b}})\citenamefont {Abbott} \emph
  {et~al.}}]{LIGOScientific:2019hgc}%
  \BibitemOpen
  \bibfield  {author} {\bibinfo {author} {\bibfnamefont {B.~P.}\ \bibnamefont
  {Abbott}} \emph {et~al.} (\bibinfo {collaboration} {LIGO Scientific,
  Virgo}),\ }\href {\doibase 10.1088/1361-6382/ab685e} {\bibfield  {journal}
  {\bibinfo  {journal} {Class. Quant. Grav.}\ }\textbf {\bibinfo {volume}
  {37}},\ \bibinfo {pages} {055002} (\bibinfo {year} {2020}{\natexlab{b}})},\
  \Eprint {http://arxiv.org/abs/1908.11170} {arXiv:1908.11170 [gr-qc]}
  \BibitemShut {NoStop}%
\bibitem [{\citenamefont {Grishchuk}\ \emph {et~al.}(2001)\citenamefont
  {Grishchuk}, \citenamefont {Lipunov}, \citenamefont {Postnov}, \citenamefont
  {Prokhorov},\ and\ \citenamefont {Sathyaprakash}}]{Grishchuk:2000gh}%
  \BibitemOpen
  \bibfield  {author} {\bibinfo {author} {\bibfnamefont {L.~P.}\ \bibnamefont
  {Grishchuk}}, \bibinfo {author} {\bibfnamefont {V.~M.}\ \bibnamefont
  {Lipunov}}, \bibinfo {author} {\bibfnamefont {K.~A.}\ \bibnamefont
  {Postnov}}, \bibinfo {author} {\bibfnamefont {M.~E.}\ \bibnamefont
  {Prokhorov}}, \ and\ \bibinfo {author} {\bibfnamefont {B.~S.}\ \bibnamefont
  {Sathyaprakash}},\ }\href {\doibase 10.1070/PU2001v044n01ABEH000873}
  {\bibfield  {journal} {\bibinfo  {journal} {Phys. Usp.}\ }\textbf {\bibinfo
  {volume} {44}},\ \bibinfo {pages} {1} (\bibinfo {year} {2001})},\ \Eprint
  {http://arxiv.org/abs/astro-ph/0008481} {arXiv:astro-ph/0008481} \BibitemShut
  {NoStop}%
\bibitem [{\citenamefont {Moore}\ \emph {et~al.}(2015)\citenamefont {Moore},
  \citenamefont {Cole},\ and\ \citenamefont {Berry}}]{Moore:2014lga}%
  \BibitemOpen
  \bibfield  {author} {\bibinfo {author} {\bibfnamefont {C.~J.}\ \bibnamefont
  {Moore}}, \bibinfo {author} {\bibfnamefont {R.~H.}\ \bibnamefont {Cole}}, \
  and\ \bibinfo {author} {\bibfnamefont {C.~P.~L.}\ \bibnamefont {Berry}},\
  }\href {\doibase 10.1088/0264-9381/32/1/015014} {\bibfield  {journal}
  {\bibinfo  {journal} {Class. Quant. Grav.}\ }\textbf {\bibinfo {volume}
  {32}},\ \bibinfo {pages} {015014} (\bibinfo {year} {2015})},\ \Eprint
  {http://arxiv.org/abs/1408.0740} {arXiv:1408.0740 [gr-qc]} \BibitemShut
  {NoStop}%
\bibitem [{Des()}]{Design-sensitivity}%
  \BibitemOpen
  \href@noop {} {}\bibinfo {howpublished}
  {\url{https://dcc.ligo.org/LIGO-T1800044/public}}\BibitemShut {NoStop}%
\bibitem [{CE-()}]{CE-psd}%
  \BibitemOpen
  \href@noop {} {}\bibinfo {howpublished}
  {\url{https://dcc.cosmicexplorer.org/cgi-bin/DocDB/ShowDocument?docid=T2000017.}}\BibitemShut
  {Stop}%
\bibitem [{ET-()}]{ET-psd}%
  \BibitemOpen
  \href@noop {} {}\bibinfo {howpublished}
  {\url{http://www.et-gw.eu/index.php/etsensitivities}}\BibitemShut {NoStop}%
\bibitem [{\citenamefont {Sesana}(2017)}]{Sesana:2017vsj}%
  \BibitemOpen
  \bibfield  {author} {\bibinfo {author} {\bibfnamefont {A.}~\bibnamefont
  {Sesana}},\ }\href {\doibase 10.1088/1742-6596/840/1/012018} {\bibfield
  {journal} {\bibinfo  {journal} {J. Phys. Conf. Ser.}\ }\textbf {\bibinfo
  {volume} {840}},\ \bibinfo {pages} {012018} (\bibinfo {year} {2017})},\
  \Eprint {http://arxiv.org/abs/1702.04356} {arXiv:1702.04356 [astro-ph.HE]}
  \BibitemShut {NoStop}%
\bibitem [{\citenamefont {Feroz}\ \emph {et~al.}(2009)\citenamefont {Feroz},
  \citenamefont {Hobson},\ and\ \citenamefont {Bridges}}]{Feroz:2008xx}%
  \BibitemOpen
  \bibfield  {author} {\bibinfo {author} {\bibfnamefont {F.}~\bibnamefont
  {Feroz}}, \bibinfo {author} {\bibfnamefont {M.~P.}\ \bibnamefont {Hobson}}, \
  and\ \bibinfo {author} {\bibfnamefont {M.}~\bibnamefont {Bridges}},\ }\href
  {\doibase 10.1111/j.1365-2966.2009.14548.x} {\bibfield  {journal} {\bibinfo
  {journal} {Mon. Not. Roy. Astron. Soc.}\ }\textbf {\bibinfo {volume} {398}},\
  \bibinfo {pages} {1601} (\bibinfo {year} {2009})},\ \Eprint
  {http://arxiv.org/abs/0809.3437} {arXiv:0809.3437 [astro-ph]} \BibitemShut
  {NoStop}%
\bibitem [{\citenamefont {Feroz}\ \emph {et~al.}(2019)\citenamefont {Feroz},
  \citenamefont {Hobson}, \citenamefont {Cameron},\ and\ \citenamefont
  {Pettitt}}]{Feroz:2013hea}%
  \BibitemOpen
  \bibfield  {author} {\bibinfo {author} {\bibfnamefont {F.}~\bibnamefont
  {Feroz}}, \bibinfo {author} {\bibfnamefont {M.~P.}\ \bibnamefont {Hobson}},
  \bibinfo {author} {\bibfnamefont {E.}~\bibnamefont {Cameron}}, \ and\
  \bibinfo {author} {\bibfnamefont {A.~N.}\ \bibnamefont {Pettitt}},\ }\href
  {\doibase 10.21105/astro.1306.2144} {\bibfield  {journal} {\bibinfo
  {journal} {Open J. Astrophys.}\ }\textbf {\bibinfo {volume} {2}},\ \bibinfo
  {pages} {10} (\bibinfo {year} {2019})},\ \Eprint
  {http://arxiv.org/abs/1306.2144} {arXiv:1306.2144 [astro-ph.IM]} \BibitemShut
  {NoStop}%
\bibitem [{\citenamefont {Buchner}\ \emph {et~al.}(2014)\citenamefont
  {Buchner}, \citenamefont {Georgakakis}, \citenamefont {Nandra}, \citenamefont
  {Hsu}, \citenamefont {Rangel}, \citenamefont {Brightman}, \citenamefont
  {Merloni}, \citenamefont {Salvato}, \citenamefont {Donley},\ and\
  \citenamefont {Kocevski}}]{Buchner:2014nha}%
  \BibitemOpen
  \bibfield  {author} {\bibinfo {author} {\bibfnamefont {J.}~\bibnamefont
  {Buchner}}, \bibinfo {author} {\bibfnamefont {A.}~\bibnamefont
  {Georgakakis}}, \bibinfo {author} {\bibfnamefont {K.}~\bibnamefont {Nandra}},
  \bibinfo {author} {\bibfnamefont {L.}~\bibnamefont {Hsu}}, \bibinfo {author}
  {\bibfnamefont {C.}~\bibnamefont {Rangel}}, \bibinfo {author} {\bibfnamefont
  {M.}~\bibnamefont {Brightman}}, \bibinfo {author} {\bibfnamefont
  {A.}~\bibnamefont {Merloni}}, \bibinfo {author} {\bibfnamefont
  {M.}~\bibnamefont {Salvato}}, \bibinfo {author} {\bibfnamefont
  {J.}~\bibnamefont {Donley}}, \ and\ \bibinfo {author} {\bibfnamefont
  {D.}~\bibnamefont {Kocevski}},\ }\href {\doibase 10.1051/0004-6361/201322971}
  {\bibfield  {journal} {\bibinfo  {journal} {Astron. Astrophys.}\ }\textbf
  {\bibinfo {volume} {564}},\ \bibinfo {pages} {A125} (\bibinfo {year}
  {2014})},\ \Eprint {http://arxiv.org/abs/1402.0004} {arXiv:1402.0004
  [astro-ph.HE]} \BibitemShut {NoStop}%
\bibitem [{\citenamefont {Finn}(1992)}]{Finn:1992wt}%
  \BibitemOpen
  \bibfield  {author} {\bibinfo {author} {\bibfnamefont {L.~S.}\ \bibnamefont
  {Finn}},\ }\href {\doibase 10.1103/PhysRevD.46.5236} {\bibfield  {journal}
  {\bibinfo  {journal} {Phys. Rev.}\ }\textbf {\bibinfo {volume} {D46}},\
  \bibinfo {pages} {5236} (\bibinfo {year} {1992})},\ \Eprint
  {http://arxiv.org/abs/gr-qc/9209010} {arXiv:gr-qc/9209010 [gr-qc]}
  \BibitemShut {NoStop}%
\bibitem [{\citenamefont {Thrane}\ and\ \citenamefont
  {Talbot}(2019)}]{Thrane:2018qnx}%
  \BibitemOpen
  \bibfield  {author} {\bibinfo {author} {\bibfnamefont {E.}~\bibnamefont
  {Thrane}}\ and\ \bibinfo {author} {\bibfnamefont {C.}~\bibnamefont
  {Talbot}},\ }\href {\doibase 10.1017/pasa.2019.2} {\bibfield  {journal}
  {\bibinfo  {journal} {Publ. Astron. Soc. Austral.}\ }\textbf {\bibinfo
  {volume} {36}},\ \bibinfo {pages} {e010} (\bibinfo {year} {2019})},\ \bibinfo
  {note} {[Erratum: Publ.Astron.Soc.Austral. 37, e036 (2020)]},\ \Eprint
  {http://arxiv.org/abs/1809.02293} {arXiv:1809.02293 [astro-ph.IM]}
  \BibitemShut {NoStop}%
\bibitem [{\citenamefont {Dorband}\ \emph {et~al.}(2006)\citenamefont
  {Dorband}, \citenamefont {Berti}, \citenamefont {Diener}, \citenamefont
  {Schnetter},\ and\ \citenamefont {Tiglio}}]{Dorband:2006gg}%
  \BibitemOpen
  \bibfield  {author} {\bibinfo {author} {\bibfnamefont {E.~N.}\ \bibnamefont
  {Dorband}}, \bibinfo {author} {\bibfnamefont {E.}~\bibnamefont {Berti}},
  \bibinfo {author} {\bibfnamefont {P.}~\bibnamefont {Diener}}, \bibinfo
  {author} {\bibfnamefont {E.}~\bibnamefont {Schnetter}}, \ and\ \bibinfo
  {author} {\bibfnamefont {M.}~\bibnamefont {Tiglio}},\ }\href {\doibase
  10.1103/PhysRevD.74.084028} {\bibfield  {journal} {\bibinfo  {journal} {Phys.
  Rev. D}\ }\textbf {\bibinfo {volume} {74}},\ \bibinfo {pages} {084028}
  (\bibinfo {year} {2006})},\ \Eprint {http://arxiv.org/abs/gr-qc/0608091}
  {arXiv:gr-qc/0608091} \BibitemShut {NoStop}%
\bibitem [{\citenamefont {Berti}\ \emph {et~al.}(2007)\citenamefont {Berti},
  \citenamefont {Cardoso}, \citenamefont {Gonzalez}, \citenamefont {Sperhake},
  \citenamefont {Hannam}, \citenamefont {Husa},\ and\ \citenamefont
  {Bruegmann}}]{Berti:2007fi}%
  \BibitemOpen
  \bibfield  {author} {\bibinfo {author} {\bibfnamefont {E.}~\bibnamefont
  {Berti}}, \bibinfo {author} {\bibfnamefont {V.}~\bibnamefont {Cardoso}},
  \bibinfo {author} {\bibfnamefont {J.~A.}\ \bibnamefont {Gonzalez}}, \bibinfo
  {author} {\bibfnamefont {U.}~\bibnamefont {Sperhake}}, \bibinfo {author}
  {\bibfnamefont {M.}~\bibnamefont {Hannam}}, \bibinfo {author} {\bibfnamefont
  {S.}~\bibnamefont {Husa}}, \ and\ \bibinfo {author} {\bibfnamefont
  {B.}~\bibnamefont {Bruegmann}},\ }\href {\doibase 10.1103/PhysRevD.76.064034}
  {\bibfield  {journal} {\bibinfo  {journal} {Phys. Rev. D}\ }\textbf {\bibinfo
  {volume} {76}},\ \bibinfo {pages} {064034} (\bibinfo {year} {2007})},\
  \Eprint {http://arxiv.org/abs/gr-qc/0703053} {arXiv:gr-qc/0703053}
  \BibitemShut {NoStop}%
\bibitem [{\citenamefont {Thrane}\ \emph {et~al.}(2017)\citenamefont {Thrane},
  \citenamefont {Lasky},\ and\ \citenamefont {Levin}}]{Thrane:2017lqn}%
  \BibitemOpen
  \bibfield  {author} {\bibinfo {author} {\bibfnamefont {E.}~\bibnamefont
  {Thrane}}, \bibinfo {author} {\bibfnamefont {P.~D.}\ \bibnamefont {Lasky}}, \
  and\ \bibinfo {author} {\bibfnamefont {Y.}~\bibnamefont {Levin}},\ }\href
  {\doibase 10.1103/PhysRevD.96.102004} {\bibfield  {journal} {\bibinfo
  {journal} {Phys. Rev. D}\ }\textbf {\bibinfo {volume} {96}},\ \bibinfo
  {pages} {102004} (\bibinfo {year} {2017})},\ \Eprint
  {http://arxiv.org/abs/1706.05152} {arXiv:1706.05152 [gr-qc]} \BibitemShut
  {NoStop}%
\bibitem [{\citenamefont {Carullo}\ \emph {et~al.}(2018)\citenamefont {Carullo}
  \emph {et~al.}}]{Carullo:2018sfu}%
  \BibitemOpen
  \bibfield  {author} {\bibinfo {author} {\bibfnamefont {G.}~\bibnamefont
  {Carullo}} \emph {et~al.},\ }\href {\doibase 10.1103/PhysRevD.98.104020}
  {\bibfield  {journal} {\bibinfo  {journal} {Phys. Rev. D}\ }\textbf {\bibinfo
  {volume} {98}},\ \bibinfo {pages} {104020} (\bibinfo {year} {2018})},\
  \Eprint {http://arxiv.org/abs/1805.04760} {arXiv:1805.04760 [gr-qc]}
  \BibitemShut {NoStop}%
\bibitem [{\citenamefont {Kamaretsos}\ \emph {et~al.}(2012)\citenamefont
  {Kamaretsos}, \citenamefont {Hannam}, \citenamefont {Husa},\ and\
  \citenamefont {Sathyaprakash}}]{Kamaretsos:2011um}%
  \BibitemOpen
  \bibfield  {author} {\bibinfo {author} {\bibfnamefont {I.}~\bibnamefont
  {Kamaretsos}}, \bibinfo {author} {\bibfnamefont {M.}~\bibnamefont {Hannam}},
  \bibinfo {author} {\bibfnamefont {S.}~\bibnamefont {Husa}}, \ and\ \bibinfo
  {author} {\bibfnamefont {B.~S.}\ \bibnamefont {Sathyaprakash}},\ }\href
  {\doibase 10.1103/PhysRevD.85.024018} {\bibfield  {journal} {\bibinfo
  {journal} {Phys. Rev. D}\ }\textbf {\bibinfo {volume} {85}},\ \bibinfo
  {pages} {024018} (\bibinfo {year} {2012})},\ \Eprint
  {http://arxiv.org/abs/1107.0854} {arXiv:1107.0854 [gr-qc]} \BibitemShut
  {NoStop}%
\bibitem [{\citenamefont {Shi}\ \emph {et~al.}(2019)\citenamefont {Shi},
  \citenamefont {Bao}, \citenamefont {Wang}, \citenamefont {Zhang},
  \citenamefont {Hu}, \citenamefont {Sesana}, \citenamefont {Barausse},
  \citenamefont {Mei},\ and\ \citenamefont {Luo}}]{Shi:2019hqa}%
  \BibitemOpen
  \bibfield  {author} {\bibinfo {author} {\bibfnamefont {C.}~\bibnamefont
  {Shi}}, \bibinfo {author} {\bibfnamefont {J.}~\bibnamefont {Bao}}, \bibinfo
  {author} {\bibfnamefont {H.}~\bibnamefont {Wang}}, \bibinfo {author}
  {\bibfnamefont {J.-d.}\ \bibnamefont {Zhang}}, \bibinfo {author}
  {\bibfnamefont {Y.}~\bibnamefont {Hu}}, \bibinfo {author} {\bibfnamefont
  {A.}~\bibnamefont {Sesana}}, \bibinfo {author} {\bibfnamefont
  {E.}~\bibnamefont {Barausse}}, \bibinfo {author} {\bibfnamefont
  {J.}~\bibnamefont {Mei}}, \ and\ \bibinfo {author} {\bibfnamefont
  {J.}~\bibnamefont {Luo}},\ }\href {\doibase 10.1103/PhysRevD.100.044036}
  {\bibfield  {journal} {\bibinfo  {journal} {Phys. Rev. D}\ }\textbf {\bibinfo
  {volume} {100}},\ \bibinfo {pages} {044036} (\bibinfo {year} {2019})},\
  \Eprint {http://arxiv.org/abs/1902.08922} {arXiv:1902.08922 [gr-qc]}
  \BibitemShut {NoStop}%
\bibitem [{\citenamefont {Maselli}\ \emph {et~al.}(2020)\citenamefont
  {Maselli}, \citenamefont {Pani}, \citenamefont {Gualtieri},\ and\
  \citenamefont {Berti}}]{Maselli:2019mjd}%
  \BibitemOpen
  \bibfield  {author} {\bibinfo {author} {\bibfnamefont {A.}~\bibnamefont
  {Maselli}}, \bibinfo {author} {\bibfnamefont {P.}~\bibnamefont {Pani}},
  \bibinfo {author} {\bibfnamefont {L.}~\bibnamefont {Gualtieri}}, \ and\
  \bibinfo {author} {\bibfnamefont {E.}~\bibnamefont {Berti}},\ }\href
  {\doibase 10.1103/PhysRevD.101.024043} {\bibfield  {journal} {\bibinfo
  {journal} {Phys. Rev. D}\ }\textbf {\bibinfo {volume} {101}},\ \bibinfo
  {pages} {024043} (\bibinfo {year} {2020})},\ \Eprint
  {http://arxiv.org/abs/1910.12893} {arXiv:1910.12893 [gr-qc]} \BibitemShut
  {NoStop}%
\bibitem [{\citenamefont {Baibhav}\ and\ \citenamefont
  {Berti}(2019)}]{Baibhav:2018rfk}%
  \BibitemOpen
  \bibfield  {author} {\bibinfo {author} {\bibfnamefont {V.}~\bibnamefont
  {Baibhav}}\ and\ \bibinfo {author} {\bibfnamefont {E.}~\bibnamefont
  {Berti}},\ }\href {\doibase 10.1103/PhysRevD.99.024005} {\bibfield  {journal}
  {\bibinfo  {journal} {Phys. Rev. D}\ }\textbf {\bibinfo {volume} {99}},\
  \bibinfo {pages} {024005} (\bibinfo {year} {2019})},\ \Eprint
  {http://arxiv.org/abs/1809.03500} {arXiv:1809.03500 [gr-qc]} \BibitemShut
  {NoStop}%
\bibitem [{\citenamefont {Baibhav}\ \emph {et~al.}(2020)\citenamefont
  {Baibhav}, \citenamefont {Berti},\ and\ \citenamefont
  {Cardoso}}]{Baibhav:2020tma}%
  \BibitemOpen
  \bibfield  {author} {\bibinfo {author} {\bibfnamefont {V.}~\bibnamefont
  {Baibhav}}, \bibinfo {author} {\bibfnamefont {E.}~\bibnamefont {Berti}}, \
  and\ \bibinfo {author} {\bibfnamefont {V.}~\bibnamefont {Cardoso}},\ }\href
  {\doibase 10.1103/PhysRevD.101.084053} {\bibfield  {journal} {\bibinfo
  {journal} {Phys. Rev. D}\ }\textbf {\bibinfo {volume} {101}},\ \bibinfo
  {pages} {084053} (\bibinfo {year} {2020})},\ \Eprint
  {http://arxiv.org/abs/2001.10011} {arXiv:2001.10011 [gr-qc]} \BibitemShut
  {NoStop}%
\bibitem [{\citenamefont {Leaver}(1986)}]{Leaver:1986gd}%
  \BibitemOpen
  \bibfield  {author} {\bibinfo {author} {\bibfnamefont {E.~W.}\ \bibnamefont
  {Leaver}},\ }\href {\doibase 10.1103/PhysRevD.34.384} {\bibfield  {journal}
  {\bibinfo  {journal} {Phys. Rev. D}\ }\textbf {\bibinfo {volume} {34}},\
  \bibinfo {pages} {384} (\bibinfo {year} {1986})}\BibitemShut {NoStop}%
\bibitem [{\citenamefont {Stark}\ and\ \citenamefont
  {Piran}(1985)}]{Stark:1985da}%
  \BibitemOpen
  \bibfield  {author} {\bibinfo {author} {\bibfnamefont {R.~F.}\ \bibnamefont
  {Stark}}\ and\ \bibinfo {author} {\bibfnamefont {T.}~\bibnamefont {Piran}},\
  }\href {\doibase 10.1103/PhysRevLett.55.891} {\bibfield  {journal} {\bibinfo
  {journal} {Phys. Rev. Lett.}\ }\textbf {\bibinfo {volume} {55}},\ \bibinfo
  {pages} {891} (\bibinfo {year} {1985})},\ \bibinfo {note} {[Erratum:
  Phys.Rev.Lett. 56, 97 (1986)]}\BibitemShut {NoStop}%
\bibitem [{\citenamefont {London}\ \emph {et~al.}(2014)\citenamefont {London},
  \citenamefont {Shoemaker},\ and\ \citenamefont {Healy}}]{London:2014cma}%
  \BibitemOpen
  \bibfield  {author} {\bibinfo {author} {\bibfnamefont {L.}~\bibnamefont
  {London}}, \bibinfo {author} {\bibfnamefont {D.}~\bibnamefont {Shoemaker}}, \
  and\ \bibinfo {author} {\bibfnamefont {J.}~\bibnamefont {Healy}},\ }\href
  {\doibase 10.1103/PhysRevD.90.124032} {\bibfield  {journal} {\bibinfo
  {journal} {Phys. Rev. D}\ }\textbf {\bibinfo {volume} {90}},\ \bibinfo
  {pages} {124032} (\bibinfo {year} {2014})},\ \bibinfo {note} {[Erratum:
  Phys.Rev.D 94, 069902 (2016)]},\ \Eprint {http://arxiv.org/abs/1404.3197}
  {arXiv:1404.3197 [gr-qc]} \BibitemShut {NoStop}%
\bibitem [{\citenamefont {Baibhav}\ \emph {et~al.}(2018)\citenamefont
  {Baibhav}, \citenamefont {Berti}, \citenamefont {Cardoso},\ and\
  \citenamefont {Khanna}}]{Baibhav:2017jhs}%
  \BibitemOpen
  \bibfield  {author} {\bibinfo {author} {\bibfnamefont {V.}~\bibnamefont
  {Baibhav}}, \bibinfo {author} {\bibfnamefont {E.}~\bibnamefont {Berti}},
  \bibinfo {author} {\bibfnamefont {V.}~\bibnamefont {Cardoso}}, \ and\
  \bibinfo {author} {\bibfnamefont {G.}~\bibnamefont {Khanna}},\ }\href
  {\doibase 10.1103/PhysRevD.97.044048} {\bibfield  {journal} {\bibinfo
  {journal} {Phys. Rev. D}\ }\textbf {\bibinfo {volume} {97}},\ \bibinfo
  {pages} {044048} (\bibinfo {year} {2018})},\ \Eprint
  {http://arxiv.org/abs/1710.02156} {arXiv:1710.02156 [gr-qc]} \BibitemShut
  {NoStop}%
\bibitem [{\citenamefont {Giesler}\ \emph {et~al.}(2019)\citenamefont
  {Giesler}, \citenamefont {Isi}, \citenamefont {Scheel},\ and\ \citenamefont
  {Teukolsky}}]{Giesler:2019uxc}%
  \BibitemOpen
  \bibfield  {author} {\bibinfo {author} {\bibfnamefont {M.}~\bibnamefont
  {Giesler}}, \bibinfo {author} {\bibfnamefont {M.}~\bibnamefont {Isi}},
  \bibinfo {author} {\bibfnamefont {M.~A.}\ \bibnamefont {Scheel}}, \ and\
  \bibinfo {author} {\bibfnamefont {S.}~\bibnamefont {Teukolsky}},\ }\href
  {\doibase 10.1103/PhysRevX.9.041060} {\bibfield  {journal} {\bibinfo
  {journal} {Phys. Rev. X}\ }\textbf {\bibinfo {volume} {9}},\ \bibinfo {pages}
  {041060} (\bibinfo {year} {2019})},\ \Eprint
  {http://arxiv.org/abs/1903.08284} {arXiv:1903.08284 [gr-qc]} \BibitemShut
  {NoStop}%
\bibitem [{\citenamefont {Bhagwat}\ \emph {et~al.}(2020)\citenamefont
  {Bhagwat}, \citenamefont {Forteza}, \citenamefont {Pani},\ and\ \citenamefont
  {Ferrari}}]{Bhagwat:2019dtm}%
  \BibitemOpen
  \bibfield  {author} {\bibinfo {author} {\bibfnamefont {S.}~\bibnamefont
  {Bhagwat}}, \bibinfo {author} {\bibfnamefont {X.~J.}\ \bibnamefont
  {Forteza}}, \bibinfo {author} {\bibfnamefont {P.}~\bibnamefont {Pani}}, \
  and\ \bibinfo {author} {\bibfnamefont {V.}~\bibnamefont {Ferrari}},\ }\href
  {\doibase 10.1103/PhysRevD.101.044033} {\bibfield  {journal} {\bibinfo
  {journal} {Phys. Rev. D}\ }\textbf {\bibinfo {volume} {101}},\ \bibinfo
  {pages} {044033} (\bibinfo {year} {2020})},\ \Eprint
  {http://arxiv.org/abs/1910.08708} {arXiv:1910.08708 [gr-qc]} \BibitemShut
  {NoStop}%
\bibitem [{\citenamefont {Forteza}\ \emph {et~al.}(2020)\citenamefont
  {Forteza}, \citenamefont {Bhagwat}, \citenamefont {Pani},\ and\ \citenamefont
  {Ferrari}}]{Forteza:2020cve}%
  \BibitemOpen
  \bibfield  {author} {\bibinfo {author} {\bibfnamefont {X.~J.}\ \bibnamefont
  {Forteza}}, \bibinfo {author} {\bibfnamefont {S.}~\bibnamefont {Bhagwat}},
  \bibinfo {author} {\bibfnamefont {P.}~\bibnamefont {Pani}}, \ and\ \bibinfo
  {author} {\bibfnamefont {V.}~\bibnamefont {Ferrari}},\ }\href {\doibase
  10.1103/PhysRevD.102.044053} {\bibfield  {journal} {\bibinfo  {journal}
  {Phys. Rev. D}\ }\textbf {\bibinfo {volume} {102}},\ \bibinfo {pages}
  {044053} (\bibinfo {year} {2020})},\ \Eprint
  {http://arxiv.org/abs/2005.03260} {arXiv:2005.03260 [gr-qc]} \BibitemShut
  {NoStop}%
\bibitem [{\citenamefont {Mourier}\ \emph {et~al.}(2021)\citenamefont
  {Mourier}, \citenamefont {Jim\'enez~Forteza}, \citenamefont {Pook-Kolb},
  \citenamefont {Krishnan},\ and\ \citenamefont {Schnetter}}]{Mourier:2020mwa}%
  \BibitemOpen
  \bibfield  {author} {\bibinfo {author} {\bibfnamefont {P.}~\bibnamefont
  {Mourier}}, \bibinfo {author} {\bibfnamefont {X.}~\bibnamefont
  {Jim\'enez~Forteza}}, \bibinfo {author} {\bibfnamefont {D.}~\bibnamefont
  {Pook-Kolb}}, \bibinfo {author} {\bibfnamefont {B.}~\bibnamefont {Krishnan}},
  \ and\ \bibinfo {author} {\bibfnamefont {E.}~\bibnamefont {Schnetter}},\
  }\href {\doibase 10.1103/PhysRevD.103.044054} {\bibfield  {journal} {\bibinfo
   {journal} {Phys. Rev. D}\ }\textbf {\bibinfo {volume} {103}},\ \bibinfo
  {pages} {044054} (\bibinfo {year} {2021})},\ \Eprint
  {http://arxiv.org/abs/2010.15186} {arXiv:2010.15186 [gr-qc]} \BibitemShut
  {NoStop}%
\bibitem [{\citenamefont {Okounkova}(2020)}]{Okounkova:2020vwu}%
  \BibitemOpen
  \bibfield  {author} {\bibinfo {author} {\bibfnamefont {M.}~\bibnamefont
  {Okounkova}},\ }\href@noop {} {\  (\bibinfo {year} {2020})},\ \Eprint
  {http://arxiv.org/abs/2004.00671} {arXiv:2004.00671 [gr-qc]} \BibitemShut
  {NoStop}%
\bibitem [{\citenamefont {Bustillo}\ \emph {et~al.}(2021)\citenamefont
  {Bustillo}, \citenamefont {Lasky},\ and\ \citenamefont
  {Thrane}}]{Bustillo:2020buq}%
  \BibitemOpen
  \bibfield  {author} {\bibinfo {author} {\bibfnamefont {J.~C.}\ \bibnamefont
  {Bustillo}}, \bibinfo {author} {\bibfnamefont {P.~D.}\ \bibnamefont {Lasky}},
  \ and\ \bibinfo {author} {\bibfnamefont {E.}~\bibnamefont {Thrane}},\ }\href
  {\doibase 10.1103/PhysRevD.103.024041} {\bibfield  {journal} {\bibinfo
  {journal} {Phys. Rev. D}\ }\textbf {\bibinfo {volume} {103}},\ \bibinfo
  {pages} {024041} (\bibinfo {year} {2021})},\ \Eprint
  {http://arxiv.org/abs/2010.01857} {arXiv:2010.01857 [gr-qc]} \BibitemShut
  {NoStop}%
\bibitem [{\citenamefont {Finch}\ and\ \citenamefont
  {Moore}(2021)}]{Finch:2021iip}%
  \BibitemOpen
  \bibfield  {author} {\bibinfo {author} {\bibfnamefont {E.}~\bibnamefont
  {Finch}}\ and\ \bibinfo {author} {\bibfnamefont {C.~J.}\ \bibnamefont
  {Moore}},\ }\href {\doibase 10.1103/PhysRevD.103.084048} {\bibfield
  {journal} {\bibinfo  {journal} {Phys. Rev. D}\ }\textbf {\bibinfo {volume}
  {103}},\ \bibinfo {pages} {084048} (\bibinfo {year} {2021})},\ \Eprint
  {http://arxiv.org/abs/2102.07794} {arXiv:2102.07794 [gr-qc]} \BibitemShut
  {NoStop}%
\bibitem [{\citenamefont {Forteza}\ and\ \citenamefont
  {Mourier}(2021)}]{Forteza:2021wfq}%
  \BibitemOpen
  \bibfield  {author} {\bibinfo {author} {\bibfnamefont {X.~J.}\ \bibnamefont
  {Forteza}}\ and\ \bibinfo {author} {\bibfnamefont {P.}~\bibnamefont
  {Mourier}},\ }\href {\doibase 10.1103/PhysRevD.104.124072} {\bibfield
  {journal} {\bibinfo  {journal} {Phys. Rev. D}\ }\textbf {\bibinfo {volume}
  {104}},\ \bibinfo {pages} {124072} (\bibinfo {year} {2021})},\ \Eprint
  {http://arxiv.org/abs/2107.11829} {arXiv:2107.11829 [gr-qc]} \BibitemShut
  {NoStop}%
\bibitem [{\citenamefont {Jaramillo}\ \emph
  {et~al.}(2021{\natexlab{a}})\citenamefont {Jaramillo}, \citenamefont
  {Panosso~Macedo},\ and\ \citenamefont {Al~Sheikh}}]{Jaramillo:2020tuu}%
  \BibitemOpen
  \bibfield  {author} {\bibinfo {author} {\bibfnamefont {J.~L.}\ \bibnamefont
  {Jaramillo}}, \bibinfo {author} {\bibfnamefont {R.}~\bibnamefont
  {Panosso~Macedo}}, \ and\ \bibinfo {author} {\bibfnamefont {L.}~\bibnamefont
  {Al~Sheikh}},\ }\href {\doibase 10.1103/PhysRevX.11.031003} {\bibfield
  {journal} {\bibinfo  {journal} {Phys. Rev. X}\ }\textbf {\bibinfo {volume}
  {11}},\ \bibinfo {pages} {031003} (\bibinfo {year} {2021}{\natexlab{a}})},\
  \Eprint {http://arxiv.org/abs/2004.06434} {arXiv:2004.06434 [gr-qc]}
  \BibitemShut {NoStop}%
\bibitem [{\citenamefont {Jaramillo}\ \emph
  {et~al.}(2021{\natexlab{b}})\citenamefont {Jaramillo}, \citenamefont
  {Panosso~Macedo},\ and\ \citenamefont {Sheikh}}]{Jaramillo:2021tmt}%
  \BibitemOpen
  \bibfield  {author} {\bibinfo {author} {\bibfnamefont {J.~L.}\ \bibnamefont
  {Jaramillo}}, \bibinfo {author} {\bibfnamefont {R.}~\bibnamefont
  {Panosso~Macedo}}, \ and\ \bibinfo {author} {\bibfnamefont {L.~A.}\
  \bibnamefont {Sheikh}},\ }\href@noop {} {\  (\bibinfo {year}
  {2021}{\natexlab{b}})},\ \Eprint {http://arxiv.org/abs/2105.03451}
  {arXiv:2105.03451 [gr-qc]} \BibitemShut {NoStop}%
\bibitem [{\citenamefont {Cheung}\ \emph {et~al.}(2021)\citenamefont {Cheung},
  \citenamefont {Destounis}, \citenamefont {Macedo}, \citenamefont {Berti},\
  and\ \citenamefont {Cardoso}}]{Cheung:2021bol}%
  \BibitemOpen
  \bibfield  {author} {\bibinfo {author} {\bibfnamefont {M.~H.-Y.}\
  \bibnamefont {Cheung}}, \bibinfo {author} {\bibfnamefont {K.}~\bibnamefont
  {Destounis}}, \bibinfo {author} {\bibfnamefont {R.~P.}\ \bibnamefont
  {Macedo}}, \bibinfo {author} {\bibfnamefont {E.}~\bibnamefont {Berti}}, \
  and\ \bibinfo {author} {\bibfnamefont {V.}~\bibnamefont {Cardoso}},\
  }\href@noop {} {\  (\bibinfo {year} {2021})},\ \Eprint
  {http://arxiv.org/abs/2111.05415} {arXiv:2111.05415 [gr-qc]} \BibitemShut
  {NoStop}%
\bibitem [{\citenamefont {Aguirregabiria}\ and\ \citenamefont
  {Vishveshwara}(1996)}]{Aguirregabiria:1996zy}%
  \BibitemOpen
  \bibfield  {author} {\bibinfo {author} {\bibfnamefont {J.~M.}\ \bibnamefont
  {Aguirregabiria}}\ and\ \bibinfo {author} {\bibfnamefont {C.~V.}\
  \bibnamefont {Vishveshwara}},\ }\href {\doibase 10.1016/0375-9601(95)00937-X}
  {\bibfield  {journal} {\bibinfo  {journal} {Phys. Lett. A}\ }\textbf
  {\bibinfo {volume} {210}},\ \bibinfo {pages} {251} (\bibinfo {year}
  {1996})}\BibitemShut {NoStop}%
\bibitem [{\citenamefont {Vishveshwara}(1996)}]{Vishveshwara:1996jgz}%
  \BibitemOpen
  \bibfield  {author} {\bibinfo {author} {\bibfnamefont {C.~V.}\ \bibnamefont
  {Vishveshwara}},\ }in\ \href@noop {} {\emph {\bibinfo {booktitle} {{18th
  Conference of the Indian Association for General Relativity and
  Gravitation}}}}\ (\bibinfo {year} {1996})\ pp.\ \bibinfo {pages}
  {11--22}\BibitemShut {NoStop}%
\bibitem [{\citenamefont {Carullo}\ \emph {et~al.}(2019)\citenamefont
  {Carullo}, \citenamefont {Del~Pozzo},\ and\ \citenamefont
  {Veitch}}]{Carullo:2019flw}%
  \BibitemOpen
  \bibfield  {author} {\bibinfo {author} {\bibfnamefont {G.}~\bibnamefont
  {Carullo}}, \bibinfo {author} {\bibfnamefont {W.}~\bibnamefont {Del~Pozzo}},
  \ and\ \bibinfo {author} {\bibfnamefont {J.}~\bibnamefont {Veitch}},\ }\href
  {\doibase 10.1103/PhysRevD.99.123029} {\bibfield  {journal} {\bibinfo
  {journal} {Phys. Rev. D}\ }\textbf {\bibinfo {volume} {99}},\ \bibinfo
  {pages} {123029} (\bibinfo {year} {2019})},\ \bibinfo {note} {[Erratum:
  Phys.Rev.D 100, 089903 (2019)]},\ \Eprint {http://arxiv.org/abs/1902.07527}
  {arXiv:1902.07527 [gr-qc]} \BibitemShut {NoStop}%
\bibitem [{\citenamefont {Isi}\ \emph {et~al.}(2019)\citenamefont {Isi},
  \citenamefont {Giesler}, \citenamefont {Farr}, \citenamefont {Scheel},\ and\
  \citenamefont {Teukolsky}}]{Isi:2019aib}%
  \BibitemOpen
  \bibfield  {author} {\bibinfo {author} {\bibfnamefont {M.}~\bibnamefont
  {Isi}}, \bibinfo {author} {\bibfnamefont {M.}~\bibnamefont {Giesler}},
  \bibinfo {author} {\bibfnamefont {W.~M.}\ \bibnamefont {Farr}}, \bibinfo
  {author} {\bibfnamefont {M.~A.}\ \bibnamefont {Scheel}}, \ and\ \bibinfo
  {author} {\bibfnamefont {S.~A.}\ \bibnamefont {Teukolsky}},\ }\href {\doibase
  10.1103/PhysRevLett.123.111102} {\bibfield  {journal} {\bibinfo  {journal}
  {Phys. Rev. Lett.}\ }\textbf {\bibinfo {volume} {123}},\ \bibinfo {pages}
  {111102} (\bibinfo {year} {2019})},\ \Eprint
  {http://arxiv.org/abs/1905.00869} {arXiv:1905.00869 [gr-qc]} \BibitemShut
  {NoStop}%
\bibitem [{\citenamefont {Capano}\ \emph {et~al.}(2021)\citenamefont {Capano},
  \citenamefont {Cabero}, \citenamefont {Westerweck}, \citenamefont {Abedi},
  \citenamefont {Kastha}, \citenamefont {Nitz}, \citenamefont {Nielsen},\ and\
  \citenamefont {Krishnan}}]{Capano:2021etf}%
  \BibitemOpen
  \bibfield  {author} {\bibinfo {author} {\bibfnamefont {C.~D.}\ \bibnamefont
  {Capano}}, \bibinfo {author} {\bibfnamefont {M.}~\bibnamefont {Cabero}},
  \bibinfo {author} {\bibfnamefont {J.}~\bibnamefont {Westerweck}}, \bibinfo
  {author} {\bibfnamefont {J.}~\bibnamefont {Abedi}}, \bibinfo {author}
  {\bibfnamefont {S.}~\bibnamefont {Kastha}}, \bibinfo {author} {\bibfnamefont
  {A.~H.}\ \bibnamefont {Nitz}}, \bibinfo {author} {\bibfnamefont {A.~B.}\
  \bibnamefont {Nielsen}}, \ and\ \bibinfo {author} {\bibfnamefont
  {B.}~\bibnamefont {Krishnan}},\ }\href@noop {} {\  (\bibinfo {year}
  {2021})},\ \Eprint {http://arxiv.org/abs/2105.05238} {arXiv:2105.05238
  [gr-qc]} \BibitemShut {NoStop}%
\bibitem [{\citenamefont {Cotesta}\ \emph {et~al.}(2022)\citenamefont
  {Cotesta}, \citenamefont {Carullo}, \citenamefont {Berti},\ and\
  \citenamefont {Cardoso}}]{Cotesta:2022pci}%
  \BibitemOpen
  \bibfield  {author} {\bibinfo {author} {\bibfnamefont {R.}~\bibnamefont
  {Cotesta}}, \bibinfo {author} {\bibfnamefont {G.}~\bibnamefont {Carullo}},
  \bibinfo {author} {\bibfnamefont {E.}~\bibnamefont {Berti}}, \ and\ \bibinfo
  {author} {\bibfnamefont {V.}~\bibnamefont {Cardoso}},\ }\href@noop {} {\
  (\bibinfo {year} {2022})},\ \Eprint {http://arxiv.org/abs/2201.00822}
  {arXiv:2201.00822 [gr-qc]} \BibitemShut {NoStop}%
\bibitem [{\citenamefont {Nollert}(1999{\natexlab{b}})}]{8marcel-grossman}%
  \BibitemOpen
  \bibfield  {author} {\bibinfo {author} {\bibfnamefont {H.-P.}\ \bibnamefont
  {Nollert}},\ }\href@noop {} {\emph {\bibinfo {title} {The Eighth Marcel
  Grossmann Meeting}}}\ (\bibinfo  {publisher} {World Scientific Pub Co Inc},\
  \bibinfo {year} {1999})\BibitemShut {NoStop}%
\bibitem [{\citenamefont {Cardoso}(2019)}]{Cardoso:amplitude}%
  \BibitemOpen
  \bibfield  {author} {\bibinfo {author} {\bibfnamefont {V.}~\bibnamefont
  {Cardoso}},\ }\href@noop {} {}\bibinfo {howpublished} {private communication}
  (\bibinfo {year} {2019})\BibitemShut {NoStop}%
\bibitem [{GW1()}]{GW150914-data}%
  \BibitemOpen
  \href@noop {} {}\bibinfo {howpublished}
  {\url{https://www.gw-openscience.org/events/GW150914/}}\BibitemShut {NoStop}%
\bibitem [{\citenamefont {Kass}\ and\ \citenamefont
  {Raftery}(1995)}]{Kass:1995loi}%
  \BibitemOpen
  \bibfield  {author} {\bibinfo {author} {\bibfnamefont {R.~E.}\ \bibnamefont
  {Kass}}\ and\ \bibinfo {author} {\bibfnamefont {A.~E.}\ \bibnamefont
  {Raftery}},\ }\href {\doibase 10.1080/01621459.1995.10476572} {\bibfield
  {journal} {\bibinfo  {journal} {J. Am. Statist. Assoc.}\ }\textbf {\bibinfo
  {volume} {90}},\ \bibinfo {pages} {773} (\bibinfo {year} {1995})}\BibitemShut
  {NoStop}%
\bibitem [{\citenamefont {Abbott}\ \emph
  {et~al.}(2021{\natexlab{f}})\citenamefont {Abbott} \emph
  {et~al.}}]{LIGOScientific:2020kqk}%
  \BibitemOpen
  \bibfield  {author} {\bibinfo {author} {\bibfnamefont {R.}~\bibnamefont
  {Abbott}} \emph {et~al.} (\bibinfo {collaboration} {LIGO Scientific,
  Virgo}),\ }\href {\doibase 10.3847/2041-8213/abe949} {\bibfield  {journal}
  {\bibinfo  {journal} {Astrophys. J. Lett.}\ }\textbf {\bibinfo {volume}
  {913}},\ \bibinfo {pages} {L7} (\bibinfo {year} {2021}{\natexlab{f}})},\
  \Eprint {http://arxiv.org/abs/2010.14533} {arXiv:2010.14533 [astro-ph.HE]}
  \BibitemShut {NoStop}%
\bibitem [{\citenamefont {Vallisneri}(2008)}]{Vallisneri:2007ev}%
  \BibitemOpen
  \bibfield  {author} {\bibinfo {author} {\bibfnamefont {M.}~\bibnamefont
  {Vallisneri}},\ }\href {\doibase 10.1103/PhysRevD.77.042001} {\bibfield
  {journal} {\bibinfo  {journal} {Phys. Rev. D}\ }\textbf {\bibinfo {volume}
  {77}},\ \bibinfo {pages} {042001} (\bibinfo {year} {2008})},\ \Eprint
  {http://arxiv.org/abs/gr-qc/0703086} {arXiv:gr-qc/0703086} \BibitemShut
  {NoStop}%
\bibitem [{\citenamefont {Bhagwat}\ \emph {et~al.}(2016)\citenamefont
  {Bhagwat}, \citenamefont {Brown},\ and\ \citenamefont
  {Ballmer}}]{Bhagwat:2016ntk}%
  \BibitemOpen
  \bibfield  {author} {\bibinfo {author} {\bibfnamefont {S.}~\bibnamefont
  {Bhagwat}}, \bibinfo {author} {\bibfnamefont {D.~A.}\ \bibnamefont {Brown}},
  \ and\ \bibinfo {author} {\bibfnamefont {S.~W.}\ \bibnamefont {Ballmer}},\
  }\href {\doibase 10.1103/PhysRevD.94.084024} {\bibfield  {journal} {\bibinfo
  {journal} {Phys. Rev. D}\ }\textbf {\bibinfo {volume} {94}},\ \bibinfo
  {pages} {084024} (\bibinfo {year} {2016})},\ \bibinfo {note} {[Erratum:
  Phys.Rev.D 95, 069906 (2017)]},\ \Eprint {http://arxiv.org/abs/1607.07845}
  {arXiv:1607.07845 [gr-qc]} \BibitemShut {NoStop}%
\bibitem [{\citenamefont {Isi}\ and\ \citenamefont {Farr}(2021)}]{Isi:2021iql}%
  \BibitemOpen
  \bibfield  {author} {\bibinfo {author} {\bibfnamefont {M.}~\bibnamefont
  {Isi}}\ and\ \bibinfo {author} {\bibfnamefont {W.~M.}\ \bibnamefont {Farr}},\
  }\href@noop {} {\enquote {\bibinfo {title} {{Analyzing black-hole
  ringdowns}},}\ } (\bibinfo {year} {2021}),\ \Eprint
  {http://arxiv.org/abs/2107.05609} {arXiv:2107.05609 [gr-qc]} \BibitemShut
  {NoStop}%
\bibitem [{\citenamefont {Gossan}\ \emph {et~al.}(2012)\citenamefont {Gossan},
  \citenamefont {Veitch},\ and\ \citenamefont {Sathyaprakash}}]{Gossan:2011ha}%
  \BibitemOpen
  \bibfield  {author} {\bibinfo {author} {\bibfnamefont {S.}~\bibnamefont
  {Gossan}}, \bibinfo {author} {\bibfnamefont {J.}~\bibnamefont {Veitch}}, \
  and\ \bibinfo {author} {\bibfnamefont {B.~S.}\ \bibnamefont
  {Sathyaprakash}},\ }\href {\doibase 10.1103/PhysRevD.85.124056} {\bibfield
  {journal} {\bibinfo  {journal} {Phys. Rev. D}\ }\textbf {\bibinfo {volume}
  {85}},\ \bibinfo {pages} {124056} (\bibinfo {year} {2012})},\ \Eprint
  {http://arxiv.org/abs/1111.5819} {arXiv:1111.5819 [gr-qc]} \BibitemShut
  {NoStop}%
\bibitem [{\citenamefont {Yang}\ \emph {et~al.}(2017)\citenamefont {Yang},
  \citenamefont {Yagi}, \citenamefont {Blackman}, \citenamefont {Lehner},
  \citenamefont {Paschalidis}, \citenamefont {Pretorius},\ and\ \citenamefont
  {Yunes}}]{Yang:2017zxs}%
  \BibitemOpen
  \bibfield  {author} {\bibinfo {author} {\bibfnamefont {H.}~\bibnamefont
  {Yang}}, \bibinfo {author} {\bibfnamefont {K.}~\bibnamefont {Yagi}}, \bibinfo
  {author} {\bibfnamefont {J.}~\bibnamefont {Blackman}}, \bibinfo {author}
  {\bibfnamefont {L.}~\bibnamefont {Lehner}}, \bibinfo {author} {\bibfnamefont
  {V.}~\bibnamefont {Paschalidis}}, \bibinfo {author} {\bibfnamefont
  {F.}~\bibnamefont {Pretorius}}, \ and\ \bibinfo {author} {\bibfnamefont
  {N.}~\bibnamefont {Yunes}},\ }\href {\doibase 10.1103/PhysRevLett.118.161101}
  {\bibfield  {journal} {\bibinfo  {journal} {Phys. Rev. Lett.}\ }\textbf
  {\bibinfo {volume} {118}},\ \bibinfo {pages} {161101} (\bibinfo {year}
  {2017})},\ \Eprint {http://arxiv.org/abs/1701.05808} {arXiv:1701.05808
  [gr-qc]} \BibitemShut {NoStop}%
\bibitem [{\citenamefont {Zimmerman}\ \emph {et~al.}(2019)\citenamefont
  {Zimmerman}, \citenamefont {Haster},\ and\ \citenamefont
  {Chatziioannou}}]{Zimmerman:2019wzo}%
  \BibitemOpen
  \bibfield  {author} {\bibinfo {author} {\bibfnamefont {A.}~\bibnamefont
  {Zimmerman}}, \bibinfo {author} {\bibfnamefont {C.-J.}\ \bibnamefont
  {Haster}}, \ and\ \bibinfo {author} {\bibfnamefont {K.}~\bibnamefont
  {Chatziioannou}},\ }\href {\doibase 10.1103/PhysRevD.99.124044} {\bibfield
  {journal} {\bibinfo  {journal} {Phys. Rev. D}\ }\textbf {\bibinfo {volume}
  {99}},\ \bibinfo {pages} {124044} (\bibinfo {year} {2019})},\ \Eprint
  {http://arxiv.org/abs/1903.11008} {arXiv:1903.11008 [astro-ph.IM]}
  \BibitemShut {NoStop}%
\end{thebibliography}%

\end{document}